\documentclass[preprint,12pt]{elsarticle}
\usepackage{amssymb,siunitx}

\usepackage{color}

\usepackage[colorlinks=true, citecolor=midblue, linkcolor=midblue, urlcolor=midblue]{hyperref}

\usepackage{ulem}

\usepackage{amsmath}

\DeclareUnicodeCharacter{2500}{\textendash}

\newcommand{\apjl}{Astrophys. J. Lett.}
\newcommand{\apjs}{The Astrophysical Journal Supplement}
\newcommand{\aap}{Astron. Astrophys.}%{Astronomy \& Astrophysics}  

\begin{document}
\begin{frontmatter}
\title{Vertical structure and dynamics of a galactic disk}
\author[1]{Chanda J. Jog\corref{corr}}
\ead{cjjog@iisc.ac.in}

\address[1]{Department of Physics, Indian Institute of Science, Bangalore 560012, India.}

\begin{abstract}
Most of the visible mass in a typical spiral galaxy is distributed in a thin disk,
 with a radial extent much larger than its thickness. While the planar disk structure, including non-axisymmetric features 
such as spiral structure, has been studied extensively, the vertical structure has 
not received comparable attention. 
This review aims to give a comprehensive, pedagogic introduction to the rich
 topic of vertical structure of a galactic disk in hydrostatic equilibrium,
 and discuss the theoretical developments in this field  in the context of recent 
observations. A realistic multi-component disk plus halo model of a galaxy has been 
developed and studied by us in detail. This takes account of both stars and 
interstellar gas, treated as isothermal components with different 
velocity dispersions, which are gravitationally coupled; further, the disk is 
in the gravitational field of the dark matter halo. This review focuses on this model
and the results from it in different physical cases.

The gas and halo crucially affect the resulting self-consistent stellar distribution 
such that it is vertically constrained to be closer to the mid-plane, and has a steeper 
profile than in the standard one-component case, in agreement with modern observations.
A  typical stellar disk is shown to flare by a factor of few within the visible radial extent of the disk. These robust results question the sech$^2$ profile and a constant scale height, routinely used in the literature for convenience.
In an important application, the observed HI gas scale height is used as a constraint on the  
model which helps determine the shape and the density profile of the dark matter halo 
for galaxies. Finally, we outline some key, open questions which can be addressed in 
the near future using the above model, and new observational data -- for example, from 
the IFU surveys and JWST. These promise to give a better understanding of this topic.
  
\end{abstract}

\end{frontmatter}

\tableofcontents

\section{Introduction \label{sec:Intro}}
\subsection{Background and motivation}

\begin{figure}
\centering
\includegraphics[height=2.5in, width=2.5in]{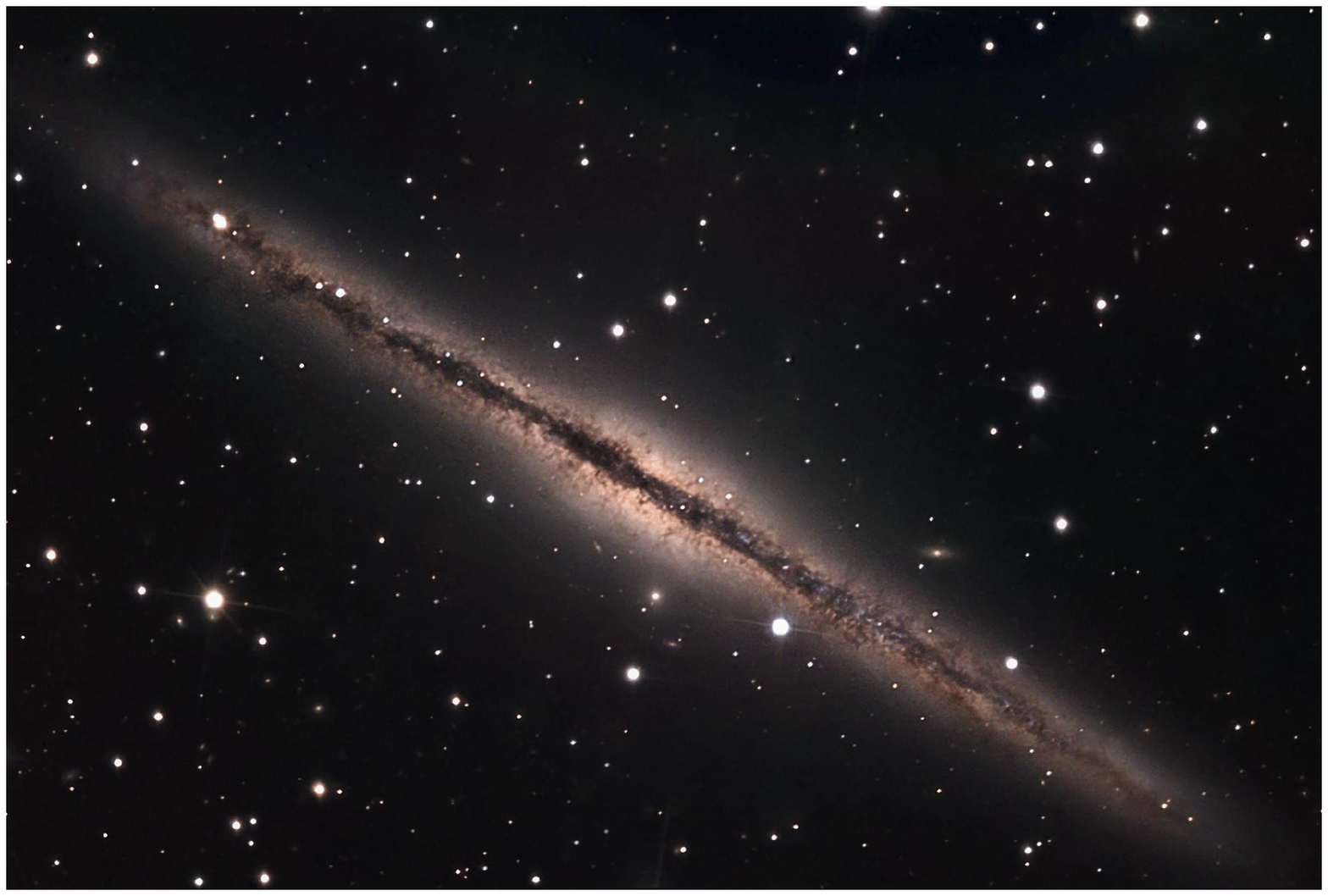}
\medskip 
\includegraphics[height=2.5in, width= 2.5in]{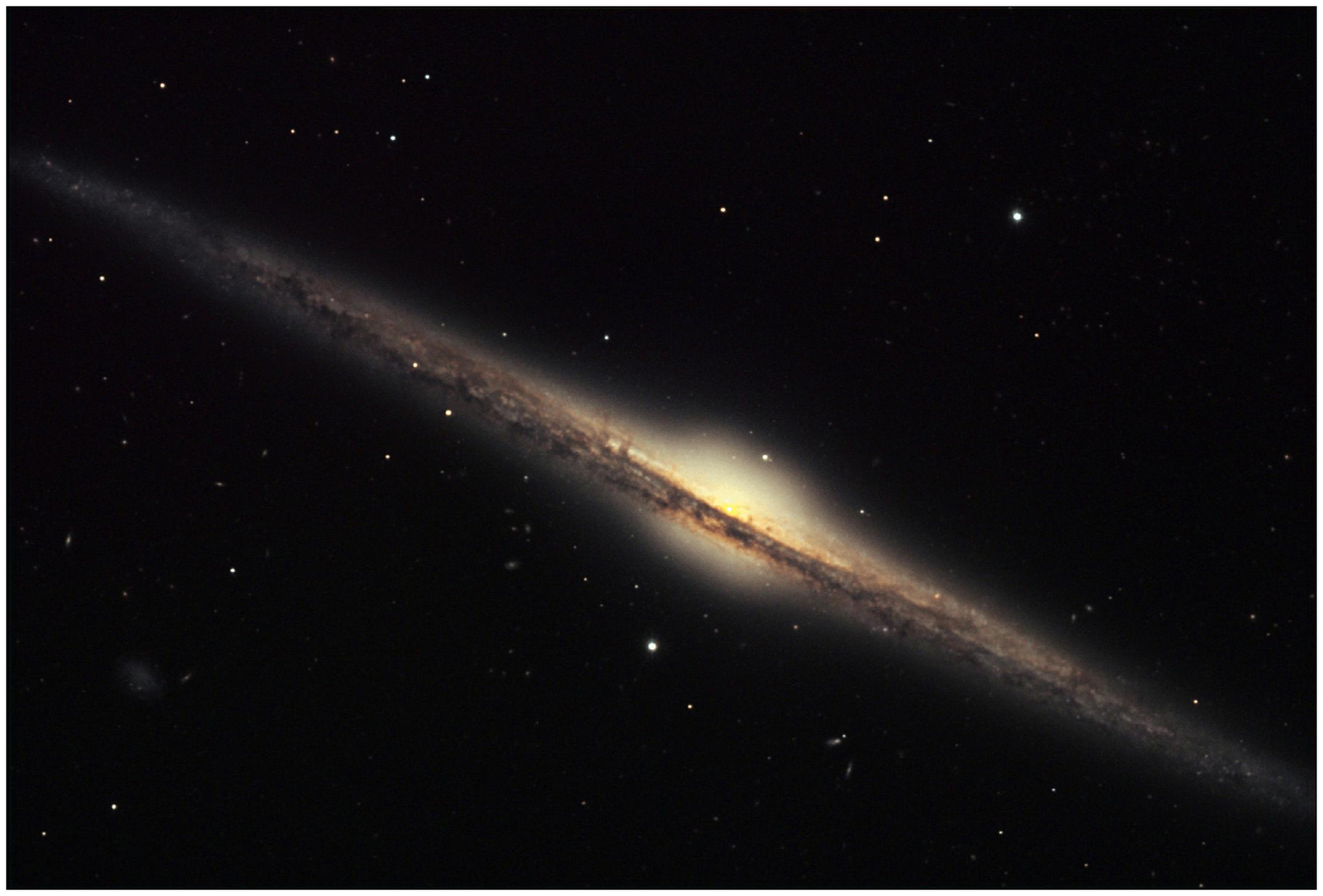} 
\bigskip
\includegraphics[height=2.5in, width=2.5in]{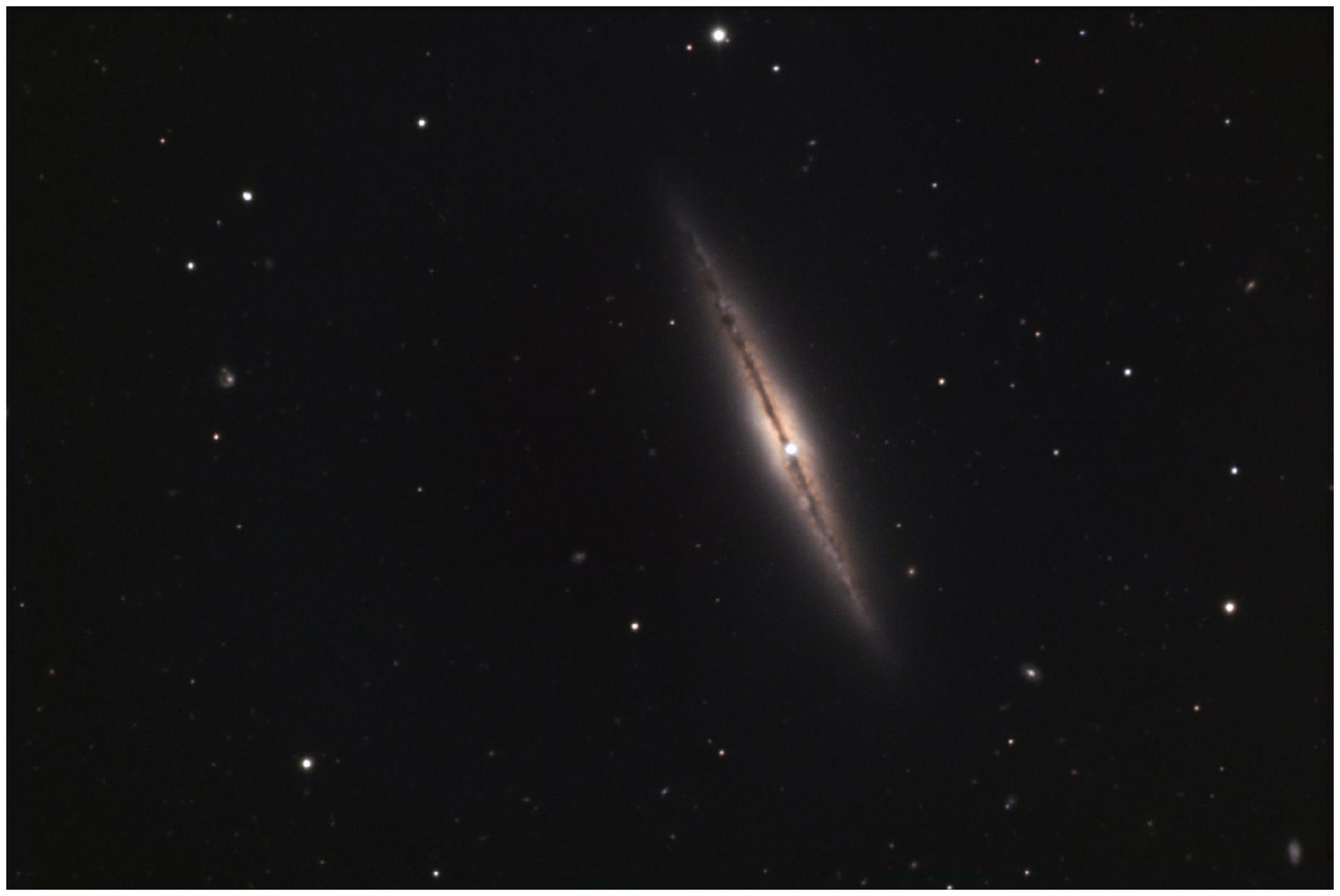}
\medskip
\includegraphics[height=2.5in, width=2.5in]{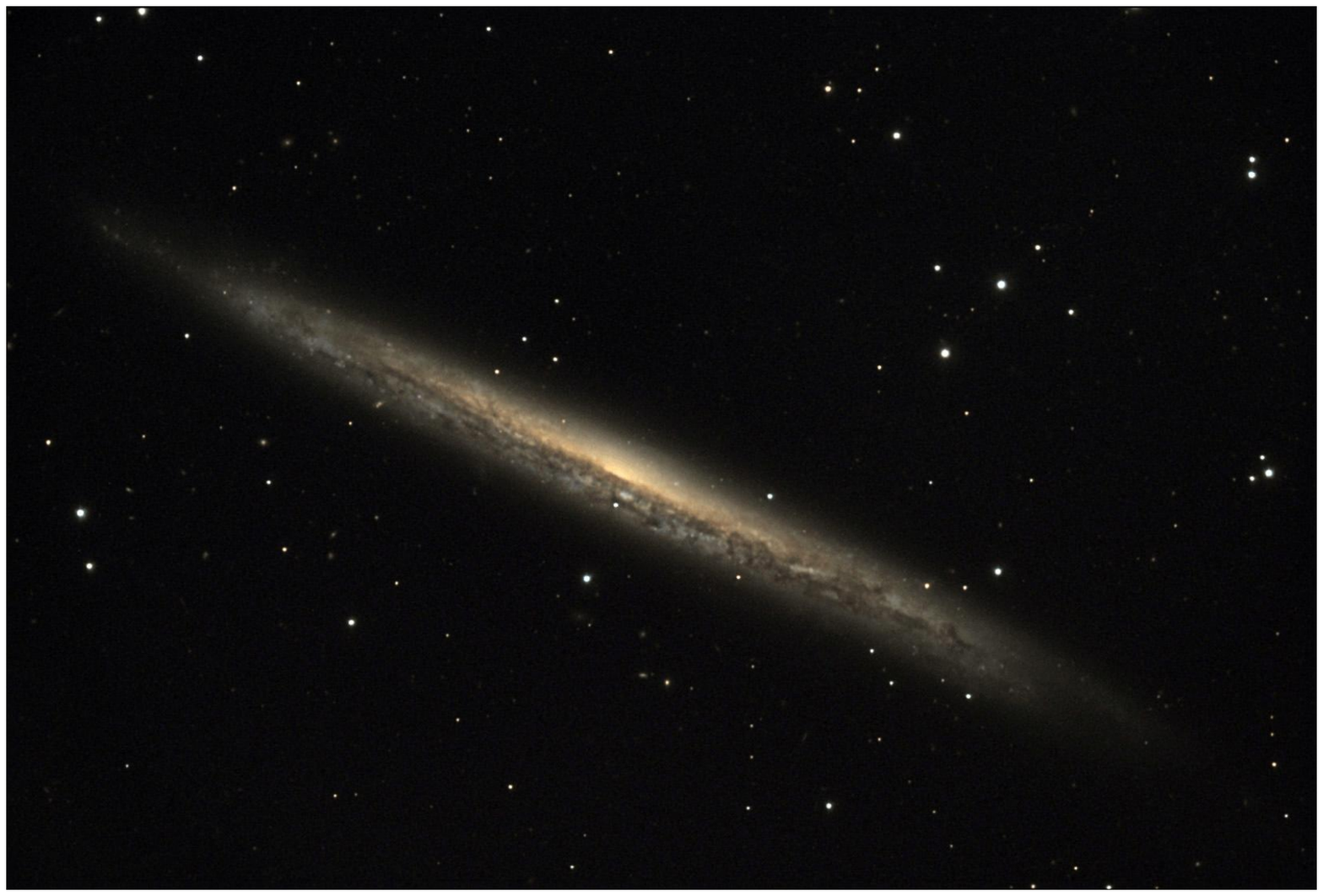}
\bigskip
\caption{A block of four typical edge-on galaxies:  
NGC 891 (top left panel); NGC 4565 (top right panel); NGC 4013 (lower left panel); and NGC 5907 (lower right panel). An edge-on galaxy is by chance  so oriented that the l.o.s. from the observer cuts across the disk plane. This brings out the thin nature of the disk.  The dark band cutting across the middle of each image is due to absorption by dust associated with the interstellar gas in the galactic disk. Edge-on galaxies provide crucial information
for the study of vertical structure of galactic disks.
(Courtesy: NOAO)}
\label{fig.1}
\end{figure}

A glance at the image of a spiral galaxy inclined close to  edge-on w.r.t. the line of sight (e.g. NGC 891) tells us that most of the visible mass of the galaxy is distributed in a thin disk, with the radial extent much larger than the vertical thickness  \cite{Sandage}. 
 For example, the ratio of the thickness to the radius of the Milky Way is $\sim 1/40$ \cite{BT1987}. 
 Most of the mass in a galactic  disk is in stars, and a small fraction ($\sim 10-15 \%$) is contained in  
 interstellar gas.
 The disk distribution is the most striking feature of a galaxy: see Fig. \ref{fig.1} for images of four typical examples of galaxies that are viewed edge-on.
Further, the rotation in a galaxy can be measured which is not strictly face-on,
this indicates an underlying disk distribution \cite{Hubble1936,SofueRubin2001}.  
The rotation in the Milky Way was first clearly measured using the 21 cm line emission from HI gas \cite {1965gast.book..455O}.
For these two reasons, spiral galaxies are also sometimes referred to as disk galaxies. 

The vertical structure of a galactic disk 
is a tracer of disk potential; hence, it is crucial to know the vertical structure in order to
understand  the disk dynamics and evolution. Edge-on galaxies play a crucial role in the study of disk vertical structure.
The thin disk distribution of matter in a galaxy simplifies its dynamics, because  the dynamics in the vertical direction can then be
 taken to be de-coupled from that in the plane (\cite{BT1987}; also, see Section \ref{sec:3.1}).{\footnote{ However, the physical origin of the thinness of the disk, and the origin of the disk itself, is not yet fully understood; although the conservation of angular momentum is believed to be the reason for this \cite{WhiteGalForm2010,Silk2014diskform}. During the formation of a galaxy,  the gas clouds cannot collapse easily in a direction normal to the direction of the angular momentum vector -- due to the conservation of the angular momentum -- but they can collapse readily along it. Hence the collapse leads to a disk.}}
 The disk is supported against gravitational collapse in the plane mainly by rotation w.r.t. the disk centre; whereas it is supported against collapse in the vertical direction by its internal pressure gradient. 
 While the planar disk
 structure, including non-axisymmetric features such as spiral structure, has been
 studied extensively; the vertical structure has not received comparable attention. This is partly due to the observational challenges of studying the disk vertical structure (Section \ref{sec:2}); and also due to the necessity of having to theoretically solve for the entire self-consistent vertical distribution at a go (Sections \ref{sec:3}, \ref{sec:4}).

In recent years, an explosion of astrometric data  with unprecedented detail for the position and kinematics of over a billion individual stars in the Milky Way has become available:
from \textit{Gaia}, LAMOST, APOGEE and other telescopes. This  gives very detailed information on the disk vertical structure, and allows us to determine the vertical distribution in situ. This has to be understood and interpreted theoretically for an optimal use of the data. 

Although isolated papers in the literature address some issues related to the  disk vertical structure, there is so far no 
cohesive review of this fast-growing, active field. Given these reasons, 
a review article
focused on the vertical structure of a galactic disk is warranted  
at this juncture.

Some noteworthy previous reviews on related topics are: about galaxy disks  in general 
 \cite {vdkF2011} ; and the 
structure and kinematics of the Milky Way  \cite{BHGARAA}. Two recent reviews  deal with the dynamics and evolution of the Milky way,  particularly in light of the \textit{Gaia} observations
\cite{Deason_Belokurov2024,Vasiliev2025}.
However, these do not specifically focus on the vertical structure of a galactic disk in detail. 

\subsection{Outline of the review}

In this review, we aim to give a comprehensive introduction to the rich topic of  vertical structure and related dynamics of a typical, thin galactic disk, in the context of recent observations.  
We will 
present this subject  starting from the early observations and theoretical work on this in the literature, followed by more recent developments, in a pedagogical fashion; to bring out the physics involved clearly.

The study of the vertical structure of a galactic disk has a long history. 
The classic work by Spitzer  \cite{Spitzer1942} considered an idealized self-gravitating,  isothermal, locally dense disk{\footnote{We discuss in Section \ref{sec:3.1} how this is equivalent to treating a thin disk 
  -- for which the result by \cite{Spitzer1942} is routinely applied.}} which was shown to obey a sech$^2$
 (or a hyperbolic secant squared) form (also, see  \cite{Camm1950}; and Section \ref{sec:3}).

However,
with the advent of R-band studies of external galaxies in early 1980's, which allowed studies of underlying old stellar mass distribution; it was already clear that the distribution close to the mid-plane showed an excess over that expected for a sech$^2$ distribution \cite{vdk1988, BD1994}. These profiles could instead be better fit by a steeper functional form, such as  a sech or an exponential.
A somewhat ad hoc 
model was proposed by  \cite{vdk1988} to fit this observed trend.
Surprisingly, no further  work was done on this problem for a long time.

In recent years, the problem of disk vertical structure has been addressed in a more physically motivated approach, starting from the work by Narayan \& Jog (2002) \cite{NJ2002}, and later followed by other studies 
\cite{Kalberla2003,BJ2007,SJ2018} .
The  multi-component disk plus halo model first proposed and worked out by 
\cite{NJ2002},
treats a realistic
galactic disk as a multi-component system, where the stars and interstellar gas (HI and H$_2$) are taken to be
 gravitationally coupled; and the disk is taken to be in the gravitational field of the dark matter halo. 
 Thus, this model constitutes a more complete physical treatment of the disk vertical structure than the stars-alone case that is usually considered; and it also naturally
 explains various observational features of the vertical density distribution of stars and gas.
 In this review, we focus
   on this model and the results from it.

In the above model
\cite{NJ2002}, 
the self-consistent vertical density distribution of each disk component is obtained simultaneously
by solving the coupled, joint Poisson-hydrostatic balance equations for all the three disk components.  
  The resulting vertical density distribution is obtained
 in terms of the physical parameters such as the surface density and velocity dispersion of each disk component, and the halo mass distribution.

The most significant finding is that  the gravitational force due to the interstellar gas and the dark matter halo 
significantly affects the self-consistent vertical stellar distribution in the disk. 
The modified stellar distribution is constrained closer to the mid-plane, such that: it has a higher mid-plane density, a smaller scale height, and is steeper than the corresponding stars-alone case. The constraining effect due to the gas and the halo dominate in the inner and the outer Galaxy, respectively.
 The resulting smaller vertical thickness for all three disk components agree well with observations for the Milky Way \cite{NJ2002, SJ2018}.
 The resulting stellar density distribution is steeper than sech$^2$, as observed 
 \cite {BJ2007, SJ2018}.

The vertical structure and dynamics of a galactic disk is a complex and challenging subject. 
We will discuss various physical cases that deal with progressively more detailed physical points, and tie these in with real astrophysical conditions.
On inclusion of some of these realistic physical effects, such as a non-isothermal velocity dispersion,
 the resulting vertical density distribution can differ substantially, by $\sim$ 30 to 40 \% , compared to the sech$^2$ distribution obtained in the standard isothermal, one-component, thin disk case \cite{SJ2020Jeans, SJ2020noniso}.

A generic, robust result obtained is that for a typical stellar disk in hydrostatic
 equilibrium, the resulting scale height increases by a factor of few within the optical disk.{\footnote{The radius of the optical disk or the visible extent of the stellar disk is typically defined   to be the radius corresponding to a limiting isophote: say, 26.5 mag {arc sec}$^{-2}$, which is known as the Holmberg radius. See Appendix A for details.}} 
The flaring stellar disk is in contrast to a constant scale height commonly assumed 
 in the literature \cite{vdkS1981a}. 
However,
 such a trend of a flaring disk is in agreement with recent observations and simulations.

Further, in an important application, the observed HI gas
 scale heights and the rotation curve have been used as two simultaneous constraints on
 the above model to determine the shape and the density profile of the dark matter
 halo. This approach has been applied to a number of galaxies, including the Milky Way and M31.
  
We mention that this review will focus on the "canonical" thin disk. We do not treat the chemically and kinematically distinct "{\it Thick disk}" component that is common in many galaxies. However, we will discuss the latter  briefly for the sake of completeness (see Section \ref{sec:7.1}; also, see Sections \ref{sec:4.5.1}, and \ref{sec:8}).

The study of disk vertical structure is an active field, with many new, important questions -- mainly triggered by  input from new observations. 
 For example, critical input parameters such as the stellar velocity dispersion in external galaxies can now be extracted from the near-IR, kinematical IFU data. Using these, more accurate model vertical density distribution for  galaxies can be determined.
The model can be also applied to understand the vertical structure of high redshift galaxies; for example, using the data from JWST (James Webb Space Telescope).
Further, the overall vertically constrained  stellar density distribution in a coupled case can have important dynamical consequences: it can make the disk more stable against distortion; and also affect dynamical features such as warps. These effects need to be studied.

Finally, apart from being interesting in itself, the vertical structure of a galactic disk is important because it holds clues to the formation and evolution of the disk \cite{Jia2024evo,Donghia2024history}. The evolution, including the increase in the disk scale height, could be  either due to internal or secular processes, such as heating by spiral arms and bars; or, due to external sources such as  tidal encounters with external galaxies (see \cite{BT1987,Jia2024evo, Donghia2024history}, and references therein). 

A section-wise break-up of the contents of this review is as follows. In Section \ref{sec:2} , we describe the observational challenges of studying the vertical structure, and the data obtained. 
Section \ref{sec:3} mainly discusses the classic work by Spitzer 
 \cite{Spitzer1942}.   
 In Section \ref{sec:4}, we present the modern
multi-component disk plus halo model, and its applications, to cases with different physical conditions: such as, a thin disk;  a thick
or a low density disk -- which  contain progressively more detailed physics in the problem.
 These cases represent different regions and types of galaxies.
Sections \ref{sec:5} and \ref{sec:6} present and explain a flaring stellar disk as a generic result; and how the shape and the density profile of dark matter halo are deduced, respectively. These two sections could
 be read as stand-alone mini-reviews on these two topics. Section \ref{sec:7} gives a brief introduction to related topics such as thick disks, bending instabilities, and phase-space spirals. Section \ref{sec:8} outlines  some specific, well-defined, open problems related to the disk vertical structure, and future trends in the field. Section \ref{sec:9}
 summarizes the conclusions from this review. In Appendix A, various physics issues related to the thickness of a self-gravitating disk and the disk vertical density profile as presented in the review are put together and 
 discussed.
 At the end, a glossary of frequently used physical terms is defined in Table 1; also,  a list of frequently used acronyms is given.

\section{Observations \label{sec:2}}

The observations of the vertical intensity profiles for an edge-on galaxy; the density distribution and velocity dispersion of stars and gas in a galactic disk,  
will be given here briefly. These provide background about the topic; also,  these are necessary to help formulate theoretical models of the disk vertical structure. This section aims to give an idea of the basic issues and complexities involved in
observations related to the disk vertical structure.

\subsection{Vertical stellar density distribution \label{sec:2.1}}

An image of any edge-on galaxy clearly shows that the  light, and hence the stellar mass distribution, in a galaxy is mainly in a thin disk. 
However, to determine  the vertical  density distribution from the observed intensity distribution  in external galaxies is problematic because of various effects: such as, the contamination due to projection effects from non-local regions in the disk along the line of sight, and dust extinction, as discussed next.
In our Galaxy, an alternative approach of actual star counts is possible in the solar neighbourhood. This allows an in situ measurement of density distribution in great detail, as will be discussed later in this Section (Section \ref{sec:2.1.3}). This approach was first used in the classic work by Oort \cite{Oort60}  to determine the mid-plane density; the latter problem  has now found resurgence with new data from \textit{Gaia} and other sources (see Sections \ref{sec:2.1.3}, and \ref{sec:3.1.1}).

\subsubsection{Analysis of intensity profiles\label{sec:2.1.1}}

The determination of vertical density distribution from observations is a challenging task.
This is why a basic theoretical model for the vertical structure of a thin, self-gravitating disk was first developed ab initio  \cite{Spitzer1942,Camm1950}. This was later used  in the literature to analyze the observed intensity profiles, so as to determine the model parameters (from \cite{Spitzer1942}) for the vertical disk density distribution, as discussed next. This is in contrast to most topics in astronomy where the observations are used as a guideline to construct a basic theoretical model of the system.

To study the vertical density profile
in a galactic disk, one needs to look at galaxies that are edge-on or tilted close to edge-on (or $90^0$) w.r.t. line of sight. 
However, due to projection effects, the intensity as observed from a point R on the major axis (at which a line of sight cuts it) has contributions from non-local regions along the line of sight. These regions are at different radial distances from the galactic centre.
Further, the light from a galactic disk suffers an extinction due to interstellar dust. This effect is proportional to the distance along the line of sight within the disk, since optical depth is a cumulative property. The dust extinction is 
important at visible wavelengths and becomes less at longer wavelengths. Therefore, the early detailed studies of edge-on studies of galaxies 
were done using R-Band or I-band observations, e.g.,\cite{Burstein1979}, to minimize the effect of dust extinction. The modern studies use near-IR data where the effect of dust extinction is further minimized,
see for example, the S$^4$G (Spitzer Survey of Spiral Galaxies) catalog \cite{Sheth2010}.

\medskip

\noindent {\bf  Intensity profiles}

Some of the pioneering work on this topic was done by van der Kruit \$ Searle (1981) \cite{vdkS1981a,vdkS1981b}. Their aim was to study the vertical structure of a set of edge-on galaxies by analyzing  the three-dimensional distribution of light in the galactic disk. 
From the observed photometric image of a galaxy, they plotted intensity profiles along z at cuts taken at different radii along the major axis (\cite{vdkS1981a}). 
This procedure was later followed by many authors, and applied for the study of other galaxies 
\cite {BD1994,deGrijsP1997}. 

To analyze the intensity profiles,  \cite{vdkS1981a}
adopted a basic theoretical model for the vertical density distribution as for an isothermal disk \cite{Spitzer1942}; which has a 
 sech$^2$$(z/z_0)$ form, where $z_0$ is an indicator of disk thickness
(to be discussed in Section \ref{sec:3.1}). Further, they
 assumed a surface density that falls off exponentially with radius (as observed, see \cite{Freeman1970}). Next, they calculated the theoretical intensity profiles, $I(R,z)$, 
where $R$ is the distance from the galaxy centre along the major axis, where the line of sight cuts the major axis.
For simplicity, all the stars were taken to form a single component.
These theoretical profiles were then compared with the observed intensity profiles to deduce the model parameters (from \cite{Spitzer1942}). This is analogous to be basic approach taken to study interstellar medium in galaxies, where also it is difficult to disentangle the local and non-local effects \cite{Spitzer1978,Draine2011}.

\medskip

\noindent {\bf Problems with the standard analysis:}

However, there are serious problems with the details of the subsequent procedure adopted by  \cite{vdkS1981a}. Hence their conclusion that the scale height is constant with radius is not correct, this is discussed next.  For the above adopted model, the luminosity density at any point (R,z) in the disk is given by

\begin{equation}
\mathrm{ L(R,z) = L(0,0) exp ({-R/R_D}) sech^2 (z/z_0)}  \label{eq2.1}
     \tag{2.1}
 \end{equation}    

\noindent where $z_0$ is a scale parameter which is an indicator of thickness of disk vertical distribution at a given radius,$R$ (see \cite{Spitzer1942,Camm1950}; also, see Section \ref{sec:3.1}). R$_D$ is the exponential scale length of the radial distribution of surface density in the disk, and $L(0,0)$ is the luminosity density at the galactic centre.
 
Next, the theoretical intensity profile for the above adopted model, say at a distance R from the galactic centre along the major axis as seen by the external observer, is obtained by integrating the contributions from the non-local regions along the line of sight which are at different radial distance w.r.t. the galactic centre. 
In a typical real galaxy, $z_0$ would be expected to vary with the radial distance from the galactic centre, 
hence integrating over this contribution is difficult. To simplify this, \cite{vdkS1981a} assume that the scale height $z_0$ is constant with radius in the disk. 
With this assumption,  it is straightforward to obtain
the net edge-on distribution of surface brightness or intensity as seen from a point at a distance R from the galactic centre,  by integrating the luminosity density along the line of sight. The resulting model surface brightness is given as: 

\begin{equation}
 \mathrm{ I(R,z) = I(0,0) \frac {R}{R_D} K_1 \left(\frac{R}{R_D}\right) sech^2 (z/z_0) }                               \label{eq2.2}
           \tag{2.2}
\end{equation}           
\noindent  where $I(0,0) = 2 R_D L(0,0)$ and $K_1$ is the modified Bessel function of the second kind.

Then by comparing this model profile with the observed intensity profile measured at a cut normal to the major axis at a given $R$, \cite{vdkS1981a} obtain the value of $z_0$.
However, note that the above expression for the the model surface brightness 
(Eq.(\ref{eq2.2})) was derived assuming the scale height $z_0$ to be constant with radius, which is not physically justified.   

Next, they stack  the intensity profiles  taken along normal cuts at different R together, 
 at an arbitrarily chosen $z'$ point; and   claim 
that these show a small spread.  Hence they conclude that these profiles are well-fit by a single composite curve
with a constant $z_0$.
Therefore they claim that their assumption of $z_0$ being constant with radius, used to obtain the model integrated surface brightness along the line of sight, 
(Eq.(\ref{eq2.2}))
from (Eq.(\ref{eq2.1})), is justified.
This procedure proposed by \cite{vdkS1981a} is routinely used by others to determine the scale height for edge-on galaxies (e.g., \cite{Kregel2002,Peters2017}).

However, there is a
 serious flaw with the  above approach proposed by \cite{vdkS1981a}, as shown next. Their claim that the profiles at different $R$  are well-fit by a single composite curve with a small spread around it --
thus indicating a constant scale height --
itself has been questioned by
 \cite{NJ2002ltr},  and \cite{Rohlfs1982}. 
 
Using the same data and the composite profile as obtained by \cite{vdkS1981a};
\cite{NJ2002ltr} show that, in fact,
there is a finite
  spread in the observed data for surface brightness around the composite profile. 
 Further, \cite{NJ2002ltr} show that since the spread is given in units 
of mag arc sec$^{-2}$, it
can actually mask an increase in scale height, of as much as a factor of 1.7-2.5 within the optical disk,  in the two galaxies they studied: namely, NGC 891 and NGC 4565.  
See section \ref{sec:4.2.7} for details of \cite{NJ2002ltr}; also, see Section \ref{sec:5} on a flaring stellar disk.
Indeed, flaring  is now shown to be a generic feature of a galactic stellar disk.

The same conclusion, namely, that the scale height $z_0$ is not constant with radius, was reached by  \cite{Rohlfs1982}. In an early and perceptive paper, \cite{Rohlfs1982} pointed out that the scale parameter $z_0$ will in general depend strongly on $R$, and a constant scale height is only possible for some specially chosen variation of stellar velocity dispersion with $R$. Hence they investigate variation of $z_0$ with $R$ in the observed data. They point out that since we see a projected system, $z_0 (R)$ cannot be measured directly; rather, we see a convolved quantity for the luminosity.
Hence for a particular trial function for $z_0 (R)$, they match the resulting model intensity profiles, obtained by directly
integrating the density profile (Eq. (\ref{eq2.1})) along the line of sight -- without making an assumption of constant $z_0$ (as done by \cite{vdkS1981a} to obtain (Eq. (\ref{eq2.2})) -- with the observed intensity profiles. 

 Further, \cite{Rohlfs1982} point out that the use of constant $z_0$  does not give a good fit to the data, which is in contrast to the claim by \cite{vdkS1981a}. Instead, \cite{Rohlfs1982} try some physically motivated, simple analytical functions for $z_0(R)$; and find that a best-fit between their models and observed profiles from \cite{vdkS1981a} is obtained when $z_0$ is taken to increase exponentially with radius with a scale length, R$_D$ 
(see Figs. 1 and 2 in \cite{Rohlfs1982}).  
This was a striking result; but unfortunately, this paper was not much noticed in the literature. 
In another early study, \cite{deGrijsP1997} showed from their data on edge-on galaxies -- by doing the fitting over two radial bins --
 that the stellar scale height increases with radius. Unfortunately, none of these important papers, showing the result that the scale height of the stellar disk increases with radius, has received much attention in the literature. 

\medskip

\noindent {\bf New, correct iterative approach proposed:}

To do the problem correctly; namely, to determine  the scale height from the observational data from external edge-on galaxies accurately, the possible scale height variation with R must be taken into account. To take account of 
the cumulative effect of the possible variation in scale height at points along the line of sight,
the fitting has to be done iteratively, 
to get the optimal choice of $z_0(R)$  which
gives the best-fit to the observed intensity profiles. In this task, different density profiles, that is, a functional form for $\rho(z)$ other than sech$^2$, could also be tried. 
This is a challenging task, and we hope this  will taken up in future to accurately determine the  true scale height, $z_0$, and its possible variation in R.
This could also give us some understanding about the physical process that gives rise to the radial variation in $z_0$.
This general approach proposed above
 is logically the next step, long overdue, in the  approach initiated along this direction by \cite{Rohlfs1982}, who had chosen a specific trial function of $z_0(R)$.

In an important recent study,  \cite{Kasparova2020} model the observed intensity profile from NGC 7572 as a sum of a thin disk and a thick disk, with the aim to study the variation of disk thickness with radius. To do this, they use model intensity profile as a sum over two sech$^2$ profiles, and let the thickness of each of these vary with $R$. Despite allowing for this variation, they find that best-fit case still left strong residues (that is, an excess of emission from the observed data) at high z compared to their  model best-fit profiles.
 They point out that this excess indicates contribution by non-local regions, that is,  from different points along the line of sight that cuts the major axis at $R$. They note that this
 excess occurs because the simplified mathematical method used
does not correctly capture the contributions from non-local regions, when doing the double sech$^2$ fits. They caution that while their results clearly show the trend of a flaring disk; in view of the approximate mathematical model used, the exact flaring values cannot be given by their work.
We stress that this again points to the necessity of doing an iterative mathematical fitting of observed intensity profiles, as we have proposed above. 

There is  motivation for taking up this problem, because theoretical studies, as well as a different line of observations: namely, the star counts data for the Milky Way; and other direct observational data for the Milky Way, as for example, from COBE, show that flaring of a stellar disk is a generic result. It would be interesting to see whether the correct fitting of observed profiles as  proposed above for external galaxies also shows such a flaring trend.
 The results for scale height as a function of radius obtained from an iterative fitting approach (as proposed above) would be a good check against the resulting model scale height values from theoretical studies done using the multi-component disk plus halo model. It is possible to obtain the model scale height values for some galaxies for which input parameters for this model are known from observations (see Sections \ref{sec:5.5}, \ref{sec:8}  for details).

This approach of integrating along a line of sight  across the entire galaxy to obtain the net intensity \cite{vdkS1981a, Rohlfs1982}
implicitly assumes that 
the optical depth effect is negligible across a line of sight in an edge-on galaxy. However, in reality, a
galaxy is optically thick, at least in the central regions, down to almost the inclination angle of 60$^0$ \cite{Xi1999}. 

It should be noted that the presence of other vertical spatial features such as warps ($m=1$ type modes, see Section \ref{sec:7.2}) makes it harder to observationally determine the galactic plane, and hence the vertical density distribution including the scale height, especially in the outer parts of the Milky Way. This analysis has been done, and the existence of flaring of  stars \cite{Vig2005,Lopez2002,Lopez2014}, and  HI gas \cite{Wouter1990} has been established from the data.

\subsubsection {Analysis of 3-D light distribution \label{sec:2.1.2}}

Another approach to extract the detailed density distribution including the
vertical structure of edge-on galaxies is to de-project the 3-D light distribution of an edge-on galaxy in terms of inverse Abel transform \cite{BT1987}, see for example \cite{Peters2017}. This method has been applied by \cite{Florido2001,Florido2006,Pohlen2007}.
However, these papers do not address the possible variation of $z_0$ with $R$; instead, these mainly focus on the radial variation of intensity including disk truncation. The effect of dust extinction has been included in the full 3-D modelling of galaxies \cite{Xi1999,Bianchi2007,degeyter2014,Savchenko2023}. 
We point out that to do this problem correctly, these studies
 must be redone so as to take account of the possible radial variation of $z_0$. For that, the fitting has to be done iteratively 
to get the variation in $z_0$ with $R$ that gives the best-fit to the data, and thus  extract the correct vertical density distribution in the stellar disk. 

\subsubsection{Analysis of star counts data in the Galaxy\label{sec:2.1.3}}

So far we have discussed observations of external
galaxies by photometry to deduce the vertical density distribution. An alternative method to obtain the vertical density distribution is possible in the Galaxy; namely,
to directly measure the star counts as a function of $z$. 
The advantage of using this approach is that it does not require any conversion from intensity to mass density that would involve some assumption about mass-to-light ratio in the disk (as pointed out, for example, in \cite{2009ApJ...705.1686H}). 

At an earlier time this approach was used by Oort  \cite{Oort60} who used the observed star counts data for K-giants in the solar neighbourhood  in the Galaxy as a constraint to obtain the total mid-plane density or the Oort limit, to be discussed later in Section \ref{sec:3}. 

In recent years, there have been great strides in observations of star counts. These permit the measurement of vertical density distribution of stars, as a function of type, age etc. over a larger radial range of several kpc: for example, the work done using 2MASS data by \cite{Momany2006}; using SDSS-SEGUE data by \cite{Lopez2014}; using SDSS data by \cite{Ferguson2017}; and using LAMOST data by \cite{Wang2018}.

In principle, the vertical density profile for stars can be obtained from this data in the Milky Way. However, in practice, it is difficult to determine the density profile. The results seem to depend on the actual choice of data cuts taken and the analysis done. For example, the analysis of \textit{Gaia} data by \cite{Bovy2017} shows a  sech$^2$ profile.  
A Bayesian analysis by \cite{Dobbie2020}, based on the SDSS data of several million K and M stars by \cite{Ferguson2017}, shows the stellar density profiles to be steeper than sech$^2$, and it shows a moderate evidence
for sech and exponential profiles compared to a sech$^2$ profile.

However, despite these caveats in the analysis of star counts data,  it is still useful to compare theoretical model results for density profile with the observations of number density from star counts (see Section \ref{sec:4.1.1}, and Section \ref{sec:8}). The observed data could in fact be used to fine-tune or improve the theoretical models.

Remarkably, new detailed astrometric and kinematical data for the line-of-sight velocities for individual stars from SDSS and \textit{Gaia}  are now available for an external galaxy, namely, the LMC (Large Magellanic Cloud). This involves about 15 million stars out of the two billion stars in the Clouds. This information has now been used to model  a 3-D mass  distribution in the LMC \cite{LMC_3Dmodel}.

\subsubsection {Stellar velocity dispersion data \label{sec:2.1.4}}

An important parameter describing a stellar disk is the typical stellar velocity dispersion. This has been measured in the Milky Way by observations along the so-called Baade's window, which is a region of low dust extinction. This permits kinematics of  stars to be measured between 1-17 kpc from the Galactic centre \cite{LewisF1989}. This study shows that the radial stellar velocity dispersion falls exponentially with radius, with a scale length,  $R_v = 8.7$ kpc. 

The velocity dispersion measurement had been done for only a few external galaxies until the  1990's \cite{Bottema1993}, that too, in the inner regions where the dispersion is high. This situation has changed for the better in the last $\sim$ ten years due to the   
availability of 2-D IFU data on stellar kinematics of external galaxies: for example, see, the DiskMass survey \cite{Bershady2010};
CALIFA (The Calar Alto Legacy Integral Field spectroscopy Area) survey  \cite{Sanchez2012, Falcon2017}; MaNGA  (Mapping Nearby Galaxies at Apache Point Observatory) survey \cite{Westfall2019,Bundy2015};  SAMI (Sydney-AAO Multi-object Integral Field Spectrograph) survey \cite {Croom2021}; also, see \cite{BHGARAA} for details of various IFU surveys. 
However, due to resolution problems, the stellar velocity dispersion measurements from many of these are still limited to
the inner regions only. Often, a single value for the dispersion in the effective radius ($R_e$) is
extracted from the data.{\footnote{An effective radius, $R_e$, of a galaxy is the radius from within which half the luminosity of the galaxy is emitted. For an exponential disk,  $\mathrm{R_e \sim 1.7 R_D}$. For a typical galaxy with a bulge, $R_e$ would be smaller than this.}}

However, CALIFA does have measurements up to 1-2 effective radii  and these are well-sampled, see the detailed work by  \cite{Falcon2017}
that gives stellar dispersion as a function of radius. Similarly, the data from the various surveys also need to be analyzed to obtain the stellar velocity dispersion values as a function of radius. These are needed as input parameters to study disk vertical structure. This will be discussed in detail in Section \ref{sec:8}. 
Given the difficulty of measuring the velocity dispersion, an alternate way has been suggested by \cite{Bershady2024} where the measured line-of-sight velocities are used to estimate the radial velocity dispersion using the concept of asymmetric drift. Here, the asymmetric drift is used as a proxy for stellar velocity dispersion (see Fig. 9 from \cite{Bershady2024}). However, in the above work, the sampling points are few, and the analysis is limited to the central two effective radii, so the results for velocity dispersion obtained are not very useful for studies of vertical structure across the disk.

For the Galaxy, using \textit{Gaia} data, very detailed information -- such as the dispersion versus stellar age -- is now available (e.g.,\cite{Mackereth2019}). For convenience, typically in the literature, starting from \cite{Spitzer1942, Camm1950}, as also in this review, all the stars in a galactic disk are taken to form a single component with the same velocity dispersion (see Section \ref{sec:4.2.1} for a discussion on this). 
Recent data gives the velocity dispersion to be non-isothermal. This information is used to build a more general theoretical model (Section \ref{sec:4.4.2}).

\subsection {Interstellar gas: Surface density and velocity dispersion \label{sec:2.2}}

Although this review mainly focuses on the stellar disk, it also shows that the interstellar gas needs to included for treating a realistic, multi-component galactic disk. Hence, we briefly describe here the properties of the interstellar gas in a typical galaxy like the Milky Way. The interstellar gas consists of two main mass components: the interstellar atomic hydrogen gas (HI), and the interstellar molecular hydrogen gas (H$_2$). Each of these is distributed 
 in a thin disk, and is characterized by different values of surface density and velocity dispersion, which vary radially. The interstellar medium (ISM) in galaxies is a vast, well-developed field  \cite{Spitzer1978,Draine2011}. The observational properties of ISM are complex and have been extensively studied: for the HI gas, for  early in-depth reviews, see \cite{KH1988,DL1990}; and for more recent reviews see 
 \cite{KalberlaARAA, McClure2023}. For the H$_2$ gas in the Milky Way,  see the reviews by \cite{SS1987,H2ARAA2015}; and for reviews for H$_2$ in external galaxies, see \cite{YS1991,Omont2007}. 
  Thus, the properties of HI and H$_2$ gas, and their radial distribution, have been extensively studied for the Galaxy as well as for external galaxies.
  
The interstellar gas is observed to be distributed in a thinner layer than the stars, with the H$_2$ distribution being thinner than the HI distribution. The three disk components are  taken to be co-spatial, concentric and coplanar; also, see Section \ref{sec:4.1}.
The ISM is a complex, multi-phase system
\cite{MckeeOstriker1977,1987ASSL..134...87K}, consisting of CM (cold medium), WNM (Warm Neutral Medium) and HIM (Hot Ionized Medium).
Despite the multi-phase nature of HI, for simplicity, we consider it to be characterized by a single value of surface density, $\Sigma (R)$ and $\sigma_{HI}(R)$, as is usually done in the literature. This represents the CM  and WNM components of HI \cite{1987ASSL..134...87K,Dickey2009}.
The contribution of HIM to the mass in the thin disk region is small, hence it is neglected for the problem of the vertical structure of the disk.

For the purpose of this review we need to consider the observed values of the global 
properties  such as  the surface density and the velocity dispersion of the HI and H$_2$ components. In the multi-component disk plus halo model to be discussed extensively in Section \ref{sec:4} onward, both HI and H$_2$ gas are treated as gravitating disk components, which interact gravitationally with stars and each other  (see Section  \ref{sec:4.1}).  The gas is supported by turbulent pressure and the corresponding velocity dominates over the thermal velocity in each component \cite{KH1988,SS1987,Draine2011}.
Typically, in the inner disk, the stellar surface density is about 10 times higher than the gas surface density; and the stellar velocity dispersion is higher by a factor of few than the HI dispersion, which in turn is slightly higher than the H$_2$ dispersion.  The total mass content in HI and H$_2$ is nearly comparable:   most of the H$_2$ gas is located inside of R= 8.5 kpc, whereas most of the HI gas is distributed in the outer Galaxy \cite{SS1987}.  Thus the radial distribution of HI and H$_2$ components is different.
The HI velocity dispersion is supported by turbulence; the pressure due magnetic fields and cosmic rays is not important for the internal support of HI gas \cite{LG1991}.

The HI dispersion values decrease with radius and then taper off to a saturation value (e.g., see the discussion in \cite{NJ2005}; also,  Section \ref{sec:4.2.3}).  
At the solar position, the observed radial stellar velocity dispersion is $\sim 18 
$ km s$^{-1}$ and decreases exponentially with radius; while the isotropic HI velocity dispersion is $\sim 8$ km s$^{-1}$ which tapers off at large radii to  $\sim 7$ km s$^{-1}$ (see Section \ref{sec:4.2.3}).
The specific values of these parameters chosen, with some variation as per the galaxy under consideration, will be introduced as necessary during the application of theoretical models in the later Sections (e.g., Section \ref{sec:4}, Section \ref{sec:6}).

In contrast to the optical studies of stars where dust extinction is a problem; for HI gas -- using the 21 cm radio observations as a tracer -- the 3-D HI data cube allows one to get detailed maps of HI distribution. Here the effect of optical depth is not  important since one sees local HI (at a particular line-of-sight
velocity bin) at each point along the line of sight. So it is possible to obtain a 3-D distribution map of HI in the Galactic disk, as well as in external galaxies. As an example of later, see the 3-D HI data cube used to study rotation curves in galaxies (e.g., \cite{Lelli2016,2023MNRAS.524.6213B}).
This point and the applications of 3-D HI data cube will be discussed further in Sections \ref{sec:4.5.2} and \ref{sec:6.5}.

\bigskip

\section{Self-consistent vertical density distribution in a galactic disk\label{sec:3}}
In this section, we develop the topic of self-consistent vertical density distribution in a galactic disk systematically. We start from the basic, classic model of a thin, one-component, isothermal disk; to be followed in Section \ref{sec:4} by  recent, more realistic multi-component disk plus halo model. 

At the outset, we point out that
at a basic level the vertical structure is more difficult to study than the planar mass distribution in a rotationally supported galactic disk. A galactic disk is mainly supported by rotation in the plane; the random velocities do not play a major role in its support. For a disk in rotational equilibrium, the rotation velocity at any point is determined by the global or non-local mass distribution in the disk. The motion of a particle in the plane can be treated as a test particle in the field of this disk. 
One can then use the observed rotation curve as a constraint to deduce the radial mass distribution in a galaxy. A galactic mass model typically is built on this idea (e.g., \cite{Bahcall1980,Mera1998,Sylos2023}.
This has in fact been used to show the important result of existence of dark matter (e.g. \cite{Rubin1980,Rubin1983}; also, see Section \ref{sec:6}), although the details of the deduced mass distribution are more complicated and depend on the actual geometry  \cite{BT1987}. 

On the other hand, for a thin disk in hydrostatic equilibrium, the vertical motion is determined locally at any radius. 
This is because the vertical gradient of the gravitational force is dominant and hence the vertical density distribution is determined by the local conditions, as shown later in this section. Thus, interestingly, the disk vertical structure can to be used to uniquely trace the local gravitational potential at a given radius; this is in contrast to the planar case. However, the entire vertical density distribution along the z direction at a given radius has to be determined self-consistently by jointly solving the Poisson equation and the equation of hydrostatic equilibrium. 
In  a  limiting case, at low z values close to the mid-plane -- corresponding to the region of nearly constant density; the motion of a single particle
can be treated as that of a test particle, and it follows a simple harmonic motion
\cite{MihalasRoutly, BT1987}.  However, that does not permit us to obtain the entire $\rho(z)$ profile. In contrast, in the planar case, using the test particle approach one can use  the observed rotation curve 
to build the entire galaxy mass model. 
This is what makes the vertical structure of a galactic disk, or the mass distribution along z, harder to study than the planar case.

\medskip

\subsection{Basic, classical model of a galactic disk \label{sec:3.1}}

The vertical density distribution of 
a galactic disk is typically obtained by treating it as a self-gravitating, isothermal
system supported vertically by its internal pressure gradient.
In a pioneering work, Spitzer (1942) \cite{Spitzer1942} showed that the vertical density distribution of such an idealized 
disk has a sech$^2$ form, as discussed next. 
We discuss this giving the details of the physics in the derivation.
As for any self-gravitating system, 
 the vertical structure and dynamics of a galactic disk are directly related through the Poisson equation.  
For an axisymmetric galactic disk, 
the $\phi$ component of the Laplacian is zero.

\cite{Spitzer1942} considered the disk to be locally dense, or that the local density of matter in the disk  is much greater than the the average density of the rest of the galaxy interior to that point R; hence the effect of the rest of the galaxy is neglected. Thus, to a first approximation, the disk distribution is treated as  a one-dimensional problem along $z$. 
Thus, in \cite{Spitzer1942} the R term in the Poisson equation is neglected and the Poisson equation is taken to reduce to:

\begin{equation}
\mathrm { \frac{{\partial}^2 \Phi (R,z)}{\partial {\mathit z}^2} = 4 \pi G \rho(R, z) }
                \label{eq3.1}
            \tag{3.1}
\end{equation}           

\noindent where both $\Phi (R,z)$, the gravitational potential and $\rho (R,z)$, the mass density are given at a local radius R. 
Thus, the force along  $z$,  and the vertical density distribution, $\rho (z)$, are determined locally, and depend only on the physical conditions  
at a given radius $R$. Thus, 
 both are taken to vary purely as a function of $z$. 

The disk is assumed to be supported against gravitational collapse along the $z$ direction by the vertical gradient of its own internal pressure. That is, the self-gravitational force is balanced by the gradient of the pressure along the vertical direction: this decides the equilibrium vertical density distribution.
 Typically, the pressure support is taken to be due to the velocity dispersion along $z$, hence the pressure is given by $\mathrm{P = \rho(z) \langle{({\mathit v}_z^2)}\rangle}$ where $ \mathrm{\langle{({\mathit v}_z^2)}\rangle }    $  is the mean square velocity dispersion 
along $z$. The disk is taken to be isothermal along $z$ for simplicity, so that the velocity dispersion is constant along $z$.
 
The force equation along the $z$-axis, that is, the equation of hydrostatic 
balance, is given by (e.g., \cite{Rohlfsdensity}):

\begin{equation}
\mathrm{\frac {{\langle{({\mathit v}_z^2)}\rangle}}{\rho} \: \frac {\partial \rho}{\partial{\mathit z}} \: = \:     {\mathit K}_z } \:     
     \label{eq3.2}
     \tag{3.2}
\end{equation}     
     
\noindent where $\mathrm{K_z \: = \: -  {{\partial}{\Phi}}/ {{\partial}z}} $ is 
the gravitational force force per unit mass along the $z$-axis. For a reasonable physical behaviour of pressure falling with $z$, 
the assumption of hydrostatic balance (Eq.
\ref{eq3.2}) is satisfied in a typical galactic disk.

The distribution is taken to be stationary or steady-state with no mass motions along the $z$ direction, hence the continuity equation is not employed to obtain the vertical density distribution.

The self-consistent vertical density distribution, $\rho(z)$, is obtained by solving the Poisson equation (eq. (\ref{eq3.1})) and the equation of hydrostatic equilibrium (Eq. (\ref{eq3.2})) together. On combining the above two first-order, 
linear differential equations, the following second-order, linear differential equation in $\rho(z)$ is obtained. This is the joint Poisson-hydrostatic balance equation for a one-component disk:

\begin{equation}
\mathrm { \frac {\partial}{\partial {\mathit z}} \: \left ( \frac {\langle{({\mathit v}_z^2)}\rangle}    {\rho} \: 
    \frac {\partial \rho}{\partial {\mathit z}} \right ) \: = \: - 4 \pi G \rho  }  
      \label{eq3.3}
       \tag{3.3}
\end{equation}

The solution for this equation was obtained by  \cite{Spitzer1942} to be:

\begin{equation}
  \mathrm{ \rho(z) = \rho_{0} \: sech^2 (z/z_0) }                    \label{eq3.4}
   \tag{3.4}
\end{equation}   

\noindent where $\rho_0$ is the mid-plane density and the parameter $z_0$ is the scale parameter which is a measure of disk thickness or scale height. 
(See Appendix A for a detailed discussion of vertical scale height for a gravitating disk.)
This solution is obtained analytically in terms of a subsidiary equation as explained in detail in \cite{Rohlfsdensity}. For small $z << z_0$, the above solution for the density, $\rho(z)$ (Eq. (\ref{eq3.4})) reduces to: $\mathrm{\rho(z)=\rho_0 \: exp -({z/z_0})^2}$; that is, a Gaussian form. At the other limit of $z >> z_0$, the above solution reduces to:   $ \mathrm{\rho(z) = 4 \: \rho_0 \: exp  ({- 2 z/z_0})}$; that is, an exponential.

The two boundary conditions used to solve the above second-order differential equation (Eq. \ref{eq3.4}) are: first,  the disk surface density, $\Sigma$, at the region is constant (as discussed later in this section); and second, 
$d \rho/dz$=0 at the mid-plane, or $z=0$ \cite{Rohlfsdensity}. The latter condition arises because for a realistic distribution, the density  is homogeneous close to the mid-plane. 
These two conditions are used to determine the two parameters $\rho_{0}$ and $z_0$.

The resulting values of  $\rho_{0}$ and $z_0$, are specified in terms of physical quantities, namely the surface density and the velocity dispersion, and are given as follows:

\begin{equation}
\mathrm{  \rho_{0} =  \frac {2 \pi G \Sigma^2} {\langle{({\mathit v}_z^2)}\rangle} ; \: \: \:  z_0 = \Sigma/\rho_{0} = \frac  {\langle{({\mathit v}_z^2)}\rangle} {2 \pi G \Sigma}= 
 \left [\frac   {\langle{({\mathit v}_z^2)}\rangle}  {2 \pi G \rho_{0}} \right ]^{1/2}    }            
         \label{eq3.5}
         \tag{3.5}
\end{equation}

\noindent On substituting $\rho(z)$ (from Eq. (\ref{eq3.4})) into Eq.(\ref{eq3.2})  gives the expression for $|K_z|$, the vertical force per unit mass, as follows:

\begin{equation}
\mathrm{ |{\mathit K}_z| = 2 {\frac  {\langle{({\mathit v}_z^2)}\rangle}  {z_0}} {tanh \frac{z}{z_0}} }
         \label{eq3.6}
         \tag{3.6}
\end{equation}         

 Physically, the scale parameter $z_0$, a measure of disk thickness, is set by the equilibrium between the local gravitational force and the internal pressure gradient as given by the equation of hydrostatic balance.

We will refer to this  as the {\it sech$^2$ model} in the rest of the review article. Recall that the assumption given above (in \cite{Spitzer1942}), namely, that the disk is locally dense, led to Eq. (\ref{eq3.1}). Hence the resulting density distribution, $\rho(z)$,  depends only on the local conditions in the disk. This classic solution for  
$\rho(z)$, the vertical density distribution along $z$ (Eq. (\ref{eq3.4})), is taken to be the standard model and routinely applied to describe the vertical structure of a galactic disk, as discussed below.

The vertical structure of an isothermal, self-consistent distribution was also studied by \cite{Camm1950} who assumed a disk of constant surface density so that there are no radial gradients in density or force. Hence, the radial component of the Poisson equation is neglected and the problem again reduces to a one-dimensional problem of distribution along $z$. \cite{Camm1950} also obtained the same solution for the vertical density distribution as given by Eq. (\ref{eq3.4}).

Historically,  it was later shown  quantitatively and more rigorously by Oort and others that the above simplified form of Poisson equation that includes only the z term in the Laplacian (Eq.(\ref{eq3.1})), is valid in the solar neighbourhood for the Milky Way.  To show this, the $R$ term of the Poisson equation was obtained in terms of the observed rotation curve and its local derivatives, and was shown to be less than $(4 \pi G \rho)$, and hence
the $R$ term could be neglected \cite{1932BAN.....6..249O,1965gast.book..455O, MihalasRoutly,Bahcall1984paper1}. 
The same result was shown to be true using modern data -- for rotation curve parameters and the Galactic constants  and the mid-plane total density -- at the solar neighbourhood, as well as at the other radii in the inner Galaxy \cite{SJ2022}.

The above simplified form of Poisson equation as given by Eq.(\ref{eq3.1}) has been shown to be valid on general physical grounds for a typical thin disk as well (see Section 2.2 in \cite{BT1987}).
Thus, the Poisson equation for the thin disk also contains only the $z$ term, 
so it has the same simplified form as used by  \cite{Spitzer1942} for a locally dense disk. Hence, the density distribution is a one-dimensional problem along $z$ for a thin disk as well. 
 The vertical force,  and the resulting vertical density distribution, for the thin disk are determined locally, 
 and depend only on the local surface density at a given point in the disk.
Thus the condition of high local density as assumed by \cite{Spitzer1942}; and the condition of a thin disk, are equivalent and give rise to the same formulation for the vertical structure problem. Eq. (\ref{eq3.1}) shows that the planar and vertical dynamics are decoupled for a thin disk, e.g., \cite{BT1987}, 
as well as for a locally dense disk considered in \cite{Spitzer1942}. We underline an important point that, for the one-dimensional, local problem considered here; the surface density is constant at a given $R$. Hence, the use of local surface density being constant as one of the boundary conditions used in solving  Eq. (\ref{eq3.3}) is justified. The hydrostatic balance equation remains the same in both approaches. Thus, the resulting density distribution for a thin disk is also specified by Eq.(\ref {eq3.4}).{\footnote{Although Eq.(\ref{eq3.1}) was shown to be valid for a thin disk by \cite{1932BAN.....6..249O,BT1987}, they
use this formulation to study the mid-plane density -- but do not obtain the solution $\rho(z)$ for a thin disk.}} 

It is worth pointing out that in the literature,  the paper by Spitzer \cite{Spitzer1942}, and the resulting 
sech$^2$ density profile (Eq.(\ref{eq3.4})) from that work, are routinely cited in the literature in the context of a thin disk. However,  \cite{Spitzer1942} did not explicitly mention the assumption of a thin disk in that paper; rather, the paper was formulated for a locally dense disk. As discussed above, the treatment and the results for $\rho(z)$ are identical for both these assumptions. Hence the use of the classic {\it sech$^2$} solution \cite{Spitzer1942} (obtained for a locally dense disk),  to also formally denote the self-consistent vertical density distribution 
for a thin, stars-alone, one-component galactic disk -- as is normally done  the literature, is justified.

Strictly speaking, while obtaining the terms involving the $z$ derivative of the potential while deriving the Poisson equation (Eq.(\ref{eq3.1})), and the equation of hydrostatic balance (Eq.(\ref{eq3.2})); the complete Jeans equation along $z$ (e.g., Eq. (4.29 c) from \cite{BT1987}) should be used. This would also include the cross terms and terms involving kinematic effects. This is not generally noted in the literature and the above simplified forms are routinely used in subsequent papers on the topic.  It has been shown that these terms are small for a typical, thin disk (e.g., \cite{BT1987}), as explained in detail next.

The complete Jeans equation along $z$ (Eq. 4.29 c from \cite{BT1987}) 
also includes terms containing the cross term $<v_R v_z>$, where the average
is taken over the velocity dispersions.
The term $<v_R v_z>$ is a component of the velocity ellipsoidal tensor whose value is
set by the tilt of the velocity ellipsoid w.r.t. the galactic plane. For any general orientation of the velocity ellipsoid that
is pointing toward the galactic center, it can be shown that the ratio of the two
cross terms involving $<v_R v_z>$ (which are dropped), to the other terms in Eq. (\ref{eq3.2}) (which are kept), is  $\sim z^2/(R R_D)$ \cite{MihalasRoutly, BT1987}. This ratio is small and hence these terms can be neglected for the thin disk in general \cite{BT1987}. For example, this ratio is $< 0.01$ for $z < 1 $ kpc for the solar neighbourhood (e.g., \cite{Bahcall1984paper1,SJ2019}).
 Hence the above simplified form for the $z$ term of Poisson equation (eq.(\ref{eq3.1})) and the equation of hydrostatic balance (Eq. (\ref{eq3.2})), as is generally done in the literature, is justified for a typical, thin galactic disk.

Similarly, the cross terms were shown to be negligible for a thin disk, while evaluating the $R$-term of the Poisson equation which was then shown to be small compared to the $z$ term for a thin disk, see \cite{1965gast.book..455O, MihalasRoutly, Bahcall1984paper1}. The cross terms being small is the reason why the $R$-term could be evaluated simply in terms of the observed rotation curve and its derivatives, as done in these papers. 
The significance of the cross terms in the context of a thick or low density disk will be discussed later in  Section \ref{sec:4.3.1} (in particular, see the discussion after Eq. (\ref{eq4.9}) and  Eq.  (\ref{eq4.11}). 

The above solution for the vertical density distribution (Eq.(\ref{eq3.4})) has been the workhorse of vertical disk structure and is taken as the default model for the vertical disk structure. However, it was  found that it  does not explain the recent observed data. This was one motivation for the formulation of a more realistic multi-component galactic disk plus halo model   (Section \ref{sec:4}).

\subsubsection{A brief comment about the Oort limit\label{sec:3.1.1}}

At around the same time as the above model was proposed; a parallel, independent line of study by Oort(1932) \cite{1932BAN.....6..249O}  focused on  
the determination of the mid-plane density in the solar neighbourhood, which is now known  
as the Oort limit in his honour. Here, the same two equations (the Poisson equation and the hydrostatic balance equation along $z$) are used; but then the observed values of number density of stars obtained by star counts are used to determine the force $K_z$ and hence the mid-plane density.

If the disk consists of several sub-components, then the equation of hydrostatic balance for  a particular population of tracer stars has the following form, where its
pressure gradient is balanced by the net gravitational force of the entire disk:

\begin{equation}
\mathrm { \frac {{\langle{({\mathit v}_z^2)_i}\rangle} }{\nu_i} \: \frac {\partial\nu_i}{\partial{\mathit z}} \: = \:
      {\mathit K}_z  }
      \label{eq3.7}
      \tag{3.7}
\end{equation}      

Here $\nu_i (z)$ is the number density of tracer  stars (denoted by the subscript i), at a given height $z$ from the mid-plane; and $<(v_z^2)>_i$ is the velocity dispersion of these tracer stars, and it is assumed to be isothermal. 
$K_z$, the gravitational force per unit mass, is due to the entire disk. The Poisson equation for the entire disk has the same form as for a one-component disk (Eq.(\ref{eq3.1}) where $\Phi$ now denotes the total potential of the disk.

Thus, combining Eq.(\ref{eq3.7}) and Eq.(\ref{eq3.1}), the joint Poisson-hydrostatic balance equation that describes the vertical density distribution of the particular tracer component is given by:

\begin{equation}
\mathrm {\frac {\partial}{\partial {\mathit z}} \: \left ( \frac   {\langle{({\mathit v}_z^2)_i}\rangle}  {\nu_i} \: 
    \frac {\partial\nu_i}{\partial z} \right ) \: = \: - 4 \pi G \rho } \:
      \label {eq3.8}
    \tag{3.8}
\end{equation}

By measuring the observed number density of a particular population of stars used as tracers  and using the above equation; one can, in principle, estimate the mid-plane density.
It should be pointed out that such an estimate of the mid-plane density would be very uncertain because it involves a triple differentiation of star counts: once to get the number density as a function of $z$, and twice more (as seen from the l.h.s. of the equation of joint Poisson-hydrostatic balance equation (Eq. (\ref{eq3.8})), see \cite{BT1987} for details. 

However, the uncertainty inherent in this method can be reduced by using different tracer stars to obtain the mid-plane density, and taking the average. This was done using F-stars and K-giants as two different set of tracers by Oort(1932) \cite{1932BAN.....6..249O}, (also, Oort (1960), \cite{Oort60}),
who concluded that $\rho_0$, the total mid-plane density, in the solar neighbourhood  is equal to 0.15 M$_{\odot}$ pc$^{-2}$. This is now called the "Oort limit". This pioneering work attempted to   
measure the total mid-plane density in the solar neighbourhood. By comparing with the observed density of gas and stars at that time, Oort concluded that $\sim 40 \%$ of this total mass density is not accounted for, or is "missing". 
It was conjectured at that time that this "missing" or unseen mass could be in the form of faint white dwarfs or interstellar gas clouds \cite{MihalasRoutly}. Later this came to be known as the local dark matter, see, e.g., \cite{Trimble}, for a discussion of the possible constituents of this local dark matter and the early searches for it. 

Later, the Oort limit problem was reworked by  \cite{Bahcall1984paper2} who aimed to reduce the uncertainty introduced by multiple differentiation of observed data. To do this, \cite{Bahcall1984paper2} considered the disk to consist of 16 different isothermal stellar components, and following a somewhat different procedure obtained the mid-plane density, $\rho_0$ to be $\sim 0.18 M_{\odot} pc^{-3}$, see \cite{BT1987} for details. At that time, the observed total mass density in the solar neighbourhood, including the stellar and gas density, and estimates of the density in white dwarfs as relics of stellar evolution,  was found to be be $\sim 0.114 M_{\odot} pc^{-3}$. Hence the missing mass or the local dark matter density was deduced to be
$\sim 0.066 M_{\odot} pc^{-3}$ or 36.7 \% of the total mid-plane density.   

This problem has been extensively worked on over the years, using more modern data: for example,  using the HIPPARCOS data (from the ESA HIPPARCOS satellite)\cite{Flynn1994}, \cite {Holmberg2000}; using RAVE (RAdial Velocity Experiment survey) data  for red clump stars  \cite{Bien2014}; and 
using the HST data for M dwarfs 
\cite{Mckee2015}. 
The analysis to determine the total mid-plane density, and the resulting value for it, have been heavily debated in the literature: 
see for example, \cite{Monibidin2012} and \cite{BovyT2012}.
More recently, 
the local total surface density as well as the local dark matter density has been extensively studied: using \textit{Gaia} DR2 and/or LAMOST data  \cite{Hagen,Buch2019,Guo2020}; and using APOGEE and \textit{Gaia} data \cite{Nit2021}. For some recent reviews on the local dark matter density, see \cite{Read2014,desalas2021}. 

This section gives a brief introduction, including the early history, of this important topic. 
The determination of the Oort limit, and the contribution of the observed or baryonic components  to it,
continues to be a subject of active research. 
The estimate of the total mid-plane density, and hence local dark matter density, is of great significance for the study of galactic dynamics. This is because it gives a local constraint on the large-scale Galactic mass model, as well as the extended dark matter halo distribution. The total local mid-plane density has interesting implications for the formation and evolution of disk galaxies \cite{Flynn1994}.

 Extensive efforts have been made to study the nature and constituents of the local dark matter (e.g., \cite{Trimble,StrigariDM}); but no clear, convincing candidate has emerged so far. 

\subsection {Gas in a galactic disk in the classical treatment \label{sec:3.2}}

The interstellar gas typically contains a small mass fraction ($\sim 10-15 \%$) of the disk mass (e.g., \cite{YS1991,BM1998}). Hence, typically, in the early work, the self-gravity of the gas was ignored even when obtaining its own vertical distribution. Instead the gas was treated to consist of massless test particles responding to the dominant gravitational potential, namely that of stars. The gas distribution, $\rho_g(z)$ is then shown to have the following Gaussian form  \cite{Rohlfsdensity}:

\begin{equation}
\mathrm{ {\rho}_{gas}(z) = \rho_{s0} \: exp-( z/z_0')^2}  
     \label{eq3.9}
     \tag{3.9}
\end{equation}   
     
\noindent where $z_0' =  {\sigma_g^2}$/$(2 \pi G \Sigma_s)$, and $\rho_{s0}$ is the total mid-plane stellar density in the disk, and $\sigma_g$ is the gas velocity dispersion. Note that the above result for  $\rho_{gas}(z)$ depends only on the stellar gravitational potential, and the gas velocity dispersion; and has a Gaussian form. This form agreed fairly well with basic form of the early observations \cite{Lockman1984,1987ASSL..134...87K,DL1990}.
 This has been the standard usage in the literature for the vertical distribution of  interstellar gas in a galactic disk, for convenience. However, it does not take account of the gas gravity, which is shown to affect the gas distribution (see Section \ref{sec:4}, and Appendix A).  

In this scenario, the atomic hydrogen gas, in presence of the stellar disk potential alone, should have a scale height which increases exponentially with radius, since the stellar surface density falls exponentially with radius while the gas dispersion is nearly constant with radius (see Section \ref{sec:4.2.3}, and Section \ref{sec:6.5}). However, what is observed is the opposite in the inner Galaxy. In fact, the observed near-constancy of HI scale height in the inner Galaxy ($< 8$ kpc) had been a long-standing puzzle Oort(1962) \cite{1962dmim.conf....3O,1987ASSL..134...87K,
DL1990}.
This led  \cite{NJ2002} to propose that the gas gravity should be included since it would decrease the gas scale height, which could better explain the observed gas thickness.
 This was indeed one of the motivations for the multi-component disk model proposed by  \cite{NJ2002}, as will be discussed in detail in the next section. As discussed in that section, the gas gravity needs to be taken into account to get the correct physical description of the vertical structure of all the disk components, including gas.

The flaring of HI gas is in fact observed and well-established at larger radii, in the outer Galaxy \cite{Wouter1990,DL1990,KalberlaD2008,Levine2006}, also see Section \ref{sec:4.2.6}, and Section \ref{sec:6}. The physical origin  of the gas flaring seen at large radii  can be explained naturally by the multi-component disk plus halo model,  
see Section \ref{sec:4.2.6}, and Section \ref{sec:5}.

\bigskip

\subsection{Vertical constraining effect of molecular cloud complex on disk\label{sec:3.3}}

In contrast to the classical treatment which ignores the gas gravity; the vertical constraining effect due to the gravitational field of a typical molecular cloud complex on the the galactic disk stars was studied
in a remarkably prescient paper by  \cite{complex}. They noted that nearly half the interstellar gas in the Milky Way is in a molecular hydrogen form, and $\sim 90\%$ of it is in the inner Galaxy \cite{SS1987}.
 This gas is contained in giant molecular clouds; 
which are further segregated into well-defined, large, discrete gravitating  features which have been variously called complexes or clusters or chains, as shown in the early work \cite{Sanders1985,Rivolo1986,Dame1987}. 
 The typical molecular cloud complex has a size of  a few 100 pc, has an oblate spheroidal shape with the full height of 120 pc, and a mass of $\sim 10^7$ M$_{\odot}$ each.    Such massive complexes are seen in other galaxies as well (see \cite{complex} for details).

 \cite{complex} pointed out that the average observed density of H$_2$ in a typical complex is $\sim 16 cm ^{-3}$ or $\sim 1 M_{\odot} pc^{-3}$. This is about six times higher than the dynamical or total mid-plane mass density in the solar neighbourhood, or, the Oort limit (Section \ref{sec:3.1.1}). Such a dense, massive cloud complex would dominate the local gravitational field in a galactic disk. 
 Motivated by this, \cite{complex} studied the dynamical effect of such a complex on the vertical distribution of stars.
 To do this, the mass distribution in a cloud complex is taken to be that of an oblate spheroid of constant density. The force per unit mass along z due to a complex is taken from \cite{Schmidt1956},  who had calculated it in the context of galaxy mass models. 

\begin{figure}
\centering
\includegraphics[height=2.6in, width= 3.1in]{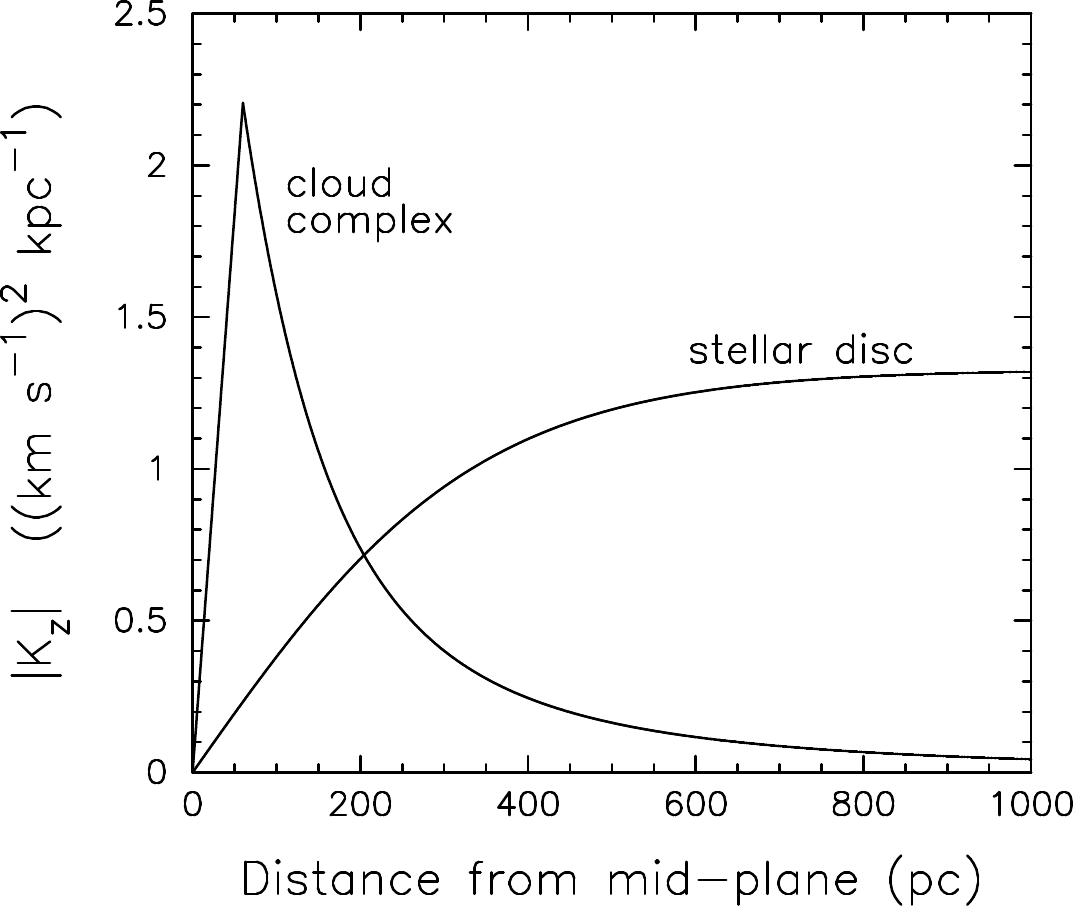} 
\bigskip
\caption{Plot of $\mathrm{|K_z|}$, the vertical gravitational force per unit mass
 due to a cloud complex, and that due to the undisturbed stellar disc, versus
 $z$, the distance from the mid-plane, at the complex centre.  For $z < 200$ pc,
 the force due to the complex dominates over that due to the disk. The ratio is
 a maximum, equal to 9.5, at the outer edge of the complex, at $z= 60$ pc. $\: $  {\it Source}: Taken from \cite{complex}} 
 \label{fig.2}
\end{figure}

To get an idea about the effect of the complex, the gravitational force per unit mass,$\mathrm{|K_z|}$, due to a stars-alone disk
(Eq.\ref{eq3.4}); and that due to the complex at the centre of the complex ($r=0$) vs. $z$ are plotted in Fig. \ref{fig.2}. Here  $r$ is the radial distance measured from the centre of the complex in 
the disk plane.{\footnote{The parameter $r$ is used later to denote the radius in the spherical polar co-ordinates, as in the usual notation, in Section \ref{sec:4.3.2}}} The centre of the complex is located at $R=8.5$ kpc. 
It can be seen that for $z < 200$ pc,  the force due to the complex
 dominates over that due to the disc; with the ratio being a maximum, equal
 to 9.5, at the outer edge of the complex ($z=60$ pc).  
A similar result is seen over the radial extent of the complex (see Fig. 2 in \cite{complex}). 
 Thus, the force due to the complex dominates over the self-gravity of the stellar disk over a large spatial range in $R$ and $z$. This has a strong effect in modifying the stellar distribution near the mid-plane, as shown next.

The equations to be solved to obtain the modified stellar mass distribution in presence of a complex are: the equation of hydrostatic balance, or the force equation along the $z$ direction: 

\begin{equation}
  \mathrm { \frac  {\langle{({\mathit v}_z^2)}\rangle}  {\rho_s} \: \frac {\partial \rho_s}{\partial \mathit z} \: = \:
      (K_z)_s \: + \: (K_z)_{complex}    }         \label{3.10}
            \tag{3.10}
\end{equation}            

\noindent and, the joint Poisson equation for the stars and the complex: 

\begin{equation}
\mathrm {\frac {\partial ({\mathit K}_z)_s}{\partial {\mathit z}} \: + \: \frac {\partial ({\mathit K}_z)_{complex}}{\partial {\mathit z}}
  \: + \: \frac {\partial ({\mathit K}_r)_{complex}}{\partial {\mathit r}} \: + \: \frac {({\mathit K}_r)_{complex}}
    {r}    \: = \: - 4 \pi G (\rho_s \: + \: \rho_{complex} )}
    \label{3.11}
    \tag{3.11}
\end{equation}

\noindent The complex being massive is assumed to be unaffected by the disk, hence the joint Poisson equation effectively reduces to that for the redistributed stellar case alone:

\begin{equation}
\mathrm {\frac {\partial (K_z)_s}{\partial {\mathit z}}   
   \: = \: - 4 \pi G \rho_s  }            
      \label{3.12}
      \tag{3.12}
\end{equation}      

\noindent where $\rho_s$ denotes the redistributed stellar density in the field of the complex.
On combining the above two coupled first-order linear differential equations (Eq. (\ref{3.10}) and Eq. (\ref{3.12}) ), gives the following  second-order, linear 
differential equation in $\rho_s$:

\begin{equation}
{<{(v_z^2)_s}>}\frac {\partial}{\partial {\mathit z}} \: \left ( \frac {1}{\rho_s} \: 
    \frac {\partial \rho_s}{\partial {\mathit z}} \right ) \: = \: - 4 \pi G \rho_s \:
    + \:  \frac {\partial (K_z)_{complex}}{\partial{\mathit z}}
 \label{3.13}
    \tag{3.13}
\end{equation}    

\begin{figure}
\centering
\includegraphics[height=2.55in, width= 3.1in]{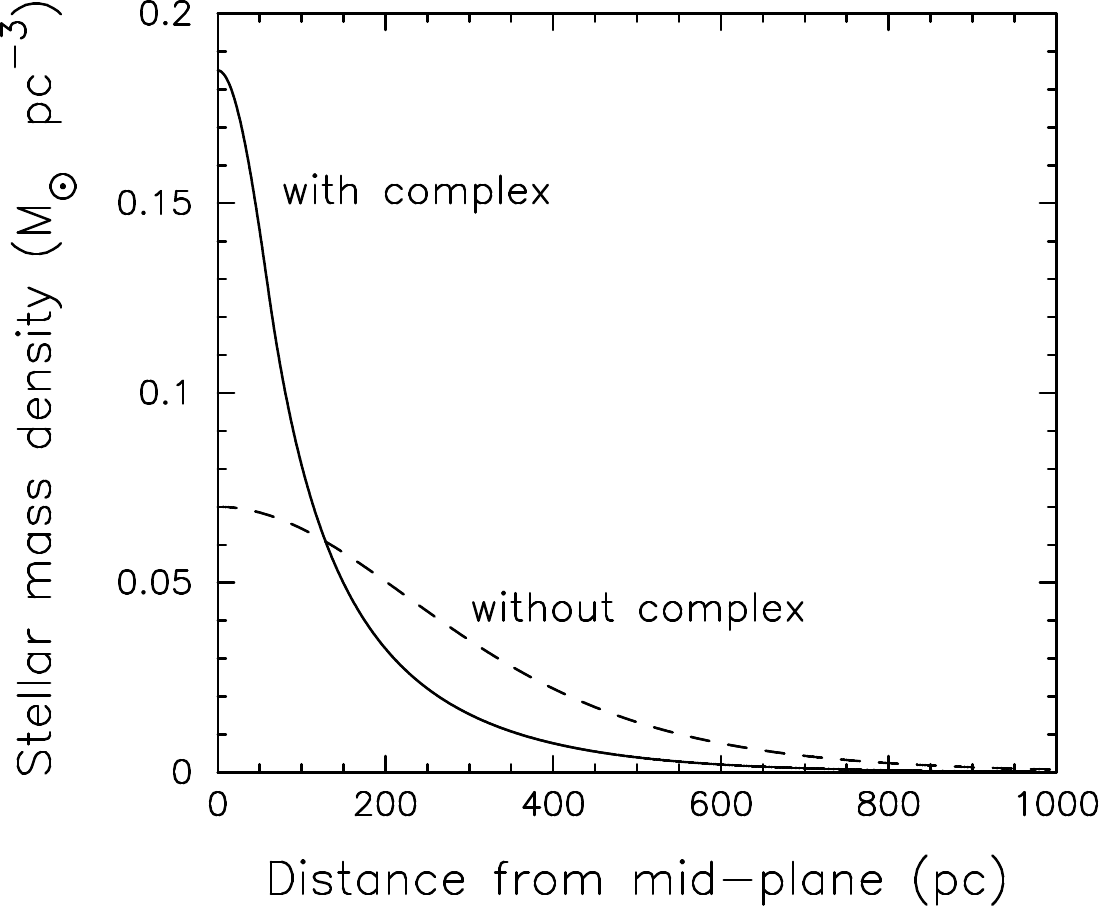} 
\bigskip
\caption{Plot of self-consistent  density for self-gravitating, undisturbed stellar disk (dashed line), and the modified stellar disk density in the field of a cloud complex (solid line) vs. $z$, at
at the centre of the cloud complex. The complex centre  is located at R=8.5 kpc. The modified mid-plane density is 2.6 times higher and the vertical scale height (HWHM) is 3.4 times smaller than the undisturbed case, showing the strong constraining or pinching effect due to the cloud complex.  
$\: $  {\it Source}: Taken from \cite{complex}
} 
 \label{fig.3}
\end{figure}

This 
equation is solved numerically as an initial-value problem from the inside-out, using the fourth-order Runge-Kutta method  \cite{NumRec1986}. The initial condition of $d \rho_s / dz =0$ at z=0 is used since that would be true for any realistic mass distribution in  a disk. The value of the redistributed density $\rho_s $ at the mid-plane ($z=0$) is not known, a priori.  Its value is obtained by trial and error, by assuming that the total column density of the undisturbed stellar disk, $\Sigma_s$, remains constant even in  the presence of a complex. This is justified because the complex is assumed to redistribute the stars only along $z$ at a given radius, since the radial force due to the complex can be neglected as argued by \cite{complex}. Hence, 
the determination of density distribution for stars in presence of a complex becomes effectively a one-dimensional problem along $z$.
To illustrate the effect of the gravitational field of the complex, consider the vertical stellar density distribution at the central position of the complex (r=0), located at $R$=8.5 kpc. On solving Eq. (\ref{3.13}) for $(k_z)_{complex}$=0, the undisturbed sech$^2$ distribution \cite{Spitzer1942} is obtained, as expected (Fig. \ref{fig.3}, dashed line); while including the effect of complex results in the modified stellar distribution (Fig. \ref{fig.3}, solid line). 
The gravitational force due to the complex strongly alters the vertical distribution of the stellar disk: the mid-plane density increases by a factor of 2.6 to be equal to $0.19 M_{\odot} pc^{-3}$,
while the scale height (HWHM)   decreases by a factor of 3.4 to be equal to 88 pc. Thus, overall the density distribution is constrained to be closer to the mid-plane. This strong "constraining" or "pinching" effect is a generic result,
arising due to the additional gravitational force of the complex acting on the stellar disk.

\begin{figure}
\centering
\includegraphics[height=2.6in, width= 3.1in]{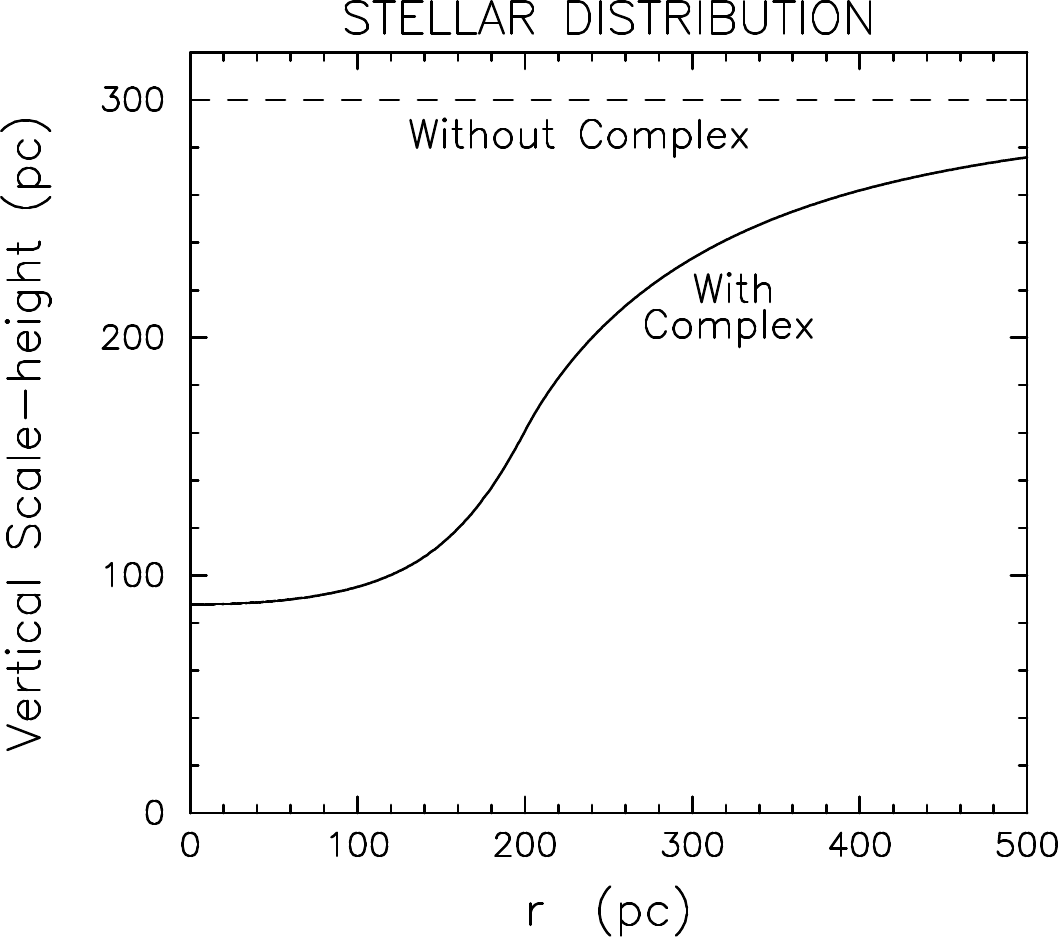} 
\bigskip
\caption{Plot of vertical scale height (HWHM) for the modified stellar distribution for stars in the field of the cloud complex vs. $r$, the radial distance from the centre of the cloud complex (solid line); and the constant scale height for the undisturbed stellar disk (dashed line). A striking result is that the vertical constraining by the complex is significant over a large radial distance $\sim 500$ pc from its centre; this is due to the extended mass distribution within a massive complex.
$\: $  {\it Source}: Taken from \cite{complex}
} 
 \label{fig.4}
\end{figure}

Repeating this procedure at different radial distances, $r$, from the centre of the complex (in the galactic plane) gives the resulting modified vertical scale height (HWHM) at each $r$. This is plotted as a function of $r$, (see Fig. \ref{fig.4}, solid line); also plotted is the constant scale height for the undisturbed stars-alone case (Fig. \ref{fig.4}, dashed line).  
The HWHM of the stellar distribution decreases by a substantial  factor of $\sim 3.4$, at $r=0$ or the central location of the complex, as already shown in Fig. \ref{fig.3}.
Another striking result is  
that the trough representing the scale height decrease is broad;  
so that the scale height reverts back asymptotically to its original undisturbed value of 299 pc only beyond 1 kpc (not shown in Fig. \ref {fig.4}). Thus the resulting scale height distribution of stars is locally corrugated. The vertical distribution of HI gas in the galactic disk also shows a similar constraining
effect due to the gravitational field of the complex.

Thus, a typical dense, massive and extended molecular cloud complex in a galactic disk extends a considerable constraining effect on the vertical distribution of stars and HI gas for a sizeable distance ($\sim 500$ pc) within its neighbourhood (see Fig. \ref{fig.4}). This result suggested that the constraining effect due to the gravitational force of the average gas distribution in the disk, and the dark matter halo, on the stellar disk could also be important. This was the theoretical motivation for proposing the multi-component disk plus halo model  \cite{NJ2002}, which will be discussed extensively in the next section.

Due to the superposition of effect of several such complexes, the gravitational potential in the disk would be distinctly non-uniform. 
This would be of interest for general galactic disk dynamics, such as for heating of stars (see \cite{complex} for details). Such complexes are seen in other galaxies as well. Hence, this topic needs to be explored further.

Recently,  \cite{Meera2024},  analyzed the high resolution archival HI data in the Galaxy from the HI4PI Collaboration \cite{Bekhti2014}; and clearly detected this constraining effect due to a molecular cloud complex in the Galaxy. This effect has been detected around the molecular cloud complex region W43-W44. This complex is located serendipitously near the  tangent point region between the Galactic longitude of $\sim 20^0 - 40^0$; so that the effect of the tangent point distance ambiguity is minimal. 
In a real galaxy, one does not see the effect of an isolated complex. Rather, the constraining effect of a set of cloud complexes of different masses, located at different locations ($R, \phi$),  
would be seen in superposition. This makes the analysis and the identification of the constraining effect due to a single complex (as predicted theoretically by \cite{complex}; see Fig. \ref{fig.4} in this Section) more challenging. However, this detection has been  achieved by taking cuts in velocity space along a given line of sight (see \cite{Meera2024} for details).

\section{Multi-component disk plus halo model\label{sec:4}}

In this section, we describe the modern theoretical study of the vertical structure of a realistic galactic disk. The disk is treated as a gravitationally coupled  multi-component disk system in the gravitational field of the dark matter halo. Each disk component is assumed to be isothermal with a different velocity dispersion. 
We first consider the simple case of a thin disk (Section \ref{sec:4.2}); followed by the thick or a low density disk case (Section \ref{sec:4.3}). Later, we consider a more general treatment that includes various kinematic effects,  and the effect of non-isothermal velocity dispersion (Section \ref{sec:4.4}).
The aim is to start with simplifying, physically motivated assumptions such as that of a thin disk; and then to
consider progressively more detailed physics in the problem, and thus to study more realistic cases. These cases were motivated by possible applications to different regions and types of galaxies.

\subsection{Motivation and basic outline of the model\label{sec:4.1}}

\subsubsection{Motivation\label{sec:4.1.1}}

The initial studies \cite{vdkS1981a,vdkS1981b} showed that the intensity profiles ($I$ vs $z$) did overall agree with a sech$^2$ distribution expected  for a self-gravitating, isothermal disk (\cite{Spitzer1942}, Section \ref{sec:3.1}). However, 
the I-band and R-band observations in 1980's began to reveal that 
the observed intensity profiles (e.g., \cite{vdk1988,BD1994})
 are generally steeper, with an excess near the mid-plane; compared to the sech$^{2}$ distribution that is expected for an isothermal, single-component disk.      The observed profiles could instead be better fitted by either a sech or an exponential function.
 A phenomenological model for an isothermal, one-component disk was proposed by  \cite{vdk1988} to explain this, but it was somewhat ad hoc and did not have a clear physical basis. 

 In fact, since a  thin, isothermal, single-component stellar disk can only result in a sech$^2$ profile (Eq. (\ref{eq3.4})); it is obvious that some additional physics has to be invoked in the theoretical analysis to explain the observed steeper profiles. This point has not been clearly recognized in the literature. It turns out that the theoretical model proposed on physical grounds \cite{NJ2002}, as discussed next, naturally results in 
steeper profiles. This will be discussed in detail in this section. 

To theoretically understand the disk vertical structure, a physically motivated model was proposed and 
developed by Narayan \& Jog (2002) \cite{NJ2002}. This model considers the vertical structure of a realistic multi-component galactic disk which includes gas gravity; and treats the stars and gas on an equal footing. The stars and gas are gravitationally coupled, and the disk is taken to be in the gravitational field of the dark matter halo. 
The aim of the model is to obtain the self-consistent vertical density distribution of each disk component in terms of the physical parameters, such as the surface density and velocity dispersion of the disk components.  
The additional gravitational force between the coupled disk components, and the effect of the halo,
naturally results in a vertically constrained, and hence a steeper density profile for each disk component, 
as we will discuss in this section.

The idea for the above model was suggested from the previous theoretical work by  
\cite{complex}. That work showed that a typical molecular cloud complex in the Milky way  significantly constrains the vertical stellar
distribution around it (see Section \ref{sec:3.3}).
It was then natural to ask how a uniform distribution of co-spatial stars and gas that interact gravitationally would affect the vertical distribution of each other. 
The other motivation to include the gas self-gravity was to see if this could explain
the puzzling near-constancy of  gas scale height observed in the inner Galaxy, which could not be explained by treating the gas as massless test particles responding to the stellar potential as done in the past (see Section \ref{sec:3.2}).

The work by  \cite{NJ2002} is the first fully self-consistent treatment in the literature to study the disk vertical structure by considering a gravitationally coupled disk in the field of the dark matter halo. This model aims to give a physically complete picture involving all the disk components and the dark matter halo; and the treatment is rigorous.
In the past,  \cite{Bahcall1984paper1} had  formulated the problem in a self-consistent fashion but the focus of the work was not on solving for the self-consistent density profile as shown here. Instead, the observed stellar density distribution (in terms of star counts data) was used as input to deduce the total mid-plane density, or the Oort limit (see Section \ref{sec:3.1.1} for details). Further, the halo was used as a perturbation in that work \cite{Bahcall1984paper1}. The importance of including gas gravity to study the gas vertical structure was pointed out by \cite{vdk1988,Olling1995}; but the effect of gas on stars was not considered, and the problem was not solved in a self-consistent way.

The multi-component disk plus halo model proposed by  \cite{NJ2002} was  subsequently investigated 
in a systematic way over the years in a series of papers. These include various detailed physics points -- as mentioned in the first paragraph in this Section. 
These cases will be reviewed methodically in this section.   
Indeed, the focus of this review is to present this model representing a realistic galactic disk and the results from it.

As we will discuss, this model explains quite well the observed disk thickness for stars and gas in the Milky Way; and the modern observational trends. These include  a departure from the sech$^2$ law, indicating a steeper profile; and wings at high $z$. These results arise naturally from this physically motivated model \cite{NJ2002,Kalberla2003,SJ2018}. 
Further, the results for $\rho(z)$ from this model can now be directly compared with the observed data for star counts in the Milky Way (see Section \ref{sec:8} for details). Interestingly, in a different but complementary approach, the results from this model could also be directly compared with the density profile values, $\rho(z)$, obtained from N-body simulations.

\subsubsection{Basic outline of the  model\label{sec:4.1.2}}

The following treatment for the multi-component disk plus dark matter halo model is from \cite{NJ2002},
who first formulated and studied this approach in detail. The treatment is general and valid for any galactic disk, although this was applied by them in the context of the Milky Way. The results from this model obtained in Section \ref{sec:4.2} to Section \ref{sec:4.6}, are generic, and the trends obtained are valid for external galaxies as well.

In this approach the stars, the interstellar atomic hydrogen gas (HI), and the interstellar molecular gas (H$_2$) are treated as three separate disk components. For simplicity, these are assumed to be concentric, co-spatial and coplanar w.r.t the galactic mid-plane ($z=0$). The steady-state disk distribution is taken to be symmetric w.r.t. the mid-plane ($z=0$). The disk is embedded within an extended dark matter halo which is taken to be concentric with the disk. 
The interstellar gas is taken to consist of two components: atomic hydrogen gas, $\mathrm {HI}$; and molecular hydrogen gas, H$_2$. This is necessary as these are known to have different velocity dispersions and surface densities; also, the values of these parameters as well as the fraction of gas in HI or H$_2$ form varies with radius (see Section \ref{sec:2}). Hence for a realistic representation of the Galactic disk, it is essential to formulate it as a three-component disk. 
The self-gravity of each disk component is taken into account. Thus, the Galactic disk is treated as a gravitationally coupled, three-component system (consisting of stars, interstellar HI and H$_2$ gas),
which is in the gravitational field of the dark matter halo.

\begin{figure}
   \centering
   \includegraphics[height=3.0 in, width= 5.0in]{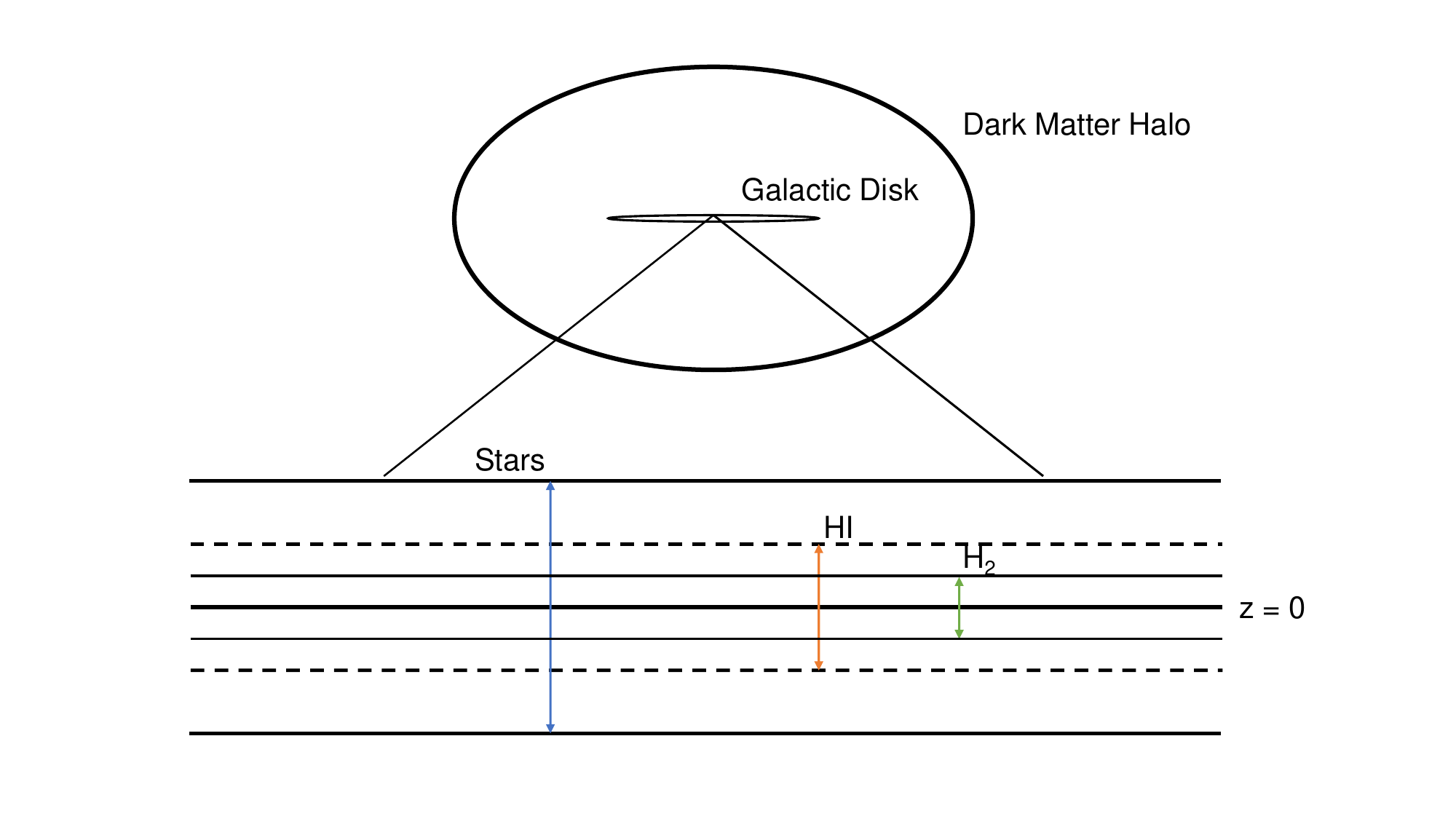}
   \caption{Schematic diagram of the  distribution of the three disk components: namely, stars, interstellar atomic hydrogen gas (HI) and molecular hydrogen gas (H$_2$), in a typical galactic disk.
The radial extent of the disk is much 
larger than its vertical thickness (not drawn to scale). The three disk 
components are taken to be axisymmetric, concentric, coplanar, and symmetrically distributed w.r.t. 
the disk mid-plane ($z=0$). The stellar scale height is higher than the HI scale height, 
which in turn is higher than the H$_2$ scale height, as observed.
These relative values are set by their observed velocity dispersion values. The disk is taken to be embedded  within an extended, massive dark matter halo which is concentric with the disk; the size and shape of the halo are not yet well-understood. (The relative sizes of the disk and halo shown are not to scale. The shape of halo is assumed to be oblate for the sake of illustration).
}
        \label{fig.5}
\end{figure}

The disk is assumed to be axisymmetric. Each disk component is assumed to be in a hydrostatic equilibrium along the vertical direction. That is, for each disk component, its internal pressure gradient balances the net gravitational force due to all the disk components and the halo, along the $z$ direction.
Each disk component is taken to be isothermal and is characterized by a different velocity dispersion. 
Due to its lower dispersion,  gas forms a thinner layer along the $z$ direction and thus the gas layers are embedded within the stellar disk of a higher thickness. Further, the molecular gas, H$_2$ is embedded within the HI gas. 
See Fig. \ref{fig.5} for a schematic diagram of the distribution of the  three disk components: namely,  stars, HI and H$_2$, embedded within the dark matter halo.

The self-consistent vertical density distribution for stars and gas  
in the joint potential of the multi-component disk and the halo is obtained. 
The additional gravitational force due to the other disk components and the halo in the coupled case
is shown to modify the vertical density distribution of each disk component, constraining it to be closer to the disk mid-plane. 
Since the interstellar gas forms a thinner layer than the stars; therefore, despite its low mass fraction in the disk, it is shown to exert a strong gravitational force  on the stars  near the mid-plane. This significantly affects the vertical density distribution of stars, as discussed later in this section.

A general point to note is that, in this model, the two gas components are distinguished from each other, and from the stellar component, only in terms of their respective vertical velocity dispersion. That is, this model does not take account of the dissipational nature of the gas components.

\subsection{Multi-component disk plus halo model: Thin disk\label{sec:4.2}}

\subsubsection{Formulation of equations: Thin disk\label{sec:4.2.1}}

The equations describing the vertical structure for the thin disk case, under  the general assumptions (described above in the basic outline of the model), were formulated by  \cite{NJ2002}, 
and are given next. The steady-state, self-consistent vertical density distribution, $\rho(z)$, for each disk component is obtained by solving together the equations for the hydrostatic balance for each disk component and the joint Poisson equation for the disk-halo system for the thin disk case.

It is necessary to take account of the effect of the dark matter halo in the study of the disk vertical structure. The inclusion of the halo complicates the formulation of the problem but also makes it richer.  
We note an important general point that for two different gravitating structural components such as the disk and the halo, the density, $\rho$, and the potential, $\Phi$, for each satisfies its own Poisson equation locally and a linear superposition describes the net system. This will be used in writing the joint Poisson equation (Eq. (\ref{eq4.2})) later.

The force equation or the Euler equation or the equation of hydrostatic equilibrium along the $z$-axis  for each disk component is given as follows, generalizing the one-component case (for the latter, see \cite{Rohlfsdensity}; or, Section \ref{sec:3.1}):

\begin{equation}
 \mathrm {\frac {\langle{({\mathit v}_z^2)_i}\rangle}{\rho_i} \: \frac {\partial {\rho_i}}
   {\partial{\mathit z}} \: = \: {({\mathit K}_z)_s} \: + \: {({\mathit K}_z)_{HI}} 
   \: + \: {({\mathit K}_z)_{H_2}} \: + \: {({\mathit K}_z)_{h}} }  
      \label{eq4.1}
      \tag{4.1}
\end{equation}      

\noindent where $\rho$ is the mass
density,  and   $ \mathrm{\langle{({\mathit v}_z^2)}\rangle }^{1/2} = \sigma_z    $ 
is the velocity
dispersion  of a disk component along $z$ at a given radius $R$, and
 $(K_{\mathrm z}) \: = \: -  {{\partial}{\Phi}}/ {{\partial}z} $ is 
the force per unit mass along the $z$-axis, where $\Phi$ is the corresponding
potential. The subscript {\it i} = s, HI, and H$_2$ denotes these 
quantities for stars, HI and H$_2$ respectively.
The last term on
the right hand side denotes the force per unit mass along the $z$ direction due to the dark 
matter  halo. The right hand side of 
Eq.(\ref{eq4.1})  denotes the total vertical force due to all the components. This can modify or redistribute the density ($\rho$) and the potential ($\Phi$) of each particular disk component.

 Thus each disk component feels the gravitational force due to all the disk components and the halo.
 This includes gas being affected by stars, even though the stellar disk thickness is higher than that of the gas.
This is possible because of the open geometry of the disk distribution (see Appendix A in \cite{JogSolomon1984}). 

Since the disk is thin and much less massive compared to the halo, its effect on the vertical distribution within the halo is neglected, so the halo is taken to remain undisturbed by the disk. Thus, the halo is treated as a reservoir that just provides an external gravitational field acting on the disk. 

The joint Poisson equation for the above system is considered next. For a thin, axisymmetric disk, only the $z$ component of the Laplacian for the disk components (stars and gas) needs to be retained in the Poisson equation
 for the disk (see Eq. (2.57) from \cite{BT1987}; also, see Section \ref{sec:3.1}).

The joint Poisson equation for the multi-component disk plus halo in the thin disk case is then given by:

\begin{equation}
\begin{split}
\frac {{\partial}^2{\Phi_s}}{\partial{\mathit z^2}} \: + \frac {{\partial}^2{\Phi_{HI}}}
   {\partial{\mathit z^2}} \: + \frac {{\partial}^2{\Phi_{H_2}}}{\partial{\mathit z^2}}
   & + 
\frac{1}{R}\frac{\partial}{\partial R}\left(R\frac{\partial \Phi_{\mathrm{h}}}{\partial R}\right) 
  \:+ 
\frac {{\partial}^2{\Phi_{h}}}
          {\partial{\mathit z^2}}  \\
  & = \:  4 \pi {\mathit G} \left ({\rho}_s \: + \: {\rho}_{HI} \: + \: 
   {\rho}_{H_2} \: +
   {\rho}_{h} \: \right ) 
   \end{split}
    \label{eq4.2}
    \tag{4.2}
\end{equation}

Since the halo is assumed to be not affected by the disk, the above equation reduces to:

\begin{equation}
 {\mathrm {\frac {{\partial}^2{\Phi_s}}{\partial{\mathit z^2}} \: + \frac {{\partial}^2{\Phi_{HI}}}
   {\partial{\mathit z^2}} \: + \frac {{\partial}^2{\Phi_{H_2}}}{\partial{\mathit z^2}}
   \: = \:  4 \pi {\mathit G} \left ({\rho}_s \: + \: {\rho}_{HI} \: + \: 
   {\rho}_{H_2} \right )}} 
   \label{eq4.3}
   \tag{4.3}
\end{equation}   

Thus, in a thin disk, the disk Poisson equation and the halo Poisson equation are effectively decoupled. However, the disk and halo are still coupled gravitationally, because of the coupling between a disk component with the other disk components and the halo  through the equation of hydrostatic balance (Eq. (\ref{eq4.1})). This leads to a redistribution of the vertical density profile for each disk component.
In the above equations (Eq. (\ref{eq4.1}), and Eq. (\ref{eq4.3})), $\rho_i(z)$ denotes the re-distributed or modified density for each disk component in the  multi-component disk plus halo case.

 From Eq. (\ref{eq4.3}) it can be seen that the force and the resulting density distribution are functions of $z$ alone, analogous to the one-component case considered in Section \ref{sec:3.1}. Hence the value of $\rho_i(z)$ depends only on the local conditions at the given radius. Thus, the problem reduces to a one-dimensional, and hence a local, problem of determining the density distribution, $\rho_i(z)$, at a given radius. Hence, the vertical density distribution of a given disk component, is only affected by the local gravitational force of itself, the other disk components, and the halo at a given $R$ (which only depend on the local values of their parameters).
Since the molecular gas is mainly confined to the inner disk, it affects the stellar vertical density distribution in the inner disk; whereas the effect of HI on the vertical stellar distribution is seen in the outer disk as well  (see Section \ref{sec:4.2.2} to Section \ref{sec:4.2.6}).

Combining Eq.(\ref{eq4.1}) and Eq.(\ref{eq4.3}) gives the following joint Poisson-hydrostatic balance equation for the coupled case, which governs the 
self-consistent vertical density distribution, $\rho(z)$, of a disk component at a given radius:

\begin{equation}
 {\mathrm {\frac {{\partial}^2{\rho_i}}{\partial{\mathit z^2}} \: = \: \frac {\rho_i}
   {\langle({\mathit v}_z^2)_i \rangle} \: \left [ - 4 \pi {\mathit G} \: 
   ({\rho_s} + 
   {\rho_{HI}} + {\rho_{H_2}}) + \frac {\partial({\mathit K}_z)_{h}}{\partial{\mathit z}} 
   \right ] \: + \: \frac {1}{\rho_i} \: \left ( \frac {\partial{\rho_i}}{\partial{\mathit z}} 
   \right )^{2} }} 
   \label{eq4.4}
   \tag{4.4}
\end{equation}   
   
\noindent where the square brackets contain the terms that arise due to the joint potential of the three disk components and the halo in the coupled system. Even though this is the same for all the disk components; its effect on the resulting density distribution of each disk component is different due to the different vertical velocity dispersion for each disk component, as shown later in this section.

\noindent {\bf  General comments}

1. It is easy to see that when there is only one gas component present, we can set $\Sigma = 0$ for the other gas component and then the above equation reduces to that for a two-component (stars plus gas) disk. 

2.  \cite{NJ2002} treated stars to be a single component for simplicity, as is routinely done in the study of stellar disk structure in the literature (e.g., \cite{vdkS1981a}). In the literature, the sech$^2$ result (Eq.(\ref{eq3.4}))  is routinely applied to a one-component stellar disk
(e.g., \cite{vdkS1981a}), so that all the stars of different types are treated in a cumulative fashion.
This is a reasonable assumption since most of the disk mass is in stars of type G-K-M  \cite{BJ2007}.
 It would be worth treating a multi-component stellar disk if the values of the surface density
 and velocity dispersion for each such stellar sub-component are known observationally. In that case, it would be straightforward to extend the above analysis to include a multi-component stellar disk (see Section \ref{sec:8} where this is listed as a possible future problem).

\medskip

\subsubsection{Numerical solutions\label{sec:4.2.2}} 

The three coupled second-order differential equations (represented by Eq. (\ref{eq4.4}) are solved simultaneously to obtain the self-consistent vertical density distribution of each disk component. Each second order differential equation is split into two first order differential equations for the sake of simplicity. These are solved numerically as an initial value problem, using the fourth-order Runge-Kutta method of integration \cite{NumRec1986}. For each disk component, the boundary conditions applied at the mid-plane, $z=0$ are: 

\begin{equation}
{\mathrm \rho_i = (\rho_{0d})_i ;   \: \:  and \: \:  \:  \frac {d\rho_i}{d {\mathit z}} = 0 } 
  \label{eq4.5}
   \tag{4.5}
\end{equation}

For a realistic distribution close to the mid-plane, the density distribution along the $z$ axis is homogeneous close to the mid-plane, hence $d\rho_i/dz=0$. Now, $(\rho_{0d})_i$, the mid-plane density  of the re-distributed matter for a given disk component $i$ is not known a priori. However, the observed surface density $\Sigma_i (R)$ at a given $R$ is known,
and can be used to constrain the value of $(\rho_{0d})_i$ as explained next. The surface density is twice the area under the curve $\rho_i(z)$ vs. $z$. Given the observed value of $\Sigma_i(R)$, the value of the mid-plane density, $(\rho_{0d})_i$ can be found by trial and error. Once this is obtained, the solution for the vertical density distribution, $\rho_i(z)$, is determined easily. In retrospect, a similar approach was taken to determine the mid-plane density, and hence the solution for $\rho(z)$, for one-component case (\cite{Rohlfsdensity}; also, see Section \ref{sec:3.1}).

As seen above (in Section \ref{sec:4.2.1}), here the distribution of matter can be treated as a one-dimensional, local problem along the z-axis. Hence the surface density, $\Sigma_i(R)$ for each disk component in the coupled case will remain unchanged even when the joint gravitational potential is considered.
This is an important point and this occurs because the gravitational coupling between the disk components causes a redistribution of matter within a component only along $z$, but not along $R$. Hence, even in the coupled case, one can still use the local $\Sigma (R)$ as a constraint to obtain the corresponding re-distributed mid-plane density value, $(\rho_{0d})_i$ at a given point.

All the three disk components: namely, stars, HI and H$_2$; affect the density distribution of each other via the coupled equations (Eq. (\ref{eq4.4})). At each R, the solutions for the three density functions are obtained simultaneously by taking account of the effect
 of the other components in an iterative fashion. See \cite{NJ2002} for the details of this iteration procedure, where the effect of the three components is added successively.  
The above iteration cycle is repeated four times until each of the 
distribution converges with a fifth decimal accuracy.
An important feature of this scheme is that the above procedure yields the solution for the
self-consistent vertical density distribution for all three disk components (stars, HI and H$_2$) simultaneously.

The density distribution of each component will be affected by the joint gravitational potential, and would no longer be give by a sech$^2$ function as for a one-component case.
\cite{NJ2002}  point out that they obtain a modified,  sech$^2$-like distribution for each component. They define the HWHM (half-width-at-half-maximum) of the distribution to be the vertical scale height. This definition was also used in the earlier paper on the disk distribution affected by a molecular cloud complex \cite{complex}.
The HWHM of the resulting distribution is refereed to as the vertical scale height or disk thickness  in the subsequent papers that studied this model. This was a pragmatic choice, and gives the actual value of the thickness measured -- that is not related to any presupposed or ad hoc functional form. 
See Appendix A for further detailed  discussion of HWHM, and how it is  
a robust indicator of disk thickness. 

The above numerical calculation to obtain $\rho_i(z)$ was repeated at different $R$ values, to obtain  the modified or redistributed density distribution, $\rho_i(z)$, and hence the resulting scale height (HWHM) vs. $R$ for each disk component in the coupled case. 

\medskip

\subsubsection{Application to the Milky Way\label{sec:4.2.3}}

The above model was first formulated in  general terms and then applied to the Milky Way in the inner Galaxy $R< 12$ kpc) by 
 \cite{NJ2002}.  The scale heights are shown to be smaller for each disk component compared to the corresponding one-component values; due to the additional gravitational force in the coupled case. These results are in a good agreement with observations for all the three disk components studied, as discussed next.
This problem was studied in greater detail using the same model by  \cite{SJ2018}, who explored the various features of the re-distributed stellar density distribution in the coupled case. Further, they covered a larger radial range (4-22 kpc), with a particular focus on the stellar distribution in the outer disk. It was found that the gas plays the dominant role in constraining the disk vertically in the inner Galaxy; while the halo has the dominant constraining effect  in the outer Galaxy (beyond R=18 kpc), as will be discussed next.
The main results from these studies are summarized below. 

\medskip

\noindent {\bf Parameters used} 

The input parameters needed to solve the coupled equations (Eq. (\ref{eq4.4})) are the surface density and velocity dispersion as a function of radius for the stars, HI and H$_2$ gas. These values are taken from observations and are briefly given here, see \cite{NJ2002} for details of the observed values. 
The mass model for the Galaxy by \cite{Mera1998} was used as being modern and consistent with other observations at that time, such as the local surface density as given by \cite{Flynn1994}. On subtracting the total observed gas density \cite{SS1987}; the local stellar surface density at the solar neighbourhood, R= 8.5 kpc, is obtained as 45 M$_{\odot}$ pc$^{-2}$. An exponential stellar disk distribution, with a radial disk scale length, R$_D$ = 3.2 kpc is adopted from \cite{Mera1998}. This gives the central stellar surface density as 640.9 M$_{\odot}$ pc$^{-2}$. 

The observed stellar radial velocity dispersion values from R= 1 to 17 kpc in the Galaxy were measured by \cite{LewisF1989} (see Section \ref{sec:2.1.4}). These fall exponentially with radius with a scale length, R$_v$ = 8.7 kpc. Assuming the same ratio of the vertical to radial velocity dispersion as in the solar neighbourhood (= 0.45), gives the vertical stellar velocity dispersion values as a function of radius. Thus, the vertical velocity dispersion is also taken to fall exponentially with radius with the same scale length, R$_v$ = 8.7 kpc. 
\cite{SJ2018} assume the same rate of fall-off with radius in the outer Galaxy, until the calculated stellar velocity dispersion is $\sim$ close to the gas velocity dispersion. Beyond this radius, the stellar velocity dispersion is taken to saturate at this value. This physically motivated assumption is based on the fact that the stellar velocity dispersion cannot be smaller than the dispersion in the gas from which stars form. This occurs at R=17 kpc in the Milky Way, for the observed HI gas dispersion of  $\sim 7$ km s$^{-1}$, as discussed next. 
This point will be discussed more in Section \ref{sec:5}, along with the observational evidence for the saturation in the stellar velocity dispersion at large radii \cite{Sharma2021} -- which confirms the above assumption by \cite{SJ2018}.

The surface density values for HI and H$_2$ are taken from \cite{SS1987}. 
In the work reported here by \cite{NJ2002} and \cite{SJ2018}, the vertical HI gas velocity dispersion is taken to be 8 km s$^{-1}$ in the inner disk (up to $R$=12 kpc). This is based on the value given by \cite{Spitzer1978}, and as measured for $\sim 200$ face-on galaxies by \cite{Lewis1984}. The HI gas dispersion is observed to slowly fall and saturate to 7 % $\pm$ 
km s$^{-1}$ at large radii \cite{Kamphuis2008,Dickey1996}. Hence, the HI dispersion is assumed to fall  gradually beyond R=12 kpc at the rate of - 0.2 km s$^{-1}$ kpc$^{-1}$. This tapers off to a value slightly higher than 7 km s$^{-1}$ at $R$=17 kpc; and is assumed to be constant beyond this radius, see \cite{SJ2018}. 
The model scale height results are compared with the observed values. The observed scale height values for HI for $R$ $<$ 8.5 kpc are taken from \cite{Lockman1984};  and those beyond  $R >$ 8.5 kpc (corrected for the Galactic warp) are taken from \cite{Wouter1990}. 

The dark matter halo is taken to have a pseudo-isothermal density profile:

\begin{equation}
\mathrm {\rho(R,z) \: = \: \frac{ \rho_{0h}}{(1 + (R^2+z^2)/{R_c}^2) }}
    \label{eq4.6}
    \tag{4.6}
\end{equation}

\noindent following the mass model of \cite{Mera1998} that was adopted. Here $\rho_{0h}$ and $R_c$ are the halo  central density and the core radius respectively, whose values are $\rho_{0h}$ = 0.35 M$_{\odot}$ pc$^{-3}$, and $R_c$= 5 kpc.

\subsubsection{Results for vertical scale heights in the inner Galaxy\label{sec:4.2.4}}

The resulting vertical scale heights for the three disk components in the inner disk ($R< 12$ kpc) and a comparison with observations (from \cite{NJ2002}) are briefly discussed next. This radial range clearly illustrates the basic features of a multi-component disk model. Although the dark matter halo was included in this work in order to be comprehensive, it does not contribute much to the determination of the stellar distribution in the radial range considered. For example, even at $R$=12 kpc, the largest value of the radius which was considered in that work; the inclusion of the halo affects the HI scale height (HWHM) by only 12\%. 
The effect of the halo on stellar distribution becomes important, even dominant, at large radii \cite{SJ2018}, as discussed later in Section \ref{sec:4.2.5} and Section \ref{sec:4.2.6}.

\begin{figure}
\centering
\includegraphics[height=2.5in, width=2.67in]{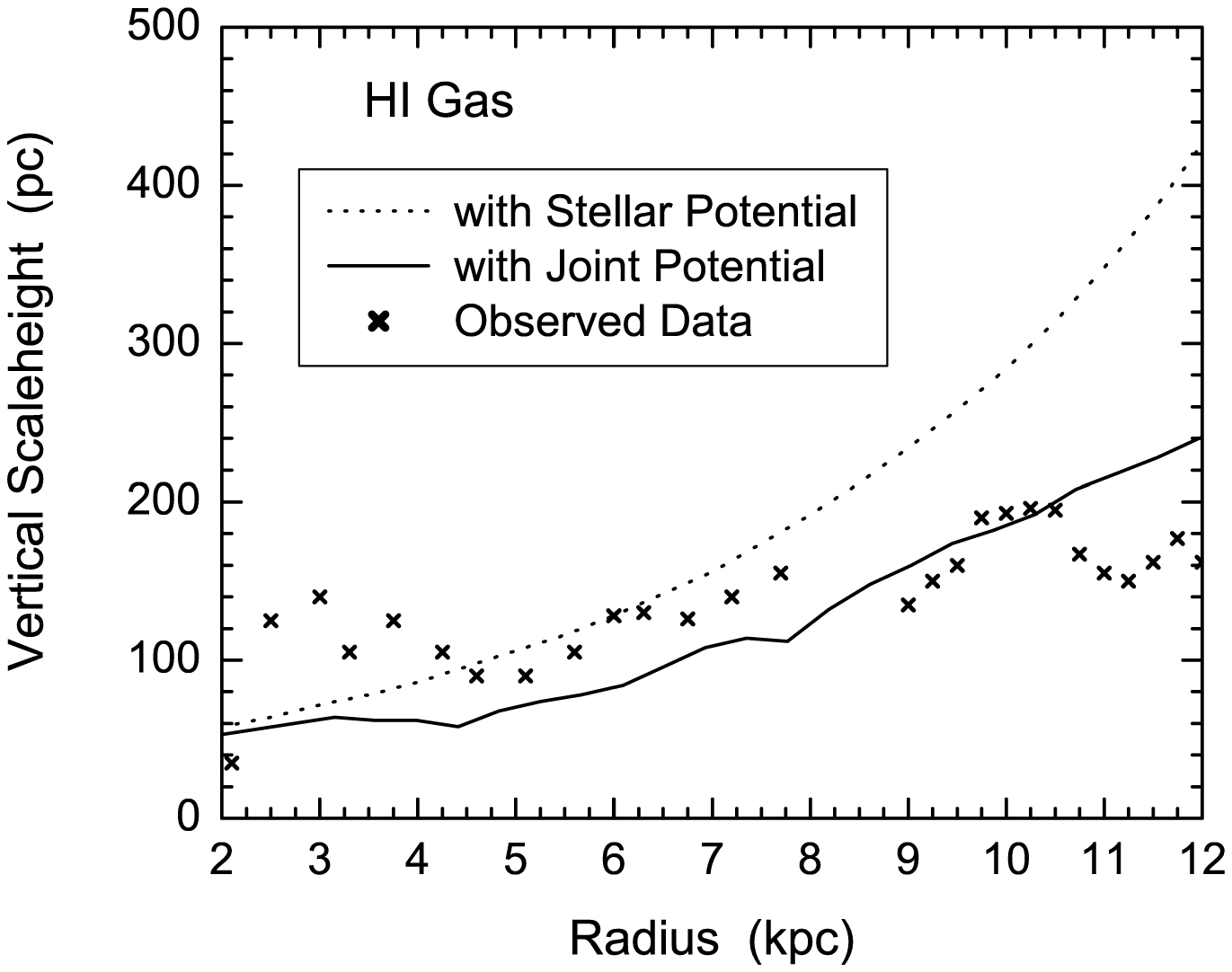}
\medskip 
\includegraphics[height=2.5in, width= 2.67in]{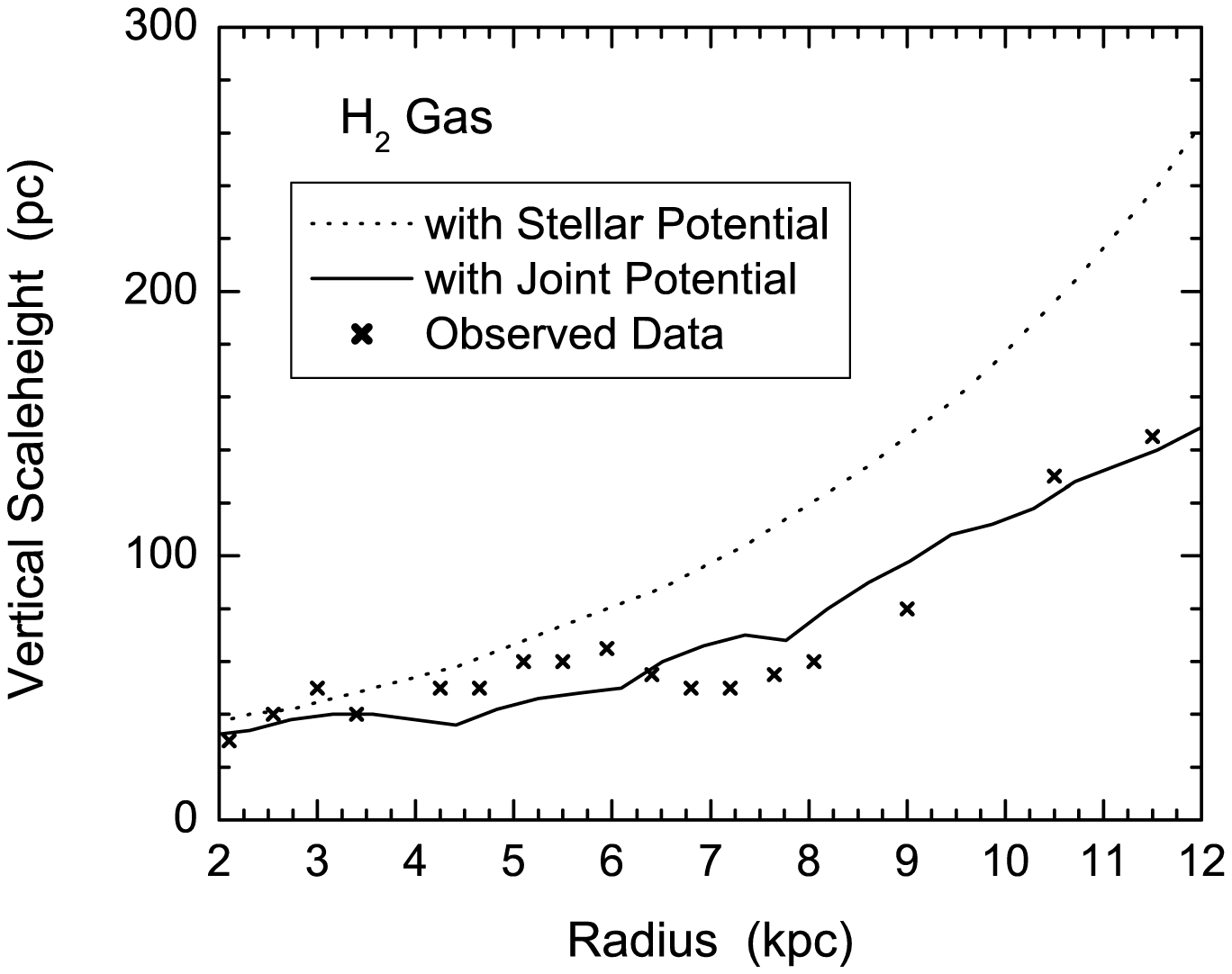} 
\bigskip
\includegraphics[height=2.5in, width=2.67in]{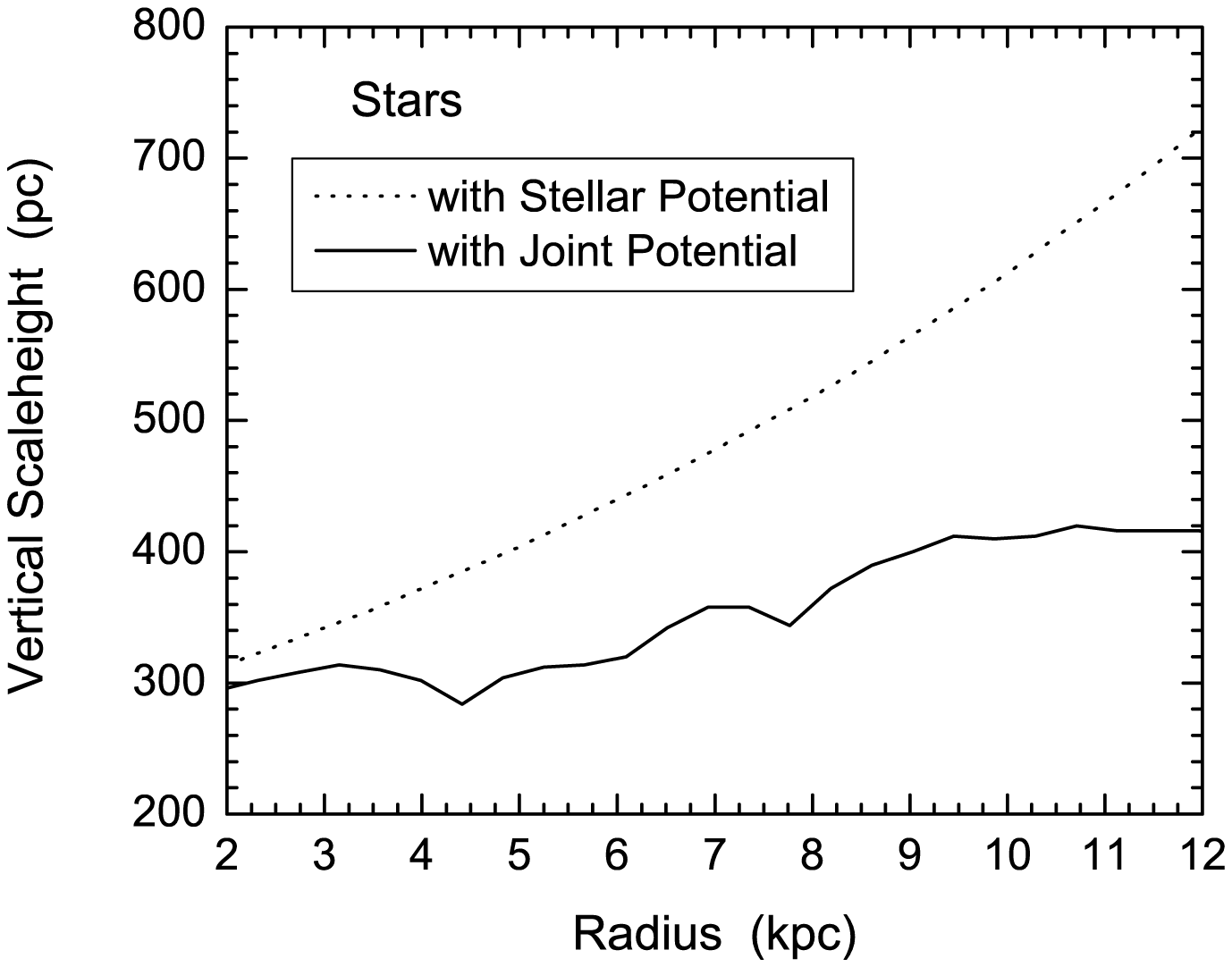}
\bigskip
\caption{The plot of scale height (HWHM) vs. R for interstellar HI, H$_2$ and stars vs. R, in the three panels respectively, at R=8.5 kpc. In each case, the
solid curve denotes the theoretical result obtained under the joint potential; and the dashed curve represents the theoretical result
obtained under the stellar potential alone, i.e.,
with no self-gravity in case of HI and H$_2$. For each disk component, the scale height values under the joint potential are significantly lower, compared to under the stellar potential alone - particularly at large radii; and show an overall better agreement with observations. In particular, the joint potential case explains the old puzzle \cite{1962dmim.conf....3O} of nearly constant HI scale height observed in the inner Galaxy. See Fig. \ref{fig.7} for an improved fit to the HI data.
$\: $  {\it Source}: Taken from \cite{NJ2002}
} 
 \label{fig.6}
\end{figure}

The resulting value of the HWHM for each disk component, responding to the joint potential in the coupled case,
is  plotted (as a solid line) in Fig. \ref{fig.6}, while the dashed line corresponds to the HWHM values when each component responds to the stellar potential alone (i.e., with no self-gravity in the case of HI and H$_2$). The three panels in
Fig. \ref{fig.6} contain results for HI, H$_2$ and stars, respectively. 

The most striking result is that for each disk component (stars, HI and H$_2$), the joint potential approach gives  a smaller scale-height at each radius than the value obtained using stellar potential alone. This effect is stronger at larger radii. Thus, including the gas self-gravity, and the  additional gravitational force due to gravitational coupling between the disk components, decreases the resulting scale heights
and brings them closer to the observed values. The detailed comparison is given next.

\begin{figure}
\centering
\includegraphics[height=2.5in, width=2.67in]{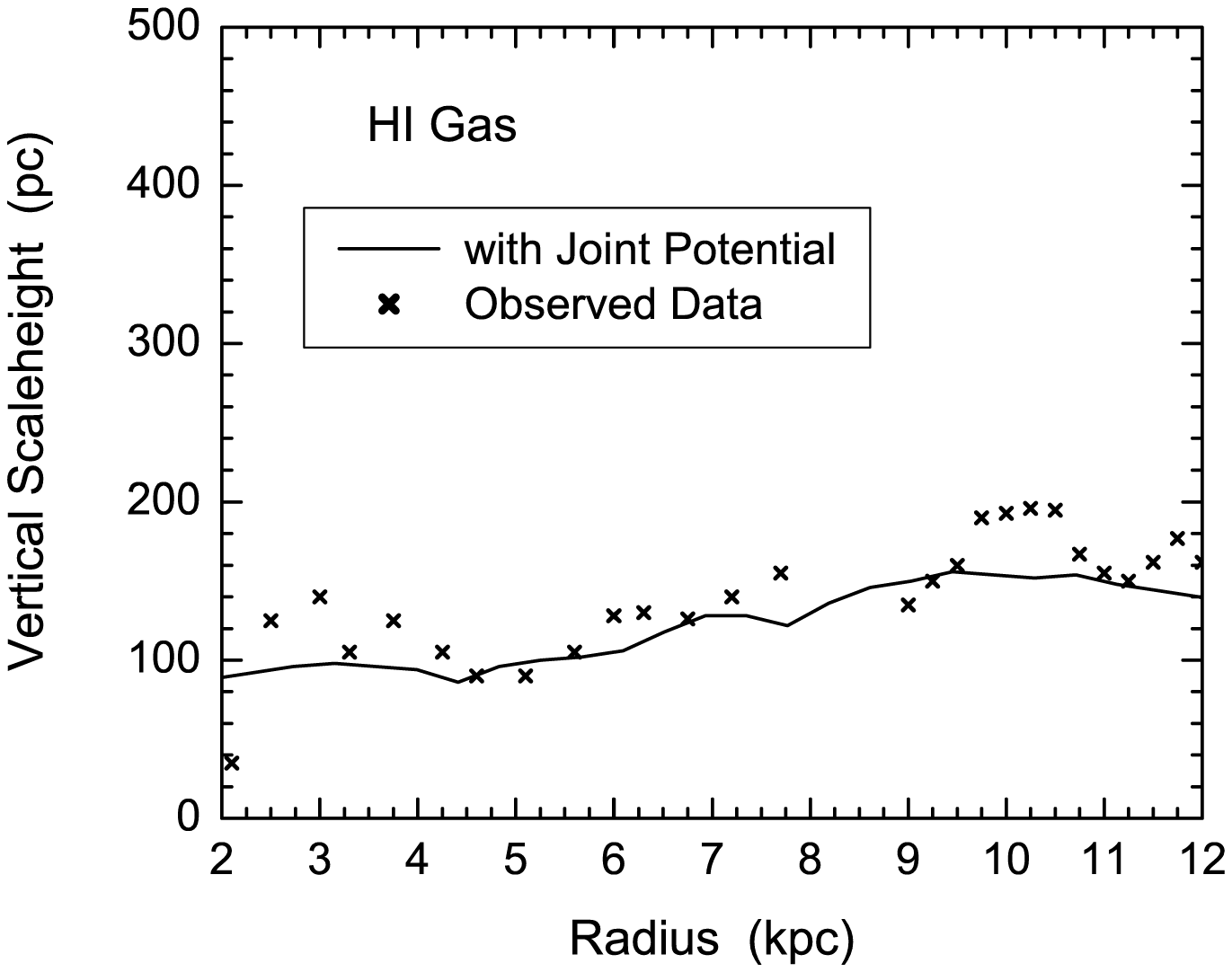}
\medskip 
\includegraphics[height=2.5in, width= 2.67in]{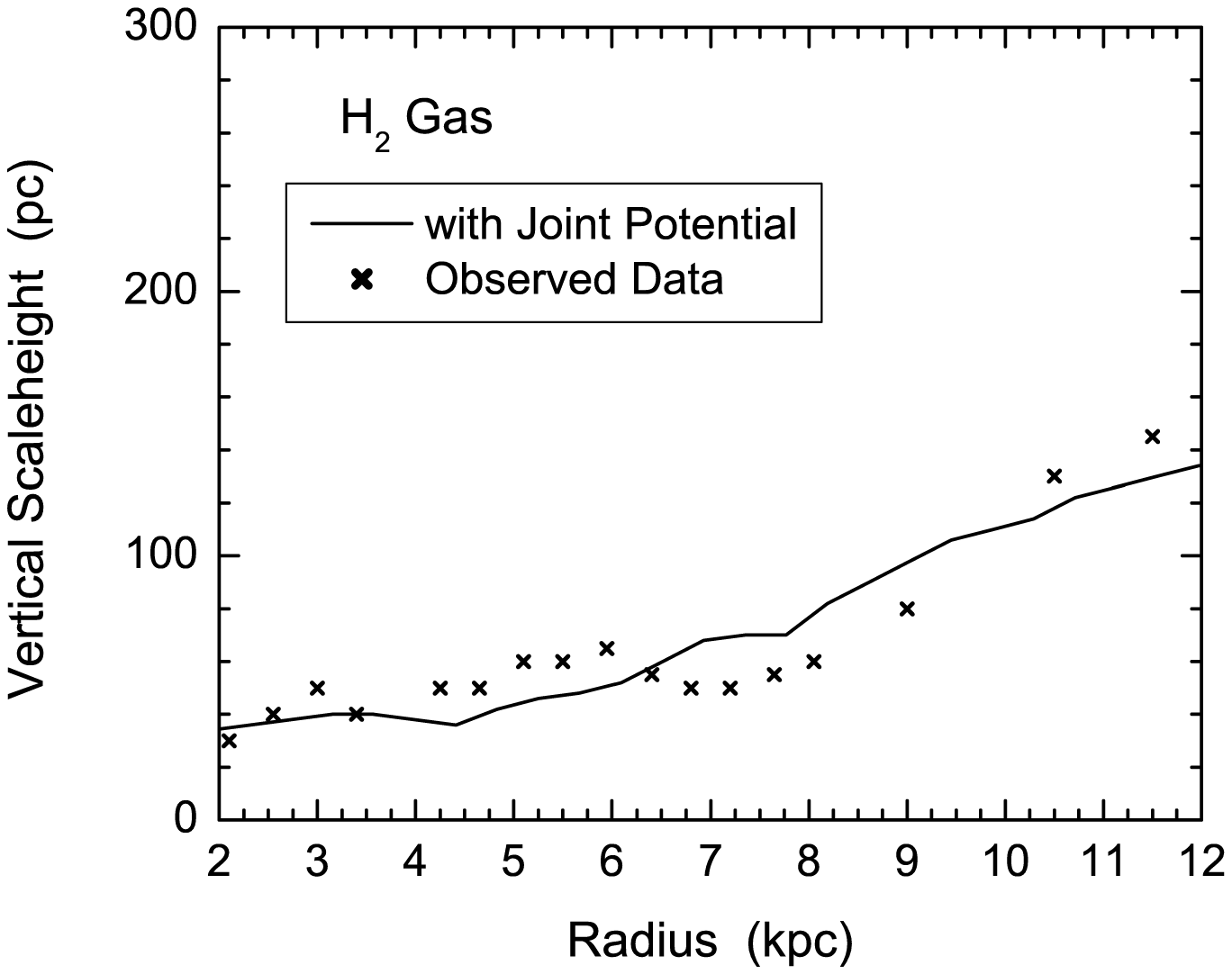} 
\bigskip
\includegraphics[height=2.5in, width=2.67in]{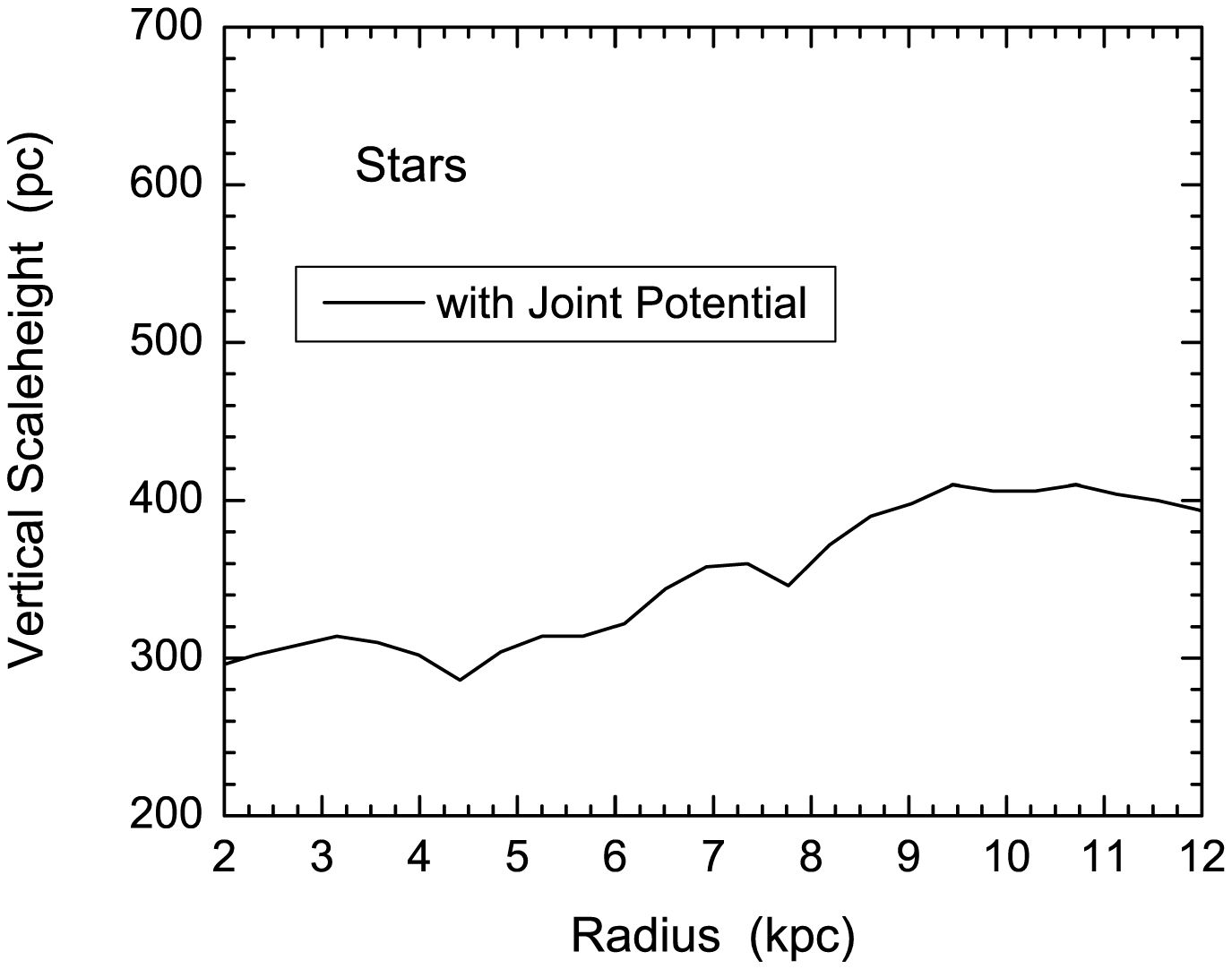}
\bigskip
\caption{The plot of scale height (HWHM) vs. R for HI, H$_2$ and stars vs. R in the three panels as shown, at R=8.5 kpc; where a radial gradient in HI velocity dispersion with a value of -0.8 km s$^{-1}$ kpc$^{-1}$ has been included in the calculations. The resulting HI scale heights are in good agreement with observations over the entire radial range studied, and the fit is better than in panel for HI of Fig. \ref{fig.6} . Interestingly, the values of scale heights for H$_2$ and stars remain nearly unchanged  on introducing the gradient in the HI velocity dispersion (compare Fig. \ref{fig.6}, panels for H$_2$ and stars with Fig. \ref{fig.7}, panels for H$_2$ and stars, respectively) -- see the text for the physical explanation.   $\: $  {\it Source}: Taken from \cite{NJ2002}} 
 \label{fig.7}
\end{figure}

Consider the HI case shown in Fig. \ref{fig.6} 
first. The dashed line representing the HWHM as a response to the stellar potential alone, increases exponentially with radius; and  thus deviates strongly from the observed values beyond 8 kpc. On using the joint potential, the scale height reduces significantly, especially at large radii; because the HI gas gravity becomes important. Hence, the model scale heights then show an overall agreement with observations, especially in the middle range of 5-10 kpc. Next, 
a small linear gradient of -0.8 km s$^{-1}$ kpc$^{-1}$ was tried, starting with a value of HI velocity dispersion of 8 km s$^{-1}$ at  $R$ = 8.5 kpc. The resulting scale height values vs. R for HI, H$_2$ and stars are given in the three panels in Fig \ref{fig.7}. 
This value of the velocity gradient was chosen since it was found  to give the best-fit between the model HWHM and the observed values for HI for the range 2-12 kpc (see the panel for HI in Fig. \ref{fig.7}). On comparing the panel for HI in Fig \ref{fig.7} with panel for HI in Fig. \ref{fig.6}, it is clear that including the HI velocity gradient 
gives a better overall agreement with the observed data.
A plausible explanation for this 
radially varying HI velocity dispersion (with a higher velocity dispersion at smaller radii) is the higher energy input due to the supernova input  \cite{MckeeOstriker1977}. This rate is expected to be higher in the inner region, where the molecular gas -- which forms the site of star formation -- is seen to dominate. Thus this model explains the old puzzle  of the nearly constant HI scale height in the inner Galaxy \cite{1962dmim.conf....3O}.

For H$_2$ as well, neglecting the gas gravity (dashed line) gives a large deviation from the observed values,  see panel for H$_2$ of Fig. \ref{fig.6}.
On including the gas gravity, and using the joint potential  in the gravitationally coupled case, the theoretical results for HWHM (solid curve) agree very well with the observed values, see \cite{SS1987}.
Thus, this model gives a physical explanation for the vertical scale height distribution of the scale heights of the H$_2$ gas in the Galaxy, which had not been studied before  in the literature.

For stars, the stars-alone potential gives an exponentially increasing curve for HWHM versus $R$ (solid line) (see the panel for stars, Fig. \ref{fig.6}). In contrast,  the stellar scale height curve in the joint case shows lower values, with a nearly flat behaviour (with a constant value of HWHM of $\sim 300$ pc) up to $R=$ 5 kpc. Between R= 5-10 kpc, it shows a moderate increase with the best-fit gradient of 24 pc kpc$^{-1}$.  Beyond 10 kpc, the curve corresponding to the joint potential saturates at 480 pc. Thus, the additional gravitational force due to gas and  the dark matter halo in the joint system  decreases the amount of flaring in the stellar disk to a moderate value. For example, at R= 8.5 kpc, the stellar potential alone gives a stellar scale height of 550 pc, but the inclusion of gas gravity reduces it to 380 pc. This reduced value is in a good agreement with directly observationally determined value of 350 pc for the Milky Way (\cite{BT1987}; also, see Appendix A).

The variation of stellar scale height with radius cannot be compared with optical observations in our Galaxy due to dust extinction. However, the observed near-IR data available at that time, namely from {\it Spacelab2} also showed a moderate increase in scale height with radius with the best-fit value of 20 pc kpc$^{-1}$ in this radial range \cite{Kent1991}. This trend is in a good agreement with the model results for the coupled case, as given above. 

It is interesting that 
the stellar disk is not strictly flat as has claimed by \cite{vdkS1981a}; rather, it shows a small but finite increase in scale height within the optical disk.
 Such a moderate flaring of the stellar disk is a generic result, and will be discussed further in Section \ref{sec:5}. 

 This approach, studied using realistic input parameters, cohesively and naturally explains the observed scale height distributions of all the three disk components: namely, stars, HI and H$_2$, in the inner region (R=2-12 kpc). This was the success of this model. 

An interesting result to note is that when the velocity gradient in HI velocity dispersion as given above is used, the corresponding resulting H$_2$ and stellar scale height distributions respectively in the coupled case do not show any noticeable difference (compare panels for H$_2$ and stars in Fig. \ref{fig.6} with the corresponding ones in Fig. \ref{fig.7} respectively).  
That is, the change in velocity dispersion of HI gas affects the HI distribution but has little effect on that of the H$_2$ or stellar disk.
This is because the Jeans equation describing the pressure-equilibrium, or the equation of hydrostatic equilibrium, of a given component only depends on its own velocity dispersion (Eq.(\ref{eq4.1})). Hence, the velocity dispersion of each component only directly affects its own scale height (see Eq.(\ref{eq4.4})). 
Therefore, a change in the velocity dispersion of HI only indirectly affects the density distribution of other coupled disk components, through the change in the density distribution of HI whose velocity dispersion has been changed (see Eq.(\ref{eq4.4})).
This is an important physical point and it comes about because the pressure support in the vertical Jeans equation only depends on the vertical velocity of that particular component, whereas each disk component feels the joint gravitational force due to all the coupled disk components. This is analogous to the study of stability of linear, axisymmetric planar perturbations in a two-fluid (stars plus gas disk), where the support in the radial Euler equation for a component
only depends on the pressure term due to itself, but each component feels the net gravitational force due to both the  disk
components \cite{JogSolomon1984}. 

\subsubsection{Physical understanding of re-distribution of stellar density\label{sec:4.2.5}} 

The results in Section \ref{sec:4.2.4}  show a decrease in the vertical scale height (HWHM) for each disk component in a joint potential compared to that in the stellar potential. The physics behind this  is illustrated clearly in the study by  \cite{BJ2007}. They considered a three component (stars, HI and H$_2$ gas) disk where the focus was to show
the dynamical effect of the low dispersion component, namely gas, on the stellar vertical density distribution in the coupled case.

\begin{figure}
\centering
\includegraphics[height=2.65in, width= 3.1in]{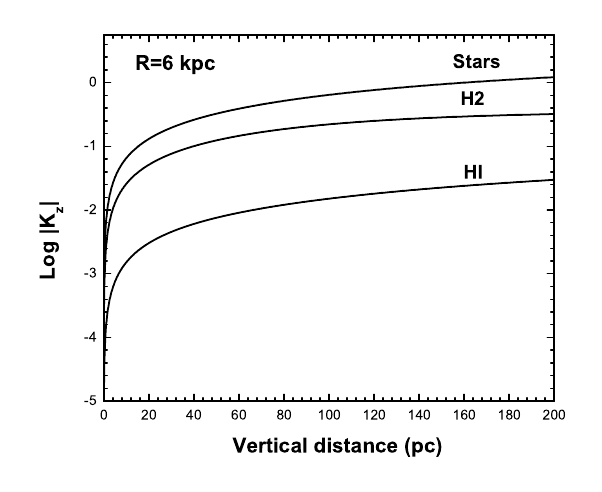} 
\bigskip
\caption{Plot of vertical self-gravitational force per unit mass, $\mathrm{|K_z|}$: for stars-alone; H$_2$-alone; and HI-alone cases; drawn on  a log scale vs. $z$, at R=6 kpc. The force due to H$_2$ gas is a significant fraction, $\sim 30 \%$, of the force for the stars-alone case; for $z$ values close to the Galactic mid-plane ($|z| < 150$ pc).
$\: $  {\it Source}: Taken from \cite{BJ2007}
} 
 \label{fig.8}
\end{figure}

First, consider each isothermal disk component separately, with its 
density distribution, $\rho(z)$, and the corresponding force per unit mass, $K_z$, as given respectively  by Eq. (5) and Eq. (6) from  \cite{BJ2007}; or, see Eq.(\ref{eq3.3}), and Eq.(\ref{eq3.5}).
Fig. \ref{fig.8}  gives a plot of $|K_z|$ vs. z shown on a logarithmic scale for stars, HI and H$_2$ at R=6 kpc. This shows that their magnitudes are comparable; and, in fact, up to $|z| < 150$ pc, the force due to gas is significant, $\sim 30 \%$ of that due to stars. 
 This is because of the low velocity dispersion of gas, which confines gas to a thinner layer; so that it is concentrated closer to the Galactic mid-plane than the stars.
The gas mass fraction at R=6 kpc is $16 \%$ of stellar surface density (see Table 1 from \cite{BJ2007}). Thus, the self-gravitational force close to the disk mid-plane is a non-linear function of the gas surface density in the disk.
 Hence despite its low mass fraction, the gas can have a significant effect on the vertical density distribution of the main mass component, namely, the stars, as shown next. 

\begin{figure}
\centering
\includegraphics[height= 2.65 in, width= 3.1in]{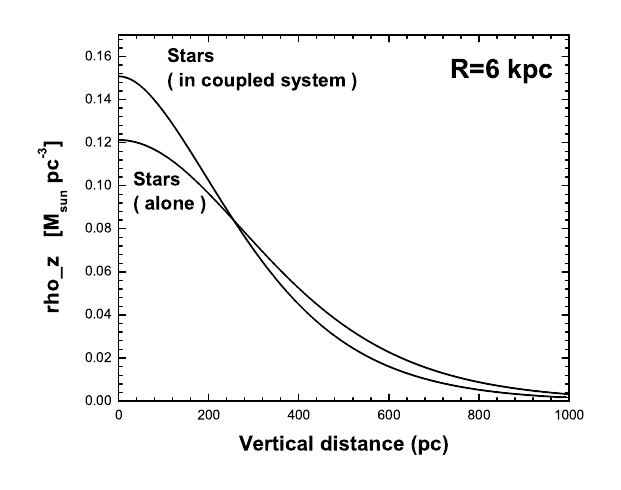} 
\bigskip
\caption{Self-consistent vertical density for stars vs. $z$: for the stars-alone case; and the stars in the gravitationally coupled case (stars, H$_2$, and HI),  at R=6 kpc. Due to the additional gravitational force of gas, the stellar distribution in the coupled case is constrained closer to the mid-plane, such that: its mid-plane density is higher; the scale height (HWHM) is smaller; and the density profile is steeper -- compared to the one-component, stars-alone case.
$\: $  {\it Source}: Taken from \cite{BJ2007}} 
 \label{fig.9}
\end{figure}

The stellar density distribution in the coupled case is obtained numerically as a solution of the coupled joint Poisson-hydrostatic balance equations (see \cite{BJ2007}). 
Fig. \ref{fig.9}  shows the model stellar vertical density distribution for the stars-alone case
 and also that in the coupled case.
This clearly shows the dynamical effect of gas gravity on the vertical stellar distribution.  In the coupled case, the central density of the stellar distribution is higher, by $\sim 25 \%$;  and the thickness (HWHM) is smaller, by $\sim 20 \%$; compared to the stars-alone case. 

This can be physically understood as follows.
The additional gravitational force due to the other disk components modifies the stellar distribution, 
 so that it is vertically constrained to be closer to the mid-plane. Thus the mid-plane density is higher compared to the stars-alone case. For a constant surface density, this therefore 
results in a smaller thickness (HWHM); and at the same time, the shape of the density profile is also altered, which becomes steeper.

Thus, the redistributed stellar distribution in the coupled case 
has a higher central density, its thickness is smaller, and the shape of its  density profile 
is steeper; compared to the stars-alone case. 
 This is the generic behaviour of redistribution of any gravitating disk component due to  additional gravitating components (disk components or halo) in the system.

While we have focused so far on  the effect of gas on the stellar density distribution; similarly, the gas density will also be redistributed, due to the gravitational force of stars and halo, and  will be constrained  closer to the mid-plane than for the gas-alone case (see Section \ref{sec:4.2.6}, and  Appendix A for further discussion on this).{\footnote {The case in \cite{NJ2002} (or Section \ref{sec:4.2.4})
is somewhat different because first the gas response to stellar potential is considered where the gas gravity is not included and then the gas distribution in a coupled (stars plus gas plus halo) case is considered that also includes gas gravity - see Fig. \ref{fig.6} panels for H$_2$ and stars.}}
Thus even though each isothermal disk component by itself would satisfy a sech$^2$ distribution; its net vertical density distribution in the coupled system is  different from sech$^2$, see Section \ref{sec:4.2.6} for further discussion on this.

\begin{figure}
\centering
\includegraphics[height=2.57in, width=2.67in]{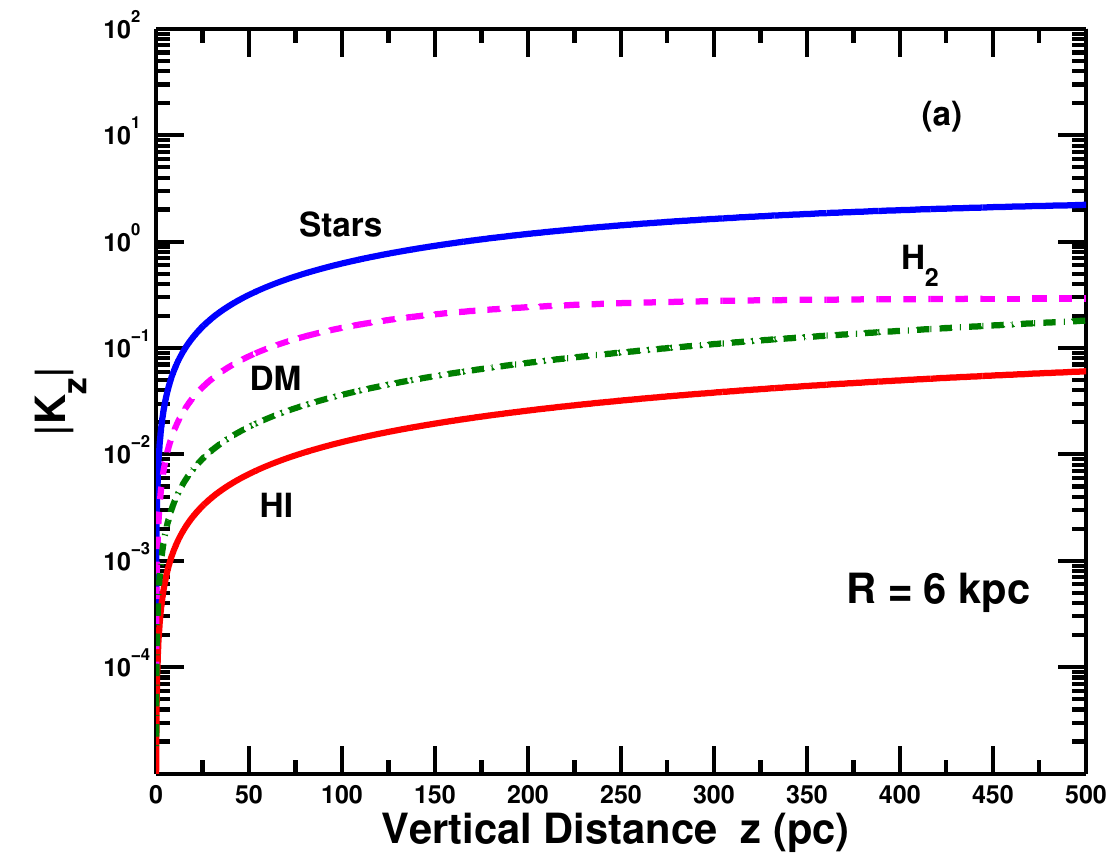}
\medskip 
\includegraphics[height=2.6in, width= 2.67in]{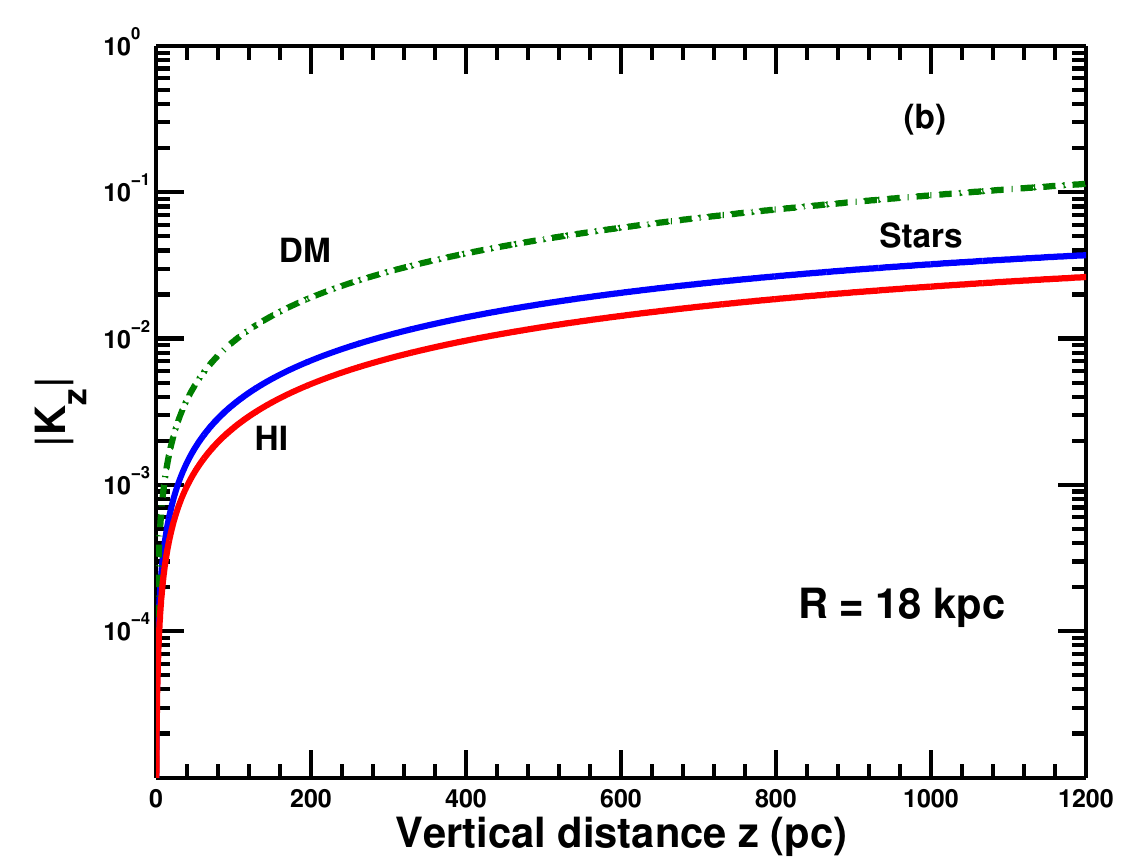} 
\bigskip
\caption{Plot of gravitational force per unit mass, $|K_z|$, exerted by each disk component (stars, HI and H$_2$) and the dark matter halo (DM) separately,
drawn on  a log scale vs. $z$, at R= 6 kpc (left panel) and R=18 kpc (right panel). This figure  shows that in the coupled case, the stellar distribution will be mainly affected by the H$_2$ gas gravity rather than by the halo or HI in the inner Galaxy, at R=6 kpc. In contrast, in the outer Galaxy, at R=18 kpc, the force due to the dark matter halo dominates over the stellar self-gravitational force itself; hence, the halo will strongly affect, and in fact, mainly determine the modified stellar distribution in the coupled case. This is confirmed by the results shown in Fig. \ref{fig.11}.
$\: $  {\it Source}: Taken from \cite{SJ2018}} 
 \label{fig.10}
\end{figure}

A detailed quantitative analysis of the redistribution of the stellar vertical density distribution, which includes the gravitational effect of gas as well as the dark matter halo; with a particular focus on the outer Galaxy, was carried out by Sarkar \& Jog (2018) \cite{SJ2018}, as briefly summarized next.
The values of $K_z$ for stars-alone and gas-alone are obtained as done above, or see \cite{SJ2018} for details.  For the dark matter halo, $K_z$ is obtained by taking the $z$ derivative of the halo potential (see Eq. 7, \cite{SJ2018}). The values for $|K_z|$ are plotted on a log scale for stars, H$_2$, HI and dark matter halo vs. $R$ in Fig. \ref{fig.10} for $R$ = 6 and 18 kpc  respectively. The choice of one radius in the inner Galaxy and one in the outer Galaxy helps bring out the dynamical result that the effect of gas and halo on stars dominates at different radial regions 
 -- namely, in the inner and outer regions, respectively.

\begin{figure}
\centering
\includegraphics[height=2.45in, width=2.74in]{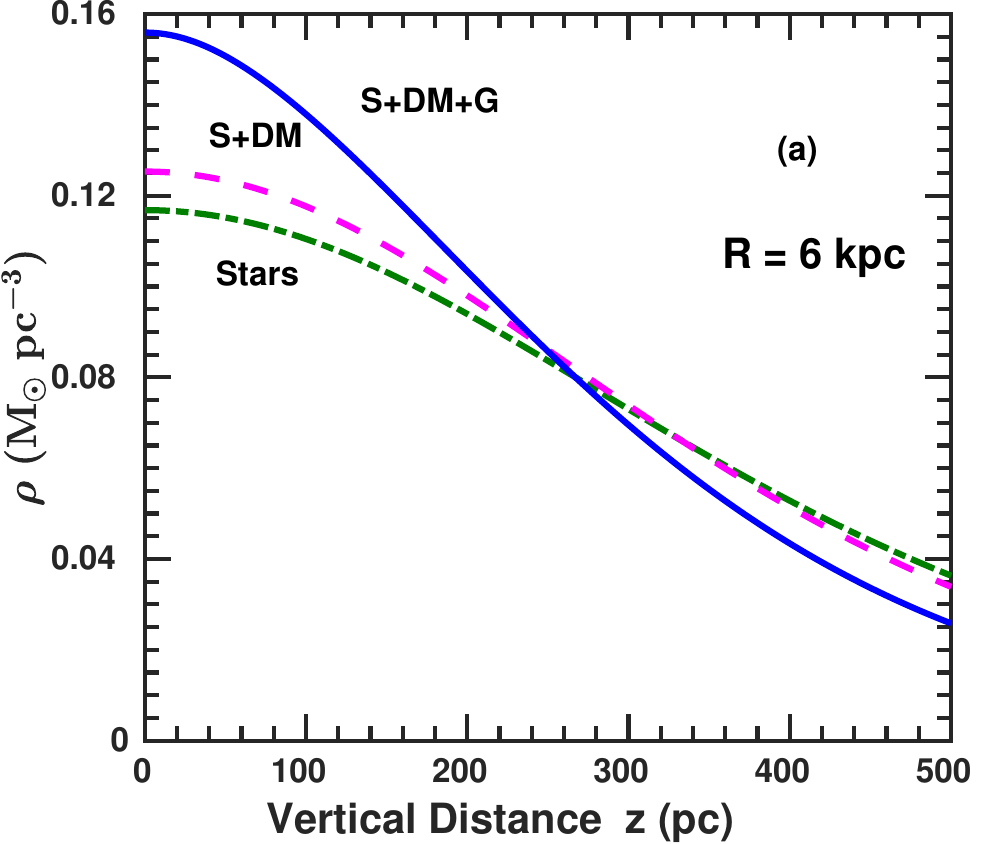}
\medskip 
\includegraphics[height=2.45in, width= 2.55in]{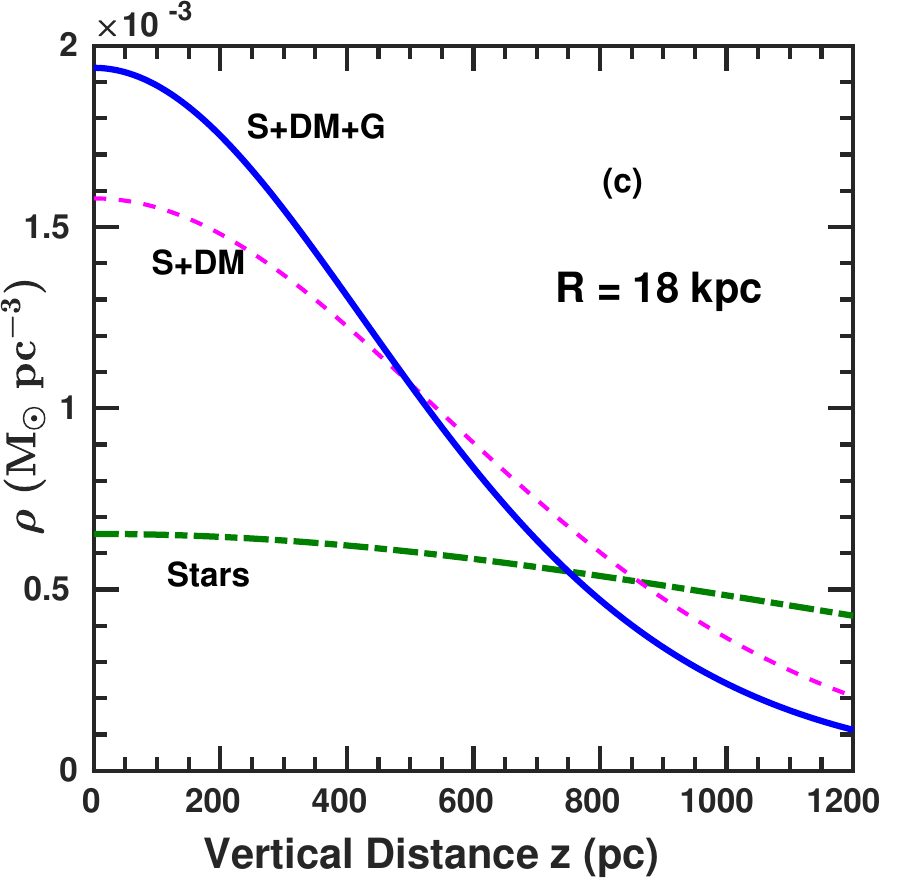} 
\bigskip
\caption{Self-consistent vertical density of stars vs. $z$ at R=6 kpc (in the inner Galaxy) (left panel), and at R=18 kpc (in the outer Galaxy) (right panel). The three curves represent the density distribution of stars in the gravitational field of: stars-alone; stars plus dark matter halo; and stars plus dark matter halo plus gas -- denoted respectively by the dashed-dot, dashed and solid curves, respectively. The inclusion of other gravitating components (gas and halo) results in an overall constraining of the stellar distribution closer to the mid-plane: such that, the mid-plane density is higher; the scale height (HWHM) is smaller; and the density profile is steeper;  compared to the results for the  stars-alone case.  
The constraining effect on stars is mainly due to gas in the inner Galaxy (left panel), while the dark matter halo has a dominant effect in the outer Galaxy (right panel).
$\: $  {\it Source}: Taken from \cite{SJ2018}} 
 \label{fig.11}
\end{figure}

Fig. \ref{fig.11} shows the resulting vertical stellar density distribution in the stars-alone case, and then  in the coupled case  (the latter obtained by solving Eq. 1 from \cite{SJ2018}) at $R$= 6 and 18 kpc respectively (see the left and right panels,  respectively in Fig. \ref{fig.11}). 
The plot at $R$=6 kpc is a generalization of Fig. 3 of \cite{BJ2007}, that is, Fig. \ref{fig.9}; since in Fig. \ref{fig.11} the effect of halo on the stellar vertical distribution is also included. 
 In the coupled case, first only the effect of dark matter halo is included; then the effect of dark matter halo as well as gas is included. This step-by-step addition  of different components allows a study of the differential effect of gas and dark matter halo on the stellar vertical distribution.

As seen in Fig. \ref{fig.10}, at $R$=6 kpc, the force $\mathrm{|K_z|}$ due to H$_2$ dominates over that due to the dark matter halo; hence the main constraining effect or the redistribution of the stars is due to H$_2$ gas,  while the halo has a small effect. 
This is confirmed by the left panel in  Fig. \ref{fig.11}  where the HWHM is reduced by $\sim 6 \%$ due to the effect of the halo as compared to the stars-alone case; whereas the addition of gas (H$_2$ and HI) reduces the HWHM by further 20\%. Thus, gas has a higher effect on the stellar distribution (see Table 2 from \cite{SJ2018} for the numerical values of HWHM) than the halo. The corresponding increase in the mid-plane density for stars  due to the gravitational effect of dark matter halo and gas is  7\% and 33 \% respectively (see Table 2, from \cite{SJ2018}).

At R=18 kpc, on the other hand, the force $\mathrm{|K_z|}$ due to dark matter halo dominates that due to the self-gravitational force of stars and HI; there is no H$_2$ gas at this radius.
Here, clearly the halo would play the dominant role in  
shaping the modified stellar vertical distribution. This is confirmed by Fig. \ref{fig.11}, right panel.  Here  the gravitational force of halo decreases the HWHM of the stellar distribution by a huge factor of $\sim 2.3$ compared to the stars-alone case; while including gas decreases it further by $\sim 20 \%$ only (see Table 2 in \cite{SJ2018}).   
In fact, beyond $R$=17 kpc, the constraining effect of halo dominates, and the halo is the main determinant of the stellar vertical distribution (see Table 2 from \cite{SJ2018})). In the outer disk, the stellar vertical density distribution is steeper than the stars-alone case mainly due to the effect of the halo. 

Thus, gas has an important constraining effect on the redistribution of stellar vertical distribution at small radii; while the halo has the dominant constraining effect in the outer Galaxy.

\medskip

\subsubsection{Detailed results for stellar density distribution in the outer Galaxy\label{sec:4.2.6}}

The quantitative results for the three features of the redistributed stellar density distribution (arising due to the constraining effect of gas and the halo):  namely, the mid-plane density, the scale heights (HWHM), and the vertical stellar profile, are given next. The focus is on the outer Galaxy. The dark matter halo plays a significant role in this radial range, confirming what was discussed above.  

\medskip

\noindent {\bf Measuring the constraining effect: Disk thickness}

\begin{figure}
\centering
\includegraphics[height= 2.7 in, width= 3.1 in]{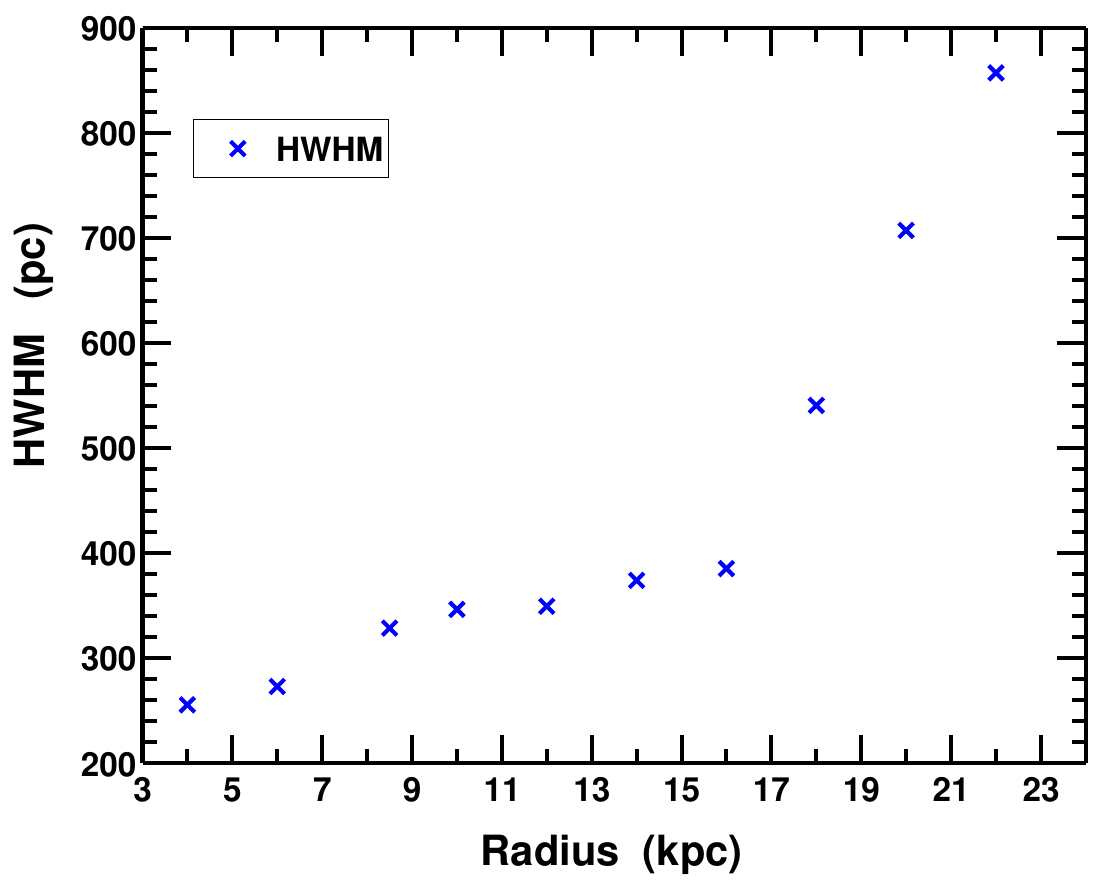} 
\bigskip
\caption{Plot of the model scale height (HWHM) of the vertical stellar density distribution in the joint potential of the disk and dark matter halo vs. R in the Galactic disk. The stellar disk thickness increases gradually until about R=17 kpc 
and then flares beyond that in the outer disk.
$\: $  {\it Source}: Taken from \cite{SJ2018}} 
 \label{fig.12}
\end{figure}

A plot of the HWHM of the vertical stellar distribution, for the multi-component disk plus halo system  vs. radius, for $R$=4-22 kpc is given in Fig. \ref{fig.12}; also, see Table 2 in \cite{SJ2018} for the numerical values. 
Fig. \ref{fig.12}  shows that in the coupled case, the stellar scale height increases by a moderate amount, by a factor of $\sim 50 \%$  from $R$=4 to 16 kpc. Beyond $R=17$ kpc, the stellar disk flares steeply because the stellar velocity dispersion saturates (by choice) to the gas dispersion
value beyond this radius (see Section \ref{sec:4.2.3}). As can be seen from Table 2, \cite{SJ2018}, the stellar disk by itself would flare by a factor of 13.6 from $R$= 4 to 22 kpc. The inclusion of gas, and mainly the halo (which is effective at larger radii),
reduces the flaring to a more moderate value of a factor of 3.3. The gas and mainly the halo restrict the thickness to be $< 1$ kpc even at $R=$ 22 kpc.
The lower disk thickness would  make the disk robust, that is, help resist distortion due to tidal perturbations (\cite{SJ2018}; also, see Section \ref{sec:4.6}). 

We point out that the various observations of the Milky Way outer stellar disk do show flaring with radius:  \cite{Momany2006} using 2MASS data for red clump and red giant star; \cite{Lopez2014} using SDSS-SEGUE data for  F8V-G5V type stars, and \cite{Wang2018} using LAMOST data on red branch stars. These agree reasonably well with the results in Fig. 12 (also, Table 1, column 4 in \cite{SJ2018}). The results of \cite{Lopez2014} show somewhat higher flaring. The higher values in \cite{Lopez2014} could be partly due to a possible contamination by thick disk stars 
(\cite{Minchev2015}; also, see Section \ref{sec:5.3}).
It is worth noting that a similar trend of flaring of stellar disks in the outer Galaxy was noted and discussed by \cite{Kalberla2014}, who had also questioned the constancy of the stellar scale height with radius.  
The detailed physics and implications of the flaring stellar disk will be discussed later in Section \ref{sec:5}.

\cite{SJ2018} also propose a new parameter: $z_{1/2}$, the half-mass scale height as measured from the mid-plane as another way of measuring the disk thickness (also, see Appendix A). This can be used as indicator of the constraining effect of gas and halo. 
The values of $z_{1/2}$ are comparable to but somewhat
smaller than the HWHM values (see Tables 2 and 3 in \cite{SJ2018}). From an observational perspective, $z_{1/2}$ can be thought of as an indicator of the half-light scale height for  a constant M/L (mass-to-light) ratio, for an edge-on galaxy. A similar parameter has also been proposed and used by \cite{Sotillo2023} to analyze the data from TNG simulations.

\medskip

\noindent {\bf Measuring the constraining effect: Mid-plane density}

The mid-plane density of each disk  component in a coupled system is higher compared to its one-component value, due to the constraining effect of the other disk components and the halo (see Section \ref{sec:4.2.5}). 
The mid-plane stellar density in the multi-component disk plus halo model  is higher by 50\% at R=8.5 kpc; while the increase is much higher, by a factor of $\sim 3-4$ at R=18, 20 kpc respectively (see Table 4, \cite{SJ2018}). This is because at large radii,   $\mathrm{|K_z|}$, the force due to the halo dominates that due to the stellar self-gravity itself. While the mid-plane stellar density is not likely to be directly observed, its higher value can have interesting dynamical consequences. 
For example, the vertical oscillation frequency normal to the plane, as given by $(4 \pi G \rho_{0})^{1/2}$ (e.g., \cite{BT1987}), would be higher by a factor of $\sim 2$ at R=18-20 kpc. This could lead to a better-mixed vertical distribution in the outer Galactic disk. A similar increase in mid-plane gas density (details are given later in this section) would help increase star formation seen in the outer Galaxy, which is otherwise hard to explain. 
 Thus, the dark matter halo has a dominant influence on the disk vertical structure in the outer disk. 

\medskip

\noindent {\bf  Measuring the constraining effect: Stellar vertical profile}

In real galaxies, the observed distribution near the mid-plane  
shows an excess over the sech$^2$ distribution, and the distribution typically obeys a sech or an exponential distribution (Section \ref{sec:4.1}).
Hence \cite{vdk1988} suggested the following family of curves that could fit the observational trends:

\begin{equation}
 \mathrm {        \rho(z) = 2^{-2/n} \: \rho_e \: sech^{2/n} (nz/2z_e) }
  \label{eq4.7}
  \tag{4.7}
\end{equation}
  
\noindent where n= 1,2 and n$\rightarrow \infty$ correspond to a density profile with a sech$^2$, sech and an exponential $z$ distribution respectively.
The parameters $n, z_e, \rho_e$ are obtained by fitting the above function to the data and doing the best-fit analysis.
This function is not based on physical principles, rather it is ad hoc as noted by \cite{vdk1988} -- who proposed it (also, Section \ref{sec:4.1.1}). 
Despite this, it has been used in many observational studies to fit the data on external galaxies so as to get the best-fit value of $n$ (e.g., \cite{BD1994,deGrijsP1997}), where $n$ is taken to be an indicator of steepness of the density profile.  

\begin{figure}
\centering 
\includegraphics[height=2.7 in, width= 3.1 in]{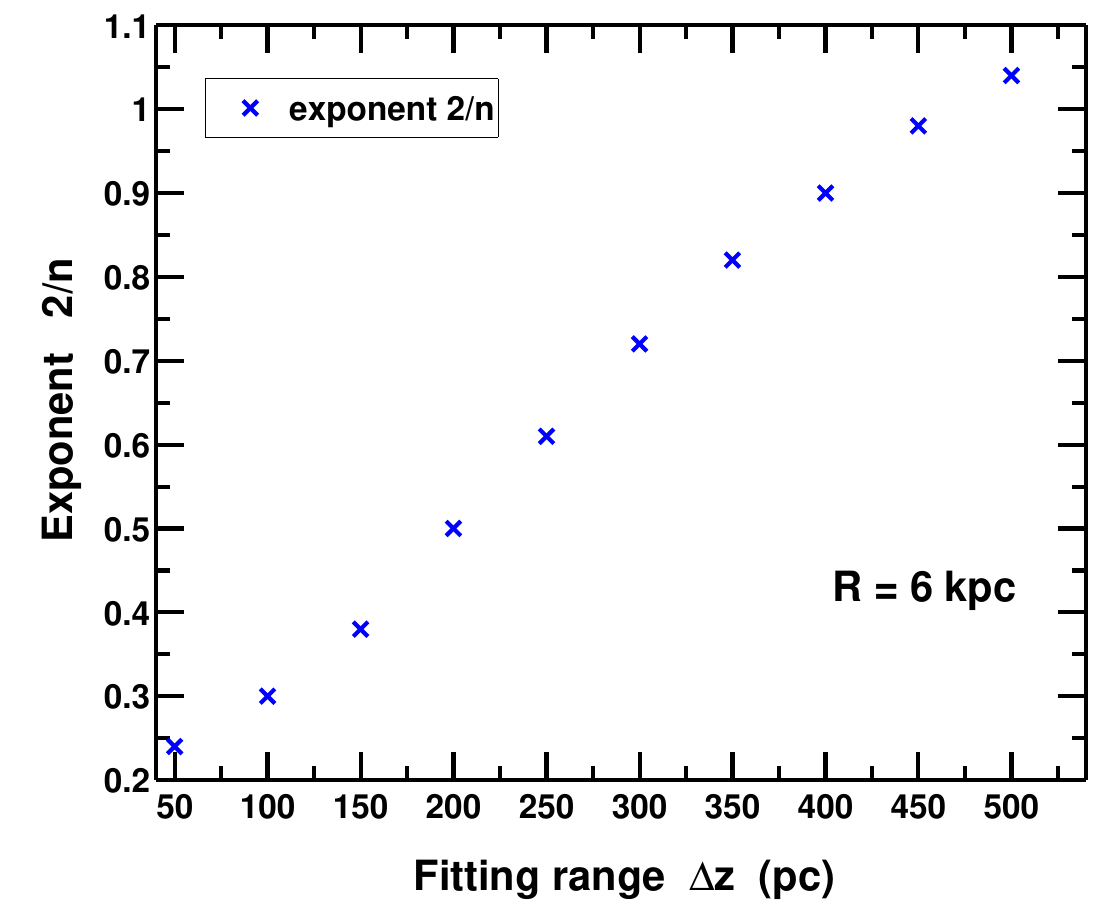} 
\bigskip
\caption{Plot of the best-fit exponent $2/n$ vs. $|\Delta z| $ for the range of $z$ values over which
  the model vertical density distribution for stars is fitted by a distribution
 of type sech$^{2/n}$; shown for R = 6 kpc. The value of $n$ varies with $|\Delta z|$, hence $n$ is not a
 robust indicator of the stellar density profile.
 $\: $  {\it Source}: Taken from \cite{SJ2018}} 
 \label{fig.13}
\end{figure}

The additional gravitational force due to the other disk components and dark matter halo results in the stellar profile being steeper than the one-component case (see Section \ref{sec:4.2.5}; and, \cite{SJ2018}). Note that even though the stellar disk is taken to be isothermal, the net stellar distribution is no longer given by a sech$^2$ profile in a coupled, multi-component disk under the gravitational field of the halo.
To quantify this effect, \cite{SJ2018} tried to fit the model results to Eq. (\ref{eq4.7}). They found
that the above expression is not adequate to explain the results for the model stellar density profiles, as discussed next. 
In the above expression, $2^{-2/n} \rho_e$ is the mid-plane stellar density which is not known from observations, but is known from the model calculations.   
\cite{SJ2018} fit the resulting model
 density distribution to the above function (Eq.(\ref{eq4.7})) to get the best-fit values of $n$ and $z_e$. They find that the best-fit value of $2/n$ is not robust; instead, it varies with $\Delta z$, the range of $z$ values chosen for the fitting -- see Fig. \ref{fig.13} which shows the results for $R$= 6 kpc. 
 Thus, the expression for the density profile, (Eq.(\ref{eq4.7})), which was proposed by \cite{vdk1988} is not physically valid for a realistic galactic disk.  

\begin{figure}
\centering 
\includegraphics[height=2.7 in, width= 3.1 in]{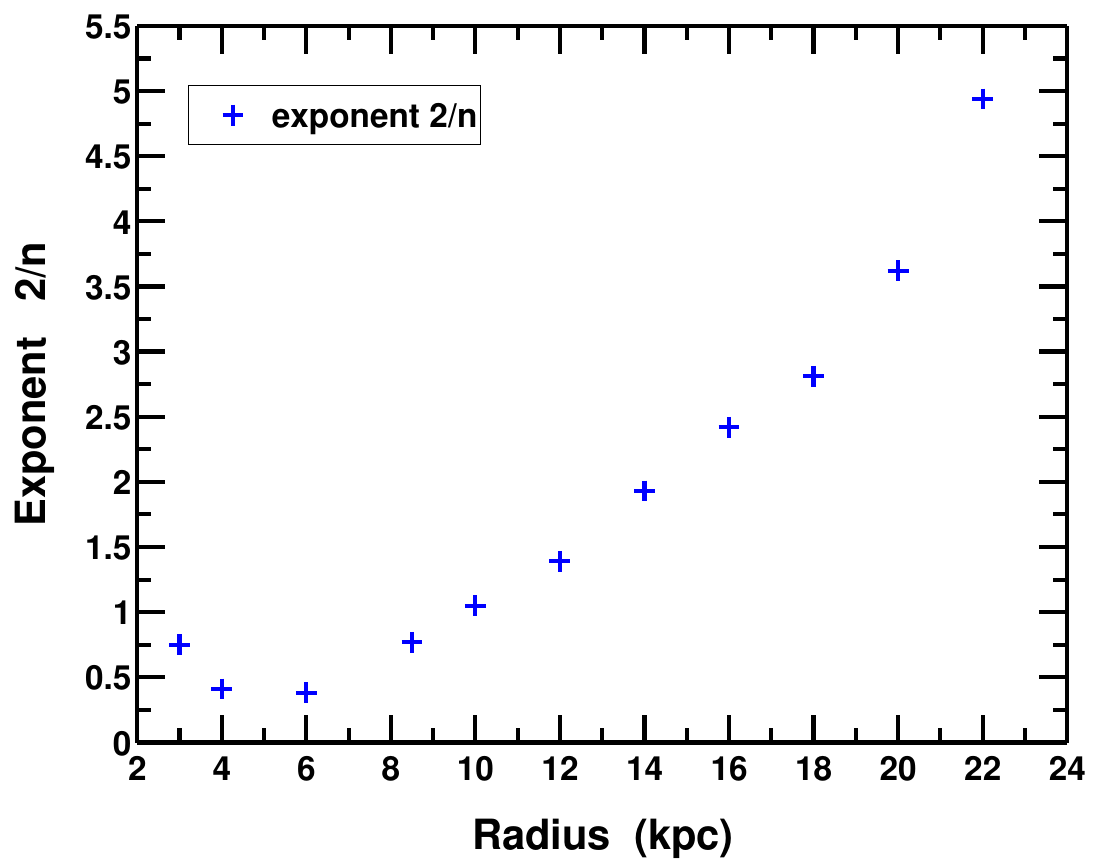} 
\bigskip
\caption{ Plot of the best-fit value of the exponent $2/n$ when the model
 density distribution is fit by a distribution of type sech$^{2/n}$ over a range of $z$ values, $|\Delta z| <$ 150 pc vs. $R$. For radii $>$ 14 kpc, a new range of $n < 1$ is obtained. This corresponds
 to the  radial region where the main constraining effect is due to the dark matter halo.
 $\: $  {\it Source}: Taken from \cite{SJ2018}} 
 \label{fig.14}
\end{figure}

Keeping in mind this caveat about $n$ not being robust; the model density distribution results are next fit for $|\Delta z| < 150$ pc with the above function (Eq.(\ref{eq4.7})) at different $R$ values; so as to get an idea of the trend, and to compare with observations where the data close to the mid-plane are studied. The resulting plot of $2/n$ vs. $R$ is given in Fig. \ref{fig.14}. 
The result gives $n >1$ up to $R < 14$ kpc. Interestingly, this is the radial range over which the gas is mainly responsible for constraining the stellar distribution (see Table 2 from \cite{SJ2018}). On the other hand, it is found that $n< 1$ in the outer Galactic disk; where the constraining is mainly due to the halo. Observations of external galaxies typically cover inner regions; which explains why the observed values give the best-fit $n$ to be $> 1$, corresponding to  the density profile between sech and an exponential. The parameter region $n < 1$ obtained here is new, and has not been studied before in the literature. The important point to stress is that, in both cases, whether $n>1$ or $n <1$, the density distribution is steeper than the distribution for the corresponding  one-component cases.

It is often stated in the literature that the index $n >1$ or $(2/n )< 2$ can be taken to be an indicator of the steepness of the profile,  with higher $n$ corresponding to a steeper profile  \cite{deGrijsP1997,BJ2007}. This is misleading since the case when $n < 1$ could also denote a steeper profile compared to the particular one-component case.  As shown above (Section \ref{sec:4.2.5}), for a constant surface density, any additional gravitational component (gas or halo) will constrain the stellar distribution closer to the mid-plane and cause a steepening of the stellar profile compared to the corresponding one-component stellar disk  of constant surface density, but $n$ could be $> 1$ or $< 1$. See for example, the two panels in Fig. \ref{fig.11} where $n>1$ for $R=6$ kpc, and $n<1$ for $R$= 18 kpc (as obtained from Fig. \ref{fig.14}).

 In reality, all the three parameters, namely, the mid-plane density, width and $n$, together decide how sharply the stellar density profile falls with
 $z$ \cite{SJ2019}. Similarly, the other disk components, HI and H$_2$, will also show a steeper profile than the corresponding one-component cases (as mentioned in Section \ref{sec:4.2.5}; also, see Appendix A for details). 
 
We  have given a detailed discussion of this topic, including the 
limitations of using the parameter $n$  as an indicator of steepness of density profile -- so as to give a complete picture; and also because Eq.(\ref{eq4.7}) has been a popular form in the literature to fit the observed excess of intensity close to the mid-plane.
As shown above,  $n$ alone is not an indicator of steepness. Moreover, it has been shown that the value of the best-fit $n$ itself is not robust; rather, it varies with $\Delta z$, the range used for fitting. Further, $n$ could be $<$ or $>$ 1 and  yet denote a steeper profile compared to the one-component case. 
Therefore, the density profile as proposed by \cite{vdk1988}, see Eq.(\ref{eq4.7}), is not physical and not  applicable to a real disk. Hence, it can only be used with some caution to  fit the data.

We highlight an important point that the $n$ value obtained from fitting the model results is not equal to 1, neither is it very large -- which would correspond to a sech$^2$ or an exponential profile respectively (see Eq. (\ref{eq4.7}); and Fig. \ref{fig.13} and Fig. \ref{fig.14} which show the model results for $n$). Thus, the actual model density profile is more complex than given by either of these limits.

Recall that, for a typical galactic disk, the determination of $\rho(z)$ is a local problem; and can be used to uniquely trace the corresponding gravitational potential (Section \ref{sec:3.1}, Section \ref{sec:4.2.1}, and Section \ref{sec:4.3.1}). This can be done numerically. However, an analytical form to represent the numerical results for the disk density $\rho(z)$ obtained from the multi-component disk plus halo model for the various cases; or, the corresponding gravitational potential at a given $R$; is not easy to obtain and has not been attempted so far.

\medskip

\noindent {\bf Effect on HI gas}

\begin{figure}
\centering
\includegraphics[height=2.57in, width=2.67in]{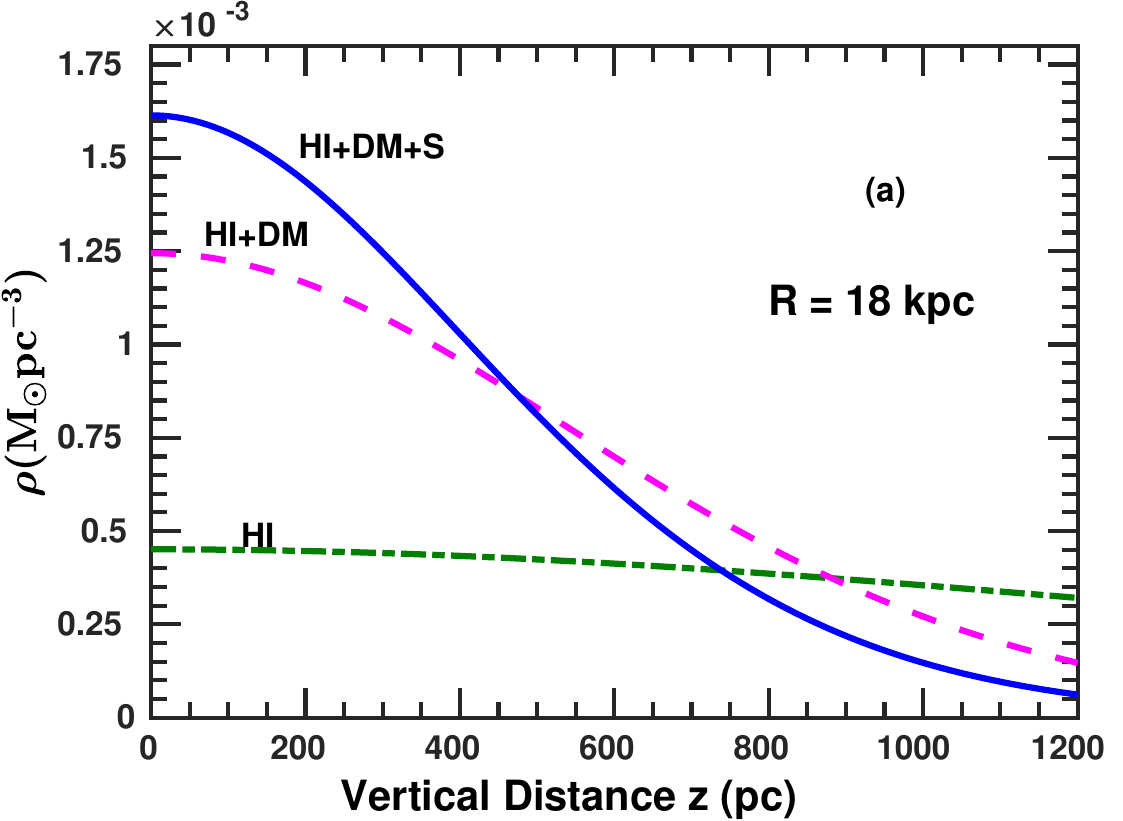}
\medskip 
\includegraphics[height=2.55in, width= 2.66in]{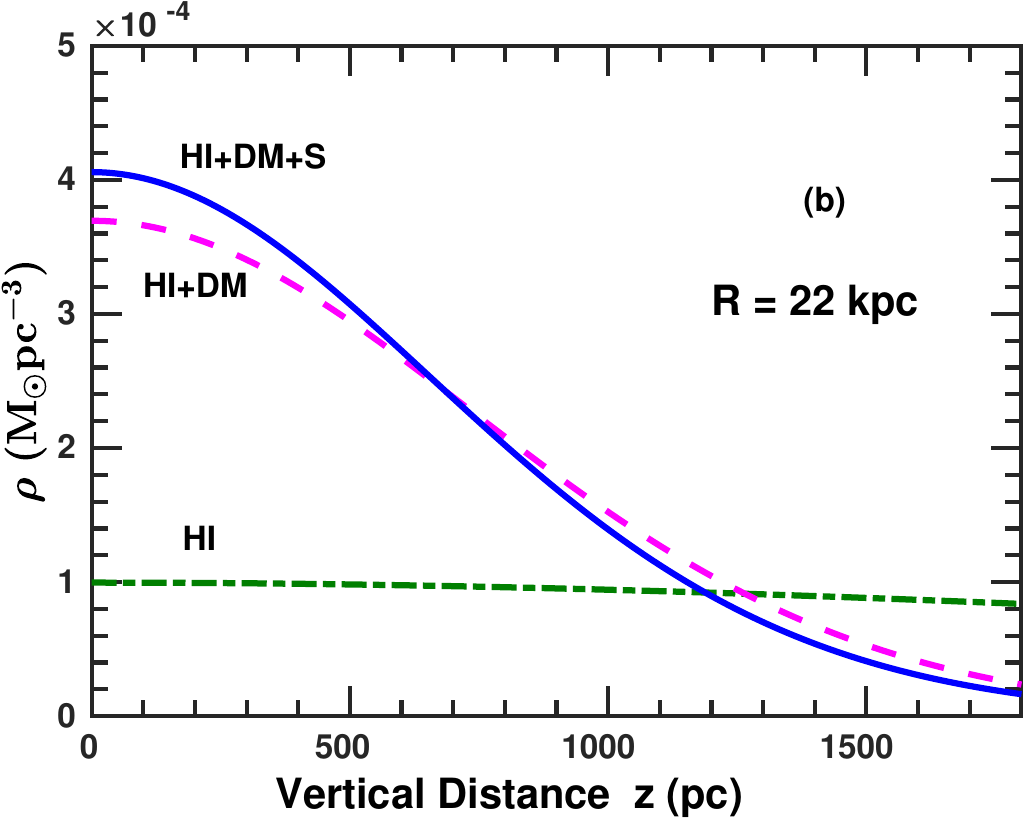} 
\bigskip
\caption{Self-consistent vertical density distribution of HI gas vs. $z$ in the outer Galaxy, at R=18 kpc (a), and R=22 kpc (b); under its own self-gravity, then also including the halo gravitational force, and then also including the gravitational force of the dark matter 
  halo and stars (shown by dashed-dot, dashed and solid curves respectively). Under its own self-gravity the HI distribution is extremely extended due to its low self-gravity in the outer Galaxy. The inclusion of halo gravitational force strongly constrains the HI vertical distribution in the outer Galaxy, decreasing the thickness (HWHM) by a factor of 3-4, and the effect is progressively higher at larger radii. The reduced HI thickness values for HI in the outer Galaxy agree well with the observed values, see the   text for details. $\: $  {\it Source}: Taken from \cite{SJ2018}} 
 \label{fig.15}
\end{figure}

The  model results for the vertical density distribution of HI gas in the outer Galaxy (from \cite{SJ2018}) are given next.
The HI vertical density distribution for the three cases: HI gas-alone, gas plus dark matter  halo, and the coupled stars plus gas in the halo potential, is given for $R$= 18 kpc and 22 kpc in Fig. \ref{fig.15}.
 At there radii, there is no observed H$_2$ gas, so the gas component consists of only HI gas. It is striking that the HI gas-alone under its own gravity is highly extended vertically; 
with a HWHM of 1.8 kpc and 3.8 kpc at R= 18 and 22 kpc, respectively.{\footnote{Note that this treatment of HI gas is different from that in Fig. \ref{fig.6} panel (b) and (c), where first the gas response to stellar potential is considered without including the gas gravity; and then the gas distribution in a coupled potential (of stars, gas and halo) that includes gas gravity is considered.}} 

 These values are much higher than the observed values  \cite{Wouter1990}. Including the effect of the gravitational field of the halo substantially reduces the HWHM, by a factor of 2.7 and 4.3 at these two radii, respectively; and the HI distribution becomes steeper in the coupled case \cite{SJ2018}. The decrease is only slightly larger on including the effect of stars, so that the net HWHM values are equal to 502 pc and 798 pc, respectively. 

  These net resulting HWHM values of HI gas showing flaring (studied up to R= 22 kpc) agree fairly well with the observed values  \cite{Wouter1990,KalberlaD2008}, while \cite{Levine2006} get higher observed values.
  An important  point to note  is that
  without the confining effect of the halo, the gas distribution would be very extended; and it would then be  more susceptible to disturbance by external perturbations, such as due to tidal encounters,  as well as gas dynamical processes. Thus the halo cushions  the outer disk against getting tidally distorted.

An interesting point is that in the outer Galaxy, beyond R=18 kpc, the surface density of stars and gas is comparable (see Table 1 in \cite{SJ2018}). Also,  their velocity dispersion values are identical, by choice (see Section \ref{sec:4.2.3}). Hence their density distributions are nearly identical.
By R= 22 kpc, the gas surface density dominates (see Table 1 in \cite{SJ2018}). In any case, beyond R=18 kpc, the main gravitational force is due to the halo -- which mainly determines the vertical density distribution of a disk component.

The corresponding HI mid-plane density at R= 18 kpc and 22 kpc increases by a factor of 3.6 and 4.1 respectively, compared to the gas-alone values. The resulting corresponding mid-plane HI density values are 1.6 $\times 10^{-3}$ and 4.1 $\times 10^{-4}$ M$_{\odot}$ pc$^{-3}$, respectively. 
 The gas density is used as an indicator of onset of star formation. The increased mid-plane density due to the constraining effect of halo, and the gas being dissipational, 
makes the gas more susceptible to star formation (also, see \cite{AbramovaZ2008}). This effect is higher at larger radii where the halo is dominant, hence it could explain the star formation seen in outer disks of some galaxies. This is an important implication from this model, and needs to be explored further.

The gas distribution is constrained towards the mid-plane, and has a steeper density profile in the coupled case (see Fig. \ref{fig.15}, and the discussion in Section \ref{sec:4.2.5}).{\footnote{The physics behind the reduction in HI scale height and the change in the gas profile in the coupled case was predicted earlier in Section \ref{sec:4.2.5}, where an analogous case of effect of gas on stars in a coupled case was considered which included gravity due to all the components as done here.}}  A quantitative measurement of this for gas distribution needs to be done, in analogy with that done for stars. It is claimed that the gas distribution in the coupled case is a Gaussian (\cite{Patra2019}, see Fig. 7 from that paper). However, an actual detailed fitting to the results has not been done.
See Appendix A for further discussion on this.

To summarise the discussion so far in Section 4, we stress that the model developed (Sections \ref{sec:4.1}, \ref{sec:4.2.1}, \ref{sec:4.2.2}) is general, although it is applied here to the Milky Way, since the parameters are the best-known for it.
The results obtained in Section \ref{sec:4.2.3} to Section \ref{sec:4.2.6} for a thin disk case are generic and the trends obtained are valid for external galaxies as well. Indeed, the model will be applied to other galaxies 
in Section \ref{sec:4.2.7} for the thin disk case. 
The results from the thin disk case are applied to obtain the self-gravitational energy of a multi-component disk in Section \ref{sec:4.6}.
The model will be developed for other physical cases and also applied to other galaxies in the following sections (see Sections \ref{sec:4.3}, \ref{sec:4.4}, \ref{sec:4.5}).

\subsubsection{Application to NGC 891, NGC 4565: Radial variation in stellar scale height\label{sec:4.2.7}}

In studies of vertical luminosity distribution of edge-on galactic disks,  \cite{vdkS1981a} measured the vertical scale height of stars and claimed that it is independent of the radius, $R$, that is, the radial distance from the centre  (Section \ref{sec:2}). They proposed that this implies a specific rate of exponential fall-off of stellar velocity dispersion ($\mathrm{R_v = 2 R_D}$) to make this possible. Here, R$_v$ is the scale length  with which the vertical velocity dispersion falls exponentially with radius. Recall that for a one-component, isothermal disk, $z_0 =  v_z^2/2 \pi G \Sigma$
 (Eq.(\ref{eq3.5})), hence, $z_0$ is constant when $\mathrm{R_v = 2 R_D}$. 
Both these claims have been accepted in the literature, although there was already observational evidence by 1990s for a moderate increase with radius in the vertical scale height \cite{deGrijsP1997,Kent1991}, and there is no compelling physical reason for the relation $\mathrm{R_v = 2 R_D}$ to be valid.
 
In an important paper,
 \cite{NJ2002ltr} 
did a careful re-examination of the analysis by  \cite{vdkS1981a} for two prototypical edge-on galaxies, NGC 891 and NGC 4565, and showed that the data actually indicates a moderate increase of scale height with radius, by a factor of $\sim 2-3$, within the optical disk.
Next, \cite{NJ2002ltr} applied the multi-component disk plus halo model in the thin disk case, 
and showed that the above variation in scale height can be used to
constrain  $\mathrm{R_V}$, which is found to be  $\mathrm{> 2 R_D}$.  
 The main points of this work are summarized next.

\noindent {\bf Re-examination of the analysis by van der Kruit \& Searle}

\begin{figure}
\centering
\includegraphics[height=3.6in, width= 2.8in]{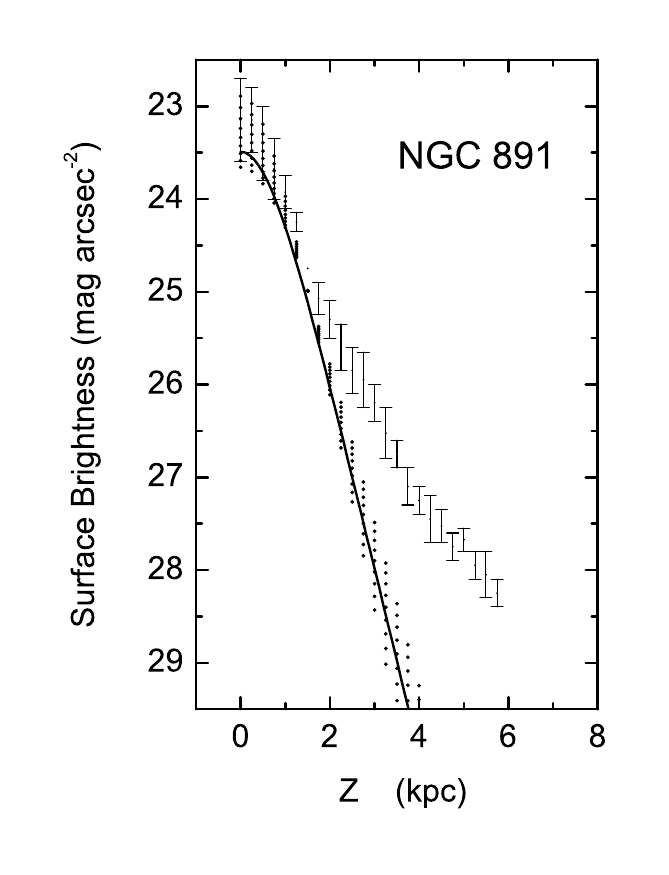} 
\bigskip
\caption{Plot of surface brightness, $I$ vs. $z$ for NGC 891 from \cite{vdkS1981a} who used their data and obtained this composite z-profile (the top curve in this figure) by vertically shifting the individual z-profiles at different R into coincidence at z'= 1.5 kpc.  The vertical bars indicate the range of the data. The solid line is the model composite curve obtained by \cite{vdkS1981a} using Eq(\ref{eq2.2}) where $z_0$ = 1 kpc, taken to be independent of radius. \cite{vdkS1981a}  claim that this model composite curve (solid line) fits the composite z-profile obtained from their data. On the other hand, \cite{NJ2002ltr} show that, when $z_0$ is allowed to increase linearly from 0.5 to 1.25 kpc within 20 kpc; the calculated surface brightness values (shown as points),  disperse over an interval of 1 mag arc sec$^{-2}$ at the mid-plane. This is exactly as seen in the original data of \cite{vdkS1981a}. Thus, \cite{NJ2002ltr} show that the observed data can allow for as much as a factor of 1.7 increase in scale height over the optical disk. The systematic deviation of observed data at high $z$ is due to the {\it Thick disk} which is not included in this study.
$\: $  {\it Source}: Taken from \cite{NJ2002ltr}} 
 \label{fig.16}
\end{figure}

In the study of edge-on galaxies by \cite{vdkS1981a},
the intensity profile, $I (R,z)$, is measured along a cut normal to the edge-on disk at a given galactocentric radius, R for the galaxy (as given in section \ref{sec:2.1.1}).
From these intensity profiles, \cite{vdkS1981a} obtained a composite z profile by vertically shifting the individual z profiles obtained at different radii and pinning them together at an arbitrary interim point $z'$. Consider the treatment for NGC 891 first. The above procedure gives the top curve in Fig. \ref{fig.16}.
Next, the model by \cite{vdkS1981a} for the intensity of an edge-on galactic disk (Eq.(\ref{eq2.2})) gives a single and well-defined model composite curve if the scale height, $z_0$, in Eq.(\ref{eq2.2}) were constant at all radii -- this is shown as the solid curve in Fig. \ref{fig.16}. 
\cite{vdkS1981a} 
 pin the data at a point $z'$ = 1.5 kpc, and claim that  the single composite model curve (solid curve)  gives the best-fit to the observed data with little scatter around it. 
  \cite{NJ2002ltr}, however, noted that the observed curves are not exactly in coincidence into a single composite curve, as would be expected if $z_0$ were constant with R as claimed by \cite{vdkS1981a}; instead, the data points show a considerable spread around this single curve. \cite {NJ2002ltr} pointed out that the spread in observed data around this composite curve corresponds to a spread in intensity, $\Delta I$, at z=0 to be equal to $\sim 1$ mag arc sec$^{-2}$. 

Further, \cite{NJ2002ltr} made an important point  that this spread being measured on a logarithmic scale, could in reality indicate a substantial variation in scale height with radius. They showed this quantitatively as follows.
\cite{NJ2002ltr} allowed $z_0$ to vary with radius and found by trial and error that   
when $z_0$ is increased linearly from 0.75 kpc to 1.25 kpc between 0-20 kpc or the entire optical disk in NGC 891;{\footnote{The radius of the optical disk is taken to be 4 R$_D$ in this work, see Appendix A for details.}}
the resulting spread in the calculated values of surface brightness for an edge-on disk (obtained using Eq.(\ref{eq2.2})), shown as dots in Fig. \ref{fig.16}, show a finite spread of intensity, $\mathrm{\Delta I \sim 1}$ mag pc$^{-2}$ at z=0 around the resulting composite
profile. This exactly matches the spread in the observed original data of \cite{vdkS1981a} (shown within error bars, see Fig. \ref{fig.16}). Thus, the spread in the observed data actually allows for a linear increase in $z_0$ from 0.75 kpc to 1.25 kpc, or by a factor of 1.7, over the optical disk.
 This is a substantial variation with radius compared to the strictly constant $z_0$ claimed by \cite{vdkS1981a}.

It should be stressed that this conclusion was reached by \cite{NJ2002ltr} using exactly the same data and the model for luminosity for an edge-on disk as in \cite{vdkS1981a}. The new point in the \cite{NJ2002ltr} study was that they noted
 the  pertinent point that the  
data actually shows a significant scatter of $\sim 1$ mag arc sec$^{-2}$ around the composite curve, and this scatter being measured
on a logarithmic scale could indicate a substantial increase in scale height with radius over the optical disk.
\medskip

\noindent {\bf Constraining the value of $\mathrm{R_v/R_d}$}

\begin{figure}
\centering
\includegraphics[height=2.8 in, width= 3.1 in]{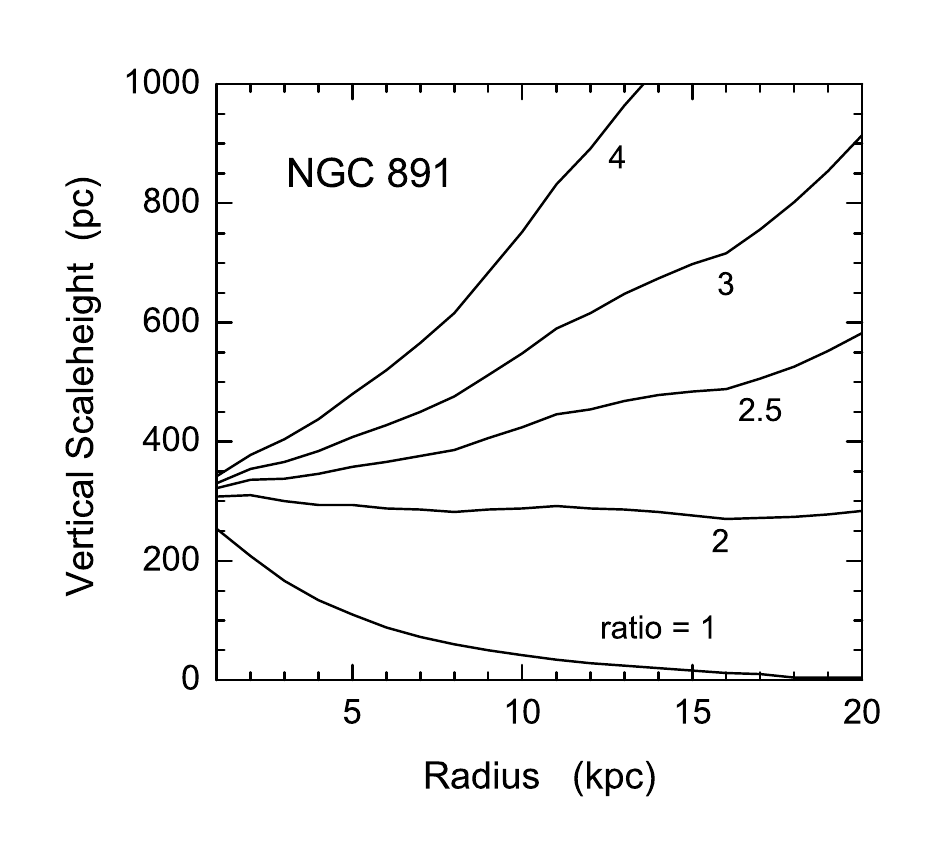} 
\bigskip
\caption{Plot of resulting stellar scale height (HWHM) vs. $R$  for NGC 891, calculated using the multi-component disk plus halo model, for different values of rate of fall-off of the vertical velocity dispersion, R$_v$. The variation in the scale height with R is found to critically depend on the value of the ratio $\mathrm{R_v/R_D}$. The scale height is found to increase by a factor of 1.8, 2.8 and 4.5  for the ratio R$_v$/R$_D$ = 2.5, 3 and 4, respectively, within the optical disk. Thus, the scale height is not constant with radius. $\: $  {\it Source}: Taken from \cite{NJ2002ltr}} 
\label{fig.17}
\end{figure}

Next, \cite{NJ2002ltr} show that the above deduced variation in scale height can be understood in terms of a disk in pressure equilibrium, and this variation allows them to constrain the value of R$_v$.
Due to the paucity of data for velocity dispersion at that time  \cite{Bottema1993}, the observed value of R$_v$ was not known. 
\cite{NJ2002ltr} apply the multi-component disk plus halo model to NGC 891, and using observed values of other input parameters for stars and gas, obtain the self-consistent vertical distribution, and hence the vertical stellar scale height as a function of $R$, for a range of values of $\mathrm{R_v/R_D}$ covering 1 to 4 (see Fig. \ref{fig.17}).

It should be pointed out that, the scale height obtained using the multi-component disk plus halo model (given as HWHM), by definition, is not identical to $z_0$
-- but the two are equivalent, and comparable, see Appendix A. So the trends in variation in these two can be compared.
This figure shows that, 
the scale height typically increases by a factor of few within the optical disk (for $\mathrm{R_v/R_D > 2}$).  
From the results in Fig. \ref{fig.17}, \cite{NJ2002ltr} conclude  that the factor of 1.7 increase in the scale height as deduced from the spread in the observed data for NGC 891, as discussed above, constrains the value of $\mathrm{R_v/R_D}$ to be between 2 and 2.5. 

A similar analysis for another edge-on galaxy, NGC 4565, done by \cite{NJ2002ltr} shows that the spread in the observed data indicates a linear increase in the vertical scale height by a  factor of 2.5 within the optical disk; which they show constrains the range of $\mathrm{R_v/R_D}$ to be between 2.5 and 3. 

Interestingly, a small change in $\mathrm{R_v/R_D}$ to 2.5-3 compared to the typical value of 2 that is used, is adequate  
to explain the observed moderate flaring of the stellar disk. 
 In Section \ref{sec:5}, we discuss the general physical significance of the value of $\mathrm{R_v/R_D} (> 2)$ and why it indicates a flaring disk. 

\noindent {\bf Discussion:}

Unfortunately,  the paper by \cite{NJ2002ltr}  and the important result from it, namely that the vertical scale height is not constant; rather, it shows a moderate radial increase, has been largely overlooked in the literature. We note that the recent observations and simulations also strongly support the above conclusion of a flaring stellar disk (see Section \ref{sec:5}). 

Finally, a few more caveats about the analysis by  \cite{vdkS1981a}  are given below. These  
contribute to uncertainties in the determination of the variation of scale height by \cite{NJ2002ltr}. 
First, the spread in the intensity profiles in the analysis by \cite{vdkS1981a} is controlled by the choice of the pinning point, $z'$, which is arbitrary. The apparent spread can be minimized by an appropriate choice of $z'$.
This point was explicitly shown by \cite{SJ2019}, 
though it was not mentioned in the paper by \cite{vdkS1981a}. 
The actual choice of $z'$ chosen  
would not matter if $z_0$ were constant at all R, since any choice of $z'$ would then yield a complete overlap and hence a single composite curve. 
Second, the spread in data is not just due to the scale height variation but also due to other effects such as  inclination, dust extinction etc (see the discussion point 1 in \cite{NJ2002ltr}).

It would be worthwhile to look at this problem  for a larger set of galaxies. However, 
in view of the caveats in the analysis of \cite{vdkS1981a} 
as described in this subsection (and in Section \ref{sec:2.1.1}), it may be advisable to follow a different approach to see if the stellar disk is flaring. For example, $z_0(R)$ could be measured directly by an iterative analysis of intensity profiles as proposed in Section \ref{sec:2}. Alternatively, if the modern IFU data gives values of stellar velocity dispersion directly, then using the multi-component disk plus halo model; the scale height as a function of radius can be calculated. This would be discussed in more detail in Section \ref{sec:5.4}.

\subsection{Multi-component disk plus halo model: Thick, or low density disk \label{sec:4.3}}

Here we relax the assumption of a disk being thin and consider a thick disk case. We next consider and compare the  various cases systematically: thin and thick disk cases -- first  a single component disk and then a multi-component disk plus halo; and see how the formulation of equations for the vertical structure is different in each case.

We stress that the term thick disk as used here and in the rest of the review is based on its general physical characteristics alone, namely a disk with a high physical thickness and/or a low density. In this case the $R$ term in the disk Poisson equation needs to be included in the formulation of equations to obtain the correct disk vertical structure. In other words, the term thick disk as used here or hereafter in this review does not specifically denote  
 the chemically and kinematically distinct structural component, which is a common feature of many galaxies including the Milky Way. In this review, the latter component will be refereed to as the {\it Thick disk}  (e.g.,  Sections \ref{sec:Intro},  \ref{sec:4.5.1}, \ref{sec:7.1}, \ref{sec:8}).
However, the results obtained in this section for a physically thick disk are general and could  be applied to study the vertical structure of the structurally and chemically distinct {\it Thick disks} in galaxies (which is outlined as a future problem, for details see Section \ref{sec:7.1} and Section \ref{sec:8}). 

As we have seen, to obtain a self-consistent vertical distribution in a self-gravitating disk, the Poisson equation and the equation of hydrostatic equilibrium have to be solved together. So far we have considered a thin disk case: for a single component disk (Section \ref{sec:3.1}); a multi-component disk in the field of the halo as applied to the Milky Way, and  also applied to model two external galaxies,  NGC 891 and NGC 4565 (see Section \ref{sec:4.2} for details). In the above studies, the radial part of the disk Poisson equation is taken to be zero. 

\subsubsection{Inclusion of radial term in the disk Poisson equation\label{sec:4.3.1}}

Now let us consider the general case when the radial term in the Poisson equation for the disk  
needs to be included in the formulation of the equilibrium vertical disk distribution. This can happen either when the disk is physically thick, or when the disk density is low as in the outer disk or at regions away from the mid-plane, so that the term representing the self-gravity in the Poisson equation, namely, ($\mathrm{4 \pi G \rho_0}$) is no longer much greater than the R-term (see for example, Eq.(2-56) in \cite{BT1987} for an expression of Poisson equation for an axisymmetric disk). In other words, the disk can no longer be taken to be locally dense. In this case, both, the radial and the vertical terms in the disk Poisson equation, need to be included for a correct formulation of the problem, see \cite{SJ2020Jeans} for details.

As a physical consequence of including both the R and z terms in the Poisson equation (in a thick or low density disk); the planar and vertical dynamics in the disk are coupled.
This is in contrast to the thin disk case where the R term can be dropped and then the planar and vertical dynamics
are decoupled (Section \ref{sec:3.1}).

\medskip

\noindent {\bf Disk-alone case: Thick or a low density disk}

Consider the disk-alone case first. The R term in the disk Poisson equation may be written in terms of the derivative of the rotation velocity, $ V_c$, and is non-zero for a non-flat rotation curve. This ignores the cross terms in the radial Jeans equation -- this is justified for $z<1$ in a galactic disk, as discussed later, after Eq.(\ref{eq4.11}), in this section.
 Thus the inclusion of the R term changes the solution for the disk vertical distribution from  the standard sech$^2$ solution (Eq. \ref{eq3.5}) obtained  for an isothermal, one-component disk. This was shown for the Galaxy (see Model A in \cite{SJ2020Jeans}) for which the resulting mid-plane disk density, $\rho_{0d}$, and HWHM are shown to be different from the values for these for the standard one-component, isothermal sech$^2$ solution by up to 10\% in the inner Galaxy (where the R term is non-zero since the rotation curve is not flat). It is easy to see how the dynamics along $R$ and $z$ is coupled here via the Poisson equation. A more general treatment for the R-z coupling was done for the one component disk-alone case by
\cite{SJ2020Jeans} where the full Jeans equations were used that contained the cross terms and also the additional terms involving the radial and azimuthal velocities (see Section \ref{sec:4.4.1} for details).

\medskip
 
\noindent {\bf Multi-component disk plus halo case: Formulation of equations for a thick disk}

Let us now compare how the joint Poisson equation behaves in the thin and thick disk cases for the multi-component disk plus halo system. Recall that for a thin disk, only the z term of the  Poisson equation for the disk needs to be retained (Section \ref{sec:3.1}). In this case, the halo drops out of the superposition or the joint Poisson equation for the disk and the halo, as discussed in  Section \ref{sec:4.2.1}.
The joint Poisson equation then reduces to that for the disk-alone case, namely Eq.(\ref{eq4.3}),
 where $\rho$ denotes the redistributed disk density. This redistribution occurs because
the two structural components; namely, the disk and the dark matter halo,
 are still gravitationally coupled, through the equation of hydrostatic equilibrium 
(see Eq. (\ref{eq4.1}), Section \ref{sec:4.2.1}). 

Next, let us consider formulation of equations for the disk vertical structure of a thick or a low density disk.
In contrast to the thin disk case, here  the R term 
 in the disk Poisson equation needs to be included for a correct formulation of the equations. In this case, the joint Poisson equation for the gravitationally coupled disk and halo system is  given by:

\begin{equation} 
\frac{1}{R}\frac{\partial}{\partial R}\left(R\frac{\partial \Phi_{\mathrm{total}}}{\partial R}\right)+\frac{\partial^{2}\Phi_{\mathrm{total}}}{\partial z^{2}} = 4\pi G\left(\rho_{\mathrm{s}}+\rho_{\mathrm{HI}}+ \rho_{\mathrm{H_2}}+ \rho_{\mathrm{h}}\right) 
        \label{eq4.8}
        \tag{4.8}
\end{equation}        

\noindent where $\Phi_{total} = \Phi_{s}+\Phi_{HI}+\Phi_{H_2}+\Phi_{h}$ is the total potential, where the first three terms are due to the three disk components considered (as defined in Section \ref{sec:4.2}).
In this case, the disk and halo 
together give the net gravitational force that keeps the disk in a rotational equilibrium. 
In this case, the net radial term in the joint Poisson equation is written in terms of the radial derivative of the net circular velocity, $V_c$, at the point R under consideration, as follows:

\begin{equation}
\mathrm { \frac{1}{R}\frac{\partial}{\partial R}\left(R\frac{\partial \Phi_{total}}{\partial R}\right) = \frac{1}{R}  \frac {\partial {V_c}^2}{\partial R}  }
      \label{eq4.9}
      \tag{4.9}
\end{equation}      

\noindent where $V_c$  includes the contribution of disk and halo. Thus, in the thick disk case, the Poisson equation for the disk and halo remain coupled through the radial, and also, the z terms (see Eq.(\ref{eq4.8})). 

We stress that  $V_c$ as defined above is taken to be equal to the observed rotational velocity, for simplicity. In writing the disk contribution to the  R term of the  Poisson equation in Eq.(\ref{eq4.9}), the contribution of the cross terms and the velocity dispersion terms in the radial Jeans equation (Eq. 4.29 a from \cite{BT1987}) is neglected. We note that this is equivalent to neglecting the asymmetric drift (e.g., \cite{BT1987}). This point is discussed in detail later, after Eq.(\ref{eq4.11}).  
Hence $V_c$ as defined by Eq. (\ref{eq4.9}) is taken to denote the observed rotational velocity in the rest of the review; with the exception of Section \ref{sec:4.4.1}, where the complete Jeans equations are used to write the R and z derivatives of the potential.

Another point to note is that, strictly speaking,
 $V_c$ is a function of both R and z; and is specified as $V_c (R,z)$. However, typically, the observed rotation velocity at each $R$ is given as the intensity-weighted average of $V_c(R,z)$ along the vertical direction, which is a function of R alone. Thus, usually for simplicity, the rotation velocity at a given $R$, $V_c(R)$, is   taken to be independent of z. 
Therefore, the term on the r.h.s. of Eq.(\ref{eq4.9}) is a function of $R$ only, and its value is determined from the gradient of the observed rotation curve. The general case where the rotation velocity, $V_c$ is a function of $z$ will be considered later in this section.

 The joint Poisson equation in terms of the observed rotation velocity is obtained by substituting Eq.(\ref{eq4.9}) into Eq.(\ref{eq4.8}), and is given to be:

\begin{equation}
\mathrm {(\frac {{\partial}^2 \Phi_{total}}{\partial z^2}) +  \frac {1}{R} \frac {\partial {V_c}^2}{\partial R}    = 4 \pi G (\rho_{s}+ \rho_{HI}+\rho_{H_2}+\rho_{h}) } 
    \label{eq4.10}
    \tag{4.10}
\end{equation}

It is interesting that the rotation curve features in the formulation of the equations in the thick disk case. In contrast, for the thin disk case, the $R$ term or effectively the rotation curve does not feature in the formulation of the equations. This is true, both, for the one-component disk-alone case (see Section \ref{sec:3.1}) because the R term in the disk Poisson equation drops out; and also for the multi-component disk plus halo thin disk case  (Section \ref{sec:4.2.1}), because the disk and halo Poisson equations are effectively decoupled. Hence the halo R term does not feature in the formulation of the equations for the vertical density distribution in the latter case.

In a multi-component galactic disk plus halo system, the gravitational force due to the other disk components as well as that due to the halo, is included in the equation of hydrostatic balance for each disk component. This can re-arrange or modify the $\rho$ and $\Phi$ of the particular disk component (Section \ref{sec:4.2}).
 
 The equation of hydrostatic balance for each disk component in the coupled multi-component disk plus halo model remains the
same as in  the thin disk case, namely:

\begin{equation}
\mathrm{
\frac{\partial }{\partial z}(\rho_{i}\langle(v_{z}^{2})_{i}\rangle)+\rho_{i}\frac{\partial \Phi_{\mathrm{total}}}{\partial z} = 0  }
   \label{eq4.11}
    \tag{4.11}
\end{equation}

\noindent where the various quantities are as defined for the thin disk case ({\ref{sec:4.2.1}}).

While writing these expressions (Eq.(\ref{eq4.9}), Eq.(\ref{eq4.10}), and Eq.(\ref{eq4.11}); truly speaking, the radial and $z$ derivative of the potential should be given in terms of the full Jeans equations (e.g., Eq. 4.29a  and  Eq. 4.29c respectively, from \cite{BT1987}). The latter also include terms containing the cross terms and the  velocity dispersion, which have been dropped here, for simplicity.
The ratio of the terms containing the cross terms and the velocity dispersion (which are dropped) and the other terms (which are kept) is  $\sim z^2 /R R_D $, which is $\sim 0.01$ so long as z $<$ 1 kpc. This has been shown
for the radial Jeans equation \cite{1965gast.book..455O,MihalasRoutly, Bahcall1984paper1}; and, also the vertical Jeans equation (\cite{MihalasRoutly, BT1987, SJ2019}; also, see Section \ref{sec:3.1}). 
Recall that, here the radial term in the disk Poisson equation has been included so as to treat a low density or a thick disk. 
But, interestingly, even when the disk is thick; so long as z $<$ 1 kpc, the cross terms can be still ignored from the radial and vertical Jeans equations, as shown above.
Hence the above simplified forms of the Poisson equation (Eq.(\ref{eq4.10})) and the equation of hydrostatic balance used (Eq.(\ref{eq4.11}))  for a thick disk case are valid for typical real galactic disks  (since these have a thickness $<$ 1 kpc).
Dropping these terms as done for the R term of the disk Poisson equation (Eq. 4.29 a from \cite{BT1987}) is equivalent to neglecting the asymmetric drift. Hence, we stress again that $V_c$ in Eq. (\ref{eq4.9}) denotes the observed rotation velocity.
 
However, as shown in  a general model (\cite{SJ2020Jeans}; also,  Section \ref{sec:4.4.1}), the inclusion of the cross terms and the terms involving velocity dispersion, and the tilt of the velocity ellipsoid;
 makes a significant difference at large $R$, or in very low density or high $z$ region. In these regimes, these additional terms need to be included for the correct formulation of the equations.

On combining  Eq. (\ref{eq4.10}) and Eq.(\ref{eq4.11}), the coupled,
joint Poisson-hydrostatic balance equation for the multi-component disk plus halo system, for the thick disk case, is obtained to be:

\begin{equation}
\mathrm{
\langle(v^{2}_{z})_{i}\rangle \frac{\partial}{\partial z}\left(\frac{1}{\rho_{i}}\frac{\partial \rho_{i}}{\partial z}\right)  = -4\pi G\left(\rho_{\mathrm{s}}+\rho_{\mathrm{HI}}+ +\rho_{\mathrm{H_2}}+\rho_{\mathrm{h}}\right) \\                                                                                                +\frac{1}{R}\frac{\partial V_c^2}{\partial R}} 
  \label{eq4.12}
  \tag{4.12}
\end{equation}

The  
solution of this coupled set of equations gives the self-consistent vertical density distribution for the thick disk case, for each of the coupled disk components (i= stars, HI and H$_2$, respectively), with the disk being under the gravitational field of the halo. The numerical solution for these is obtained following  the same procedure, including  the boundary conditions at the mid-plane; as done for the multi-component, thin disk (as in \cite{NJ2002}; also, see Section \ref{sec:4.2.2}, and Eq.(\ref{eq4.5})).

\noindent {\bf One-dimensional, local calculation}

A somewhat subtle technical point is that despite the inclusion of the radial term, one can still treat the surface density, $\Sigma(R)$, to be  constant at a given $R$ to obtain $(\rho_0)_i$, the modified mid-plane density. This is, in fact, one of the boundary conditions (see Eq.(\ref{eq4.5})) while obtaining the numerical solution to Eq. (\ref{eq4.12}).
For a disk in rotational equilibrium, the net $R$ term in the Poisson equation is a constant at a given R and its value depends on the gradient of $V_c$ at $R$ alone (although the rotational equilibrium takes into account the mass distribution at non-local regions inside of $R$.)
Thus, even though the $R$ and the $z$ terms in the Poisson equation are coupled, the inclusion of the radial term only changes the effective gravity, 
and hence the $z$ motion, locally at a given $R$.  
Thus, despite the inclusion of the radial term, 
 the surface density at a given $R$ remains unchanged.
Hence, the constant surface density  can still be used as a constrain to determine one of the boundary conditions, namely,  $(\rho_0)_i$.
The other boundary condition remains as $d \rho/dz=0$ at the mid-plane (or, $z=0$), as before (\cite{NJ2002}; also,  Section \ref{sec:4.2.2}).

Here too, the problem reduces to a one-dimensional, local one of the determination of $\rho$ along $z$; as in the thin disk case  (see section \ref{sec:4.2.1}). Hence, in the thick disk case, the vertical density distribution of a given disk component in the coupled case is only affected by the local values of the parameters of the other disk components, at a given $R$ -- as was also shown to be true for the thin disk case (see Section \ref{sec:4.2.1}).

\noindent {\bf Effect of non-zero rotation velocity on scale height}

Note that the presence of the last term on the r.h.s of Eq.(\ref{eq4.12}) opposes the effect of the net gravitational force due to the local mass density if the velocity gradient is positive.
Thus, in the rising part of the rotation curve, the last term opposes the disk gravity and this results in a puffed up distribution or a higher scale height. In contrast, in a region of falling rotation curve, the opposite is true; namely, the  inclusion of the last term 
adds to the self-gravity term. Hence, it reduces the resulting scale height. This will be illustrated by example in case of real galaxies (see Section \ref{sec:4.3.2}).

\medskip

\noindent {\bf Application to galaxies with a rotational lag:}

A careful study by various authors has revealed that there is a variation with $z$ in the rotation velocity in some galaxies; also known as the rotational lag, with a lower rotational velocity at higher $z$ values (e.g., \cite{Swaters1997, Fraternali2002,MatthewsW2003,Fraternali2005,Boomsma2008,Kamphuis2007}). 
Here the gas rotational velocity has a negative vertical gradient with $z$. In this case, Eq.(\ref{eq4.9}) is modified as follows. Assume that the lag in the rotational velocity can be specified as $\mathrm{V_c(R,z)= V_c(R,0) + C_{lag}z}$; where $\mathrm{C_{lag}= dV_c/dz}$ ($<0$) is the linear decrease, or lag in the rotation velocity with $z$. Further,  if the lag is the same at all radii, that is, $\mathrm{dC_{lag}/dR=0}$; then the r.h.s. of Eq.(\ref{eq4.9}) is modified and contains one extra term, $\mathrm{2(z/R) C_{lag} (dV_c/dR)}$. The corresponding modified coupled, joint Poisson-hydrostatic balance equation (Eq.(\ref{eq4.12}))  can be  solved to obtain the
resulting density distribution. It can be shown that the typical observed lag of a few 
km s$^{-1}$ kpc$^{-1}$, results in very small ($<$ few $\%$) change in the vertical density distribution.{\footnote {S. Sarkar, private communication. I thank S. Sarkar for checking this.}} However, we point out that this discussion is based on the observations for HI gas and ionized gas. It is not clear from observations whether the stars obey any extra-planar rotational lag, and whether its magnitude is the same as that observed for the HI gas.

\subsubsection{Thick, or low density disk case: Applications\label{sec:4.3.2}}

The above formulation of the  multi-component, thick disk plus halo case was general. In this subsection, it is applied to three specific cases:
namely, 
the Milky Way outer disk; the dwarf galaxies; and
a low surface brightness galaxy, UGC 7321. These are physically distinct cases and cover important situations of interest in real galaxies. 
The details of the equations are slightly different in each case
due to the different physical conditions in these three systems (e.g., the flat rotation curve in the Milky Way).  
Further, the results obtained cover different aspects studied in each case  (e.g., halo profile in the Milky Way). 
These three cases are presented next.

\medskip

\noindent {\bf Application to the Milky Way outer disk: Halo density profile}

The formulation involving a thick disk for the multi-component disk plus model  was studied for the outer disk of  Milky Way by Narayan et al (2005) \cite{NJ2005}, to obtain the HI gas scale heights.
The aim of this paper was to use the observed HI scale heights and the observed rotation curve as two independent constraints to determine the best-fit halo parameters. This approach will be discussed in more detail in Section \ref{sec:6}.
This study involves one of the early applications  of the general thick disk case, as developed in Section \ref{sec:4.3.1} (Eq. (\ref{eq4.12})). 
Here, the net radial term in the Poisson equation due to the disk and halo is zero 
(Eq.(\ref{eq4.9})), because the observed rotation curve in the Milky Way is flat (see \cite{NJ2005} for details). 

Recall that for the thin disk, on the other hand, the radial term in the disk Poisson equation is zero (\cite{NJ2002}, Section \ref{sec:4.2.1}).
In that case, therefore, the disk and halo Poisson equations are decoupled, as discussed earlier (Section \ref{sec:4.2.1}); while in the thick disk case, even when the R term is zero, as for a flat rotation curve; the disk and the halo Poisson equations remain coupled (see Eq. (\ref{eq4.10})). Therefore, 
the resulting coupled, joint Poisson-hydrostatic balance equations (Eqs. 4 and 5 in  \cite{NJ2005}), and the resulting solutions, are different in these two cases.
The thin disk
assumption would overestimate the scale height by as much as 10\%. Since the resulting HI scale heights are used to constrain the halo parameters, it is important to obtain accurate model values for the
scale heights. Hence \cite{NJ2005} use the general approach that considers a thick disk.

For a thick disk, and a flat rotation curve, the net coupled, Joint Poisson-hydrostatic balance equations for a thick disk, Eq. (\ref{eq4.12}), reduce to:

\begin{equation} 
\mathrm{\langle(v^{2}_{z})_{i}\rangle \frac{\partial}{\partial z}\left(\frac{1}{\rho_{i}}\frac{\partial \rho_{i}}{\partial z}\right)  = -4\pi G\left(\rho_{\mathrm{s}}+\rho_{\mathrm{HI}}+ +\rho_{\mathrm{H_2}}+\rho_{\mathrm{h}}\right) \\ }
      \label{eq4.13}
      \tag{4.13}
\end{equation}

For the given set of input parameters for the Galaxy, the 
 solution of these three coupled equations is obtained numerically using the same iterative procedure and the boundary conditions as in \cite{NJ2002}; also, see section \ref{sec:4.2.2}. This gives the self-consistent density distribution $\rho_i(z)$, and the corresponding HWHM denoting the thickness or vertical scale height at a given radius for each disk component. These results are compared with the observed HI scale height values.

To obtain the best-fit halo parameters, \cite{NJ2005} try a large range of values for the halo parameters while solving Eq.
(\ref{eq4.13}). They assume a four-parameter halo model described by the following density profile \cite{deZP1988}:

\begin{equation}
\mathrm{  \rho(R,z) = \frac {\rho_{0h} (q)}{(1+ \frac {m^2}{R_c^2(q)})^p} } 
   \label{eq4.14}
   \tag{4.14}
\end{equation}

\noindent where $\rho_{0h}$ is the central halo density, $R_c(q)$ is the core radius, $q$ is the axis ratio and $p$ is the index. By definition, $m^2= R^2+ z^2/q^2$ represents the surfaces of concentric ellipsoids. 
Note that $ q= 1$ gives a spherical halo, while $q=c/a < 1$ and $q=c/a >1$ gives an oblate and a prolate halo respectively. Here $a$ and $c$ are the semi-major axis in the disk plane, and that along the vertical direction, respectively. Note that for $p=1$ and $q=1$,  Eq. (\ref{eq4.14}) reduces to a pseudo-isothermal density profile (Eq. (\ref{eq4.6})). 

By varying the density index $p$ one can get different halo density profiles.
The choice $p=1$ corresponds to a screened isothermal or a pseudo-isothermal halo, with a density falling off as $r^{-2}$ at large radii ($r >> R_c$), this results in a flat rotation curve at large radii. Here $r$ is the radius in spherical co-ordinates. For $p =1.5$ and $p=2$ the density falls off more steeply: as $r^{-3}$ as in the NFW (Navarro, Frenk \& White) case\cite{NFW1996}; and as $r^{-4}$,  respectively. At large radii, the mass for $p=2$  to a finite value unlike the other two cases.

First, different shapes for the isothermal ($p=1$) case are tried. On varying the shape, $q$, with $p=1$, the values of the parameters ($\rho_{0h} (q), R_c (q)$) to be used in Eq.(\ref{eq4.12}) change and can be obtained in terms of the spherical case values, by imposing two physical conditions: namely, that the mass within a thin spheroidal shell, and the terminal velocity of the halo should be independent of $q$ (see Eq.7, and Fig. 1 in \cite{NJ2005}).  For the fiducial spherical halo, the pseudo-isothermal model (Eq. (\ref{eq4.6})) is used for which $R_c(1)= 5$ kpc and $\rho_{0h} (1)$ = 0.35 $M_{\odot} pc^{-3}$ (as given by \cite{Mera1998}).

\begin{figure}
   \centering
   \includegraphics[height= 3.2 in, width= 3.4 in]{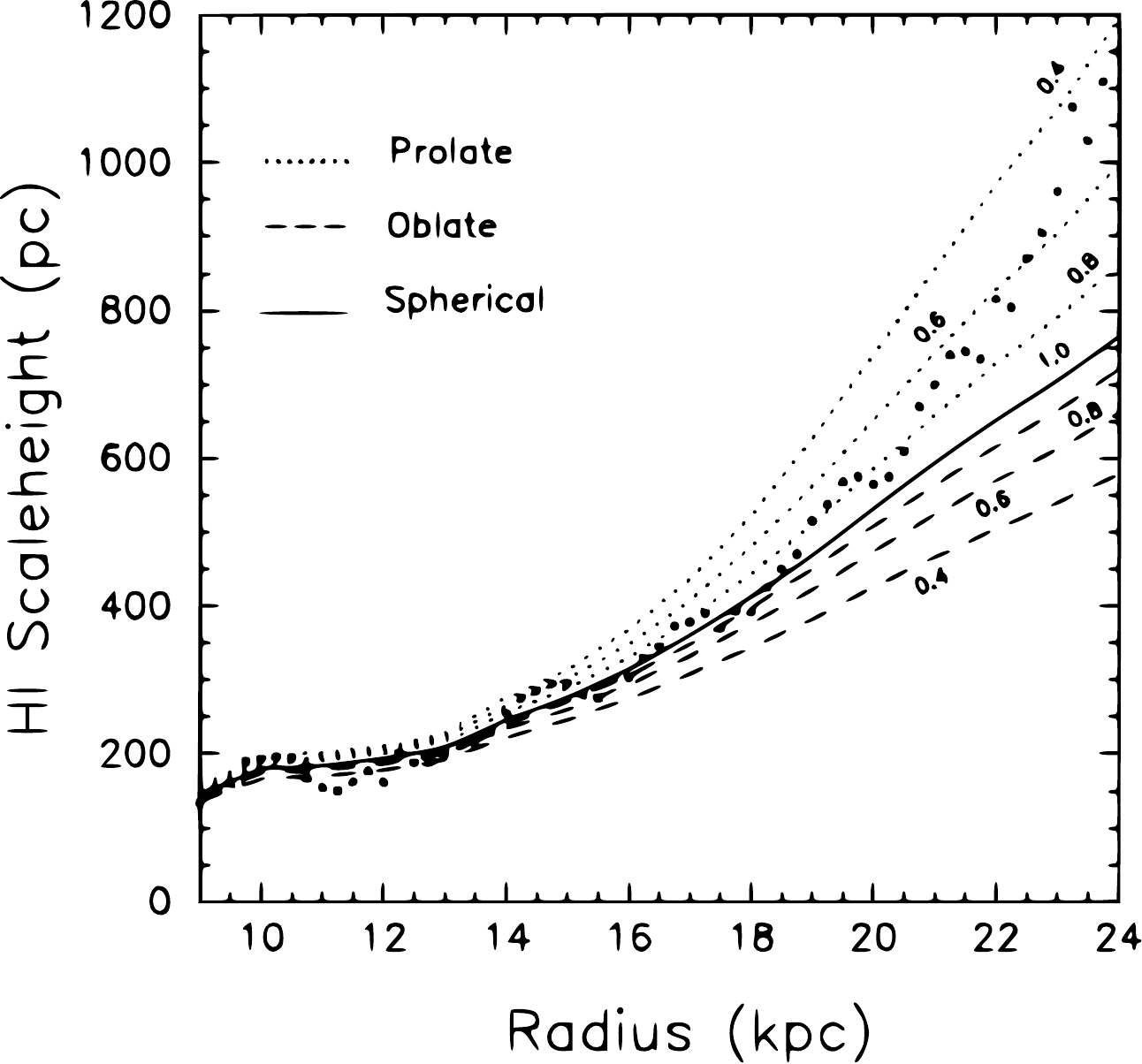}
   \caption{Plot of calculated HI scale height (HWHM) vs. R in the outer Galaxy. The shape of an initially spherical isothermal halo (with  $\rho_0 = 0.035$ $M_{\odot}$ pc$^{-3}$, and $R_c = 5$ kpc) is changed keeping its mass constant.  
    The solid line results for the spherical shape ($q = c/a $= 1); the dashed lines are  for oblate halos ($c/a$ = 0.8, 0.6, 0.4), and the dotted lines are for prolate halos ($a/c$ = 0.8, 0.6, 0.4). The points show the observed values. Note that neither the oblate nor the prolate-shaped halos are clearly favoured by the data. Beyond R= 20 kpc, a prolate halo is favoured, but no single value of prolate shape fits the data over the entire radial range.
    $\: $  {\it Source}: Taken from \cite{NJ2005}} 
 \label{fig.18}
\end{figure}

For a given radius R, assuming a certain density profile for the halo (that is, for a choice of $p$ and $q$), the corresponding $\rho_{0h} (q)$ and $R_c(q)$ are obtained as described above. Further, using the input parameters for stars and gas from observations (as given in Section \ref{sec:4.2.3}); the solution for the three coupled equations (Eq.(\ref{eq4.13})) yields the self-consistent density $\rho_i(z)$, and the HWHM, for each disk component. 
Repeating the same procedure at different radii gives the model scale height curve. The HI scale height results depend sensitively on the choice of the HI velocity dispersion (see the discussion at the end of Section \ref{sec:4.2.4}). A detailed discussion is given in \cite{NJ2005} regarding the choice of the HI dispersion used and its variation with radius. 

The resulting HI scale height curve is shown for the outer region of the Galaxy
in Fig. \ref{fig.18}. 
The solid line is due to spherical shape ($q=1$), the dashed lines are due to oblate halos,
and the dotted lines are due to prolate halos. The observed HI scale height values 
 taken from \cite{Wouter1990} -- given as points in the figure -- are used to fit the results from the model.  The \cite{Wouter1990} data do not have error bars with the data points.  Hence, to estimate a goodness of fit, 
\cite{NJ2005} compute least-square of the model generated curve.
Note that the solid line corresponding to the scale height curve for a spherical halo gives a good agreement with observations up to about $R$=20 kpc. Beyond that the calculated values fall below the observed values, thereby giving a poor fit.

If the halo shape is changed to an oblate shape while keeping the mass constant, physically it is easy to see that the gravitational force normal to the plane will be higher and hence will result in a smaller HI scale height. This would make the fit worse at all radii, and especially at large radii; since the HI thickness is observed to steeply flare with radius in the outer disk, beyond 20 kpc. 
Conversely, a prolate halo would give a better fit. However, given the steeply flaring HI disk, no single prolate shape is found to give a good fit over the entire radial range considered (see Fig. \ref{fig.18}).
A halo that is progressively more prolate with radius 
could explain the observed sharp flaring of HI scale heights in the outer Galaxy, as shown by \cite{2011ApJ...732L...8B}. This will be discussed later in Section \ref{sec:6.3.1}.

\begin{figure}
   \centering
   \includegraphics[height= 2.8 in, width= 3.2 in]{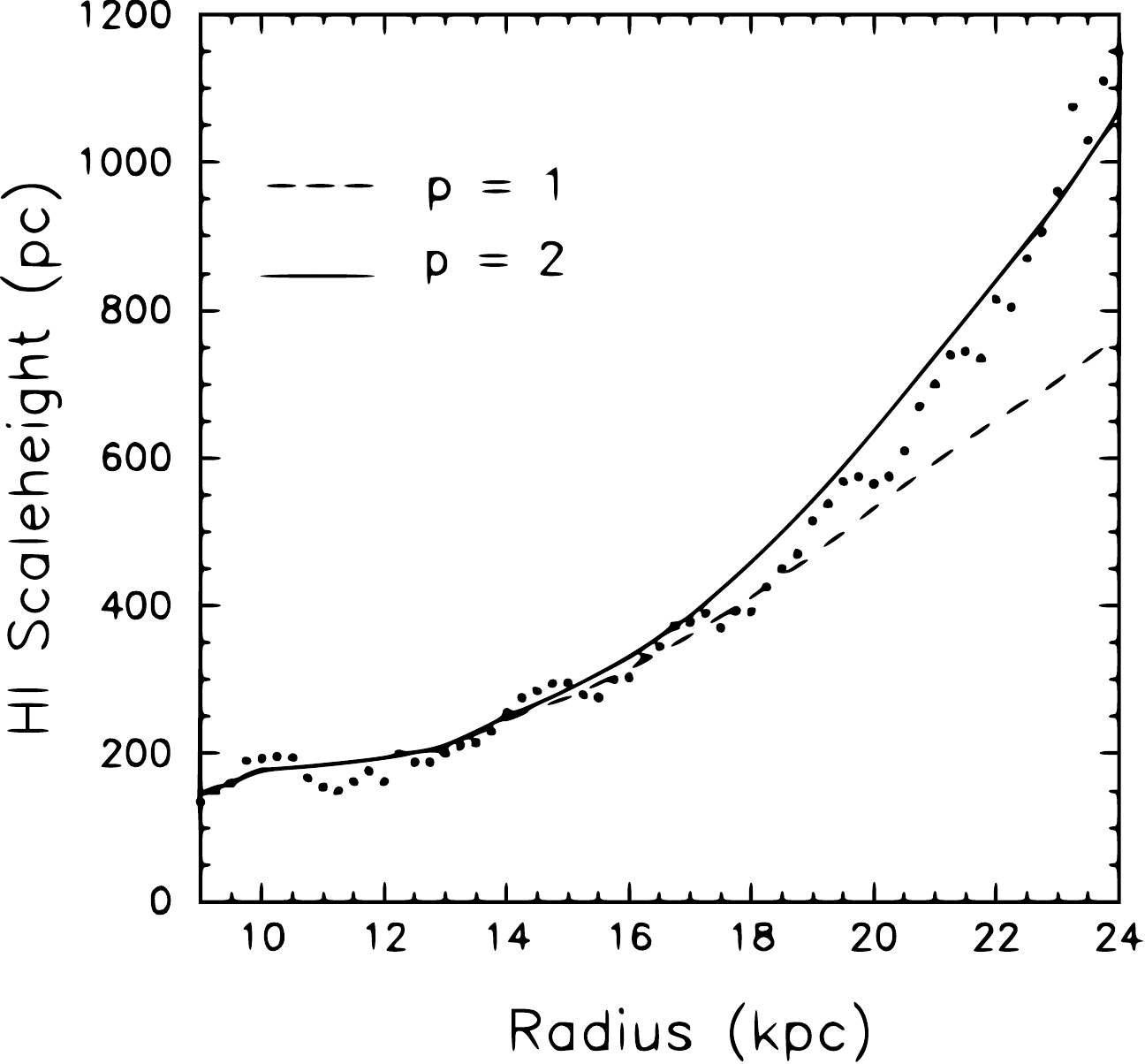}
   \caption{Plot of HI scale height  vs.R in the outer Galaxy. The best fit to the data (shown as points) is obtained for a model with a halo with $p=2$ (solid line) where $p$ is the index in the halo density profile.
For comparison, results for a model with typical isothermal halo ($p=1$) are also shown (dashed line). See the text for details.
$\: $  {\it Source}: Taken from \cite{NJ2005}}    
       \label{fig.19}
\end{figure}

Next, \cite{NJ2005} tried the halo density profiles falling steeper than $r^{-2}$  in the outer region and showed that these provide 
a better fit to the observed HI flaring, and these also fit the observed rotation curve of our Galaxy. To check this,
density profiles for index values of $p=1.5$ and $p=2$ were tried, while the shape was kept spherical for simplicity, and also so as to isolate the effect of density variation of the halo. For each value of $p$, a large range of realistic central density, $\rho_{0h}$,  and core radius, $R_c$, values were chosen to form a grid of $(\rho_{0h}, R_c)$ values. In each case, the model rotation velocity and the HI gas scale height were obtained at each R. This was repeated for the entire range of R values. By simultaneously comparing the model results with the observed rotation curve and the gas scale heights following the detailed procedure as given in \cite{NJ2005}, a good fit to the observed data was found for halos with the central density and the core radius in the range of $\rho_{0h} =  0.035-0.06 M_{\odot} pc^{-3}$  and $R_c = 8- 9.5$ kpc, respectively; with $p=2$ providing a better fit. Of this range, the best fit was obtained for $p=2$, and $\rho_0 =  0.035 M_{\odot} pc^{-3}$  and $R_c = 9.4$ kpc, see Fig. \ref{fig.19}.  
Recall that at large radii ($>>$ $R_c$), $p=2$ corresponds to a density falling off as $r^{-4}$. This steeper fall-off  helps explain the steep flaring in HI scale height with radius in the outer Galaxy, between R = 20-24 kpc. This density profile gives rise to "truncated" or "finite-sized" halos with about 95\% of the total mass within a few 100 kpc. These results were argued to be in a good agreement with the observed data (e.g., from SDSS), and numerical simulations of cosmological evolution of galaxies, as available at that time.

We point out that given the tremendous growth in the observational data as well as studies of numerical simulations, this subject is ripe for a re-look, while adopting the same formalism as in \cite{NJ2005}.

\medskip

\noindent {\bf Application to Dwarf galaxies:}

Another example of an application of the multi-component disk plus halo model in the thick disk case is the study of dwarf galaxies by \cite{BJBrinks2011}. The aim of this study was to theoretically obtain the vertical scale height of atomic hydrogen gas (HI) in four dwarf galaxies, DDO154, HoII, IC 2574, and NGC2366.
The dwarf irregular galaxies are low-mass, low-metallicity, gas-rich objects and constitute the largest number of galaxies in the present-day day observable universe \cite{Mateo1998}. These are of interest for early galaxy evolution in the paradigm of hierarchical galaxy formation.
These are believed to have large HI scale heights \cite{Brinks2002}. The actual value of the gas scale height has implications for different topics: including star formation, and the growth of HI holes in these galaxies.  
Hence, to study the vertical density distribution; the general, thick disk case is applied here Eq.(\ref{eq4.12}). 
The treatment as a thick disk serendipitously allows for the inclusion of the effect of the radial variation of the rotation curve with $R$ (see the related discussion in Section \ref{sec:4.3.1}). 
In contrast, if the disk is taken to be thin, the rotation curve does not feature in the formulation of the equations (see the discussion after Eq.(\ref{eq4.10})).

 In dwarf galaxies, over most of the observed region, the rotation curve is seen to be rising, hence the radial gradient of observed rotational velocity is positive. Thus, the last term on the r.h.s. of Eq.(\ref{eq4.12}) opposes the self-gravity term. Therefore, the inclusion of the radial term here results in slightly higher disk scale heights than in the absence of it, as already explained earlier in Section \ref{sec:4.3.1}.
The molecular gas content is negligible in dwarf galaxies. Hence, the coupled, joint Poisson-hydrostatic balance equation, as 
given by Eq.(\ref{eq4.12}) is applied to a two-component disk; with i= 1, 2 corresponding to stars and HI, respectively. These two joint second-order coupled differential equations for stars and HI are solved simultaneously;  following the same numerical procedure, including the boundary conditions, as in \cite{NJ2002}. The resulting solution gives the self-consistent vertical density distribution for each disk component. 

To do this, the observed input parameters for stars and gas are used (see \cite{BJBrinks2011} for details). 
To solve Eq.(\ref{eq4.12}), the dark matter halo parameters are also needed. The halo is assumed to be spherical with a pseudo-isothermal density profile (Eq. (\ref{eq4.6}). The values of the parameters describing it, namely the central density $\rho_{0h}$ and the core radius, $R_c$, were obtained in the literature by fitting the model rotation curve to the observed rotation curve.
Using the above input parameters,  the self-consistent solution for the vertical density distribution, $\rho(z)$, and the associated HWHM,  were obtained for HI gas. This process is repeated at different radii to get the resulting scale height variation with radius.

Interestingly, this study employs an inverse approach to the one taken in \cite{NJ2005} to study the Milky Way 
   -- as seen earlier in this subsection.
In \cite{NJ2005}, the
 observed 
HI gas scale height values were used  to constrain the best-fit halo parameters, using the same basic model -- except there the observed rotation curve and HI scale heights were used as two independent constraints to obtain the halo parameters.
\begin{figure}
\centering
\includegraphics[height=2.55in, width=2.67in]{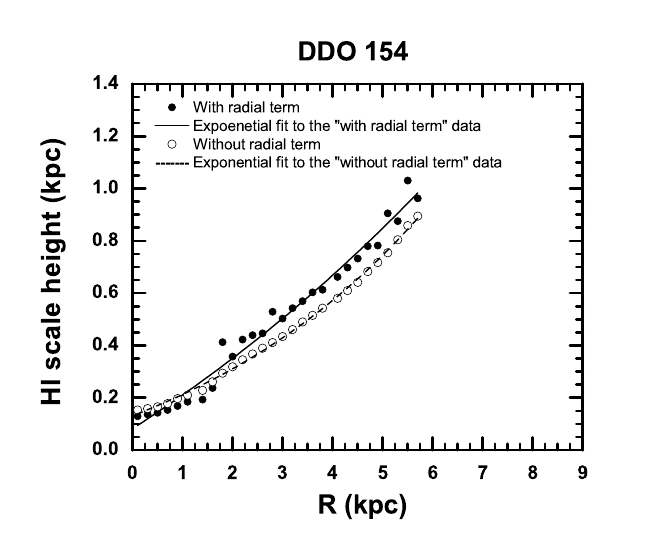}
\medskip 
\includegraphics[height=2.55in, width= 2.67in]{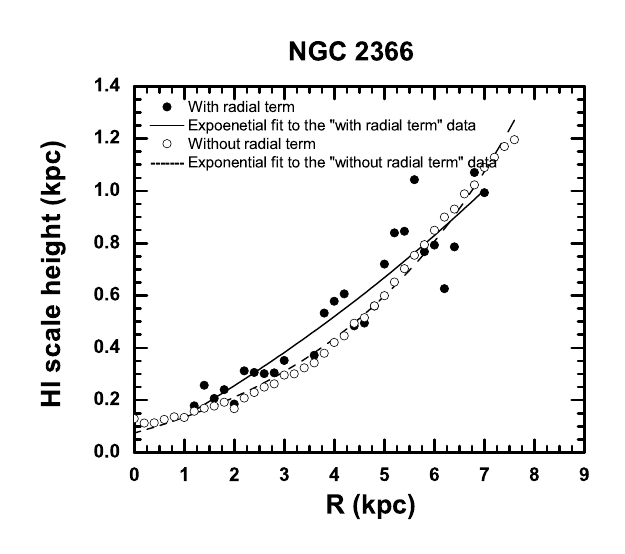} 
\bigskip
\caption{Plot of results for the HI scale height (HWHM) vs. R for two dwarf
 galaxies studied: DDO 154 and NGC 2366. The solid line and a dashed line in each case correspond to the thick disk treatment: that includes the variation of rotation
 curve with radius, and one that assumes a flat rotation curve, respectively. The two show a small difference -- see the text for details. The scale height increases
 steadily with radius up to the last point for which data are available. $\: $  {\it Source}: Taken from \cite{BJBrinks2011}}
 \label{fig.20}
\end{figure}

Fig. \ref{fig.20}  shows the resulting HI scale height (HWHM) as a function of $R$ for two of the galaxies studied, DDO 154 and NGC 2366. 
In each case, the results are shown as a solid line and a dashed line: both of which correspond to the thick disk treatment,  one that includes the variation of rotation curve with radius, and the other assuming a flat rotation curve,
respectively (obtained using Eq.(\ref{eq4.12}) and Eq.(\ref{eq4.13}), respectively). 
In each case, the scale height increases steadily with radius 
up to  the last point for which data are available, thus flaring by a factor of few within several disk scale lengths. 
Since the rotation curve is rising over most of the disk,
the resulting scale height values for the thick disk treatment which includes the variation in the rotation curve are slightly higher (by $\sim 10-20 \%$), than those obtained assuming a flat rotation curve, as expected (see the discussion above). 

The  values of HI scale height obtained are in the range of $\sim 200-400$ pc in the inner disk region, increasing up to $\sim$ 600 pc-1 kpc out to the last measured point.
The inclusion of gas gravity and the gravitational effect of halo as done here gives more reliable results for the scale heights - with smaller resulting values, 
than  obtained in the previous studies in the literature (which ignored gas gravity). This is important because the gas scale height value, in turn, is used for deriving other physical quantities, as discussed above.

Measuring the gas scale heights in galaxies is difficult: due to possible contamination by flaring and warps 
(\cite{SA1979,War1988,Wouter1990}; and, 
 due to inclination effects \cite{SA1979}, also see Section \ref{sec:6.5}). As an independent check, \cite{BJBrinks2011} 
compared the derived scale heights with the observed size distribution of type 3 HI holes (roughly spherical cavities that are contained within the gas layer), and found these to be in a good agreement. This is an indirect confirmation of the model scale heights obtained for these galaxies.

However, it should be pointed out  that dwarf irregular galaxies show a large variation of parameters \cite{KormendyF2004}. 
Hence, while the results for the HI gas scale height from this paper can be taken as clear trends, these need to be confirmed for a larger sample of irregular galaxies in a follow-up work.

\medskip

\noindent {\bf Application to a LSB galaxy, UGC 7321:} 

An interesting application of Eq.(\ref{eq4.12}) (which was meant for a thick or a low density disk) is to an edge-on low surface brightness (LSB)
galaxy, UGC 7321 \cite{SJ2019}.  The LSBs form a special class of galaxies that lie at the faint end of the galaxy luminosity function, and are unevolved.  LSBs have a low mass disk, and a  
dark matter halo that dominates from the innermost regions. Thus the LSBs are structurally, and hence dynamically, different from the typical high surface brightness (HSB) galaxies, like the Milky Way. Hence it is of interest to study the vertical distribution of the stellar disk of a LSB galaxy and see how it compares with, say, that of the Milky Way disk. 
 
Interestingly, here, in absolute terms the disk is thin due to the constraining effect of the dominant halo \cite{BJ2013}. Yet, because of the low disk mass and density,  the term denoting self-gravity in the disk Poisson equation does not dominate over the $R$ term. Hence, the $R$ term in the disk Poisson equation has to be included to correctly formulate the problem of the vertical density distribution in the disk (Section \ref{sec:4.3.1}).
 Hence, here too, the self-consistent vertical density distribution is determined by the coupled, joint Poisson-hydrostatic balance equation given by Eq.(\ref{eq4.12}).

UGC 7321 was chosen for this detailed study to obtain the  stellar scale heights because it is one of the LSBs that has been well-studied; so that the various input parameters, such as the stellar and HI gas surface density, and the HI velocity dispersion, were known observationally.  
The central stellar velocity dispersion value  has been estimated assuming the stellar disk to be in pressure equilibrium \cite{Matthews2000}; see \cite{SJ2019} for details of the parameters. Further, this is one of the few LSB galaxies whose vertical stellar structure has been studied observationally. Its vertical scale height has been measured up to $R$= 5.6 kpc,    
and it is found to be increasing with radius \cite{Matthews2000}. Here, the disk scale length, R$_D$ is 2.1 kpc \cite{MatthewsETAL1999}. 

The dark matter halo is assumed to have a pseudo-isothermal density profile (Eq.(\ref{eq4.6})). The net rotation curve obtained by quadratically adding the stellar disk and halo contributions to it, is fitted with the observed rotation curve
from \cite{UsonM2003}{\footnote{L.D. Matthews, personal communication (2009). I thank L.D. Matthews for tabular form of this data.}}.
The best-fit halo parameters are obtained to be $\rho_{0h} =0.126 M_{\odot} pc ^{-3}$ and $R_c$ = 1.4 kpc,  where $\rho_{0h}$ and $R_c$ are the central density and the core  radius
of the halo, respectively. These values indicate a dense and compact halo (as defined by \cite{BJ2013}). 
The molecular gas content in LSBs is negligible. Using the above parameters, Eq.(\ref{eq4.12}) is solved for a two-component case to obtain the self-consistent vertical density distribution  for stars and HI. To do this, the  same numerical procedure, including the boundary conditions, as in \cite{NJ2002} is used.

\noindent {\bf Flaring stellar disk}

The value of R$_v$, the rate at which the stellar velocity dispersion falls exponentially with radius, is obtained by fitting the resulting model scale height values to the observed HWHM vs. R for stars (known up to R=5.6 pc) used as a constraint. This procedure gives the best-fit value for R$_v$ = 3.2 R$_D$ \cite{SJ2019}. The same rate of fall-off is assumed to hold good for radii beyond 5.6 kpc until R=8 kpc when the stellar velocity dispersion falls to the gas dispersion value. Beyond this radius, the stellar velocity dispersion is assumed to saturate because of the physically motivated argument (see \cite{SJ2018}), that the stellar dispersion cannot be less than the dispersion in the gas from which the stars form. The 
model scale height is obtained using the above velocity dispersion values as input. The resulting scale height continues to increase beyond 5.6 kpc, the increase is gradual up to 7 kpc and then it flares rapidly (see Fig. 5 from \cite{SJ2019}). The stellar disk thickness increases (from $\sim$ 200 pc to 680 pc) or by a factor of $\sim 3$ from R = 0 to 10 kpc (see Fig. 5 from \cite{SJ2019}). The resulting HI disk thickness increases (from 130 pc to 820 pc); that is, by a factor of 6 from R= 0 to 10 kpc (see Fig. 6 from \cite{SJ2019}), in good agreement with observations.

\cite{SJ2019} note that the best-fit value of $R_v = 3.2 R_D$) obtained here is at variance with the choice of ($R_v = 2 R_D$) routinely used in the literature. The latter was proposed by \cite{vdkS1981a}; also, see Section \ref{sec:2}.
Therefore, \cite{SJ2019} emphasise that the choice of $\mathrm{R_v = 2 R_D}$  is not physically justified in dynamical modeling of galaxies, and could give erroneous results. This point, along with the relation between the value of rate of fall-off of velocity dispersion, or $\mathrm{R_v/R_D}$, to flaring, is discussed in detail in Section \ref{sec:5}. 

This work clearly shows that although UGC 7321 is dominated by dark matter halo at all radii, the stellar disk shows flaring. Thus, a flaring stellar disk appears to be a generic phenomenon in LSB galaxies as well.

\medskip

\noindent {\bf Density profile of stellar disk and wings at high $z$}

Next, \cite{SJ2019} fit the results from their theoretical model for the self-consistent vertical density distribution for UGC 7321 at different radii, to the function proposed by \cite{vdk1988},  (Eq.(\ref{eq4.7}).
The best-fit value for the parameter $(2/n)$ is obtained; first, as a function of R (over the interval taken for fitting, $\Delta$ z= 200 pc); and then as a  function of $\Delta$ z at a given R.  It is found that the best-fit value of $2/n$ fitted over $\Delta z= 200$ pc changes with $R$ and starts to be close to 2 or greater than 2 (corresponding to n $<$ 1)
from $R$= 5 kpc (or, $\sim 2 $ R$_D$) onward; that is, in the outer disk. This region is entirely dominated by dark matter halo gravity. A similar trend was found for the Galaxy beyond R=18 kpc  by \cite{SJ2018}, who pointed out that this range of $n$ is new and was not considered earlier in the literature (see Section \ref{sec:4.2.6}).
This range of ($n < 1$) corresponds to a broad distribution at high $z$.
Further, at any radius, the exponent $2/n$ increases  with increasing $\Delta z$ values; so that, a single $n$ does not provide a good fit over the entire range of $\Delta z$ considered. Thus $n$ is not a robust parameter.
 This result was also seen for the Galaxy \cite{SJ2018}.
At large $R > 5$ kpc and high z, $n$ is $<1$   (see Fig. 8 from \cite{SJ2019}). Thus, at large R and z,  the stellar    
distribution would be broader than sech$^2$ and would appear as broad wings in the intensity profiles.

Observations show that in UGC 7321 there is an excess emission or wings at high z, that deviate from the distribution at low z. To explain this,
a second disk, with a higher scale height was invoked earlier \cite{Matthews2000}. 

However, as discussed above, \cite{SJ2019} showed that a physically motivated multi-component (single) disk plus halo model  
naturally gives a broader  distribution at high z and high $R$; without the necessity of invoking a second, thicker disk as done by \cite{Matthews2000} for UGC 7321; and also for another LSB galaxy, FGC 1540, by \cite{Kurapati2018}.  \cite{SJ2019} therefore argue that such a second, thicker disk would, in fact, be 
redundant. 

\subsection{General models\label{sec:4.4}}

In this section, we discuss more general models, which allow one to explore even more realistic cases. 
First is  a general model for a thick, isothermal,one-component disk using the complete Jeans equations which effectively includes various kinematical effects.
The second is the
thin, non-isothermal, multi-component case in the field of the halo.  
In both, the idea was to to highlight the effect of a special new physics point considered,
while keeping the problem tractable.
Both these points are shown to significantly affect the vertical density distribution in a galactic disk. 

\medskip

\subsubsection{Model for thick, or low density disk, using complete Jeans equations\label{sec:4.4.1}}

Sarkar \& Jog (2020a)\cite{SJ2020Jeans} consider the general, complete model for an isothermal, one-component, thick or low density disk,  that includes both the
radial and vertical terms in the Poisson equation. These are then written in terms of the full radial and vertical Jeans equations which
 take account of the non-flat observed rotation curve, the random motions, and the cross term
 that indicates the tilted stellar velocity ellipsoid. This is a more general and complete treatment for a thick disk than given in Section \ref{sec:4.3}; where the radial term in the disk Poisson equation was taken into account, and this term was expressed  in terms of the rotational velocity only.
 
This complete approach is applied to the Milky Way, and it is found that these additional kinematical 
effects result in a density distribution that is significantly different from the standard sech$^2$ law.

The cross term  $\overline{v_R v_z}$
where the average is taken over the velocity dispersion, is a component of the velocity ellipsoid tensor whose value is set by the tilt of the velocity ellipsoid w.r.t. the disc plane.
It also indicates the coupling between the radial and vertical motions. 
The terms containing the cross terms are smaller than the other terms in the R and Z Jeans equations  by a factor of $\sim z^2/(R R_D)$, which has a value of a few $\%$ for 
$z < 1$ kpc in the solar neighbourhood ({\cite{Bahcall1984paper1,BT1987}}; and the discussion in Section \ref{sec:4.3.1}). Hence,
typically, these terms have been
neglected in dynamical studies 
for a thin disk -- this is justified near the mid-plane. 
However, recent kinematic data from various observations such as RAVE (Radial Velocity Experiment), SDSS (Sloane digital sky survey), and \textit{Gaia} shows that the stellar velocity ellipsoid is tilted in the meridional plane in the stellar disk \cite{Hagen} and is observed to have an orientation such that it aligns with the spherical polar co-ordinate system centred at the center of the Galaxy. Motivated by this,  \cite{SJ2020Jeans} included the cross terms; and to cover all aspects, also included  the other terms  in the Jeans equations. The inclusion of all these terms is shown to significantly affect the vertical density distribution. 
Some earlier studies had used the full Jeans equations to determine: the vertical force field, the shape of the dark matter halo, and local estimate of dark matter density \cite{BovyT2012,Hagen,Salcedo2016,Wegg2019};
but the effect on the vertical distribution of stars had not been studied.

The formulation of the equations in \cite{SJ2020Jeans} is briefly described next. The complete Poisson equation is used that includes both radial and vertical terms, which is given by:

\begin{equation}
 \mathrm{ {\frac {1}{R}} {\frac{\partial}{\partial R}} ( R  \frac {\partial \Phi}{\partial R}) + \frac {\partial^2 \Phi}{\partial z^2} = 4 \pi G \rho(R,z)}
    \label{eq4.15}
    \tag{4.15}
\end{equation}

To calculate the radial and the vertical gradient of the potential, the complete radial and vertical Jeans equations are used, given as follows (from \cite{BT1987}):

\begin{equation}
\mathrm { \frac{\partial \Phi}{\partial R} = -\frac{1}{\rho}\frac{\partial}{\partial R}(\rho \overline{v^{2}_{R}})-\frac{1}{\rho}\frac{\partial}{\partial z}(\rho \overline{v_{R}v_{z}})-\frac{(\overline{v^{2}_{R}}-\overline{v^{2}_{\phi}})}{R}}  
   \label{eq4.16}
   \tag{4.16}
\end{equation} 

\begin{equation}
\mathrm { \frac{\partial \Phi}{\partial z} = -\frac{1}{\rho R}\frac{\partial}{\partial R}(R \rho \overline{v_{R}v_{z}})-\frac{1}{\rho}\frac{\partial}{\partial z}(\rho \overline{v^{2}_{z}})}
     \label{eq4.17}
      \tag{4.17}
\end{equation}

Here $v_R, v_{\phi}, v_z$ denote velocities along the three axes.
Assuming that there is no net streaming motion along radial and vertical directions in the Galaxy, one can write $\overline{v_{R}^2}=\sigma^{2}_{R}$ and $\overline{v_{z}^2}=\sigma^{2}_{z}$ where $\sigma_{R}$, $\sigma_{z}$ represent velocity dispersions of stars along $R$ and $z$ directions respectively.{\footnote{In the rest of the review, the quantity
    $ {\langle{({\mathit v}_R^2)}\rangle}$ 
    (=$\sigma^{2}_{R}$)
    has been used to denote the mean square velocity dispersion
    along $R$. In this subsection, the notation $\overline{v^{2}_{R}}=\sigma^{2}_{R}$ is used 
    to denote this same quantity. This is for convenience, so as to be able to use the same
notation for the equations formulated by \cite{SJ2020Jeans}, which are used here. Similar 
notation is employed for the $z$ case as well. In this subsection, the velocity dispersions are taken to denote that for the stars (and are given without the subscript $s$).
}} The quantity $\overline{v_{\phi}^{2}}$ can be written as $\overline{v_{\phi}^{2}}=\sigma^{2}_{\phi}+\overline{v_{\phi}}^{2}$, where $\overline{v_{\phi}}$ represents observed mean rotation velocity of stars in the disk and $\sigma_{\phi}$ is the random velocity dispersion along the azimuthal direction.

When the velocity ellipsoid is aligned with the spherical co-ordinate axes centred at the centre of the Galaxy,  then the cross term $\overline{v_R v_z}$
 is given as $(\sigma_R^2 - \sigma_z^2) (z/R)$ \cite{MihalasRoutly, BT1987}. Substituting these in Jeans equations, the latter are simplified. Further, as usual, the surface density  
and the radial stellar velocity dispersion
 are assumed to fall exponentially with radius, with a scale length R$_D$ and R$_v$, respectively (Section \ref{sec:2}).
The ratio of the vertical and azimuthal dispersions in terms of the radial velocity dispersion is defined as
$b_z = \sigma_z^2 / \sigma_R^2$ and $b_{\phi} = \sigma_{\phi}^2 / \sigma_R^2$, respectively, and these are obtained in terms of observed quantities. These ratios are taken to be constants; which effectively means that  R$_v$, the rate of fall-off in velocity dispersion is assumed to be the same for radial and vertical velocity dispersions (Section \ref{sec:4.2.3}).

Using these assumptions and on substituting the Jeans equations in the Poisson equation,  gives the net second-order differential equation describing the self-consistent vertical density distribution, $\rho(z)$, to be:

\begin{equation}
\begin{split}
\frac{{\partial}^{2}\rho}{\partial z^{2}}  & = \frac{-4 \pi G \rho^{2}}{\sigma_{z}^{2}}+\frac{1}{\rho}\left(\frac{\partial\rho}{\partial z}\right)^{2}+\frac{\rho}{\sigma_{z}^{2}}\frac{2\overline{v_{\phi}}}{R}\left(\frac{\partial\overline{v_{\phi}}}{\partial R}\right)\\
                          &   -\frac{z(1-b_{z})}{b_{z}R}\left(\frac{2}{R_{D}}-\frac{2}{R_v}\right)\frac{\partial\rho}{\partial z} \\		
                          &   +\frac{\rho}{b_{z}}\bigg[\frac{2}{RR_v}-\frac{4}{{R_v}^{2}}+\frac{1}{RR_{D}}-\frac{2}{R_{D}R_v}+\frac{4(1-{b_{z}})}{RR_v} \\
                          &   \qquad +\frac{2(1-b_{\phi})}{RR_v}+\frac{(1-{b_{z}})}{RR_{D}}\bigg]  
                          \end{split}
                          \label{eq4.18}  
                          \tag{4.18}          
\end{equation}

The resulting self-consistent density distribution along z can be obtained by solving this equation at a given $R$; following the numerical procedure as in \cite{NJ2002}. Next, the radial variation of $\rho(z)$ is obtained.

It should be pointed out that if only the first two terms on the r.h.s. of Eq.(\ref{eq4.18}) are considered; it reduces to the usual thin, isothermal case with a solution proportional to sech$^2$.
The third term on the r.h.s. arises due to the radial term in the Poisson equation when expressed only in terms of the observed rotation velocity (as in \cite{SJ2019}). Thus the rest of the terms on the r.h.s. of this equation denote the additional terms that arise due to the inclusion of the terms involving velocity dispersion and cross term in the radial and vertical Jeans  equations. The three cases described above have been named as the sech$^2$ law, model A, and model B respectively. These allow a systematic comparison of the most general, complete model considered here (model B) with the other two cases. For the detailed results for these cases, see Table 1 in \cite{SJ2020Jeans} -- a few highlights are given next.

\noindent {\bf Results:}

Consider case A first. The joint Poisson equation-Hydrostatic balance equation for model A is given as follows:

\begin{equation}
\mathrm{
 \frac{\partial^{2}\rho}{\partial z^{2}}   = \frac{-4 \pi G \rho^{2}}{\sigma_{z}^{2}}+\frac{1}{\rho}\left(\frac{\partial\rho}{\partial z}\right)^{2} \\
+\frac{\rho}{\sigma_{z}^{2}}\frac{2\overline{v_{\phi}}}{R}\left(\frac{\partial\overline{v_{\phi}}}{\partial R}\right)}
  \label{eq4.19}
  \tag{4.19}
\end{equation}

The rotation curve, $v_{\phi}$, is taken to be flat at all radii starting from 8.5 kpc and beyond.
On comparing the results from model A with the sech$^2$ case, the two are found to differ by $\sim 10 \%$ upto R $<8.5$ kpc, and are identical beyond that due to the assumption of a flat rotation curve.
The difference at low radii  could be of either sign, which depends on the sign of the gradient of the observed rotation velocity. As expected, a rising rotation curve effectively reduces self-gravity (Section \ref{sec:4.3.1}, see the discussion after Eq.(\ref{eq4.12})); and hence, gives slightly puffed up disks with a lower mid-plane density. Hence, the difference between results from Model A, which has a rising rotation curve, and sech$^2$, differ by a few \%  value (see Table 1 in \cite{SJ2020Jeans}).

On the other hand, the results from the most general, complete case (Model B) differ from the sech$^2$ law at all radii, and the difference is large --
by as much as 30-40 \% at $R = 18-22$ kpc.
This may be explained as follows. In this radial range, the disk density is low, and the vertical distribution is extended \cite{SJ2018}; thus, the radial term in the Poisson equation may not be negligible. Hence, the inclusion of the $R$ term in the disk Poisson equation, and the other kinematic terms in the Jeans equations, has a substantial effect on the density distribution.
Recall that in this region, the rotation curve is taken to be flat. Hence, the difference between the results from these two models at large radii is purely due to the kinematic terms involving the velocity dispersion and the cross terms in the radial and vertical Jeans equations. 

At radii less than 7 kpc, the difference between the mid-plane densities in model B and sech$^2$ case is again quite significant $\leq 20 \%$. This is attributed to the non-flat rotation curve and the stellar  velocity dispersion. 
At an intermediate radial range of 7-12 kpc, the change as seen in Model B from sech$^2$ is small $\sim$ a few \%. By sheer coincidence, the use of sech$^2$ as is routinely done
for the stellar disk in the literature; happens to be valid reasonably  well in the intermediate radial range, including in the solar neighbourhood
($R$ = 8.5 kpc). The changes from sech$^2$ results shown in this radial range do not show any clear pattern with radius. It is the complex interplay among the various kinematical terms present in Eq.(\ref{eq4.18}) that sets the value of the mid-plane density and the difference with the sech$^2$ model results, including the sign difference.

\begin{figure}
\centering
\includegraphics[height=2.5in, width=2.67in]{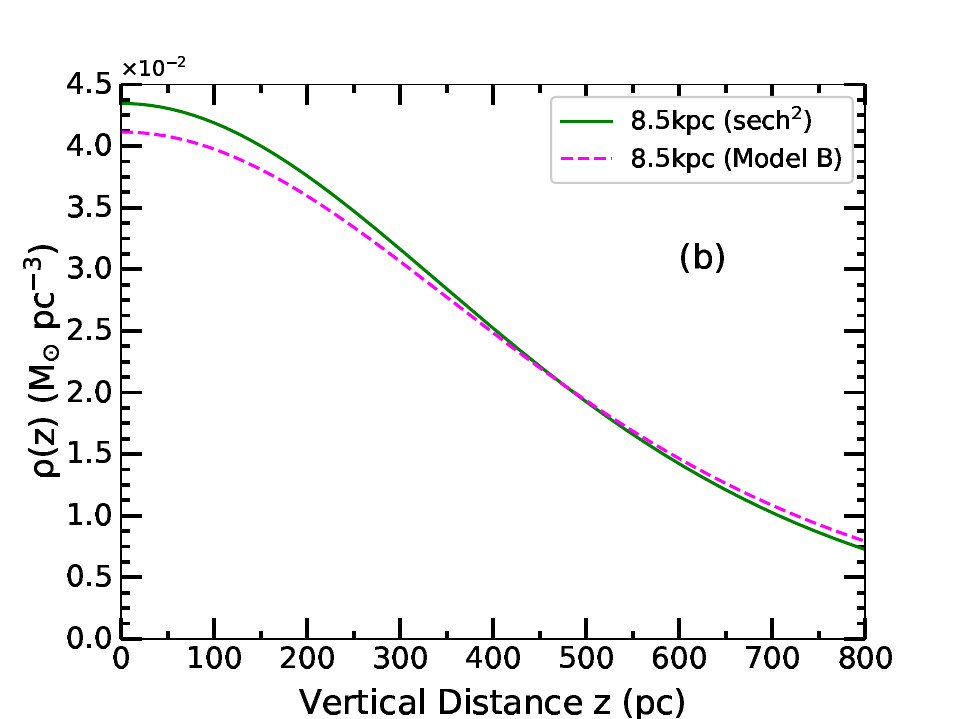}
\medskip 
\includegraphics[height=2.5in, width= 2.67in]{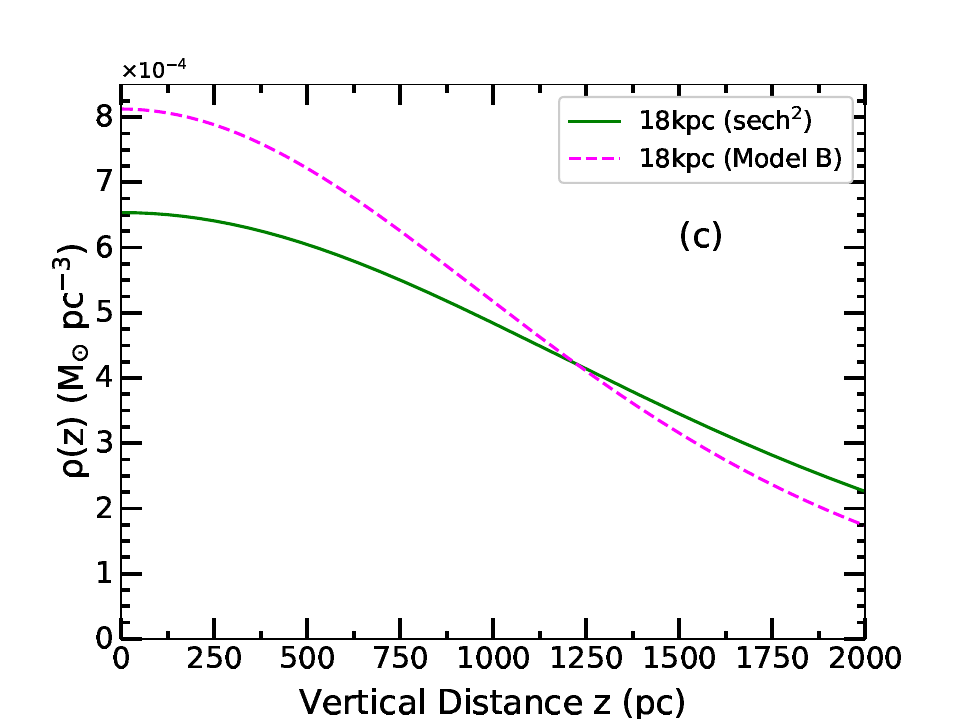} 
\bigskip
\caption{Plot of resulting stellar vertical density, $\rho(z)$ vs. $z$ at $R=8.5$ kpc and $R=18$ kpc, respectively. 
 The solid curve
represents sech$^2$ distribution; and the dashed curve represents the density distribution obtained using the complete, general model (Model B) -- see the text for details. 
The difference between the two distributions is substantial in the outer Galaxy (R=18 kpc), while it is  small in the solar neighbourhood (R=8.5 kpc). This shows the necessity of using complete Poisson and Jeans equations to obtain $\rho(z)$ for stars, in the outer Galaxy or in low density regions.
$\: $  {\it Source}: Taken from \cite{SJ2020Jeans}} 
 \label{fig.21}
\end{figure}

As an example, in Fig \ref{fig.21}, the resulting density distribution for Model B and sech$^2$ model are shown for R=8.5 kpc and 18 kpc (left and right panels, respectively).  Note that 
the density distribution for the general, complete model (Model B) shows a substantial difference from the standard 
sech$^2$ model in the outer Galaxy, at R=18 kpc, while the difference is small at R=8.5 kpc. 
This shows
the importance of using the complete Poisson and Jeans equations while solving for $\rho(z)$ for stars, in the outer Galaxy or in region of low density. 

The conclusion from this study is that
at larger radii and in regions of low density or a thick disk; the above complete, general model should be employed in future studies  to get 
accurate results for the vertical disk distribution. 
The predicted changes in the vertical density profile due to various kinematical effects considered in the model are possible to be checked against the new, accurate data, e.g. from \textit{Gaia} DR3 and DR4 or LAMOST.

A few general comments are as follows. 
First,  
the resulting density profiles for the general case (model B) can be fit  to a function of type sech$^{2/n}$, but then $n$ is fund to vary with $\Delta z$. This trend was
also found for the 
multi-component thin disk case studied by \cite{SJ2018}, also see Section \ref{sec:4.2.6}.
Thus, sech$^{2/n}$ is not valid for the density profile in this case as well.
See Appendix A for a detailed discussion on the vertical density profiles, and scale heights. 
Second, in the general case 
(model B), the disk shows flaring and the flaring over the radial range R=2 to 22 kpc is smaller  by a factor of 2 (reducing from a factor of $\sim$ 16 to 8) compared to that for the standard sech$^2$ law resulting for a thin, stars-alone disk  (see Table 2 in \cite{SJ2020Jeans}).
Thus, flaring of the stellar disk in the outer disk region appears to be a generic result (also, see Section \ref{sec:5} for further details).
                    
\subsubsection{Model for a non-isothermal galactic disk\label{sec:4.4.2}}

The vertical density distribution of stars in a galactic disk is traditionally obtained by assuming an isothermal velocity dispersion of stars, namely dispersion that is constant with $z$, for simplicity.
Only a few papers have examined the dynamical consequences of considering a non-isothermal velocity dispersion \cite{Camm1950,Perry1969}. One reason could be that observationally the velocity dispersion
was not known in detail at different heights. Interestingly, \cite{Camm1950,Perry1969} had studied the vertical density distribution for the case when the velocity dispersion increases with height from the mid-plane, and showed that a stable solution was possible.

Recently, high-quality, high-resolution data: such as obtained using Sloan Digital Sky Survey (SDSS); Large sky area multi-object fiber spectroscopic telescope (LAMOST); Radial Velocity Experiment (RAVE); and  \textit{Gaia}  shows that the vertical velocity dispersion, $ \sigma_z$, for stars 
 increases with $z$, the distance from the mid-plane.{\footnote{In this subsection, $\sigma_z$ is taken to denote the vertical velocity dispersion for stars, in keeping with the notation used in \cite{SJ2020noniso},  and is given without the subscript $s$.}} This is shown for various tracer stars; and even over different metallicity bins; as seen in the solar neighbourhood, and also at larger radii (see \cite{SJ2020noniso}, and references therein).
This shows that the non-isothermal velocity dispersion is a genuine physical feature of the stellar disc. A few of the above papers discuss the vertical variation of the velocity dispersion, particularly for the thin disc of stars \cite{Jing2016,Xia2016,Hagen,Guo2020,Sun2020}.

Although a few papers, based on the observed data, study the effect of non-isothermal velocity dispersion on the measurement of dynamical quantities -- such as, the local dark matter estimate or the total mid-plane density (i.e., the Oort limit), see e.g.,\cite{Garbari2011,Hagen}; they do not study the effect of the non-isothermal dispersion on the vertical stellar density distribution.

\noindent {\bf Formulation of Equations}

 \cite{SJ2020noniso} study the dynamical effect of such non-isothermal velocity dispersion on the self-consistent vertical distribution of thin disk stars in the Galaxy -- both for a single-component stellar disk,  as well as for stars in the multi-component disk plus halo system. In both cases, they find that the inclusion of non-isothermal dispersion has a significant effect on the vertical density distribution, as summarised next. 
Although the numerical results are obtained for the input parameters for the Galaxy; the formulation as well as the trends in results obtained in this work
 are general, and are applicable for a typical galactic disk.

\cite{SJ2020noniso} note that they model the stellar disk to consist of a single component, for simplicity. That is, they treat stars of different ages or metallicity values of the thin disc cumulatively, as is the standard practice in studies of vertical disk structure (e.g., \cite{vdkS1981a}; also, see discussion in Section \ref{sec:4.2.1}).

For the stars-alone disk,     when the vertical velocity dispersion, $\sigma_z$,  is a function of $z$; the joint Poisson-hydrostatic balance equation (Eq.(\ref{eq3.3})) is modified to:

\begin{equation}
\mathrm{
\frac{\sigma^{2}_{z}}{\rho}\frac{\mathrm{\partial}^{2}\rho}{\mathrm{\partial} z^{2}}-\frac{\sigma^{2}_{z}}{\rho^{2}}\left(\frac{\mathrm{\partial}\rho}{\mathrm{\partial}z}\right)^{2}+\frac{1}{\rho}\left(\frac{\mathrm{\partial}\rho}{\mathrm{\partial}z}\right)\frac{\mathrm{\partial}\sigma^{2}_{z}}{\mathrm{\partial}z}+\frac{\mathrm{\partial}^{2}\sigma^{2}_{z}}{\mathrm{\partial}z^{2}}  = -4 \pi G \rho  } \label{eq4.20}
\tag{4.20}
\end{equation}

\noindent 
 
Next, $\sigma_{z}$ is taken to increase linearly with $z$, based on most of the observed data in literature, and is given as:

\begin{equation}
\mathrm{ \sigma_{z}=\sigma_{z,0}+Cz }   
  \label{eq4.21}
  \tag{4.21} 
\end{equation}

\noindent where $C$ is the linear gradient ($\mathrm{d}\sigma_{z}/\mathrm{d}z$) in velocity dispersion along $z$ and $\sigma_{z,0}$ is the dispersion at $z$=0, i.e, the galactic mid-plane. Substituting this expression into the joint Poisson-hydrostatic balance equation (Eq.({\ref{eq4.20})), it gives:

\begin{equation}
\begin{split}
\frac{\mathrm{\partial}^{2}\rho}{\mathrm{\partial}z^{2}} & = \frac{-4\pi G \rho^{2}}{\left(\sigma_{z,0}+Cz\right)^{2}}+\frac{1}{\rho}\left(\frac{\mathrm{\partial}\rho}{\mathrm{\partial}z}\right)^{2} \\
		         &  -\frac{2C}{\left(\sigma_{z,0}+Cz\right)}\left(\frac{\mathrm{\partial}\rho}{\mathrm{\partial}z}\right)-\frac{2C^{2}\rho}{\left(\sigma_{z,0}+Cz\right)^{2}}
\end{split} 
        \label{eq4.22}
        \tag{4.22}
\end{equation}

\noindent The solution of this equation gives the self-consistent, non-isothermal vertical density distribution $\rho(z)$ of a stars-alone disc for a linearly increasing vertical velocity dispersion. 
It is assumed that such a stable solution exists. 
Eq. (\ref{eq4.22}) is solved using the same numerical procedure as in \cite{NJ2002}.
The corresponding isothermal case is solved numerically by setting the gradient $C=0$. Alternately the solution for the isothermal case is simply given by the sech$^2$ law (Eq.(\ref{eq3.4})).

In an analogous way, the joint Poisson-hydrostatic balance equation that governs the vertical distribution of stars 
for a
 multi-component disk plus halo model for the non-isothermal case, is obtained to be:

\begin{equation}
\begin{split}
\frac{\mathrm{\partial}^{2}\rho_{\mathrm{s}}}{\mathrm{\partial}z^{2}} & = \frac{\rho_{\mathrm{s}}}{\left(\sigma_{z,0}+Cz\right)^{2}}\left[-4\pi G\left(\rho_{\mathrm{s}}+\rho_{\mathrm{HI}}+\rho_{\mathrm{H_{2}}}\right)+\frac{\mathrm{\partial}(K_{z})_{\mathrm{h}}}{\mathrm{\partial}z}\right]  \\
					& +\frac{1}{\rho_{\mathrm{s}}}\left(\frac{\mathrm{\partial}\rho_{\mathrm{s}}}{\mathrm{\partial}z}\right)^{2} \\
					& -\frac{2C}{\left(\sigma_{z,0}+Cz\right)}\left(\frac{\mathrm{\partial}\rho_{\mathrm{s}}}{\mathrm{\partial}z}\right)-\frac{2C^{2}\rho_{\mathrm{s}}}{\left(\sigma_{z,0}+Cz\right)^{2}}
\end{split}
         \label{eq4.23}
         \tag{4.23}
\end{equation}

The gas components, HI and $\mathrm{H_{2}}$ gas, both have isothermal dispersions along $z$, as supported by observed data. Therefore, the following equation governs their vertical distribution (as was given by Eq.(\ref{eq4.4}) for a thin disk case, see Section \ref{sec:4.2.1}):

\begin{equation}
\begin{split}
\frac{\mathrm{\partial}^{2}\rho_{i}}{\mathrm{\partial}z^{2}} & 
= \frac{\rho_{i}}{\sigma^{2}_{z,i}}\left[-4\pi G\left(\rho_{\mathrm{s}}+\rho_{\mathrm{HI}}+\rho_{\mathrm{H_{2}}}\right)+\frac{\mathrm{\partial}(K_{z})_{\mathrm{h}}}{\mathrm{\partial}z}\right] \\
                           &   
+\frac{1}{\rho_{i}}\left(\frac{\mathrm{\partial}\rho_{i}}{\mathrm{\partial}z}\right)^{2} 
\end{split}
        \label{eq4.24}
        \tag{4.24}
\end{equation}

\noindent for $i$= HI \& $\mathrm{H_{2}}$ respectively.  Eq.(\ref{eq4.23}) \& Eq.(\ref{eq4.24}) are solved together
 for a  coupled system, following the approach as in \cite{NJ2002}; to obtain $\rho_i(z)$ for $\mathrm{i=s,HI,H_2}$, corresponding to stars, HI and H$_2$, respectively.

The corresponding case when the stellar dispersion is isothermal, is treated
 by setting $C=0$ in Eq. (\ref{eq4.23}) and this equation is solved  together with Eq. (\ref{eq4.24}) to obtain the density distribution for the three disk components in the coupled case. 

\medskip

\noindent {\bf Input parameters and results}

An important input parameter for this work is the non-isothermal velocity dispersion. The mid-plane value of the vertical stellar velocity dispersion,
$\sigma_{z,0}$, is obtained from observations of \cite{LewisF1989}, see the discussion in \cite{SJ2018,SJ2020noniso}. The observed stellar velocity dispersion shows a vertical velocity gradient, as determined by different surveys: such as, TAGS, RAVE \cite{Hagen};
LAMOST DR2, and \textit{Gaia} \cite{Guo2020}; and SDSS and LAMOST surveys \cite{Jing2016},  for the thin disk stars. The observed velocity profiles of thin disk stars in the solar neighbourhood are found to be similar irrespective of the tracer or the survey; hence, the measured linear gradient values are expected to lie in a small range. 
\cite{SJ2020noniso} note that 
the observed gradient values thus obtained cover thin disk populations with a finite range of metallicities and ages; thus, the 
net average velocity gradient value is applicable to the whole thin disk treated as a single component in their work.

In view of these points,
 \cite{SJ2020noniso} adopt the value for the  gradient along z  for the stellar velocity dispersion to be
 $+ 6.7$ km s$^{-1}$ kpc$^{-1}$ as determined from observations for the radial range of 6-10 kpc \cite{Hagen} as the standard value to explore the effect of the non-isothermal velocity dispersion. 
They also consider a higher value of $+ 10$ km s$^{-1}$ kpc$^{-1}$ later for a few cases to illustrate the quantitative variation with the gradient, and state that such higher values are likely to occur at large radii due to external tidal encounters. 
The other observed input parameters for the Galaxy are as given by \cite{SJ2018,SJ2020noniso}.
An interesting point is that during the integration, $\sigma_z$, the $z$ velocity dispersion of stars is kept constant at its value at $z$ =1.5 kpc (the last observed point) and beyond for all higher z points. For comparison, the corresponding isothermal cases are also solved, as described above.

\medskip

\noindent {\bf Results}

The resulting vertical stellar density distribution vs. $z$ for the stars-alone case is plotted for the isothermal case and the non-isothermal case, at R= 8.5 kpc, for a vertical velocity gradient of 6.7 (the observed value) and 10 km s$^{-1}$ kpc $^{-1}$, respectively (see Fig. \ref{fig.22}, left panel).
\begin{figure}
\centering
\includegraphics[height=2.5in, width=2.67in]{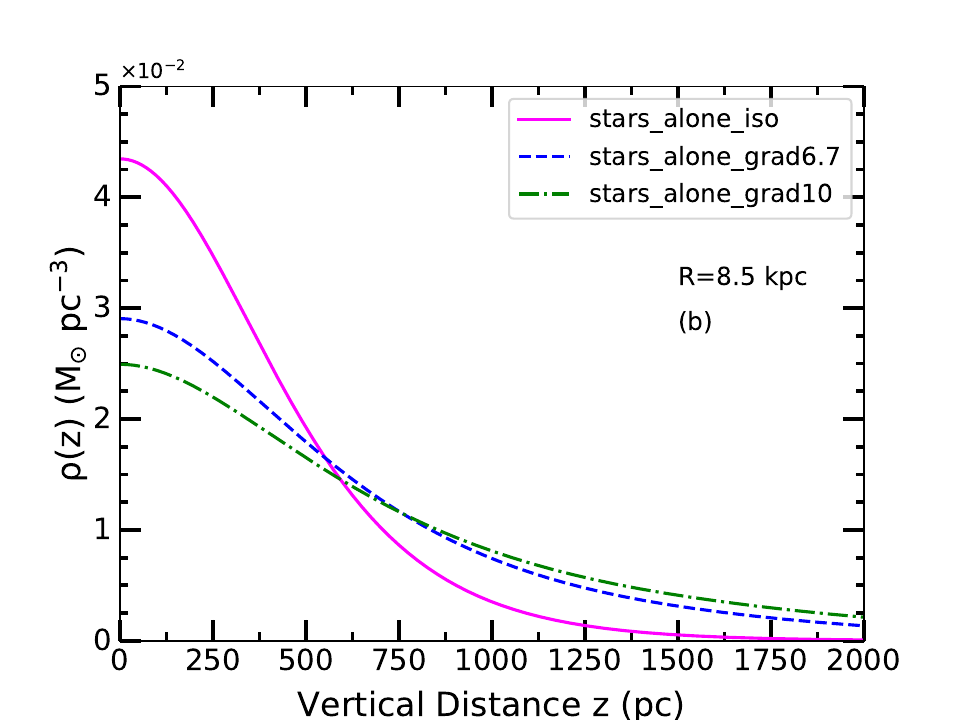}
\medskip 
\includegraphics[height=2.5in, width= 2.67in]{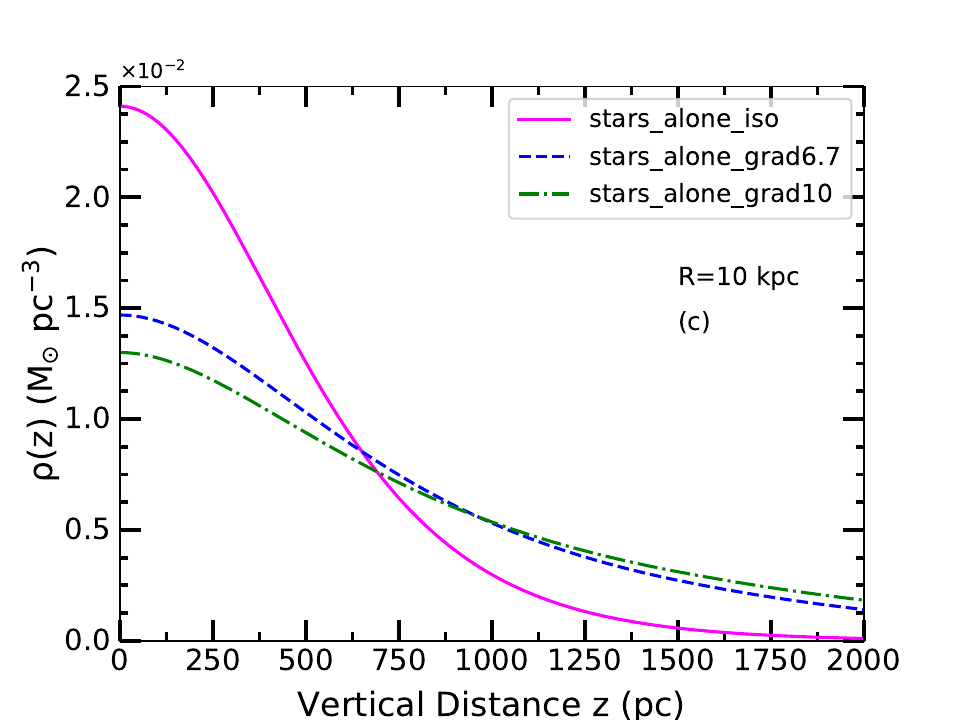} 
\bigskip
\caption{Plot of resulting self-consistent vertical density distribution, $\rho(z)$ versus $z$ for a stars-alone disk, for the isothermal case (shown as a solid curve); and for the non-isothermal cases for a vertical velocity gradient of 6.7 $km s^{-1} kpc^{-1}$ (the observed value) and 10 $km s^{-1} kpc^{-1}$ (shown as dashed and dashed-dot curves, respectively). The results for R=8.5 kpc and 10 kpc are shown in the left and right panels, respectively. The higher vertical pressure in the non-isothermal case results in a lower mid-plane density,$\rho_0$, and a higher scale height (HWHM) compared to the isothermal case. This gives rise to a distribution that is more extended along the vertical direction, or, it is flatter. This effect is stronger for a higher velocity gradient (due to higher pressure), and also at larger radii (due to lower disk self-gravity). $\: $  {\it Source}: Taken from \cite{SJ2020noniso}}   \label{fig.22}
\end{figure}
A similar plot for R= 10 kpc is given in the right panel of Fig. \ref{fig.22}.  

 At each radius, the non-isothermal density distribution, $\rho(z)$, is found to have a lower mid-plane density value, and a higher scale height (HWHM) value, than for the corresponding isothermal $\rho(z)$ distribution; and is thus more extended along the vertical direction, or, is flatter. For example, at R=8.5 kpc, the mid-plane density is lower by 32.6 percent, and the HWHM is higher by 37.1 percent, than the isothermal case for the observed gradient of 6.7 km s$^{-1}$ kpc$^{-1}$.

This effect  
 is  stronger for a higher velocity gradient at a given radius -- compare the dashed and the dashed-dot curves in each of the two panels in Fig. \ref{fig.22}. The density 
distribution for a gradient of 10 km s$^{-1}$ kpc$^{-1}$ is flatter than for the lower gradient value; so that the mid-plane density at R=8.5 kpc is lower by 41.9 percent and the HWHM value is higher by 53.7 percent than the isothermal case, compared to the smaller corresponding changes for the smaller velocity gradient as given above.

The physical reason for the extended distribution is that in the non-isothermal case
with a positive velocity gradient, the  
vertical pressure is higher than the isothermal case at each $z$. Therefore the hydrostatic balance between the self-gravity and the non-isothermal vertical pressure gradient results in a density distribution that is vertically more extended than the isothermal case.
This effect is stronger as $z$ increases, since the dispersion is higher,
hence the density distribution becomes flatter with increasing $z$. Obviously, this effect is stronger for a higher value of the velocity gradient.

Further, for a given velocity gradient, this effect is more prominent at larger radii. This is because  the self-gravity of the stellar disk decreases with increasing radii hence the non-isothermal velocity dispersion can affect the vertical density distribution more (compare the two panels in Fig. \ref{fig.22}). Since the HWHM is higher for the non-isothermal case, and its value increases for a higher radius; the stellar disk shows flaring, which is higher than seen in the isothermal case. Further, this flaring increases with the velocity gradient (see Fig. 2 in \cite{SJ2020noniso}).

The density profile for the single-component non-isothermal case thus shows a deviation from the standard sech$^2$ profile; and this deviation increases at high $z$. Similar deviation is seen at high $z$ in the observed intensity profiles, and is known as the "wing".  This will be discussed more later in this subsection.

 Next, consider the effect of non-isothermal velocity dispersion of stars in a more realistic case, namely on stars in a multi-component disk plus halo system. Here too, on inclusion of the non-isothermal velocity dispersion,
 the results show the same trend;  but the effect of non-isothermal dispersion is reduced due to the opposite, constraining effect of gas and dark matter halo gravity. Fig. \ref{fig.23} shows the resulting  density distribution of stars in a multi-component disk plus halo case vs. z, at R=8.5 kpc, for the non-isothermal case with the observed velocity dispersion gradient of 6.7 km s$^{-1}$ kpc$^{-1}$. Also shown are results  for stars in a coupled multi-component disk plus halo case with isothermal velocity dispersion, and for a single-component stars-alone, non-isothermal case. Interestingly, although the gas and dark matter halo constrain the non-isothermal stellar vertical distribution in the multi-component case (compared to the non-isothermal, stars-alone case), the distribution remains more extended that the isothermal stellar solution of the coupled case. Thus the effect of non-isothermal dispersion is opposite to the constraining effect of the gas and halo gravity, and dominates over it. 
\begin{figure}
\centering
\includegraphics[height=2.8in, width=3.1in]{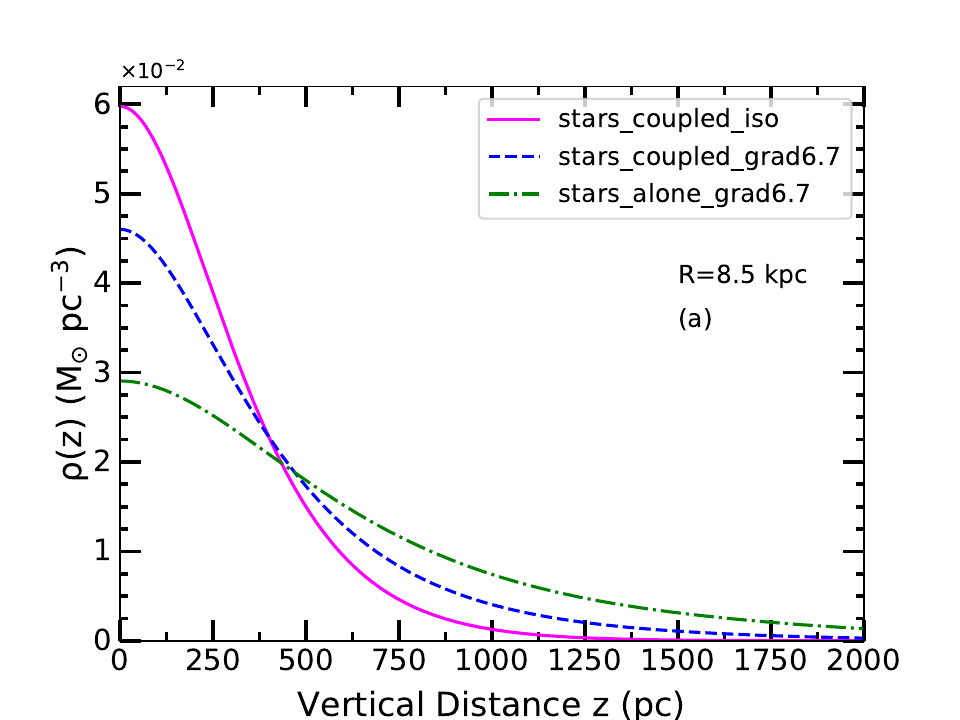}
\caption{The non-isothermal density vertical density distribution of stars versus $z$  is shown at R=8.5 kpc: for the stars-alone case (dashed-dotted curve); for the multi-component disk plus halo case (dashed curve)- both the above obtained using the dispersion gradient of $+ 6.7$ km s$^{-1}$ kpc$^{-1}$; and the stellar distribution versus $z$ for the isothermal case for the 
multi-component disk plus halo model (solid curve). The non-isothermal stellar distribution in the coupled case is more constrained towards the mid-plane, due to the gravity of gas and dark matter halo; compared to the non-isothermal stars-alone case. But it still remains more extended than the isothermal stellar distribution in the coupled case. 
$\: $  {\it Source}: Taken from \cite{SJ2020noniso}} 
 \label{fig.23}
\end{figure}

\cite{SJ2020noniso} show that the reduced mid-plane density of the stellar distribution due to the effect of non-isothermal velocity dispersion, will be reflected in a lower value of the theoretical estimate of the total mid-plane density, or the Oort limit, by $\sim 15 \%$ compared to a similar theoretical estimate obtained in the isothermal case  \cite{SJ2018}.

\medskip

\noindent {\bf Observed "wings" at high z due to non-isothermal velocity dispersion}

The vertical profile of an external galaxy is often found to deviate from the single sech$^2$ function (Section \ref{sec:2}), especially at high $z$. The  excess flux at high $z$ makes the observed distribution look like a wing w.r.t. the fitted sech$^2$ function. To account for this wing, another sech$^2$ profile representing a thicker disk is typically added during the fitting to the observed data to obtain the parameters of the two disks (e.g.,\cite{Yoachim2006}).
\begin{figure}
\centering
\includegraphics[height=2.5in, width=2.67in]{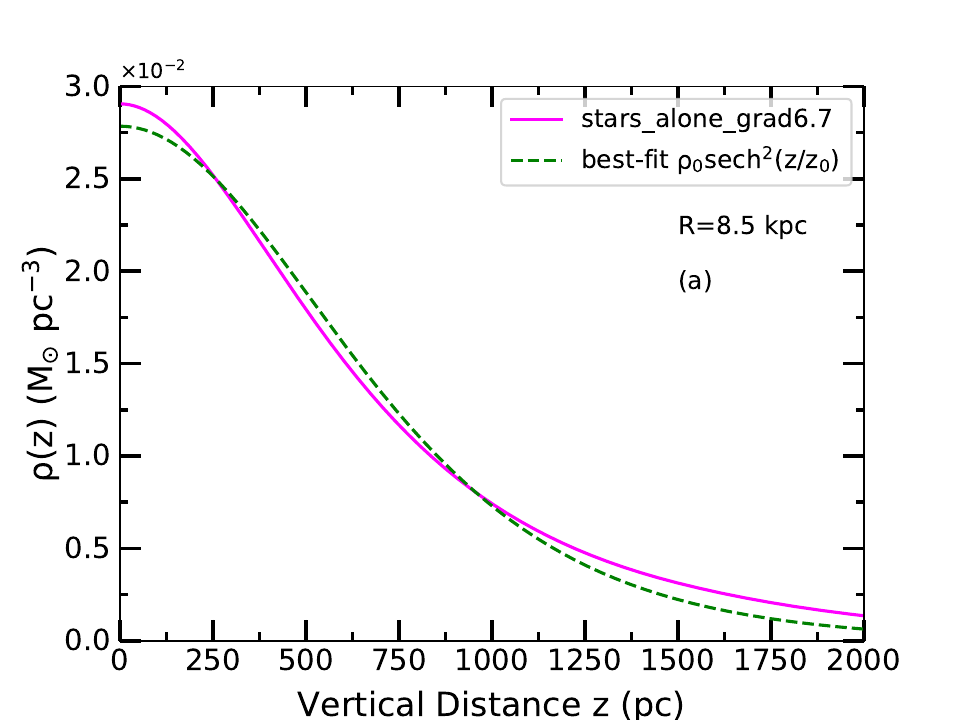}
\medskip 
\includegraphics[height=2.5in, width= 2.67in]{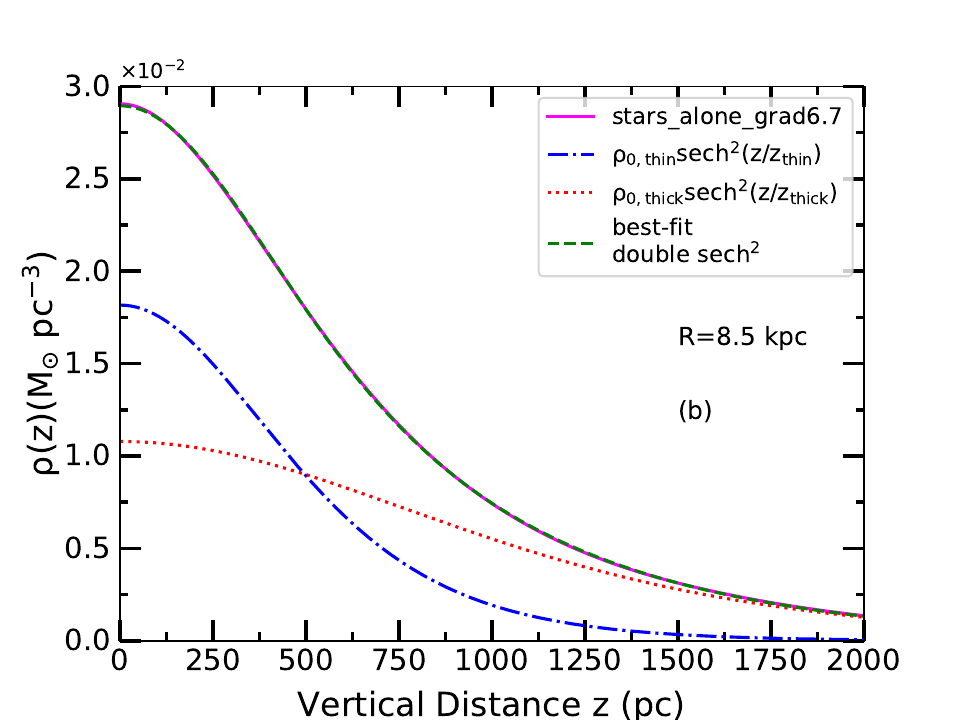} 
\bigskip
\caption{The resulting non-isothermal 
vertical stellar distribution obtained for a velocity gradient of 6.7 km s$^{-1}$ kpc$^{-1}$ for stars-alone case at R= 8.5 kpc: as fitted by a single and double sech$^2$ profiles (left and right panels), respectively. The single sech$^2$ function (left panel)  gives a poor fit to the density distribution at all $z$ values, especially at the "wing" or high $z$ part of the distribution. The right panel shows the best-fitting double sech$^2$ profiles consisting of "a thin and a thick disk" (shown as dashed-dot and dotted curves, respectively) which together give a good fit to the density distribution of a non-isothermal, stars-alone case (solid curve) at all $z$. This shows how an observer might misinterpret a non-isothermal single-component stellar disk density distribution as a superposition of two separate sech$^2$ or isothermal disks. $\: $  {\it Source}: Taken from \cite{SJ2020noniso}}    
 \label{fig.24}
\end{figure}

\cite{SJ2020noniso} note that the non-isothermal $\rho(z)$ distribution for a single stellar disk shows a similar deviation as seen in the observed profiles; and this could well be the cause of the observed "wings" in the intensity profile at high z.
To check this idea, they fit the model non-isothermal $\rho(z)$ distribution obtained at R= 8.5 kpc using the observed velocity gradient value of 6.7 km s$^{-1}$ kpc$^{-1}$ --  first, by a single, and then by double sech$^2$ profiles, see Fig. \ref{fig.24}. 

It is clearly seen that even the best-fit single sech$^2$ function gives a poor fit to the model $\rho(z)$ distribution. On the other hand, the double sech$^2$ profile fits the model results for the density distribution very well at all $z$ (see Fig. \ref{fig.24}).
Hence \cite{SJ2020noniso} caution that if an observed luminosity profile shows a wing at high $z$ and a single sech$^2$ function is found to give a poor fit to the data, this may well be due to a non-isothermal origin of the $\rho(z)$ distribution of stars of a single stellar disk. Thus, evoking a second, thicker disk -- as is routinely done in the literature -- to fit the observed profile that shows a wing at high z, could be redundant. 
A similar conclusion also holds for a non-isothermal stellar disk in the coupled case (see Fig. 5 in \cite{SJ2020noniso}).

Further, even when a genuine distinct {\it Thick disk} is present  -- as evident from its distinct chemical and kinematical properties --  see Section \ref{sec:Intro}; a part of the deviation from the sech$^2$ distribution could be due to a non-isothermal velocity dispersion in the disk. Hence, \cite{SJ2020noniso}
stress that this should be taken into account while obtaining the properties of the two disks by the usual double sech$^2$ fitting procedure.
           
An interesting point is that at large R, such observed wings or a broader distribution at high z, could also arise in the isothermal case for the multi-component disk plus halo model; as was shown for UGC 7321 (\cite{SJ2019}; see Section \ref{sec:4.3.2}). In that case too, it was argued that evoking a second disk to explain the wings could be redundant.

\medskip

\noindent {\bf  Limit of validity of the isothermal assumption:}

To determine the limit of validity of the isothermal assumption normally assumed, \cite{SJ2020noniso} try varying the values of the gradient, and show that  even a small gradient of $+ 2$ or $+ 3 $ km s$^{-1}$ kpc$^{-1}$ is found to cause a change of -9.3 \% and -14\% in the mid-plane density value of the stellar distribution compared to the isothermal case, for the stars-alone case. The corresponding changes are somewhat smaller, namely -8.3\% and -11.7\% due to the opposite, constraining effect of gas and halo in the the multi-component disk plus halo case compared to the corresponding isothermal, stars-alone case. Such small gradients are likely to occur in real galaxies, at least locally. In fact, a strictly constant  -- or isothermal -- velocity dispersion that is typically assumed, may be an idealized case. Hence, \cite{SJ2020noniso} alert that the isothermal assumption, or the resulting sech$^2$ profile, that is routinely used to describe the stellar vertical distribution in the literature, may not be strictly valid even if there is a small velocity gradient of $\sim$ 3 km$^{-1}$ kpc$^{-1}$. This warning is of practical importance since the resulting small changes in
mid-plane density are now possible to be detected by means of accurate measurements of star counts data, e.g. using \textit{Gaia}.

The above results show that the observed non-isothermal stellar vertical velocity dispersion has an important dynamical effect on the vertical disk structure. That is, it results in a self-consistent density distribution that is substantially more extended along the vertical direction, or it is flatter compared to the isothermal case. Hence this effect should be 
included in the future dynamical modelling of a galactic disk. 

\medskip

\subsection{Applications to observational data and theoretical studies in the literature\label{sec:4.5}}

The multi-component disk plus halo model developed above, for the thin disk and the thick disk cases (Section \ref{sec:4.1} to Section \ref{sec:4.4}), has been extensively applied in the literature to theoretical studies, 
and also to analyze observational data. We highlight a few examples here. Because of the physical rigour of the model, and the complete treatment that involves the disk components (stars and gas) and the halo; this model has been preferred in accurate studies of disk vertical structure -- over the usual sech$^2$ model -- even though this involves more work.

\subsubsection{Applications to analyze intensity profiles\label{sec:4.5.1}}

 In an important paper, Comeron et al (2011) \cite{Comeron2011} applied the model by \cite{NJ2002} to analyze the 3.6 $\mu$ images of a sample of 46 edge-on galaxies from the S$^4$G survey \cite{Sheth2010}. The aim was to constrain the thin and the {\it Thick disk} properties of external galaxies. Here a {\it Thick disk} is kinematically and chemically distinct from a thin disk. 
\cite{Comeron2011} point out that it is necessary to follow this rigorous 
model which gives physically motivated density profiles that are more accurate. Next, they match the corresponding model luminosity profiles with observations. They follow this procedure rather than taking the simpler approach of decomposing the observed stellar profiles in terms of a linear superposition of two sech$^2$ disk components, each with a  scale height that remains constant with radius. The latter approach is
usually used in the literature because the trial density functions then can be taken as simple analytical functions. \cite{Comeron2011} multiply their model density profiles with trial values of mass-to-light ratios for the thin and thick disks respectively, and match the total with the observed luminosity profiles from the S$^4$G data, so as to obtain the disk parameters for both the disks. The best-fit to the data indicated
that the thick disks were more massive than was generally believed till then, and could be comparable in mass to the thin disk. From this they proposed a galaxy formation scenario where the thick disk is formed first.

\cite{Comeron2011} stress that an accurate determination of the relative masses of thin and {\it Thick disks} -- which  was made possible because of the use of physically meaningful density profiles obtained -- allowed them to identify the formation mechanism for the thick disk. The formation mechanism of the thick disk had heretofore been a challenging problem.

It should be noted that the halo contribution is not included in this work for the sake of simplicity, and because the uncertainties in the mass-to-light ratio values used are stated to be comparable to the effect of neglect of the dark matter halo in the problem \cite{Comeron2011}. 

In another, earlier application of the model,  \cite{NJ2002ltr}  considered real
 data (from \cite{vdkS1981a}) for intensity profiles in two external galaxies.
By analyzing these correctly (See Section \ref{sec:2.2}  and Section \ref {sec:4.2.7}), \cite{NJ2002ltr}  showed that the disk scale height increases by a factor of 2-3 within the optical disk. Further, by applying the multi-component disk plus halo model, \cite{NJ2002ltr} showed that this flaring in the disk
constraints the parameter R$_v$, to lie in the range of  $\mathrm{2-3 \ R_D }$.

\subsubsection{Applications to the study of gas distribution\label{sec:4.5.2}}

Although this review has mainly focused on the stellar disk; the multi-component disk plus halo model simultaneously gives the resulting self-consistent vertical density distribution including the disk thickness for each disk component, including gas (Section \ref{sec:4.2.2}). 
Getting accurate gas scale height values is important for a number of physical processes, such as star formation.
We give a few examples 
where this model has been applied  so as to obtain the accurate vertical density distribution of gas and the scale height:

 \cite{Patraspiral} has applied the above model, and employed a new, iterative  approach, to simultaneously obtain the best-fit values of the HI velocity dispersion and the gas thickness in a sample of spiral galaxies, by making use of the 3-D HI data.
First, the model results for the vertical density profiles of HI and H$_2$ gas as a function of radius are obtained, using the multi-component disk plus halo model in the thick disk case (Section \ref{sec:4.3}). 
To do this, the  observed values of the stellar and gas surface densities in the literature  as well as the H$_2$ velocity dispersion are used as input parameters for the model. 
 Then, by fitting the model results with the 3-D HI data cube in an iterative fashion,
 the best-fit HI velocity dispersion and the best-fit HI 
scale height as a function of radius are obtained simultaneously. 
The best-fit HI dispersion shows a variation with radius (see \cite{Patraspiral} for details). 

In another application, a  self-consistent, multi-component disk analysis was done to obtain the vertical distribution and hence the disk thickness (HWHM) of the HI gas,
as a function of radius in the ultra-diffuse galaxy AGC 242019 \cite{Li2022}. The formation mechanism for ultra-diffuse galaxies are not yet well-understood.
\cite{Li2022} found the HI scale height values in AGC 242019 to be moderate, and comparable to that for a sample of dwarf galaxies. Hence they  conclude that stellar feedback has not been important in the formation of this ultra-diffuse galaxy.

\subsubsection{Applications to theoretical studies\label{4.5.3}}

The model has been applied to a number of theoretical studies in the literature to get a more accurate description of the vertical disk structure and stability; 
star formation; as well as to understand the halo properties, as discussed next. The change in the density profile including the thickness in the multi-component case, compared to the standard sech$^2$ solution for the one-component case; and the associated change in the gravitational potential, can affect 
various dynamical processes in the disk. Some examples of application of this model for theoretical studies  are: 

In an interesting application, \cite{Wang2010sta} use the model to develop the theory of stability of 3-D gas disk in a galaxy.  To do this, \cite{Wang2010sta} use the multi-component disk plus halo model in the thin disk case (\cite{NJ2002}; also, see Section \ref{sec:4.2.2} here) to set up a gas disk in hydrostatic equilibrium. 
 This allows them to treat the problem locally at a given radius, $R$ (see their Section 4.2); which simplifies the set-up for numerical simulations as no radial exchange of information is needed. Their analysis highlights this subtle point. Next, \cite{Wang2010sta}  show that using this initialization condition for their numerical simulations allows them to study the onset of gravitational instabilities in a 3-D gas disk more accurately. This has implications for resulting star formation rates, and formation of spiral arms. 

In another study, the multi-component disk plus halo model was applied \cite{AbramovaZ2008} to data for a set of spiral galaxies to obtain the redistributed or enhanced mid-plane density. Assuming the star formation to depend on a power, $n'$, of mid-plane density,
and comparing with the observed star formation rate, they find the best-fit value of $n'=1.5$. Hence \cite{AbramovaZ2008} conclude that the Schmidt law for star formation rate ($\propto \rho^n$, where $n=2$), as given by \cite{SchmidtLaw}, is not universal.

 \cite{OstrikerKim2022} point out that
an accurate stellar scale height as obtained by the multi-component disk plus halo model would affect the net pressure on gas and hence the resulting star formation. Hence they advise using this model rather than assuming a constant scale height and the ad hoc value of the stellar scale height taken to be equal to 
 0.27 R$_D$/5, as is routinely used, for simplicity, in the literature.

\cite{Khrapov2021} have obtained 
the vertical distribution assuming multi-component disk plus halo model, and taking account of the non-isothermal stellar velocity dispersion
as in \cite{SJ2020noniso}; and have applied these results in the numerical simulations study of spiral structure in galaxies.

In another application, the multi-component disk plus halo model in the thin disk case has been applied to study the superthin, LSB galaxies. This indicates a dense, compact halo in these galaxies that vertically constrains the stellar disk at all radii. This explains their superthin nature in  a generic way \cite{BJ2013}.

In an interesting, new approach, the observed HI scale height and rotation curve values are compared with the model results; to constrain the dark matter halo parameters such as the density profile and shape of the dark matter halo in a number of galaxies (see Section \ref{sec:6} for details).

\subsubsection{Comments on simplified, non-self-consistent prescriptions for stellar and HI scale heights in the literature\label{sec:4.5.4}} 

We warn that in the literature, a variety of simplifying, ad hoc assumptions are routinely made, for the sake of convenience,  to obtain the vertical scale heights for stars and gas in a galactic disk; instead of using a rigorous treatment to obtain a self-consistent density distribution for coupled stars and gas in a galactic disk as presented in this review (see Sections \ref{sec:4.2}-\ref{sec:4.4} for the rigorous treatment).

We point out that the resulting values of scale heights obtained from the simplified models are not accurate;  and when these are used in other, subsequent analysis, say to study star formation, this could lead to erroneous results. We mention this point to be comprehensive and illustrate it by two examples here. Also, see Section \ref{sec:6.4} for more examples of a non-self-consistent treatment. 

\medskip

\noindent {\bf Gas disk scale height and star formation}

The gas scale height, apart from being of interest for its own sake, is also of importance for other physical processes such as star formation, and containment of HI holes. 
Hence, there  is a general interest in obtaining the gas scale height as a function of radius, and in different types of galaxies. For this reason, there are several papers where the scale heights for gas have been estimated, but in a non-self-consistent fashion.

For example, often the gas gravity is neglected and the gas response to the stellar potential (see Section \ref{sec:3.2}, Eq.(\ref{eq3.8}))
is taken to determine the gas scale height. Even when the gas gravity is included  \cite{Olling1995,Peters2017,Bacchini2019}, the treatment is not self-consistent. For example, consider  a recent detailed work  \cite{Bacchini2019} which does include many physical points, such as gas gravity and the effect of finite thickness of gas on the gas potential. 
However, their formulation of the problem is not rigorously correct, and the resulting density profile is not self-consistent. Some of the simplifying assumptions in their work are given next. \cite{Bacchini2019} assume
a sech$^2$ stellar density profile; and further assume ad hoc values of the model parameters ($z_0= R_D/5$) to write the profile. The stellar potential is taken to be external and constant (that is, not affected by gas or halo).  
 Thus their formulation of the problem is not self-consistent, and this will affect the resulting gas scale heights obtained by them.

A correct treatment would involve taking account of the gas and halo. The resulting self-consistent stellar profile 
would be steeper than sech$^2$ due to the constraining effect of gas and halo (see \cite{NJ2002,SJ2018}; or, Section \ref{sec:4.2.5}),
and hence would result in smaller (true) gas scale height values. 
 Since this effect is not included by \cite{Bacchini2019}, their resulting HWHM values for gas would be higher than the true values. We caution that this would then result in  smaller predicted star formation rate in their work.
This discrepancy will be seen more at larger radii where the effect of halo on stellar profile becomes more important (Section \ref{sec:4.2.5}; also, \cite{SJ2018}). This error will be carried over in subsequent papers where the non-self-consistent model by \cite{Bacchini2019} has been applied.

\medskip

\noindent {\bf Stellar scale height:}

Similarly, in many studies in the literature,  
instead of treating the proper coupled system in a self-consistent way; often a  number of simplifying assumptions are made so as to get a simple, analytical expression for the stellar scale height as modified due to the presence of gas.  However, as we show next, these simplifications do not give accurate results and do not represent a realistic disk.

To give an example, a number of authors (e.g., \cite{Forbes2012,Aniyan2018,Tsukui2024})
 use the following relation between the stellar scale height, $h_z$ (in their notation), and the stellar vertical dispersion,$\sigma_s$, and the surface densities of stars and gas, $\Sigma_s$ and $\Sigma_g$ respectively:

\begin{equation}
\mathrm {h_z = \sigma_s^2/ [ 2 \pi G (\Sigma_s + \Sigma_g)] }
 \label{eq4.25}
 \tag{4.25}
\end{equation}

This expression was obtained  \cite{Aniyan2018} for a razor-thin gas disk embedded in a stellar disk. It was obtained by them as a solution to the problem 4.22 in \cite{BT2008}.

This calculation makes a number of serious, simplifying assumptions; and does not constitute  a  self-consistent treatment, as discussed next. First, \cite{Aniyan2018} do not take the gas to be in a hydrostatic balance. 
They do not take account of the effect of stars on the gas, as was done in the true coupled case (\cite{NJ2002,SJ2018}; also, see Section \ref{sec:4.2.4}, in particular Fig. \ref{fig.6} and Fig. \ref{fig.7}). Thus, the treatment by \cite{Aniyan2018} has not treated the gas component correctly.
Second, the above simplified expression is obtained assuming a sech$^2$ distribution for stars, hence it does not reflect the steeper  stellar distribution profile, or the corresponding smaller scale height, that is obtained in the coupled case in the rigorous, self-consistent calculation (see Section \ref{sec:4.2.6}; also Appendix A). 

Thus, the above expression is not valid for a true coupled system. The stellar scale height values for the Milky Way that result from it will differ from the results obtained in the self-consistent calculation by \cite{NJ2002,SJ2018}. The latter results agree well with observed values, as shown by  \cite{NJ2002}.

Therefore, the use of this simplified, analytical relation Eq. (\ref{eq4.25}) as done in the literature, e.g. by \cite{Aniyan2018} and other papers which cite this work, is not physically justified. This treatment gives incorrect result for the density profile of stars, and does not treat the $z$ distribution of gas at all. Thus, using Eq. (\ref{eq4.25}) in subsequent papers will give incorrect results.

Moreover, the application of this result from \cite{Aniyan2018} to a {\it Thick disk} and a thin disk as done by \cite{Tsukui2024}  is questionable, since the thin disk has a finite height (which is just a factor of 2 smaller than the {\it Thick disk} scale height  - see Section \ref{sec:7.1}). Hence, the thin disk cannot be treated as being razor-thin.

Another serious problem is that the effect of the dark matter halo is not included in Eq. (\ref{eq4.25}). \cite{Forbes2012} claim that this is justified because the halo affects scale height by only 12 \% in the inner Galaxy, as shown by \cite{NJ2002}. However, in the outer Galaxy, the dark matter halo dominates the dynamics; and the stellar scale height decreases by a factor of 2.3 and 3.3  at $R$= 18 and 20 kpc, respectively (see Table 2 in \cite{SJ2018}; also, see Fig. 11, panel for R=18 kpc here). Thus ignoring the effect of halo on the stellar disk as done in the above papers
\cite{Forbes2012,Aniyan2018, Tsukui2024}  is a serious error;  and this will give spuriously large stellar scale height values compared to the "true" stellar scale height values. The latter are as given by a self-consistent treatment, as in \cite{NJ2002,SJ2018}. 
We caution again that the above expression (Eq. (\ref{eq4.25})) is incorrect, and adopting it in subsequent papers would affect the accuracy of the results obtained.

\subsection{Gravitational potential energy of a multi-component galactic disk\label{sec:4.6}}

A generic result from the multi-component disk plus model is that the vertical disk distribution is constrained closer to the mid-plane (See \cite{NJ2002,SJ2018}; or, Section \ref{sec:4.2.5}). To understand the implications of this result for the energetics of the disk, the gravitational potential energy of a multi-component disk was studied by Sarkar \& Jog (2022) \cite{SJ2022}.

The gravitational potential energy of spherical and spheroidal/ellipsoidal gravitating objects has been extensively studied in the literature \cite{Chandra1969,BT1987}.
In contrast, the potential energy of a thin disk system has not received much attention, even though it is of direct relevance for realistic systems, 
namely the disk galaxies. One reason for this is that the geometry being "open" in this case makes it a harder problem, than say, the spheroidal case that is usually considered. This problem was studied by Camm (1967)\cite{Camm1967} who considered an isothermal, one-component, self-gravitating thin disk of constant surface density and infinite extent in the plane. By using analytical techniques and potential theory, \cite{Camm1967}
obtained the gravitational potential energy per unit area of such a stars-alone disk (see Eq.(\ref{eq4.26})), as briefly discussed below.

Normally the potential energy of a mass distribution in a finite volume is defined as the energy released in assembling the finite system from an infinitely dispersed state, and it is finite. 
For a thin disk distribution, the state of collapse of the disk mass on the $z=0$ plane is defined to be the state of zero potential energy. The work done that is required to build the disk of finite height, from that state, is considered to be the potential energy stored in the disk, so that it has a finite value -- for a detailed discussion see \cite{Camm1967,SJ2022}.
Therefore, the potential energy of the disk is defined in terms of the energy contained in a column of unit cross section, that is, as potential energy per unit area of the disk.

 The mathematical expression for the gravitational potential energy of a thin, single-component, isothermal disk of constant surface density was derived by \cite{Camm1967} and is given in terms of the integral over $\rho(z)$, the mass density; $K_z$, the force per unit mass due to the self-gravity of the disk; and $z$; and is equal to{\footnote {In writing this result from \cite{Camm1967}, the standard notation for the Poisson equation is used, which is  
$\mathrm{{\partial^{2}\Phi}/{\partial z^{2}} = 4 \pi G \rho}$. See Section \ref{sec:3.1}, and \cite{SJ2022} for details.}}
:

\begin{equation}
\mathrm{  W= \int_{-\infty}^{+\infty} \rho(z)\frac{\partial \Phi}{\partial z}z \ dz =  - \int_{-\infty}^{+\infty} \rho(z)K_z z \ dz } 
  \label{eq4.26}
  \tag{4.26}
\end{equation}

Using the symmetry of $\rho(z)$ w.r.t. z, this can be written as:

\begin{equation}
\mathrm{
W=  2 \int_{0}^{+\infty}  \rho(z)\frac{\partial \Phi}{\partial z}z
 dz  }
   \label{eq4.27}
   \tag{4.27}
\end{equation}   

\subsubsection{Motivation and Formulation\label{sec:4.6.1}}

We stress that a real galactic disk is a multi-component system of gravitationally coupled stars and gas  embedded in the potential of the dark matter halo (\cite{NJ2002,SJ2018}; also, see Section \ref{sec:4.1}-Section \ref{sec:4.3}).
\cite{SJ2022} follow the method developed by \cite{Camm1967} for a self-gravitating one-component disk; and systematically extend it to a gravitationally coupled, multi-component disk, and thus explicitly derive the expression for the potential energy per unit area for the multi-component case. We emphasize that this is not a trivial extension, see \cite{SJ2022} for the detailed steps of the extensive calculation.  
For simplicity, the dark matter halo is not included, hence their calculation is mainly applicable to the inner Galaxy. The main points regarding the calculation are summarized next.
It should be pointed out that this calculation as well as the results from this section are valid for any general galactic disk, although the numerical results are obtained for the Milky Way since the input parameters are known for it.

While deriving Eq.(\ref{eq4.26}), \cite{Camm1967} had used only the $z$ component of the Poisson equation, since an infinite disk of constant surface density was treated. 
\cite{SJ2022} show that this assumption is also justified for a general, thin galactic disk (for which the $z$ term of the Poisson equation is $>>$ the $R$ term - see \cite{BT1987}; also, Section \ref{sec:3.1} of this review). Hence the vertical density distribution, force and energy become only $z$-dependent quantities, and depend on the local value of the disk surface density, $\Sigma$. Hence the calculation is local, and the potential energy obtained at a given radius is independent of the value of the surface density at other radii.

Therefore, \cite{SJ2022} could follow the same approach as in \cite{Camm1967}, while deriving the potential  energy for a realistic, multi-component galactic disk; with  stellar and gas disks with surface densities that vary with radius.
The potential energy per unit area for the coupled case is then obtained to be:

\begin{equation}
\mathrm{
W_{\mathrm{coupled}} = \int_{-\infty}^{\infty} z~ \frac{\partial\mathrm{\Phi_{coupled}}}{\partial z}(\rho_{s}+\rho_{g})  \ dz }
   \label{eq4.28}
   \tag{4.28}
\end{equation}

\noindent 
Thus, analogous to the one-component case, the potential energy for the two-component case is obtained to be an integral over the total mass density, the total gravitational force, and the vertical distance $z$. 

The above equation can be separated into two terms; and using the symmetry of $\rho(z)$  about $z=0$, these can be further simplified as:

\begin{equation}
\mathrm{
W_{\mathrm{coupled}} = -2\int_{0}^{\infty}z~ K_{z,\mathrm{coupled}} ~\rho_{s} \ dz - 2\int_{0}^{\infty}z~K_{z,\mathrm{coupled}} ~\rho_{g} \ dz   }
   \label{eq4.29}
\tag{4.29}
\end{equation}

\noindent where $\mathrm{k_{z,coupled}}$ is the joint gravitational force per unit mass in the coupled system. The integrals in this equation can be considered to represent the potential energy per unit area of the stellar disk and the gas disk, respectively, in the coupled stars plus gas system. 
This separable form results because  the two disk components are now being  built from z=0 to their corresponding vertical distributions simultaneously, against the same  coupled force. 
Note that in the limit of $\mathrm{\rho_g \rightarrow 0}$, or $\mathrm{\rho_s \rightarrow 0}$, Eq.(\ref{eq4.29}) goes over to the corresponding  stars-alone and gas-alone cases, respectively.

The above formulation can be extended in a similar fashion to any n-component disk system (for $n>2$). This has been  done for a three-component disk consisting of stars and two gas components (HI and H$_2$) in the Milky Way (see \cite{SJ2022} for details). Also, the above formulation could be extended to a multi-component stellar disk (n $>$ 3) 
for the many stellar sub-components that have now  been identified from \textit{Gaia}. The study of the vertical distribution of the latter system has been mentioned as a possible future problem in Section \ref{sec:8}. 

\subsubsection{Potential energy  for a disk in hydrostatic equilibrium\label{sec:4.6.2}}

For a disk in a vertical hydrostatic equilibrium, $\rho(z)$ and $K_z$ used in the above expression for the potential energy, are related to each other 
by the equation of hydrostatic equilibrium.
The self-consistent solution for the vertical density distribution, $\rho(z)$, and its derivative
 have to be obtained numerically by solving the coupled, joint Poisson-hydrostatic balance equation (see Eq.(16) from \cite{SJ2022}). The joint force, $\mathrm{K_{z, coupled}}$ in the coupled case at each $z$ can be obtained from the equation of hydrostatic balance for any one disk component in the coupled case (see Eq. 14 from \cite{SJ2022}), in terms of the values of $\rho$ and $\mathrm{d \rho/dz}$ for that disk component obtained numerically, as described above.
Similarly, for the one-component case, the values of $K_z$ and $\rho$ at any $z$ are obtained using the analytical solutions (see Eq.(6) and the discussion following that in \cite{SJ2022}).
\begin{figure}
\centering
\includegraphics[height= 2.8 in, width= 3.1in]{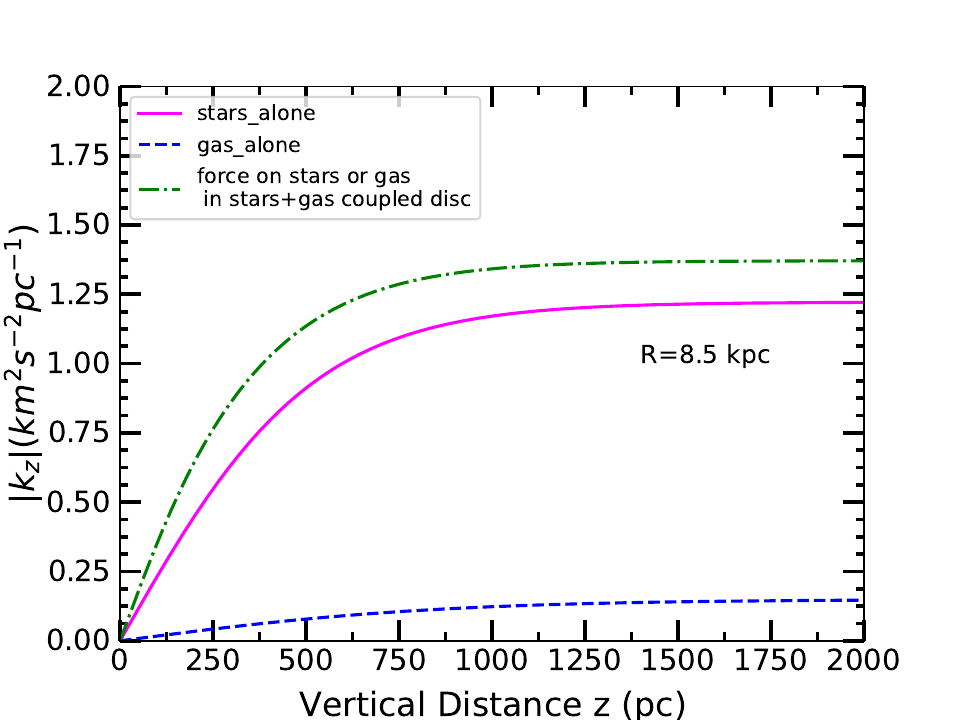}
\caption{Plot of the vertical force per unit mass, $|K_z|$, versus $z$  at R=8.5 kpc, for stars-alone and gas-alone under their own self-gravity (denoted by solid and dashed curves, respectively). The force due to gravitationally coupled stars and gas -- felt by both stars and gas  (shown by the dashed-dotted curve) is larger than the individual one-component, gravitational force for stars and gas, at each $z$.
 $\: $  {\it Source}: Taken from \cite{SJ2022}}
\label{fig.25}
\end{figure}

Using the above approach, and the observed input parameters for the Milky Way, the values of the force $K_z$ for stars and gas respectively and for the coupled force $\mathrm{K_{z,coupled}}$ are obtained, and their magnitudes are plotted in Fig.\ref{fig.25}. At each $z$, the joint gravitational force 
$\mathrm{|K_{z,coupled}|}$ is larger than the one-component values.{\footnote{Note that, we have seen the effect of this larger force  in the coupled case in constraining the disk components in the earlier sections;  but the fact that the magnitude of the force is higher in the coupled case was not explicitly shown in any of the papers discussed in  Section \ref{sec:4.1} to Section \ref{sec:4.5}.}} 

Next, the
corresponding 
integrands of the potential energy for the one-component case (solid line) and the same component in the coupled case (dashed line), obtained using Eq.(\ref{eq4.27}) and Eq.(\ref{eq4.29}) respectively, are plotted in Fig. \ref{fig.26} (panel a for stars and panel b for gas). In each case, twice the area under the curve gives the potential energy per unit area. 
\begin{figure}
\centering
\includegraphics[height=2.5in, width=2.67in]{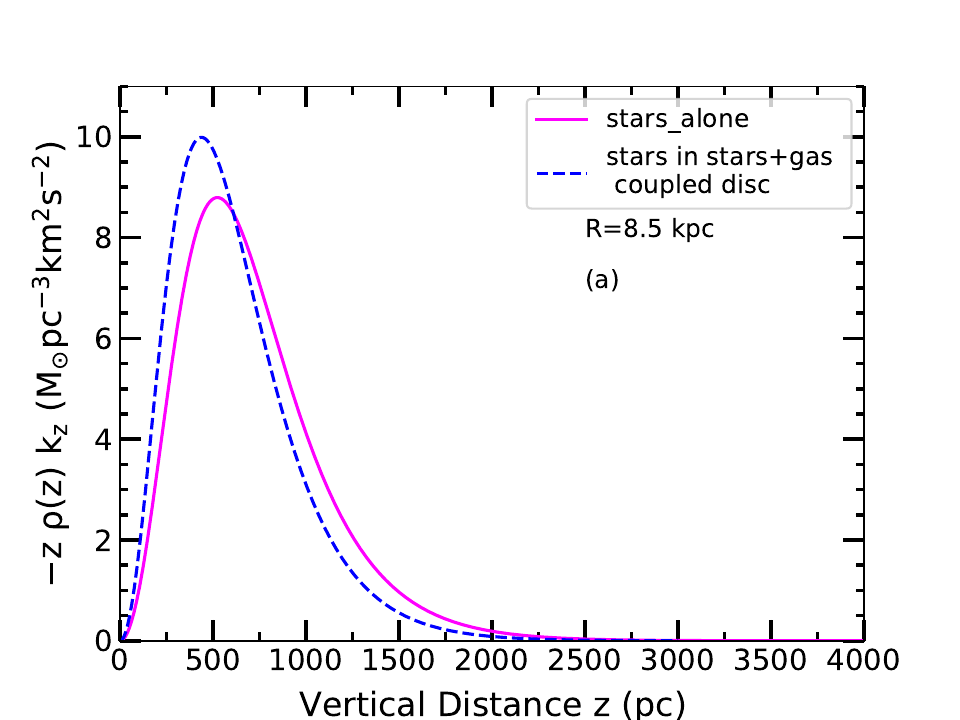}
\medskip 
\includegraphics[height=2.5in, width= 2.67in]{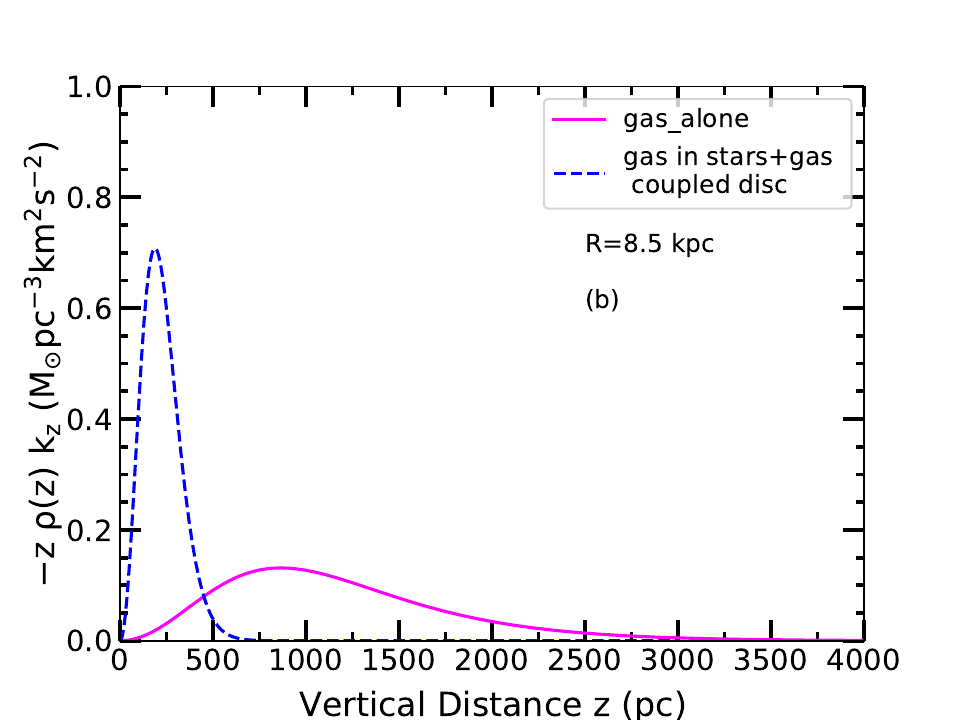} 
\bigskip
\caption{Plot of the integrand $- z \rho(z) K_z$  of the gravitational potential energy per unit area versus $z$ for stars and gas (HI) at R= 8.5 kpc; obtained for a one-component case and then for the coupled case. 
The left panel shows the plot of energy integrand for stars-alone (solid curve) and in the coupled case (dashed curve).
Twice the area under the curves gives the gravitational potential energy per unit area in the two cases; which, surprisingly, is obtained to be the same. The right panel shows a similar result for gas. Thus, the net energy per unit area for each of the two disk components remains unchanged. See the text for the physical explanation for this surprising result.
$\: $  {\it Source}: Taken from \cite{SJ2022}}
\label{fig.26}
\end{figure}

Interestingly, despite being in the higher gravitational force in the coupled system, the values of the gravitational potential energy per unit area for both stars and gas turn out to be the same as their corresponding  values for the single-component, self-gravitating cases.
This is because in the coupled case, while $|K_z|$ is higher, the density distribution is constrained to be closer to the mid-plane, hence the integrand peaks at a smaller $z$. Therefore the energy integrand is now redistributed along $z$ while conserving the total energy.
 Thus, the work done required to build up a stars plus gas disk turns out to be the same as the sum of energies required to build separate one-component, self-gravitating stellar and gas disks of same parameter values.

\medskip

\noindent {\bf Analytical simplification for potential energy per unit area of a disk:}

To investigate the physical reason for this surprising result, the expression for the potential energy above is simplified analytically by writing $K_z$ in terms of the equation of hydrostatic balance. 
On substituting $\mathrm{K_{z, coupled}}$ obtained 
in terms of the equation of hydrostatic balance for any one of the disk components in the coupled case (see Eq. (14) from \cite{SJ2022}) 
in the expression for the net potential energy per unit area in the coupled case (Eq.(\ref{eq4.29})), the latter reduces to:

\begin{equation}
\mathrm {
W_{\mathrm{coupled}}= {\sigma^2_{z,s}} \Sigma_{s} + \sigma^{2}_{z,g}\Sigma_{g} } 
          \label{eq4.30}
          \tag{4.30}
\end{equation}

\noindent Here the notation as in \cite{SJ2022} has been used to denote $\sigma_{z,s}$ and $\sigma_{z,g}$ as the velocity dispersion along $z$ for stars and gas, respectively. The expression 
for the energy, $\mathrm{W_{coupled}}$ (Eq.(\ref{eq4.30})),
 only depends on the intrinsic parameters for each component; namely, the surface density and the velocity dispersion. This is equal to the sum of potential energy per unit area for the two components taken individually (see \cite{SJ2022} for details).
This can be understood physically as follows.  
Due to the higher gravitational force in the coupled case (Fig. \ref{fig.25}), the self-consistent vertical distribution in each component is now constrained closer to the mid-plane (or, more of the mass is at a smaller height), hence the integrand in the expression for the potential energy (Fig. \ref{fig.26}) now peaks at a smaller $z$ value, so as to conserve the net potential energy per unit area. The joint gravity works like an internal force within the system, and therefore it can just redistribute the energy (along $z$) within each of the two components without changing the total value of the potential energy in each disk component.

\subsubsection{Gravitational potential of a coupled disk\label{sec:4.6.3}}

 Next, \cite{SJ2022}  consider another important physical quantity, namely, the work done or the energy required to raise a unit mass from the mid-plane
  to a certain height $h$ in  a single-component disk, $E_z,i$ (i= stars or gas); and in the coupled, two-component (stars plus gas) 
disk, $\mathrm{E_{z,i, coupled}}$ (i=stars or gas). The work has to be done against the self-gravity of the stars or gas (in the single component case); and the joint gravity of stars and gas (in the coupled case), respectively. Note that this is  the measure of gravitational potential at any height in these cases \cite{Bahcall1984paper1,Bahcall1984paper2}.

\begin{figure}
\centering
\includegraphics[height=2.8in, width=3.1in]{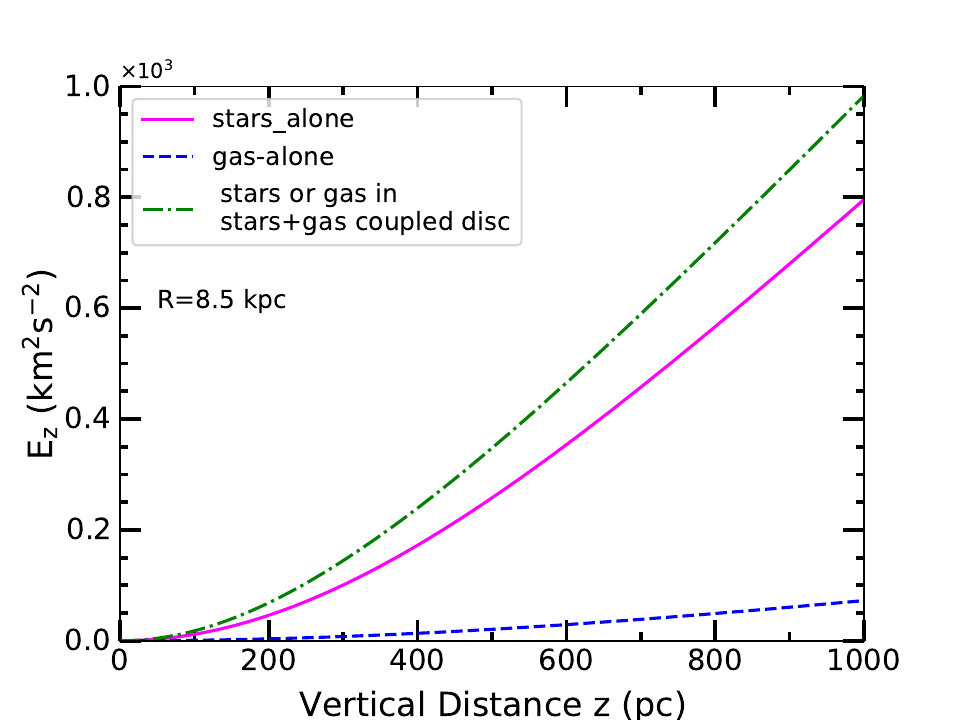}
\caption{Plot of work done, $E_z$, to raise a unit test mass from the mid-plane to a vertical height $z$ versus $z$, at R=8.5 kpc: for the stars-alone case against its self-gravity (solid curve); for gas-alone case against its own self-gravity (dashed line); and for stars or gas in the stars plus gas coupled system, against their coupled gravity (dash-dotted curve). The work  required to be done to raise a unit mass to any vertical height is higher  in the coupled case, than in the corresponding one-component cases. This is due to the higher force in the coupled case (see Fig. \ref{fig.25}). This shows that the stars or gas in the coupled system are more strongly bound to the mid-plane than the corresponding single component cases. $\: $  {\it Source}: Taken from \cite{SJ2022}}
  \label{fig.27}
\end{figure}

For a single-component disk (i = s or g), this work done is given as: 

\begin{equation}
\mathrm{
E_{z,i}=-\int_{0}^{h} K_{z,i} \ dz } 
   \label{eq4.31}
   \tag{4.31}
\end{equation}   

\noindent where $\mathrm{K_{z,i}}$ represents the self-gravity of the disk.

For the coupled case, it is given by

\begin{equation}
\mathrm{
E_{z,i,\mathrm{coupled}}=-\int_{0}^{h} K_{z,\mathrm{coupled}} \ dz } 
  \label{eq4.32}
   \tag{4.32}
\end{equation}

\noindent where i= stars or gas. In the coupled case, both stars and gas experience the same coupled force; hence, the same work has to be done to raise a unit mass of stars or gas to a certain height. Note that in each case, the energy is positive. Thus, work has to be done to raise mass to a given height. 

Using the input values for the Milky way, and obtaining the $K_z$ values as discussed above (see Section \ref{sec:4.6.2}), 
the work done as a function of $z$ for the above cases is calculated, for the results see Fig. \ref{fig.27}.   This shows that to raise a unit mass to a certain height, higher amount of work has to be done in the coupled case than in the corresponding single-component cases. 
This is due to the higher gravitational force  at each $z$ in the coupled case (Fig. \ref{fig.25}).
 
To summarize, while the vertical constraining of a disk component in a coupled case does not indicate higher gravitational potential energy, it does mean that each disk component in the coupled case is more tightly bound to the mid-plane. Hence, each disk component is better able to resist distortion; say, due to a given external, tidal encounter. A detailed N-body simulation needs to be done to confirm these predictions.

\section{Flaring stellar disk in a galaxy: A generic result\label{sec:5}}
 
At various points in the previous sections, we have reported  that the scale height of the vertical stellar distribution in a galactic disk increases with radius. The increase is moderate, a factor of $\sim$ 2-3, within the first few radial disk scale lengths, and then increases rapidly beyond that.
Thus the disk thickness or scale height is not constant; rather, the stellar disk flares with radius. 
There is clear and convincing evidence for a flaring stellar disk in observations; as well as in theoretical studies for a typical galactic disk in hydrostatic equilibrium, when observed input parameters are used.
In this section, we discuss and highlight this striking result and show that a flaring stellar disk is a generic feature, and discuss its dynamical implications.{\footnote{This section can be read as a largely stand-alone mini-review on this topic.}}

Note that the flaring stellar disk is in contrast to a flat distribution or a radially constant scale height that is commonly assumed in the literature. The latter was proposed by \cite{vdkS1981a} from analysis of their data.  
To explain this claim of a constant scale height, 
\cite{vdkS1981a} further proposed  
a specific rate of radial fall-off in dispersion;  namely, $\mathrm{R_v/R_D =2}$, where R$_v$ and R$_D$ are the scale lengths for the exponential
fall-off in the vertical velocity dispersion and surface density with radius, respectively (as defined in Appendix A). Both these claims by  \cite{vdkS1981a}: namely, that the scale height is constant, and $\mathrm{R_v/R_D =2}$; have now shown to be invalid in real galaxies, as shown by observations as well as theory, as will be discussed in this section.

\subsection{Physical basis for the radial variation in vertical thickness\label{sec:5.1}}

In order to see the significance of a flaring disk, let us first consider some general theoretical ideas on what determines the
vertical scale height of the disk at a given radius.

First, consider a single-component disk.
A self-gravitating, isothermal disk in pressure equilibrium  obeys a vertical density distribution of form sech$\mathrm{^2 (z/z_0)}$ (\cite{Spitzer1942}; also, Section \ref{sec:3}), where
$\mathrm {z_0 = [\langle(v_z)^2\rangle/ 2 \pi G \Sigma]}$. Here, the parameter $z_0$ is a measure of the disk scale height of the sech$^2$ distribution. 
The value of $z_0$ is set by the balance between the self-gravitational force and the vertical pressure gradient (Section \ref{sec:3.1}).
For a general density distribution (as for example, resulting for a multi-component disk plus halo case), the HWHM of the stellar vertical distribution is taken to define the disk thickness (\cite{NJ2002}; also, Section \ref{sec:4.2.2}, and Appendix A). For a sech$^2$ distribution, the disk thickness (HWHM) is given by = $\mathrm{z_0 \times sech^{-1} (0.5)}$ = $\mathrm{z_0 \times 0.88}  $ (Appendix A). Thus, for a one-component case, the disk thickness (HWHM) is proportional to  $ {\langle{({\mathit v}_z)^2}\rangle} /\Sigma$. This also indicates the formal functional dependence of the disk thickness (HWHM) for a general density distribution, $\rho(z)$.

\subsubsection{Dependence on radial variation of stellar velocity dispersion\label{sec:5.1.1}}
It is clear that for a one-component stars-alone disk, if $\mathrm{R_v = 2 R_D}$, then the scale height or the disk thickness for stars (which is $\propto  {\langle{({\mathit v}_z)^2}\rangle} 
/\Sigma$, as shown above, here both these parameters are for stars) remains constant with radius. Physically, here the fall-off in pressure precisely compensates for  the fall-off in the  disk gravity, hence the disk scale height remains constant with radius.  In the literature, the relation 
$\mathrm{R_v/R_D=2}$ is routinely assumed, following the claim by \cite{vdkS1981a}; although there is no physical justification for this relation to be valid. See, for example, the discussion in \cite{deGrijsP1997} about this point, and also the discussion in Section \ref{sec:5.4}. 
Indeed, for the Milky Way stellar disk, the observed value of the ratio $\mathrm{R_v/R_D = 2.8 (> 2)}$ (see Section \ref{sec:5.3.1}).

Note that if  the velocity dispersion falls off more slowly than the above required rate, namely if $\mathrm{R_v /R_D > 2}$,  the pressure support falls slower with radius than the gravitational force, hence the disk flares. That is, the pressure support is higher at a given radius - than required for a constant scale height, and this results in a flaring disk. This can be checked as follows:

\begin{equation}
\mathrm {HWHM \propto \frac {exp(- 2 R/ n R_D)}{exp(-R/R_D)} =   exp \left[{\frac {R}{R_D} (1-\frac {2}{n})}\right]}
    \label{eq5.1}
    \tag{5.1}
\end{equation}
    
\noindent where $\mathrm{R_v = n R_D}$. For $n >2$, the HWHM increases exponentially with radius. 
Thus, for a single-component self-gravitating disk, any value of $\mathrm{R_v/R_D > 2}$ results in a flaring disk. Physically, for a one-component, self-gravitating disk, for $\mathrm{R_v/R_D > 2}$, the pressure support falls slower with radius than the gravitational force, hence the disk flares.
 The higher the value of $\mathrm{R_v/R_D (>2)}$, the higher is the flaring. 

In a realistic multi-component disk plus halo system, however, no such simple relation exists between R$_v$ and R$_D$ that would result in a constant thickness,  or  otherwise a value that 
would give a flaring disk.
In the coupled case, the gas and the halo exert a considerable gravitational force on the stellar disk.
In this case, the rate of variation of stellar disk thickness would depend crucially on the value of  $\mathrm{R_v/R_D (>2)}$
 \cite{NJ2002ltr,SJ2019}; 
and also on the surface density of gas and the amount of halo mass, and their radial variation. 

Thus, it is not straightforward to predict the limiting value of $\mathrm{R_v/R_D}$ above which flaring would be seen in a multi-component disk plus halo model. However, 
it is easy to see that because of the additional gravitational force due to gas and the halo, a higher pressure gradient in the stars is required at a given radius, R, to keep the disk in pressure equilibrium with the same resulting vertical scale height. This is because in the coupled case, the additional gravitational force due to gas and halo results in a smaller vertical stellar scale height for the disk in pressure equilibrium; than in one-component stars-alone case, at each radius,
as shown for example for the Milky Way (see Fig. \ref{fig.6}, panel for stars).
 Further,
the additional gravitational force due the gas and 
halo is more important at larger radii. Hence,  the pressure has to fall more slowly with radius than in the one-component case to get a constant scale height. Hence, in general in a multi-component disk plus halo system, one needs  higher dispersion at large radii, corresponding to $\mathrm{R_v / R_D > 2} $ to even get a flat distribution.
 This was shown
in the context of modeling of two galaxies: NGC 891 and NGC 4565 by \cite{NJ2002ltr} (see Fig. \ref{fig.16}) for NGC 891).
For the same reason, 
to get flaring in a multi-component disk plus halo case, an even higher value of $\mathrm{R_v /R_D} (> 2)$ is required than in the one-component case. 

In other words, for the same value of $\mathrm{R_v/n R_D > 1}$ where $n >2$, the flaring seen in the stars in the multi-component disk plus halo system is less than in the stars-alone case (as shown for the Milky Way, see Fig. \ref{fig.6}).
Because of the constraining effect of gas and halo, the stellar scale height does not flare exponentially at large radii as it does in a single-component, stars-alone case; instead, it shows a controlled linear increase with radius, for the same observed $\mathrm{R_v/R_D = 8.7/3.2= 2.7} >$ 2, for the Milky Way (see Fig. \ref{fig.6}, panel for stars).
The stellar scale height values are lower for the multi-component disk plus halo case than for the one-component stars-alone case. These lower values agree with the trend from observational data from COBE \cite{NJ2002}.  

A related point shown from the modeling of NGC 891 and NGC 4565 \cite{NJ2002ltr} is that a higher value of $\mathrm{R_v/R_D}$ ($>$ 2) results in a higher flaring (see Fig. \ref{fig.16}). This is because, other parameters being the same for a given galaxy,
 a higher value of $\mathrm{R_v/R_D} (> 2)$  means a slower fall-off in pressure. This results in a higher pressure support at a given radius and hence results in
higher flaring (see Fig. \ref{fig.17}). This is also true for for a one-component case, as can be seen from Eq.(\ref{eq5.1}).

\subsection{Origin of sharp flaring in the outer disk\label{sec:5.2}}

In addition, there is another factor that can cause a sharp flaring of stellar disk at large radii; namely, when the stellar dispersion falls sufficiently and saturates at the gas dispersion values, as argued on physical grounds (\cite{SJ2018}; also see \cite{Sharma2021}), as discussed next.

In the Milky Way, the stellar vertical velocity dispersion falls exponentially with radius, with a scale length, R$_v$ = 8.7 kpc (\cite{LewisF1989}, see Section \ref{sec:4.2.3}).  
 The velocity dispersion in external galaxies is difficult to observe and had been measured observationally  for only a few galaxies \cite{Bottema1993}. This situation is now changing for the better over the last $\sim$ ten years due to the wide availability of IFU 2-D data on stellar kinematics in galaxies (see Section \ref{sec:2.1.4}, and Section \ref{sec:8}). 

In their study of the Milky Way focusing on the outer disk, \cite{SJ2018} proposed that beyond a certain radius when the stellar velocity dispersion falls close to the value of the HI velocity dispersion,  the stellar dispersion does not fall any further; rather, 
it  saturates to the gas dispersion value (see Section \ref{sec:4.2.3}). This is based on the physical argument that the velocity dispersion of stars cannot be lower than the dispersion within the gas from which they form.
Beyond this radius where the stellar dispersion saturates, the stellar surface density still continues to fall exponentially. Hence, the resulting vertical stellar scale height would show a sharp flaring. 

This physically motivated saturation in the stellar velocity dispersion was proposed and employed in the theoretical model by \cite{SJ2018}. This work showed that for the observed input parameters for the Milky Way, the saturation in the stellar velocity would occur at $R$ $\sim$ 17 kpc. Further, their model results indeed show that the resulting stellar scale height flares sharply beyond this radius, as argued above, see Fig. \ref{fig.12}.

Interestingly, using recent observations from \textit{Gaia} and APOGEE, \cite{Sharma2021}    
confirmed that 
the vertical stellar velocity dispersion saturates  to $\sim 10$ km s$^{-1}$ in the outer Galaxy (see Fig. 1 in \cite{Sharma2021}).
In fact they predicted that consequently the stellar disk would flare beyond this point.  The
 physical argument used by \cite{Sharma2021} to explain the observed saturation of the stellar velocity dispersion was identical to the one proposed by \cite{SJ2018}, namely that the stellar velocity cannot fall below the gas dispersion value,
and hence saturates to the non-zero velocity dispersion of stars at birth, which corresponds to the dispersion within the gas clouds. Note that,
the flaring of stellar disk beyond the point of flattening in stellar dispersion, as predicted by \cite{Sharma2021},
was  exactly already shown by the theoretical model results of \cite{SJ2018}.

Applying the same physical argument of saturation of stellar velocity dispersion at large radii in the theoretical study of a  low surface brightness (LSB) galaxy, UGC 7321, \cite{SJ2019} found that in this galaxy the stellar velocity dispersion saturates at $\sim$ 7 kpc. Their results showed that, indeed, the model stellar scale height shows sharp flaring beyond this radius (see Fig. 5 from \cite{SJ2019}; also, see Section \ref{sec:5.3.2} for details).

We predict that such onset of sharp flaring of stellar disk beyond the transition radius
is expected to be seen in all galaxies. Recall that typically the gas velocity dispersion tapers off and is constant at $\sim 7$ km s$^{-1}$ at large radii in most galaxies (Section \ref{sec:4.2.3}). The transition radius is set by the central stellar velocity dispersion, the rate of fall-off in stellar velocity dispersion, R$_v$, and the gas velocity dispersion, in the outer parts in a galactic disk (as discussed by \cite{SJ2019}). 

 The HI velocity dispersion as a function of radius has been measured for a large number of galaxies in the THINGS (The HI 
 Nearby Galaxy Survey)  \cite{2009AJ....137.4424T}. The stellar velocity dispersion values including the central dispersion can be extracted from the various IFU surveys although it is a challenging task (see Section \ref{sec:8} for details). 
In principle, therefore, it would be straightforward to apply this result to more galaxies, and thus check if sharp flaring of a stellar disk in the outer regions of disks, as predicted here, is indeed a common result. 

\subsection{Application to galaxies: Theoretical modeling and observational results\label{sec:5.3}}

Here we apply the above ideas about $\mathrm{R_v/R_D}$, and the radial variation in stellar scale height, to real galaxies, to  better understand the results from observations and theoretical models given in the previous sections. We also give  additional evidence for flaring of stellar disks from observations and simulations.

\subsubsection {The Milky Way \label{sec:5.3.1}}   Consider the Milky Way first, for which the input parameters for the multi-component disk plus halo model are well-known observationally. In fact, of all the galaxies studied so far; the input parameters, including the variation in the stellar velocity dispersion with radius, are best known for the Milky Way. In the Milky Way, the vertical stellar velocity dispersion is known to fall exponentially with radius, with a scale length, R$_v$= 8.7 kpc (Section \ref{sec:4.2.3}).
Using this and the other input parameters, a single-component stellar disk model in pressure equilibrium shows an exponentially flaring behaviour (\cite{NJ2002}; also Fig. \ref{fig.6}, panel for stars, dashed line). Inclusion of gas and halo gravity results in a more moderate, linear increase  up to R=12 kpc (Fig. \ref{fig.6}, panel for stars, solid line).
Yet even in the latter, more realistic case, the scale height clearly shows flaring. Thus, flaring of a stellar disk is a robust result. The resulting model rate of a
  moderate rise in scale height with radius, between the radial range R= 5-10 kpc,  is 24 pc  kpc$^{-1}$. This was shown to be in good agreement with the observations known at that time; namely, the linear gradient
  deduced from COBE and SPACELAB 2 data (see Section \ref{sec:4.2.4}).
 The flaring behaviour is now confirmed by more modern observations, as discussed later in this subsection. 

In retrospect, given that the observed value of $\mathrm{R_v/R_D=8.7/3.2 = 2.7}$ is $ >$ 2, a flaring disk would be expected in the Milky Way; as  seen from the discussion given at the beginning of this section.
  However, to our knowledge,
this cause and effect correlation has not been  generally noted in the literature. 
It is not often realized that the claims by \cite{vdkS1981a}; namely, that the scale height  is constant, and $\mathrm{R_v/R_D =2}$, are not valid for a stars-alone disk even in the Milky Way, for which the observed values of these parameters are known. In fact, because the observed value of $\mathrm{R_v/R_D }$ for the Milky Way is 2.7 $>2$, the stars-alone disk would flare exponentially. It is the presence of additional 
gravitational force due to the gas and the halo in the coupled case that decreases the stellar flaring to a moderate value (for $R < 12$ kpc), as seen above.

In the outer Galaxy, the physically motivated assumption 
of saturation of  stellar velocity dispersion (as discussed above), occurs at $\sim 17$ kpc. 
The theoretical model results indeed show a sharply flaring stellar disk beyond this radius, as discussed above (\cite{SJ2018}; also, see Fig. \ref{fig.12} here).
The scale height increases by a factor of 2.2 between R= 16 kpc and 22 kpc (see Table 2, column 4 from \cite{SJ2018}).
The overall phenomenon of flaring in the outer Galaxy as predicted by \cite{SJ2018} is  confirmed by observations of number counts of stars (see Section \ref{sec:4.2.6} for details).   
The flaring of the stellar disk in the Milky Way has  recently been further confirmed by studies of number counts of different tracers, such as: red clump stars \cite{Sun2024MWflaring,2024MNRAS.527.4863U};  young O-B stars 
\cite{Yu2021}; and supergiants \cite{Chroba2022}.
In fact, all these papers, including the ones mentioned in Section \ref{sec:4.2.6}, simultaneously detect a flare and a warp in the Milky Way stellar disk (also, see \cite{Vig2005}). 

The value of flaring measured in these papers (for $R< 16$ kpc) is somewhat higher than the model results. This discrepancy could arise because of the incorrect deduction of warp, or contamination by thick disk stars -- as was argued by \cite{Minchev2015} for a similar excess in the measurement of stellar flaring  seen by \cite{Lopez2014}.
Interestingly,
using the vertical metallicity gradient of stars in the Milky Way with the RAVE and \textit{Gaia} data, and comparing with simulations; \cite{Ciuc2018}  conclude that these stars have formed in a flaring stellar disk.

Thus, a flaring disk appears to be a generic result for a typical galactic disk in hydrostatic equilibrium, and when the observed parameters are used.
It should be stressed that the flaring is not due to the multi-component treatment; in fact, the reverse is the case. Keeping all the other parameters constant (including $\mathrm{R_v/R_D}$ to be equal to the observed value);
the additional gravitational force due to gas and halo in the coupled case actually decreases flaring. That is, it leads to a more moderate rise with radius in the stellar scale height. This was shown earlier for the Milky Way (see Section \ref{sec:4.2.4}, and the discussion above, as well as the next paragraph). 

The flaring of the stellar disk is also seen in  theoretical models which take account of more detailed physics: such as, the treatment with complete Jeans equations as applied to stars-alone case (Model B in \cite{SJ2020Jeans}); and the treatment that takes account of the observed non-isothermal stellar velocity dispersion for the multi-component disk plus halo case \cite{SJ2020noniso} -- both applied to the Milky Way.
Interestingly,
 a stars-alone, thin, isothermal disk in pressure equilibrium  for the observed parameters  shows a flaring of scale height (HWHM) by a factor of 13.6 between R= 4 to 22 kpc (see Table 2 from \cite{SJ2018}); while the complete, Jeans treatment (Model B from \cite{SJ2020Jeans}) shows a smaller increase in scale height of a factor of 6.2. In comparison to these two cases; 
  in the multi-component disk plus halo case, the flaring is further reduced to a factor of 3.4, due to the gravitational force of gas and halo. In this case, the stellar disk flares by a smaller factor of 3.4 between R= 4 to 22 kpc (Table 2 from \cite{SJ2018}).

The recent observations for the Milky Way provide a different picture of the radial variation of stellar kinematics, which nevertheless strengthens the feature of a flaring disk.
\cite{Mackereth2019} analyze the
 detailed information about the motion of a large sample ($\sim 65000$) of stars, located within a galactocentric radial range of 4 to 13 kpc, as observed from APOGEE and \textit{Gaia}. They show that  $z$ velocity dispersion for stars 
 shows a clear increase with radius beyond 9 kpc, for all ages of stars (see Fig. 8 from \cite{Mackereth2019}).
This is surprising, and  opposite to the exponential fall with radius, as observed in earlier studies for the Milky Way \cite{LewisF1989}; and,
also in external galaxies \cite{Bottema1993}. The physical reason for the increase of stellar velocity dispersion with radius is not well-understood. It could be due to a tidal encounter with a satellite galaxy. In any case, it is easy to see that in this case, the scale height would increase with radius. This prediction is easy to check for a one-component stellar disk, where the thickness $z_0$ would be $\mathrm{\propto (\sigma_z)^2/\Sigma}$ (both quantities for stars) -- which would steadily increase with radius. At each radius, the resulting scale height would be higher  than the values obtained earlier (using the previously observed radial fall-off in velocity dispersion with R$_v$=8.7 kpc =2.7 R$_D$). The new kinematic data indeed shows that the disk flares considerably \cite{Bovy2016haloshape,2017MNRAS.471.3057M}, which supports the above prediction.

\subsubsection {UGC 7321 \label{sec:5.3.2}} 

The scale heights of the stellar disk have been measured observationally in the edge-on LSB galaxy, UGC 7321 up to 6 kpc. This shows a moderate flaring of a factor $\sim 2.5$ upto 6 kpc ($\sim 3 R_D$). Using this information on scale heights as a constraint to the multi-component disk plus halo model, the value of the rate of fall-off, $\mathrm{R_v/R_D}$, was obtained to be =3.2 \cite{SJ2019}. The same value of $\mathrm{R_v/R_D}$ was applied to theoretically predict the scale height beyond this radius, and the model scale height was shown to be rising. 
Again, the physically motivated idea of saturation of the stellar velocity to the HI dispersion (as proposed by \cite{SJ2018})  was applied here. The saturation occurs at $\sim 7$ kpc in this case. As expected (see the discussion in Section \ref{sec:5.2}), the resulting model stellar scale height shows sharp flaring beyond this point (see Fig. 5  in \cite{SJ2019}; also, see the discussion in Section \ref{sec:4.3.2}). Thus, both the Milky Way and UGC 7321 show a similar pattern of flaring of the stellar disk, with the onset in sharp flaring occurring where the stellar velocity dispersion saturates  to the value of the gas dispersion. 

\medskip

\subsubsection {NGC 891, NGC 4565 \label{sec:5.3.3}}

Two other galaxies, NGC 891, and NGC 4565, were also  shown to have a flaring 
stellar disk. This was shown from the correct interpretation of the data of these galaxies (\cite{NJ2002ltr}; also, see Section \ref{sec:2.1.1} and Section \ref{sec:4.2.7}). On a careful 
re-examination of the analysis by \cite{vdkS1981a}, \cite{NJ2002ltr} found that the data by \cite{vdkS1981a} actually shows a significant scatter. A correct interpretation of this scatter indicates a moderate increase in the stellar scale height with radius, corresponding to a factor of $\sim 2-3$  within the optical disk.
\cite{NJ2002ltr} next
applied the multi-component disk plus halo model to these galaxies,
and used this deduced variation in scale height to constrain the rate of radial fall-off of stellar velocity dispersion. The ratio $\mathrm{R_v/R_D}$ was thus shown to lie between 2-2.5 for NGC 891 and between 2.5-3 for NGC 4565. 
An independent study, by \cite{Rohlfs1982}, also showed that for these two galaxies, a flaring disk gave a better fit to the intensity profile data of \cite{vdkS1981a} (for details, see Section \ref{sec:2.1.1}).  

\medskip

\subsubsection {Other galaxies \label{sec:5.3.4}}

It was already pointed out by \cite{deGrijsP1997}, based on a sample of 48 galaxies, that the data shows a small variation in scale height with radius; but this result was largely ignored. More recently, there has been further observational evidence for flaring of stellar disk in a number of galaxies: see \cite{Kasparova2016,Rich2019}; including an ultra-thin Galaxy UGC 11859 \cite{2023ApJ...951..149O}. 

Recently, \cite{Ranaivo_flaring_2024} have analyzed the observed data for 46 spirals from the Spitzer Survey of Stellar Structure in Galaxies (S$^4$G) catalogue. By fitting the data by an exponential and sech profile for the radial and vertical mass distribution respectively, and following the procedure as in \cite{vdkS1981a}, it has been
been shown
that the stellar disk flares by a factor of few within the optical disk in all the galaxies considered \cite{Ranaivo_flaring_2024}.

However, there are several caveats regarding this
analysis and the result. First,
 the above method uses the integrated intensity along the line of sight, given by (Eq. (\ref{eq2.2})) as proposed by \cite{vdkS1981a}. This derivation implicitly assumes the scale height to be constant with radius, which is not warranted, as pointed out in Section \ref{sec:2.1.1}.
The work by \cite{Ranaivo_flaring_2024} employs two internally inconsistent assumptions: of a constant scale height, as well as a 
scale height that varies with $R$.
Thus the scale height values, obtained by 
\cite{Ranaivo_flaring_2024}), by fitting the image along cuts at different radii will carry over this error (see the general discussion in Sec \ref{sec:2.1.1}; also, see \cite{Kasparova2016}). Thus the values of scale heights as obtained by \cite{Ranaivo_flaring_2024}; 
are not accurate; however,
 the trend of a flaring disk obtained in all galaxies by them will  probably remain valid.

Another problem is that the approach by \cite{Ranaivo_flaring_2024} involves a simple fitting of sech or sech$^2$ functions unlike the more accurate approach by \cite{Comeron2011}. \cite{Comeron2011}  had used the multi-component disk plus halo model to obtain physically motivated disk vertical density profiles for thin and {\it Thick disks}, and thus extract more accurate disk scale heights from the same S$^4$G 
data (Section \ref{sec:4.5.1}). 

Finally, another issue with their approach is that
the results given by \cite{Ranaivo_flaring_2024}
are for a single disk fitted to the data;
 so the scale height obtained represents a cumulative fit to the thin and {\it Thick disks} in each galaxy.

\medskip

\subsubsection {Evidence of flaring from numerical simulations  \label{sec:5.3.5}}

There is a growing evidence for a flaring stellar disk from results from simulation studies of galaxies,
especially when the artificial condition of constant thickness used earlier is relaxed. Note that simulations data is amenable to easily being fit by a given functional form of the density distribution to see how well the fit is; and to obtain the corresponding HWHM values.
Earlier, the flaring effect may have been missed in the simulations results; perhaps because it was not looked for, or the resolution in the earlier simulations was not sufficient.

One early evidence for a flaring stellar disk comes from simulations of galactic dynamics of the Milky Way by \cite{Kawata2017}, which showed a small increase in stellar scale height with radius. 
Recently, in an important, systematic study of the TNG50 simulations data, \cite{Sotillo2023}  have fitted  the  TNG galaxies 
by a function of the form $sech^{(2/n)} (z/z_0)$ (proposed by \cite{vdk1988}) to determine the scale height. They show that the flaring effect is robust and is seen whether the data is fitted by vertical density profile that has a sech or a sech$^2$ form. It is also seen when a half-mass scale height (also defined by \cite{SJ2018}) was used. 
However, a word of caution about the results by \cite{Sotillo2023} is in order. Recall that the self-consistent density distribution in a realistic multi-component disk plus halo model as applied to the Milky Way \cite{SJ2018}, and to UGC 7321 \cite{SJ2019}, shows that the parameter $n$  obtained is not unique. Rather,  it is a function of the range of $z$, $\Delta z$, over which the fit is made. Thus, $n$ obtained this way is not a well-defined, unique number (see Section \ref{sec:4.2.6}; and Fig. \ref{fig.13}). 
Since \cite{Sotillo2023} fit a single function to the entire $z$ range; therefore, the resulting $n$ as well as $z_0$ they obtain 
are not well-defined.
However, the trend of a flaring stellar disk obtained by them probably remains valid.

\medskip

\subsection {Discussion  \label{sec:5.4}}

To conclude from the above applications, a flaring stellar disk appears to be a generic feature in galaxies in both the High surface brightness (HSB) galaxies, such as the Milky Way; and the Low surface brightness (LSB) galaxies. 
As shown above, the stellar disk flaring is directly tied up with the value of  $\mathrm{R_v/R_D}$, which has to be $>2$ for the disk to show an increasing scale height with radius. However, what decides the value of R$_v$ is not understood; and in fact, this question has not been discussed much in the literature and needs to be explored. The value of R$_v$ would be decided by how the heating of stellar velocity dispersion varies with radius: the various sources of heating being the gravitational encounters with gas clouds, spiral arms and a bar within a galaxy; and  tidal encounters with passing satellite galaxies (see Section \ref{sec:Intro}). For recent studies of this topic, see \cite{Jia2024evo,Donghia2024history}.  
Given the diverse ways in which the value of R$_v$ could be affected, it is clear that there is no physical justification for using $\mathrm{R_v /R_D =2}$ 
as is routinely done in the literature. Also, as discussed above, a few galaxies for which R$_D$ and R$_v$ have been measured observationally, give a value of $R_v/R_D >2$.

\noindent {\bf Flaring of HI gas disk} 

For the sake of comparison, consider the scale height variation with radius for the HI disk. It is easy to see that the gas component would show flaring. For the gas responding to the stellar potential alone (Eq. \ref{eq3.9}), the gas scale height, 
$\mathrm{(z_0)_g \propto ( {\sigma_g^2}/\Sigma_s)}$,
where ${\sigma_g}$ is the gas velocity dispersion and $\Sigma_s$ is the stellar disk surface density (see Table \ref{table:1}).
Since the stellar surface density would fall with radius for any realistic disk, and the gas velocity dispersion is nearly constant with radius (see the discussion in Section \ref{sec:4.2.3}, we expect the gas to flare with radius (see Section \ref{sec:3.2}). The HI gas in the Milky Way indeed shows flaring at large radii (\cite{Wouter1990,Levine2006,KalberlaD2008}; also, see Section \ref{sec:4.2.6}, and Section \ref{sec:6}). At large radii, the gravitational force of the halo dominates, which decreases the gas flaring to a moderate value (\cite{SJ2018}; also, see Section \ref{sec:4.2.6}).  
The HI gas disk flaring at large radii is a common feature of spiral galaxies. This feature can be used to constrain the dark matter halo parameters, by theoretical modeling using a multi-component galactic disk plus halo system (see Section \ref{sec:6} for details).

 On the other hand, in the inner regions of the Galaxy, at $R < 8.5$ kpc, the gas scale height is observed to be nearly constant \cite{1962dmim.conf....3O,DL1990}: this was a long-standing  puzzle. To solve this problem,  gas gravity was brought into the picture; and  this was one main motivation to propose the multi-component disk plus halo model (\cite{NJ2002}; also, see the discussion in Section \ref{sec:3.2}, and Section \ref{sec:4.1}). 

\medskip

\noindent {\bf Disk Flaring: Why it should be included in future studies}

Thus, there is clear and growing evidence, both observationally and theoretically, that the stellar scale height is not strictly constant; rather, it increases moderately, by a factor of few, within the optical disk. Despite this,
the assumption of a constant scale height is commonly used in the literature, 
  perhaps because it simplifies the treatment. For example, the assumptions of a constant scale height and $\mathrm{R_v/R_D=2}$   
are often used in model building in the literature, including in numerical simulations, as the basic state of a galaxy (e.g., \cite{Haines2019,Donghia2015}); as well as in interpreting observational data, as in the analysis of 2-D kinematic data  \cite{Bershady2010}.  We stress that adopting these wrong assumptions is not justified physically and could lead to erroneous results.
For example, the early and much-cited work on 2-D kinematics using IFU data by \cite{Bershady2010}  has a constant disk scale height as a built-in assumption. However, since the actual scale height increases radially, the surface density values they deduce from their measurement of vertical dispersion would not be accurate. The true surface density would be lower than the values they deduce.

We urge that future theoretical studies of vertical structure of galaxies including N-body simulations, and also analysis of observations, should explicitly incorporate a flaring stellar disk 
for a correct understanding of the dynamics and evolution of a galactic disk. 

\subsection{Dynamical implications, and future directions\label{sec:5.5}}

The finite height of a galactic disk can affect many dynamical features, including the in-plane dynamics of a disk: such as, the criterion for stability against axisymmetric perturbations of a disk of finite height \cite{KimO2007,Wang2010sta,Behren2015}, and star formation \cite{OstrikerKim2022}. 
The vertical thickness of a disk also tends to dampen or quench the swing amplification of non-axisymmetric perturbations in a disk \cite{GhoshJog2018}. That is,  a finite  disk thickness  leads to suppression of spiral arms. A finite disk thickness also leads to  decrease the lifetime of a spiral density wave because it increases the group transport velocity \cite{GhoshJog2022}. A finite height of the disk thus makes it more stable against instabilities. The actual value of height will affect the effectiveness of these phenomena. 
Due to flaring of the stellar disk, these effects of a finite disk thickness would be more important at large radii. For example, a disk would resist the onset of non-axisymmetric instabilities at large $R$, and the spiral arms would be shorter-lived in the outer disk. 
Thus a flaring disk makes it harder for a disk to support instabilities within it at large radii. 

 On the other hand, a flaring disk itself is more susceptible to being tidally distorted at large $R$. The multi-component system is harder to be disturbed than the stars-alone disk \cite{SJ2022}; this needs to be checked by numerical simulations. As discussed earlier,  the constraining effect of gas and dark matter halo reduces the flaring of the stellar disk. Thus, overall, a multi-component disk and especially a dark matter halo, tend to stabilize a stellar disk against distortion by controlling the flaring to be of a moderate level. 
 
Future theoretical studies, including N-body simulations, should take account of the flaring stellar disk, in order to portray the correct dynamical picture.

\noindent {\bf Future directions for study of a flaring stellar disk}

Firstly, and most importantly; new and accurate observational data, and better analysis of the data, are needed to quantitatively better define the phenomenon of a flaring disk.
To study the stellar scale height as a function of radius; a proper, iterative modeling of the observed intensity profiles, 
$I(z)$, that takes account of the variation in thickness with radius is needed (see Section \ref{sec:2.1.1}). This will directly give the measured stellar scale height as a function of $R$. 

Ironically, even when a radial variation in scale height is claimed to be  measured, as in the recent work by \cite{Ranaivo_flaring_2024} involving analysis of S$^4$G data on galaxies, the equation used by them for fitting the data (Eq. (\ref{eq2.2}), which was derived by \cite{vdkS1981a}) implicitly assumes a radially constant scale height.
Hence the values of scale heights as measured by \cite{Ranaivo_flaring_2024}  are not strictly correct, as discussed above in Section \ref{sec:5.3.4}.  

Alternatively, the stellar scale height values can be calculated theoretically, if the necessary input parameters are available from observations. The
 wide availability of IFU kinematic data for 2-D stellar kinematics in external galaxies (Section \ref{sec:2.1.4}) can now allow for a direct measurement of the stellar velocity dispersion as a function of $R$. Note, however, that it is a painstaking task to extract this information from the data (e.g., \cite{Falcon2017};
 see Section \ref{sec:8}).
 If other physical parameters such as disk surface density are known from observations, then the multi-component disk plus halo model could be applied to theoretically predict the radial variation of stellar vertical scale height. This can check if flaring  is indeed a general feature of a galactic stellar disk.  

In addition, the various dynamical implications of a flaring disk, as discussed above, need to be studied in future.

\section{Density profile and shape of dark matter halo\label{sec:6}}

In this section, we consider an important application of disk vertical structure as a diagnostic to determine the parameters of the dark matter halo in a spiral 
galaxy.{\footnote { This section can be read as a largely stand-alone mini-review on this topic.}
We highlight how the observed gas vertical scale height can be used as a complementary constraint, along with the observed rotation curve, to quantitatively determine the parameters such as the density profile and shape of the dark matter halo.
 This approach has been applied to a number of galaxies, including the Milky Way, as discussed here.
 
The shape of the dark matter halo is an important parameter to understand its formation and evolution (e.g.
\cite{Natarajan2002} for an early discussion on this); and also because it can strongly affect the dynamics of the disk embedded in it, including features such as warps and bars in the disk. However, the properties of the halo shape are not yet well-understood. The deduction of both, the density profile and the shape of the halo, using observations of tracers is an important problem; but so far it has not received the focused attention that it deserves.

\subsection{Background on dark matter halos in galaxies\label{sec:6.1}}

First, we give a brief background on the dark matter halos in  spiral galaxies, to put this topic into a proper context.
It is now well-accepted that a typical spiral galaxy is located in an extended, massive dark matter halo (e.g., \cite{BT1987,ARAADM2018}). Within the extent of the optical or stellar disk of a galactic disk, most of the mass is in baryons. The fraction of dark matter increases with radius; such that at large radii, beyond the optical disk, 
most of the galaxy mass is in  dark matter  and not baryons (see \cite{Combes2002}; and, also the discussion in the next paragraph). 
The dark matter halo is believed to play an important role in the formation, evolution and dynamics of a galaxy \cite{ARAADM2018}. 

Despite being the dominant mass component of a galaxy, even the basic physical parameters such as the mass, size, radial density profile, and the shape of the dark matter halo are not yet well-understood. The dark matter halo thus remains one of the most important, unsolved problems in the study of galaxies. Since the dark matter halo interacts gravitationally with  the disk, 
the halo parameters could in principle be determined by its effect on the disk, which can be observed. In fact, this was precisely how the existence of dark matter halos in galaxies was deduced from  the observed, extended, nearly-flat rotation curves that did not show a Keplerian fall-off \cite{Rubin1978,Rubin1980,RobertsMS1975,Rubin1983,Bosma1978,vanalbada1985}.
This implied the existence of more gravitating mass than can be accounted for by the visible matter.  
These studies of the rotation curves of galaxies showed the existence of a dark matter halo whose fractional mass content increases with radius, and which dominates in the outer parts \cite{vanalbada1985,Kent1986,Broeils1997,
Combes2002}. For a summary of dark matter halos in galaxies as deduced from the observed rotation curves in galaxies, see \cite{SofueRubin2001,Ashman1992}. 

For  a disk in rotational equilibrium, the rotational velocity is set by the gravitational force due to the disk, plus halo, (and the bulge) balancing the  centrifugal force. In these studies, the halo is assumed to be spherical  or a slightly flattened oblate spheroid, for simplicity.  
Thus, the rotational velocity is a good tracer of mass distribution, and hence the potential in the plane of the disk. Interestingly, the rotation curve is weakly dependent on the halo shape \cite{SSparke1990}, so 
it  is not a sensitive indicator of the mass distribution normal to the galactic plane.
One needs to consider a tracer distribution extended normal to the galactic plane to study the mass distribution normal to the galactic plane. 
As discussed next, the HI gas scale height is an excellent tracer of the gravitational potential gradient, and hence the shape of the halo, normal to the plane, in the region close to the mid-plane.

\subsection{Motivation for using HI gas scale height as a constraint\label{sec:6.2}}

In a typical spiral galaxy, the HI gas extends several times  ($\sim 1.5-2$ times) farther out than the optical or stellar disk (e.g., \cite{GioHaynes1988}). 
The optical or stellar disk size as observed is typically $\sim 4-5$ R$_D$ \cite{vdkS1981a,vdkF2011,NJ2003}. It is well-known than the HI gas scale height flares in the outer Galaxy (\cite{Wouter1990,Levine2006,NJ2005}; also, see Sections \ref{sec:3.2},\ref{sec:4.3.2},\ref{sec:5}). For HI gas in hydrostatic equilibrium, a possible cause of HI flaring is that the net gravitational force acting perpendicular to the disk decreases with radius, while the velocity dispersion of HI gas is observed to remain nearly constant with radius (Section \ref{sec:4.2.3}). 
The contribution to the net vertical gravitational force comes from stars, gas and the dark matter halo; with the halo dominating in the outer disk. Thus the halo plays a major role in determining the vertical disk structure in the outer disk, especially beyond the optical disk. The HI gas vertical scale height is a sensitive indicator of the 
the vertical gradient of the potential close to the mid-plane, and hence is the best tracer to study the vertical mass distribution, and the shape of the halo, close to the mid-plane. We focus on this new approach in this section.

The model HI scale height depends critically on the choice of the HI velocity dispersion (Section \ref{sec:4.2}). Typically this is observed to be a constant at $\sim 8$ km s$^{-1}$ in the inner parts of a galactic disk, then tapering off to 7 km s$^{-1}$ in the outer parts -- see the discussion in Section \ref{sec:4.2.3}); also, see \cite{2009AJ....137.4424T} for a measurement for the THINGS (The H I Nearby Galaxy Survey) sample of galaxies. The actual value of gas velocity dispersion used in the model calculations is taken from the observed value for each galaxy under consideration, as discussed later in this Section. 

The rotation curve is only weakly dependent on the halo shape  \cite{SSparke1990}. Therefore, 
 the observed HI gas vertical scale height, especially in the outer disk, can be used as an additional and complementary constraint -- in addition to the observed rotation curve that mainly traces the planar mass distribution -- to determine the density profile and the shape of the halo. 
 Using the above two independent criteria on an equal footing allows one to better constrain the 3-D mass distribution of the 
dark matter halo; and hence the shape of the halo.
This is a novel approach and has been used to determine the halo parameters (e.g., \cite{Olling1996,BecqCombes1997,NJ2005}). This is an important application of the disk vertical structure, as we discuss  in this Section.
In contrast, normally only the rotation curve is used to determine the 
radial mass distribution of the dark matter halo halo in a galaxy (e.g., \cite{vanalbada1985,deBlok2001}).

Measuring the HI vertical scale height is difficult (\cite{SA1979}; also,  Section \ref{sec:6.5}), so the few galaxies for which the HI scale height values are known, have been studied so far to obtain the halo parameters using this approach.

We note that other tracers have also been used in the literature to study the halo parameters, we will discuss these briefly  in Section \ref{sec:6.7}.

\medskip

\noindent {\bf Radial density profile and shape of dark matter halo: values tried}

 The halo parameters, namely its radial density profile and shape, can be obtained from numerical simulations, or  from analysis of observations of visible tracers such as stars or gas. Generally the rotation curve alone is used to trace the halo parameters. 
For a particular choice of radial density profile of the halo adopted, the de-composition of the observed rotation curve in terms of the contributions from the disk, bulge and halo allows one to constrain the halo parameters for the halo profile adopted. 

Numerical cosmological simulations give a centrally peaked, NFW (Navarro-Frenk-White) radial density profile for a dark matter halo \cite{NFW1996}. This is often used as a choice of halo density profile to model the observed rotation curve in terms of the contributions by the stellar disk, gas disk, and the dark matter halo, to constrain the halo parameters \cite{deBlok2001,Sofue2015,LiLelli2020}. Once high resolution rotation curves became available and it was possible to model the rotation curves close to the centre, it became clear that the dark matter halo is actually better fitted with a pseudo-isothermal distribution with a flat, constant-density core \cite{deBlok2001}. This is particularly true in the case of late-type, or low surface brightness (LSB) galaxies. In the literature, both choices are often used to model rotation curves 
\cite{KormendyF2004,Sofue2015}. In fact,  some studies \cite{Jimenez2003} have shown that both give equally good fits. Studies of dwarf galaxies often use the density distribution given by \cite{Burkert1995} for the halo. 

Interestingly, studies in the literature that use the HI gas scale heights and the rotation curve together as tracers for the halo potential - which is the case we will focus on -  have mostly used the pseudo-isothermal density distribution with a flat core, as it is found to give a better fit to the 
observations.

\medskip

\noindent {\bf Shape of the dark matter halo}

\medskip

Cosmological N-body simulations typically give triaxial-shaped halos, with a $\neq$ b $\neq c$, where a and b are the semi-major and semi-minor axis in the galactic plane and $c$ is the semi-major axis along the vertical axis (or, the vertical direction w.r.t. the plane). In other words, a,b,c are measured along the Cartesian axes X, Y and Z respectively. Typically, the disk lies in the plane defined by $a$ and $b$ axes. In simulations, the axis ratio of the halo, or the shape, is found to vary with radius, and the shape is also seen to evolve with time 
\cite{Bailin2005,Bett2007,Vera-Ciro2011,Chua2019}. 
Simulations with baryonic feedback from the disk give a spherical or an oblate-shaped halo in the inner regions  (\cite{Bailin_etal_2005,Allgood2006, Prada2019}; also, see the discussion in \cite{Palau2023}).

In most studies of real galaxies, the triaxial shape is not taken into account, with some exceptions (e.g., \cite{Dubinski2009, Law2009triaxial}). 
For most theoretical studies dealing with real galaxies, as well as for analysis of observational data of tracers; typically, a spherical ($a=b=c$), or a spheroidal shape is adopted for simplicity:  where $c <a$ denotes an oblate spheroid, and $c >a$ denotes a prolate spheroid. 
Here onwards, we will refer to the quantity, $q = c/a$, the ratio of the vertical-to-planar axes, as the halo shape (normal to the galactic plane); so that $q<1$ and $q>1$ corresponds to an oblate and a prolate spheroid, respectively.
This is the quantity we want to trace using the HI gas scale heights.

The shape measured by analysis of observations appears to depend on the tracer used, and may be sensitive to or reflect the physical region where the tracer is located, as will be discussed in Section \ref{sec:6.7}. Due to the limitation of resolution of data, typically a constant shape across the entire halo is used in the literature, for simplicity.

In the work discussed in Section \ref{sec:6.3}, the dependence of rotation curve on the shape (albeit small) is taken into account to match with the observed rotation curve. For this work, the generalized four-parameter halo density profile (see Eq.\ref{eq4.14}) as given by \cite{deZP1988} is used, see Section \ref{sec:4.3.2} for details.

The shape of the halo in the plane ($\mathrm{=b/a}$) will not be discussed, though we mention it here briefly for a complete picture. The planar shape of the halo can be determined as follows: by Fourier analysis of the observed isophotal shapes of galaxies in the near-IR \cite{RZ1995,Bournaud2005,Zaritsky2013}; by Fourier analysis of HI surface density maps \cite{Eridanus2006,JanineMorph}; by Fourier analysis of the HI velocity fields of galaxies \cite{Schoen1997}; and, by measuring the asymmetry in the HI rotation curves \cite{Janinekinemat}.
 These studies show that, typically,  dark matter halo in a spiral galaxy shows azimuthal asymmetry of type $m=1$ (lopsidedness)  and $m=2$ (ellipticity)
 of a few percent amplitude each, in the disk plane.

\subsection{Dark matter halo traced using HI gas scale height: Self-consistent approach\label{sec:6.3}}

The pioneering idea on obtaining the halo parameters including the halo shape, $q$,  using the observed rotation curve and HI gas vertical scale heights as two independent and simultaneous constraints was proposed in a model by \cite{Olling1995}.
The above model was applied to NGC 4244 \cite{Olling1996} and NGC 891 \cite{BecqCombes1997} which gave highly flattened oblate halos with the axis ratios in the range of 0.2-0.4.
However, the model by \cite{Olling1995} is not rigorous, and does not use a  self-consistent approach to obtain the vertical distribution of stars and gas in the halo potential, as will be discussed later in  Section \ref{sec:6.4}.
 
The above idea of using the observed gas scale height and the rotation curve as two independent constraints to determine the halo parameters was extended; and applied in a systematic and accurate way to the Milky Way by \cite{NJ2005}. They obtained the self-consistent vertical distribution and hence the HI scale height by using the multi-component disk plus halo model, and explored a large range of density profiles and shapes (spherical, oblate and prolate) in a systematic way. For this, the halo was modeled as a general, four-parameter model by \cite{deZP1988} (see \cite{NJ2005} for details; also, see Eq. (\ref{eq4.14})).
A similar approach has also been applied to obtain the dark matter halo parameters of M31 and UGC 7321. These are the galaxies for which observed HI scale height values were available as constraints, as will be discussed later in this section.

\subsubsection{Application to the Milky Way\label{sec:6.3.1}}

The above careful treatment by \cite{NJ2005} has already been described in detail in Section \ref{sec:4.3.2}, as an example of application of the multi-component disk plus halo model for a thick disk case. The details of dark matter halo parameters scanned and the best-fit values are given in the same section. 
\cite{NJ2005} found that while a 
prolate-shaped halo gave an overall better fit to the observed  HI scale heights, no single prolate shape was adequate to 
explain the observed  flaring 
over the entire radial range considered ($R$=8 to 24 kpc) (see Fig. \ref{fig.18}). 

\medskip

\noindent {\bf Progressively more prolate halo} 

\medskip

Instead, a halo that is increasingly more prolate with radius was indicated that could explain the observed steep flaring of HI disk in the outer disk (Section \ref{sec:4.3.2}).
This idea was taken up and worked out in detail by  \cite{2011ApJ...732L...8B} in a subsequent paper. This work is based on the ingenious use of the fact that the mass of a spheroidal shell bounded by two surfaces $m$ and $m +\delta m$ (where the parameter $m$ defines an isodensity contour), is independent of the shape of the spheroidal shell ($q_R$) at $R$ \cite{BT1987}. Thus, if an original spheroidal shell with radius $R$ is distorted to be a prolate spheroidal shell by construction; then the condition of constancy of mass of the shell gives the following condition:

\begin{equation}
\mathrm{ q_R \rho_R(q_R) =  \rho_R(q=1)}  
  \label{eq6.1}
  \tag{6.1}
\end{equation}

\noindent where $\rho_R(q_R)$ is the density along a prolate isodensity contour through $R$ and $\mathrm{\rho_R(q=1)}$ is the density along the corresponding spherical contour with radius $R$.  
The quantity $q_R$ denotes the vertical to planar axis ratio, which is $>1$ for a prolate-shaped halo (Section \ref{sec:6.2}).  
Thus, the density of the prolate shell at $R$ will be lower by a factor $q_R$. Hence, the vertical force near the mid-plane will be lower. Recall that the self-consistent $\rho(z)$ is determined locally at a given $R$ (see Section \ref{sec:4.2.1}, and Section \ref{sec:4.3.1}). The resulting thickness of the disk in pressure equilibrium is strongly dependent on the vertical gravitational force near the mid-plane at a given $R$ (\cite{BJ2007}; also, see Section \ref{sec:4.2.5}). Therefore, the resulting thickness for a prolate shell will be higher. 
A small change in the shape of a shell of a given mass has striking effect on vertical scale height since the mid-plane density decreases linearly as $1/q_R$. This 
could in principle explain the observed steep flaring of HI  if $q_R$ were taken to increase with radius in the outer disk, where the halo contribution to the gravitational force becomes important. This was shown by 
 \cite{2011ApJ...732L...8B}  who proposed and applied this idea to work out the variation of halo shape with radius.
\begin{figure}
\centering
\includegraphics[height=3.2in, width=3.4in]{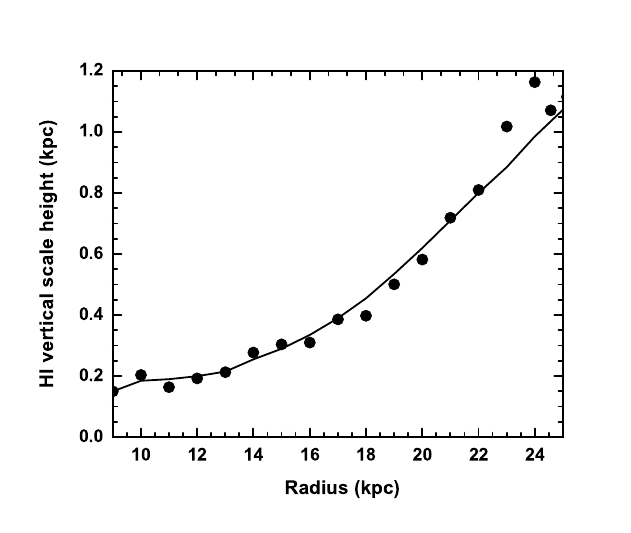}
\caption{Calculated vertical scale height for the atomic hydrogen gas (HI) (solid line) and the observed values (Wouterloot et al. 1990) (filled circles) vs. the radius R in the outer Galaxy. The theoretical curve is the best-fit case and corresponds to a dark matter halo which is increasingly more prolate. In the radial range studied, the halo is found to be most prolate with the vertical-to-planar axes ratio, $q_R=2$, at R=24 kpc. $\: $  {\it Source}: Taken from \cite{2011ApJ...732L...8B} }
  \label{fig.28}
\end{figure}

To start with, \cite{2011ApJ...732L...8B}  consider a spherical, pseudo-isothermal halo as from the Galactic mass model by \cite{Mera1998}, with the density profile $\rho(R)$ as given by Eq.(\ref{eq4.6}), at $z=0$. The density at a given $R$  multiplied by $1/q_R$ -- the reduction in density due to distortion of a shell  to a prolate shape, as given by Eq.(\ref{eq6.1}) -- gives the net reduced density in a prolate shell at a radius $R$. 
This reduced value is used as the halo density in the coupled, joint Poisson-hydrostatic balance equation (Eq. (\ref{eq4.12})) for the multi-component disk plus halo model, in the thick disk case, to obtain the HI scale height at a given $R$. Recall that the vertical structure is determined at each $R$. 

Each shell acts independently, and $q_R$ is a local property of each shell.
 The radial variation of $q_R$ is modeled as a simple second-order polynomial in radius.  See Fig. \ref{fig.28} for the best-fit 
 to the observed HI scale height data  obtained. This procedure constrains the variation in halo shape and gives $q_R$ as a function of $R$. 
The halo shape is found to vary from a spherical shape at R=9 kpc (by assumption), to an increasingly prolate halo with $q (R) =2$ at the maximum radius of $R =24$ kpc studied -- up to which the HI scale height data is available.
This method employing the shell-by-shell reconstruction 
of the halo makes it possible to  study the radial variation in halo shape.
\begin{figure}
\centering
\includegraphics[height=3.2in, width=3.4in]{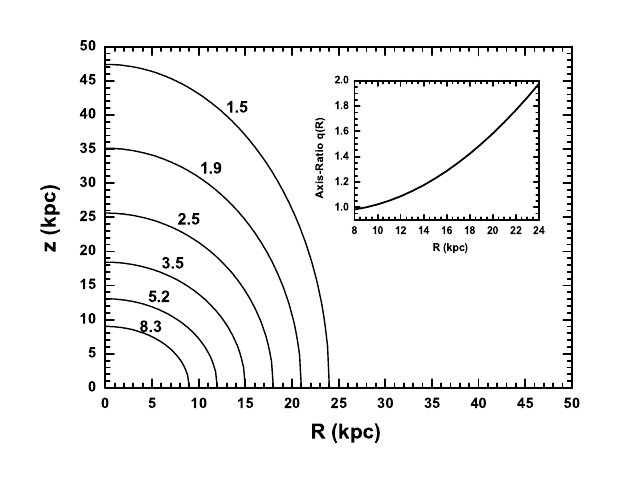}
\caption{Resulting prolate-shaped isodensity contours (on the R-z plane) of the best-fit dark matter halo; with values of density (as one moves outwards) being 8.3, 5.2, 3.5, 2.5, 1.9 and 1.5 (in units of $10^{-3}$ M$_{\odot}$ pc$^{-3}$). These contours denote a progressively more prolate halo in the outer Galaxy. This corresponds to $q_R$, the vertical-to-planar axis ratio, increasing with $R$ as shown in the in the inset. This brings out the increasingly prolate shape of the halo. 
$\: $  {\it Source}: Taken from \cite{
    2011ApJ...732L...8B}}
  \label{fig.29}
\end{figure}
In contrast, most other work in the literature until then had mostly modeled the dark matter halo as a spherical or oblate halo of constant shape (see \cite{2011ApJ...732L...8B} for a detailed discussion and the related references).

Fig. \ref{fig.29} 
gives the isodensity contours (on the R-z plane) corresponding to  the above increasingly prolate best-fit halo in the outer Galaxy; and the inset shows the resulting $q_R$ vs. $R$.
As shown above, an increase in $q_R$ from 1 to 2 corresponds to a decrease in mid-plane density of a factor of 2 (Eq. (\ref{eq6.1})).   Interestingly, since the resulting HI scale height depends non-linearly on the mid-plane density;  this moderate radial variation in the halo shape is sufficient to explain the observed steep flaring of HI scale heights 
by a factor of 6 between $R$= 9 to 24 kpc.

 Interestingly, the resulting prolate-shaped nature of halo agrees with the general trend seen in cosmological simulations of that time, which show a preference for a prolate-shaped halo \cite{Bailin2005,Bett2007}; except that, the simulations show a smaller vertical-to-planar axis ratio. However, in simulations, the axis ratio values are measured over a larger spatial scale of $\sim$ 100 kpc \cite{Bailin2005}; hence a quantitative comparison between these two is not meaningful. For the current status of halo shapes in simulations, see Section \ref{sec:6.2}. These give an oblate shape as the preferred shape; while the observations give a spherical or a slightly prolate shape as the preferred shape \cite{Palau2023}.

In another study, \cite{Kalberla2007} also tried to fit the HI flaring with a prolate halo by treating a multi-component disk plus halo model. They used  the same halo profile (Eq.(\ref{eq4.14})). They found that a constant-shaped prolate halo, even with an axis ratio, $q$, as high as 64 does not fit the steeply flaring HI observed in the outer disk.
They state that this confirms the trend found by
\cite{NJ2005}, namely that the results from a single-shape halo do not fit the scale heights over the entire radial range.  Instead, \cite{Kalberla2007} show that the best-fit to the HI scale height data up to $R$ = 40 kpc       is obtained by considering a spheroidal dark matter halo, and a dark matter ring in the Milky Way disk plane. 
There is a caveat, however, about the scale heights obtained by \cite{Kalberla2007}. For a constant mass halo, the halo central density, $\rho_{0h}$, and the core radius, $R_c$, are not independent as \cite{Kalberla2007} have assumed; rather, these two parameters are  related to each other through the halo shape, $q$, as shown by \cite{NJ2005}. Hence the results for scale heights obtained by \cite{Kalberla2003} for a particular halo shape tried are not accurate. However, their conclusion about mismatch with HI scale heights obtained for a single halo shape would probably remain valid. 

 An important point to note is that the actual value of the scale height and hence the resulting shape, $q_R$, obtained depends critically on the choice of the gas velocity dispersion; which is not  well-known  \cite{NJ2005,2011ApJ...732L...8B}. Hence, the value of $q_R$ as constrained by \cite{2011ApJ...732L...8B} may not be robust. However, given the sharply flaring HI scale height values observed beyond 18 kpc in the outer Galaxy, they expect that the qualitative trend of a halo that is increasingly more prolate with radius, will still hold good.

While the observed HI scale height constrains the shape of the halo to be increasingly more prolate with radius, what gives rise to such a halo shape
 is not understood.  
For that matter, in general  the physical origin of the halo shape in a real galaxy has not been understood or even discussed much (see Section \ref{sec:6.7}). Irrespective of the origin of the halo shape that is more prolate with increasing radius, it is interesting to ask whether such a configuration can even be stable. This was 
studied by \cite{Rathul2013} who calculated the potential for a progressively prolate halo analytically, by applying the potential theory for ellipsoids \cite{Chandra1969,BT1987}.
They calculated the differential acceleration due to  consecutive shells of different eccentricities - corresponding to the best-fit halo values obeying the $q_R$ vs. R as obtained by \cite{2011ApJ...732L...8B}. They showed that the net force due to this is small; hence, such a system is stable over the Hubble time.

A prolate halo, in particular one that is increasingly more prolate at larger radii, could have interesting implications for galaxy dynamics and evolution, which needs to be explored further (also, see Section \ref{sec:6.6}). 

\noindent {\bf Asymmetric HI scale height distribution and halo shape}

A somewhat subtle point is that the observed HI scale height distribution shows an azimuthal asymmetry \cite{Levine2006,KalberlaD2008}, being higher by a factor of two in the northern Galactic hemisphere than in the south. This has been modeled by \cite{Saha2009}  in terms of a multi-component disk plus halo model where the halo potential has a built-in azimuthal asymmetry in the plane of type $m=1$ (denoting lopsidedness) and $m=2$ (denoting bars and spiral arms). Lopsidedness is observed to be common in disk galaxies \cite{RZ1995,JogCombes,Zaritsky2013}; with a magnitude comparable to the observed values for the $m=2$ features.  The origin of lopsidedness is not yet well-understood. However, one generic mechanism that has been suggested is that the halo is distorted to have a lopsided shape due to an interaction with a satellite galaxy
\cite{Weinberg1995,JogCombes}; and the disk then responds to it \cite{Jog1997,Jog1999,Schoen1997,Jog2002vlop}.
The observed asymmetric HI scale height distribution can be modeled in terms of the self-consistent vertical disk distribution as a response to an external halo potential that is asymmetric, as shown by \cite{Saha2009}. This gives the best-fit halo potential to be lopsided  with a fractional  
amplitude 
of 0.18 (for $m=1$); on which is superposed a $m=2$ potential of amplitude 0.19 -- which is out of phase with the lopsided perturbation potential. This choice explains the observed N-S (North-South) asymmetry in the HI scale heights naturally. Such halo distortions can be generated during galaxy encounters or mergers  \cite{Weinberg1995,JogCombes,Ghoshmerger2022}. \cite{Saha2009} argue that the best-fit asymmetric halo they obtain  
is likely to be produced during hierarchical merging of galaxies in the $\lambda$CDM cosmology; and the halo being collisionless, such asymmetry could be long-lasting. 

We point out that in the work by \cite{Saha2009}, the asymmetry in the HI scale heights in the two halves of the Galaxy is modeled in terms of the response of HI to the asymmetry of the potential (of type $m=1,2$) in the disk plane. This is a different tracer compared to the planar isophotal shapes which are also used to deduce the planar 
asymmetry of the halo potential (Section \ref{sec:6.2}).

Alternatively, the azimuthal asymmetry in the HI scale heights may be explained in terms of
intergalactic accretion flows, as proposed by \cite{Lopez2009}.

 Due to better data being available for it, the Milky Way is the best-studied galaxy regarding its halo properties, and using a variety of tracers   (see Section \ref{sec:6.7}). 

\subsubsection{Application to M31 (Andromeda)\label{sec:6.3.2}} 

The above technique of using observed HI scale heights and rotation curve as two simultaneous constraints to study the halo parameters, 
was applied to the Andromeda galaxy (known as M31 or NGC 224) by \cite{BJ2008M31}. This problem is treated using the multi-component disk plus halo model, in the thick disk limit (see Section \ref{sec:4.3}; also, see \cite{NJ2005}).
M31 being an Sb galaxy, the bulge forms a substantial mass component in the central region, and can affect the rotation curve even in the intermediate radial range. Hence the bulge needs to be included in the correct formulation of the equations. 
This modifies the joint Poisson-hydrostatic balance equation used (Eq.(\ref{eq4.12})) slightly, in that the r.h.s. also includes the term corresponding to the density of the bulge, $\rho_b$.
The observed rotation curve is flat or gently falling, hence the net $R$ term in the combined Poisson equation is taken to be zero, as in \cite{NJ2005}. The corresponding three  coupled, joint Poisson-hydrostatic balance equations (Eq(\ref{eq4.13}), modified by the inclusion of $\rho_b$); are solved together numerically, using the same procedure as in \cite{NJ2002}. The input parameters for the disk and bulge are taken from observations  \cite{BJ2008M31}.
For the halo density, the four-parameter dark matter halo model is used (\cite{deZP1988}; or, see Eq.(\ref{eq4.14})). The halo parameter-space values are scanned
systematically to get the best-fit to the rotation curve and the HI scale height data in the outer disk.
In analogy with the Milky Way, the region outside of 3R$_D$ ($\sim 16$ kpc for M31) is treated to be the outer disk and used for this study.
The model results are matched with the observed HI scale height values from \cite{Braun1991} and the observed rotation curve  (see \cite{BJ2008M31}) in this radial range, to obtain the best-fit values for the halo parameters.
A flattened isothermal halo ($q=0.4$), with a  central density, $\rho_{0h}$ = $0.011 M_{\odot} pc^{-3}$; and a core radius, $R_c$ = 21 kpc is found to provide the best-fit.

However, the paucity of data on the HI scale height values -- with only three data points for HI scale height in the outer disk -- and the large error bars on these, means that 
the above result for the best-fit halo parameters is not robust.
Additional scale height data, especially in the outer disk, is needed for this technique to constrain the halo parameters better.
For comparison, in the Milky Way, the HI data was available up to 24 kpc \cite{Wouter1990}, or $\sim 8 R_D$ (that is, eight disk scale lengths); while in M31, the HI data does not extend beyond 26 kpc (or, $\sim 5 R_D$).
It is well-known, and surprising, that the two large, nearby galaxies show such different HI gas distributions, with M31 being overall gas-poor.

\cite{BJ2008M31} show that the best-fit halo obtained is highly flattened, with a shape $q=0.4$; and that such halos lie at the most oblate end of halos seen in cosmological simulations at that time \cite{Bailin2005,Bett2007}. \cite{BJ2008M31} state that the observed flat HI scale height distribution in M31 indicates a high  gravitational force; which constrains the best-fit halo as being oblate. 
They point out, however, that a moderate variation in HI gas velocity dispersion  
 -- that is well within the observational limits -- can result in the best-fit halo being less flattened. For example, a smaller velocity dispersion of 7 km s$^{-1}$ (instead of 8 km s$^{-1}$ adopted in the paper), gives a lower pressure support and hence a smaller model scale height. To explain the observed scale height, the halo then needs to be less oblate; or, be more spherical ($q=0.8-0.9$). Alternatively, assuming a small radial fall-off with radius in the gas dispersion, from 8 km s$^{-1}$ to 7 km s$^{-1}$ in the outer disk, as observed in the Galaxy; also results in a lower pressure support and lower model HI gas scale height.
Hence a less oblate halo is implied, with $q \sim 0.5-0.6$, to explain the observed HI scale height distribution. 
However, \cite{BJ2008M31} find that in this last case, the resulting $q$ values are not constrained well.
This is because the small value of model HI scale height is now mainly sought by a change in the gas velocity dispersion, hence the dependence on the halo shape is weak.

\subsubsection{Application to UGC 7321\label{sec:6.3.3}}

Another application of the above technique was to UGC 7321, which is a low surface brightness, superthin galaxy \cite{BJM2010}. The aim was to study the dark matter halo properties of the 
 low surface brightness (LSB)  galaxies. These are known to be dominated by dark matter halo starting from the smallest radii within their stellar disks, as seen from fitting of rotation curves alone \cite{deBlok1997}. 
The two simultaneous constraints of rotation curve and HI vertical scale heights were used \cite{BJM2010}; the observed values for both of which are known from observations \cite{UsonM2003}. 
See Section \ref{sec:4.3.2} for a brief discussion about low surface galaxies and the observed parameters for UGC 7321.
This galaxy has insignificant amount of molecular hydrogen gas.   
The rotation curve is assumed to be flat for simplicity.  
The corresponding coupled, joint hydrostatic-Poisson equations  for a multi-component disk plus halo model in the thick disk limit (Eq.(\ref{eq4.13}),  with $\rho_{H_2} = 0$), are solved together numerically, using the procedure as in \cite{NJ2002}. This gives the self-consistent vertical density distribution for stars and HI. 

For solving the above coupled equations, the four-parameter halo potential is used (\cite{deZP1988}), or see Eq.(\ref{eq4.14})). The halo parameters are scanned systematically and the resulting model rotation curve and the HI scale height values are matched with the observed values, so as to get the best-fit values of the halo parameters.

We point out that, in reality, however, the rotation curve is rising over most of the disk in UGC 7321  \cite{BJM2010}. Taking this into account affects the resulting $\rho(z)$ distribution and leads to a slightly puffed up disk (\cite{SJ2019}. Hence, ignoring the rising rotation curve as done by \cite{BJM2010} can affect the density distribution including the HI scale height values; this in turn, will also affect the halo parameters deduced by \cite{BJM2010}.

The choice of the gas velocity dispersion value is critical for the resulting HI vertical scale height. \cite{BJM2010} first choose 9 km s$^{-1}$ so as to explain the high observed HI scale height values; and then add a small gradient of $\mathrm{-2 km s^{-1} kpc^{-1}}$ to improve the fit. Here the resulting vertical scale height results were fitted with the observed data over the entire radial range (2-12 kpc), since the halo is important from a small radius.
 For the various gas parameters tried, the best-fit halo was found to be: isothermal, spherical, with the central density, $\rho_{0d}$, in the range of 0.039-0.057  M$_{\odot}$ pc$^{-3}$,   and the core radius, $R_c$, in the range of 2.5-2.9 kpc. The core radius, $R_c$, is just slightly greater than $R_d = 2.1$ kpc. This is in contrast to the high surface brightness (HSB) galaxies where the halo core radius is typically a few times that of the disk scale length, R$_D$ \cite{BJ2013}.
 
It should be pointed out that the high value of velocity dispersion used, namely 9 km s$^{-1}$; and even higher values as suggested by the above choice of velocity gradient, is problematic. This is because the LSB galaxies do not have internal sources of energy input by supernovae (due to lack of star formation), or heating by external encounters since LSB galaxies are typically isolated.  Both these processes could have otherwise explained these implied high HI dispersion values required to fit the observed HI scale height data.
Taking account of a rising rotation curve leads to a puffed-up disk, as shown by \cite{BJBrinks2011,SJ2019}.
This would imply a higher HI velocity dispersion for pressure support. This would make the discrepancy between the velocity necessary for pressure support, and that available in a LSB setting, even harder to explain.

Thus the self-consistent approach, and the application of two simultaneous constraints indicate a dense, compact halo for the  LSB galaxy, UGC 7321 \cite{BJM2010}. 
This result by \cite{BJM2010} supports and confirms the previous results in the literature, obtained by fitting the observed rotation curve alone, namely, that LSB galaxies are dominated by dark matter halo from the innermost regions (e.g., \cite{deBlok1997}),
and that the stabilization of a superthin disk indicates a massive, dominant halo \cite{Zasov1991}. The dense, compact halo deduced by \cite{BJM2010} is shown to explain the superthin nature of this galaxy \cite{BJ2013}.

 We point out that using the same approach of two simultaneous constraints 
except using a non-self-consistent model (Section \ref{sec:6.5}), \cite{Obrien2010} deduce a spherical-shaped halo for UGC 7321.  On the other hand, using the same data  and the same non-self-consistent model as in \cite{Obrien2010},   \cite{Peters2017} obtain a prolate-shaped halo for UGC 7321.

To summarise this subsection: it is surprising 
that the three galaxies studied using this technique, and a self-consistent treatment,
show a wide diversity of shapes. We  will discuss (in Section \ref{sec:6.5}) the various uncertainties involved in using the HI scale height constraint itself, as one possible cause of some of this variation.

\subsection{Dark matter halo traced using HI scale heights: Non-self-consistent models in the literature\label{sec:6.4}}

There are several papers in the literature that have used the HI scale height as a tracer for  dark matter halo; but these papers make ad hoc assumptions, and further they do not obtain a self-consistent vertical distribution. Thus, the resulting scale heights, or the deduced halo properties, from such models are not accurate. We discuss some of these here for the sake of completeness.

For example, the pioneering study on this topic was carried out by \cite{Olling1995}, and was applied to NGC 4244 \cite{Olling1996} (see Section \ref{sec:6.3}). However,  there are several problems with the details of the treatment used: it is not rigorous, and the vertical distribution of stars and gas is not obtained in a self-consistent fashion.
To treat the hydrostatic equilibrium for HI gas, the gravitational force due to stars and halo are included; however,  the effect of gas on stars is not included. Hence the vertical distribution of 
 stars and gas is not obtained in a self-consistent fashion as solutions of the equations describing the coupled system. 
Instead, an ad hoc choice of a Gaussian is assumed for the vertical distribution HI gas; and an exponential distribution with a constant scale height is assumed for the vertical distribution of stars.  
The shape of the halo is assumed to be oblate.  
The  equations are further simplified by various approximations so as to arrive at the shape of the halo. 
Even in the so-called multi-component treatment where the stars and gas are treated to be coupled
(see Appendix C in \cite{Olling1995}), the equations solved  are not general. Rather, these are simplified and solved analytically using the method by \cite{Bahcall1984paper1}, which assumes  the halo to be a perturbation. \cite{Bahcall1984paper1} had developed this approach to solve for vertical distribution for the multi-component stellar system in the inner disk, where this assumption is valid. However,  it is certainly not correct to treat the halo as a perturbation in the outer disk region -- as done by  \cite{Olling1995,Olling1996} -- where the vertical force due to the halo dominates over that due to the disk  \cite{SJ2018}.

Thus, the resulting value of the shape, namely, the oblateness $q$, obtained for  NGC 4244 by \cite{Olling1996} ($q=0.2$); and as obtained for NGC 891 ($q=0.2$) by \cite{BecqCombes1997}, who adopted the method by \cite{Olling1995}; are not accurate. The halo deduced in these two papers is extremely flat. 

A similar criticism applies to the studies to determine the halo shape 
for UGC 7321 \cite{Obrien2010}, and for a number of galaxies by \cite{Peters2017}; who also use an approach similar to \cite{Olling1996}, and make a number of ad hoc assumptions to obtain the scale height. Thus the solution for density distribution obtained  is not self-consistent.
These papers preselected a flattened or oblate halo by choice, and did not even consider prolate-shaped halos -- with the exception of \cite{Peters2017}
who considered a prolate halo as a possibility for UGC 7321. 

Finally, we mention a recent study for a set of face-on galaxies by \cite{Das2023DM}. Their analysis has many 
ad hoc assumptions, a few of which are mentioned here. First of all, unlike the other papers mentioned above in this subsection, here a purely one-dimensional problem is treated, so  that
only the pressure equilibrium of HI is studied, including the vertical force due to stars and the halo.
 The halo is assumed to be oblate by choice and obeys a logarithmic potential. 
The treatment is not self-consistent; instead, an exponential vertical distribution is assumed for stars and gas, with constant and different scale heights for stars and gas.
Further, a constant
gas scale height is assumed; and its value is chosen to be 0.5 kpc or 1 kpc for a particular galaxy. This assumption of constant gas scale height across the disk is a serious error,  and is in direct conflict with the observed flaring of HI gas seen in many galaxies. In fact, in such galaxies, the observed flaring of HI  is a sensitive indicator of the halo shape, as we have seen above. 
Given the various ad hoc assumptions and errors, the analysis by \cite {Das2023DM} is unphysical, and the values of halo flattening as obtained by them are not reliable.

To summarize, the treatment in the papers mentioned in this sub-section is not rigorous nor self-consistent; hence, the resulting values of the halo shape, $q$, obtained are not reliable. Moreover, it is not clear how the resulting value of the halo shape, $q$, is affected given the many assumptions and approximations made in the treatment. 
This has been explicitly mentioned by \cite{Peters2017}.
 
\subsection{Discussion and Future directions\label{sec:6.5}}

From the above discussion, it is clear that the few galaxies that have been studied by this technique; whether using a rigorous approach involving the self-consistent multi-component disk plus halo model (the Milky Way, M31, and UGC 7321; see Section \ref{sec:6.3}); or, using the non-self-consistent models (NGC 4244, NGC 891, UGC 7321, ES0 274-G0001, see Section \ref{sec:6.4}); both show a diverse range of halo shapes. That is, the galaxies studied do not show a preference for a particular shape (also, see
\cite{Obrien2010} for a discussion on this point).
This could represent a genuine variation in shape, although it difficult to see why  halos would form with such a wide range of shapes. 

\medskip

\noindent {\bf Caveats in using gas scale height as a constraint}

Alternatively, the range in halo shapes deduced could also be partly attributed to the shortcomings inherent in the determination of the shape by this method itself. 

The first, and the most serious point is that, the halo extends vertically much beyond the disk and has a spheroidal (non-disk) distribution. 
 Hence, one cannot use the effect of its gravitational force on gas at low z to trace or extrapolate the halo potential or its shape at large z uniquely{\footnote{D. Pfenninger 2005, personal communication}}. Thus the uniqueness of the halo shape deduced this way is not well-established. In particular, the shape at high z could well be different. An ellipsoidal shape for the halo, even if it changes with R, has been assumed for simplicity in the literature; and because this choice is indicated physically for a self-gravitating system. In reality, the shape may not even be pure ellipsoidal, especially when taking account of the back-reaction of the disk.
This point has not been noted in the various papers that have used this technique so far.

Second, the model gas scale height obtained theoretically depends sensitively on the choice of the velocity dispersion and its variation with radius \cite{NJ2002,NJ2005}, both these quantities are not known accurately observationally  \cite{NJ2002,NJ2005,Peters2017}. This will affect the resulting model values of HI gas flaring. A small decrease in the gas velocity dispersion, as observed in the Milky Way (for details, see \cite{NJ2005,SJ2018}) -- if also applied to other galaxies such as NGC 4244, NGC 891 or M31 -- could lower gas pressure support and hence would indicate slightly less flattened (or more rounded) dark matter halo, as was pointed out  by 
(\cite{BecqCombes1997,BJ2008M31,Obrien2010}; also, see Section \ref{sec:6.3.2}).  Thus, the results for $q$ are sensitively dependent on the gas dispersion and its radial variation.

Third, the measurement of HI disk thickness is a challenging task. For example,  even a small deviation in the inclination angle from edge-on (or, 90$^0$), and the HI optical depth, can affect the values of the thickness measured as shown in the classic work by \cite{SA1979}.
A small deviation in inclination angle could also affect  the rotation velocity measured and this will then affect the estimated mass of the halo; see \cite{Peters2017} for a discussion of this point and other problems inherent with the technique of using HI scale height as a constraint. 
Note, however, that the availability of 3-D HI data cube now overrides these problems to some degree. This will be further discussed next.

Fourth, here two independent constraints are being used: the rotation curve to trace the halo potential in the plane, and the HI gas vertical scale height, to trace the halo potential in the vertical direction. This approach would, in principle, give an overall better fit to the 3-D density distribution of the halo by matching the model results with the data, than using either one of the two criteria singly. However, it is also true that the best-fit  parameters; namely, the core radius and the central density, obtained this way  
give a poorer fit to either the rotation curve or vertical scale height, compared to  if the halo parameters were determined by fitting a single constraint alone.
The best-fit parameters obtained using two independent criteria are in a parameter range that is the overlap from that due each of the constraints; 
so that the final best-fit parameter value obtained this way  was not the best-fit obtained using either of the two constraints alone (see the discussion in \cite{BJ2008M31,BJM2010}). 

\noindent {\bf  Future Trends:}

Despite these caveats, the approach using the HI gas scale height and the rotation curve still remains the best way to determine the vertical gradient of the potential and hence the halo shape on a small scale, near the mid-plane (also; see Section \ref{sec:6.7}). 

In the past, the main limitation of applicability of this technique has been the difficulty of measuring the HI scale heights \cite{SA1979}. 
These measurements were known for only a handful of galaxies which could be used to study the dark matter halo using this approach (Section \ref{sec:6.3}). This problem may be now mitigated to a large degree by
the availability of modern 3-D HI data cube for galaxies.  There has been tremendous growth in the availability of such data  
in the past 10-15 years, since the work reported in the  papers in Section \ref{sec:6.3} and Section \ref{sec:6.4}. This new data can allow us to map a complete 3-D distribution of HI including an accurate measurement of the HI scale heights and rotation curves for external galaxies, 
 without the errors and uncertainties inherent in the older data and its analysis. 
(see e.g., \cite{Teodoro2015,Lelli2017,Patraspiral,2023MNRAS.524.6213B}). 

We urge that it is now time for a systematic  study of the halo shape using this technique 
 for a larger number of galaxies;  based on the modern data, and adopting the self-consistent approach that uses the multi-component disk plus halo model (see Section \ref{sec:6.3}).
Such a study would give a systematic understanding of halo profile and shapes for a larger number of galaxies. 
It would tell us 
if the results show a preference for any particular halo shape; or if there is any dependence on galaxy type; and if such a result is statistically significant.

\subsection{Shape of halo and its effect on disk dynamics\label{sec:6.6}}

The shape of the dark matter halo is an important physical parameter. First, it plays an important role in galaxy dynamics and evolution \cite{Ryden1990,Bekki2002}, and traces the evolution of a galaxy \cite{Natarajan2002,Alimi2024}. The  shape could depend on its constituents \cite{Combes2002,versic2024DM}. Second, it can have important implications for the disk dynamics, as discussed next. The dark matter halo and its radial distribution has been well-recognized as being important for the formation and evolution of a galaxy (see e.g., Section \ref{sec:6.1}; also, see \cite{ARAADM2018}).

First of all, it is worth stressing that the dark matter halo has a stabilizing effect on the disk, which makes it difficult to form planar instabilities
\cite{Jog2014Q,Ghosh2014LSB,GhoshJog2018}. Further, the dark matter halo vertically constrains the disk \cite{NJ2002,SJ2018}. This effect is stronger in the outer disk where the halo dominates. Hence the halo can make the disk resist distortion by a tidal encounter \cite{SJ2022}. Although the above dynamical studies considered a spherical halo for simplicity, the trends in the results obtained would be true irrespective of the halo shape.

The halo shape, and the orientation of the disk in it, are believed to have a strong impact on the disk dynamics, such as the effect on warps and bar
(see Section \ref{sec:7.2} for a discussion about warps).
 The origin of warps has been attributed  to: a misaligned angular momenta of the disk and the halo \cite{DebaWarp1999};   to a disk in a tumbling triaxial halo \cite{Dubinski2009}; or a disk that is tilted w.r.t. the halo mid-plane \cite{Han2023warp}. A prolate halo shape has been indicated to explain long-lived warps \cite{Ideta2000,Ryden1990}. 
However,
\cite{Rathul2013} calculated analytically the gravitational potential for the progressively more prolate halo as deduced for the Milky Way by \cite{2011ApJ...732L...8B}; and showed that the shape of the halo has little effect on the longevity of the warp.
Similarly, the disk response, and hence the radius of onset of warps, is shown to be nearly independent of the shape of the halo \cite{Pranav2010}. On the other hand, the bar evolution is shown to be affected by the shape of the halo \cite{AnkitDMshape}.
Thus the role, or even the possible importance, of halo shape on disk dynamics is not well-understood; this topic needs to be explored more.

\subsection{Other tracers and complementarity\label{sec:6.7}}

Here we briefly discuss the other tracers that have been used to measure the halo shape, so as to give a broad coverage of the topic. It is difficult to study the halo shape, in particular the shape normal to the galactic plane; since one needs sufficient number of luminous tracers, that are distributed normal to the galactic plane, up to a large distance from the mid-plane. 
In the literature, various tracers have been used to study the halo shape. For example, stellar streams in the Milky Way have been used as a tracer; and these
indicate the halo to be oblate, or such studies  have  used  an oblate halo as a preselected choice while analyzing the data \cite{Ibata2005,Helmioblate,Vera-ciro2013,Ibata2024stream}. On the other hand, some studies using stellar streams indicate a prolate halo \cite{Helmiprolate,Palau2023}. For a recent review of stellar streams as a tracer of dark matter halo, in the context of \textit{Gaia}, see \cite{Bonaca2025stream}. Globular cluster distribution in the Galaxy indicates a spherical halo in the central 20 kpc \cite{PostiHelmi}. 

There are, however, some cases where a prolate-shaped halo is strongly indicated, as in polar ring galaxies \cite{CombesArnaboldi,Khoperskov2014}. In some cases, the halo shape can vary with radius within a 
single galaxy, as for the Milky Way (\cite{2011ApJ...732L...8B,Bovy2016haloshape}; see Section \ref{sec:6.3.1}); and in external galaxies \cite{Khoperskov2014}.

The shape of the dark matter halo measured by different tracers seems to depend on the choice of the tracer used,
and it may also be sensitive to the
physical region where the tracer is located.
It could also depend on the properties of the constituents of the dark matter halo \cite{Combes2002,versic2024DM}. Due to the limitation of resolution of data to be used as a constraint, typically a constant shape across the entire halo is used, for simplicity.
The paucity of some of these tracers (e.g. globular clusters) means that using these tracers, the potential cannot be sampled uniformly. Hence, these are less reliable as tracers of the halo shape 
than, say, the HI vertical scale height as a tracer.
For a good summary of different tracers used, and the resulting shape of the halo deduced, see Section 6 in \cite{Palau2023}.

Most of these other tracers are located at a large distance from the mid-plane; hence they trace the halo potential at high $z$ values. This is complementary to the information on the halo shape deduced from HI distribution, which is distributed in a thin disk close to the mid-plane.  

In summary, the halo shape as deduced from various tracers seems to show a bewildering variety of shapes and a range of axes-ratio values, even within large spiral galaxies.  The physical origin for this diversity has not been clearly understood or discussed much in the literature. This needs to be addressed.
We suggest that the variety of halo shapes deduced reflects to some degree the physical region occupied  by the tracer and/or the uncertainty with the technique using a particular tracer. The last point was discussed in Section \ref{sec:6.5}, to show that this could be partly responsible for the variety of halo shapes seen in different galaxies using the same tracer, namely the HI vertical scale height.

To conclude, a systematic and comprehensive study of the halo profiles and shapes  using the HI scale heights as well as other tracers is required, to get a full understanding of the halo parameters.

\subsection{MOND (MOdified Newtonian Dynamics) as an alternative to Dark matter halo\label{sec:6.8}}

In this review, we have taken the standard $\lambda$CDM approach; where the dark matter halo  has been evoked to mainly explain the nearly flat rotation curves observed in the outer regions of spiral galaxies (Section \ref{sec:6.1}). 
It is interesting to ask whether at low accelerations as seen at large $R$ and large $z$ away from the mid-plane, an alternative theory of gravity, namely, MOdified Newtonian Dynamics (MOND), could be relevant.
MOND, an alternative to Newtonian gravity, was proposed by \cite{Milgrom1983}; meant to be applicable in regions of low accelerations, such as the outer disks of galaxies. This was meant to explain the observed, nearly flat rotation curves without the necessity of evoking a dark matter halo. Since then much work has been done on the various forms, and tests, of MOND (see, e.g., \cite{Milgrom2014,Famaey2012,BanikZhao}). The effect of MOND on the flaring of HI gas in the outer disk of the Milky Way, was studied by \cite{ChaitraMOND2008}. They showed that the observed HI flaring
between 17-40 kpc could be explained reasonably well by MOND; but in the range of 10-16 kpc, the HI data showed deviations from the prediction of MOND.

We note, however, that stars constitute the main mass component of the disk. Hence it would be worth checking whether MOND can explain the scale height values, including the flaring of the stellar distribution that is seen within the optical disk of Milky Way and other galaxies (see \cite{NJ2002,NJ2002ltr}; and Sections \ref{sec:4}, \ref{sec:5}). It has been shown (see \cite{NJ2002}; or, Fig. \ref{fig.6}) that for the observed parameters, under Newtonian dynamics, a stars-alone disk in hydrostatic equilibrium shows exponential flaring. The inclusion of gas gravity in the coupled system brings down the stellar scale heights which match well with the observed values for $R< 12$ kpc. In this radial range, the dark matter halo contribution to gravity is negligible, and it does not affect the scale heights of stellar distribution significantly \cite{NJ2002}. 
MOND gives an excess effective gravitational force compared to $|K_z|$, the Newtonian vertical force per unit mass (e.g., \cite{BanikZhao}).
Since Newtonian gravity already gives a good agreement with observed scale heights within R$ < 12$ kpc \cite{NJ2002}; therefore, on the face of it,  applying MOND to this radial region would give a mismatch between predictions from MOND theory and the observed data for stellar scale heights. Hence, MOND does not appear to be favoured to explain the observed stellar scale heights in the inner Galaxy ($R < 12$ kpc). However, 
 this should be checked more quantitatively to confirm whether MOND can be ruled out from  the study of disk vertical structure; or, whether it does play some role in the determination of the stellar vertical scale heights,
 including flaring, in the Milky Way stellar disk. 

As discussed above, MOND predicts a higher $|K_z|$ than the Newtonian value. It has been shown that the MOND effects are expected to be significant when the disk surface density falls below the critical surface density of 137 M$_{\odot}$ pc$^{-2}$ \cite{BanikZhao}. We show that in the the Milky Way, with central surface density of 649 M$_{\odot}$ pc$^{-2}$ and R$_D$ = 3.2 (see Section \ref{sec:4.2.3}); the above critical surface density will occur at R$\sim 5$ kpc. Hence, the MOND effects should become apparent at radii beyond R=5 kpc. 
 Thus the radial range of 5-12 kpc  
is the ideal radial range where the future tests for MOND versus Newtonian dynamics should be carried out.
Other possible indicators of the vertical dynamics,  such as the tilt of the velocity ellipsoid, do not yet clearly favour either $\lambda$CDM, or MOND, as discussed by \cite{BanikZhao}.

\section{Related topics\label{sec:7}}

The focus of this review is on the equilibrium vertical structure of a typical "thin" galactic disk. 
In this section, for the sake of completeness, we briefly discuss a few other important topics related to the study of disk vertical structure. These include  the kinematically and chemically distinct {\it Thick disk} component (see Section \ref{sec:Intro} for the
definition) in the Milky Way; and various dynamical features and instabilities along the vertical direction. These are well-developed and active topics of research in their own right. 
For the sake of keeping the scope of this review within limits,  
we do not cover the above topics in detail.
Instead, here  we briefly mention the salient points, 
and unresolved questions about these related topics, and point to a few important references.
The material covered in the earlier Sections (Section \ref{sec:4.2}, Section \ref{sec:4.3}, Section \ref{sec:4.4}) has direct relevance  for some of these topics, in particular Section \ref{sec:7.1}, as will be pointed out in this section.

\subsection{Thick disk\label{sec:7.1}}

The Milky Way, as well as other galaxies,  contain a kinematically and chemically distinct, metal-poor {\it Thick disk} component; in addition to the thin disk component which we have focused on in this review. The topic of {\it Thick disks} is a vast, well-developed topic that is evolving rapidly, triggered by new data and studies involving numerical simulations.
The aim of this subsection is to give a brief glimpse into this topic.

Starting from the work of \cite{1983MNRAS.202.1025G}  in 1980's, over the past forty years, it has been clearly established that the Milky Way has a {\it Thick disk}, which is thicker than the canonical thin disk. However, its properties including the thickness and the fraction of mass in it compared to the thin disk, or its origin, are yet not well-understood  \cite{vdkF2011,BHGARAA,Vieira2022}. The surface density distribution of the {\it thick disk} falls exponentially with radius. The typical scale length for this
fall-off is measured to be $\sim 2$ kpc  \cite{Cheng2012,Bovy2012b753} vs. the typical scale length of $\sim 3$ kpc for the thin disk (e.g., \cite{BT1987,Mera1998}). Thus, most of the mass in the {\it Thick disk} is in the inner Galaxy, well inside R=8.5 kpc. 

 As the name suggests, the {\it Thick disk} component is vertically more extended and kinematically hotter than the thin disk component. The thickness of the vertical density distribution of the {\it Thick disk} is in the range of 0.5-1.2 kpc \cite{vdkF2011} -- for example, a value of 750 pc is obtained using SEGUE photometry \cite{deJong2010}; versus a thickness of 300 pc determined for the thin disk \cite{BT1987}.
The {\it Thick disk} mass is now believed to be comparable to that of the thin disk \cite{Comeron2011,Haywood2013,Snaith2014,Snaith2015,Vieira2022}. The rotational velocity of the {\it Thick disk} lags  w.r.t. the thin disk by $\sim 50 -100 $ km s $^{-1}$ \cite{Wyse1986,Ojha1994,Fuhr2008,Bensby2014}.

 The Milky Way {\it Thick disk} may, in fact, contain several sub-components \cite{Bovy2012b753}. Also, there could be structural continuity between the thin disk and the {\it Thick disk} \cite{Bovy2012a751,Bovy2012b753}. 
Thus dividing a galactic disk into two discrete components -- a thin disk and a  {\it Thick disk} -- may be an oversimplification \cite{Bovy2012a751,Ghosh2023}. However, for the sake of simplicity, most studies still use two discrete components: namely, a thin disk and a {\it Thick disk} (see the discussion in \cite{Ghosh2023}).

A  {\it Thick disk} is common in external galaxies as well  \cite{DalcantonB2002,Yoachim2006,Comeron2011,Kasparova2016,Pinna2019,Comeron2019,Scott2021}.
{\it Thick disks}  were first discovered in external galaxies as an excess of light over the thin disk in edge-on galaxies (\cite{Tsikoudi,Burstein1979,Comeron2011}).
Typically, a sech$^2$ function is used for the density profile to analyze data to study {\it Thick disks} \cite{Yoachim2006,Yoachim2008}. However, this choice of density may not be correct for a real disk.
As already shown by \cite{Comeron2011}, a correct, physically motivated multi-component treatment and the resulting self-consistent density profiles -- which are different from sech$^2$ -- affect the deduced values of mass contained in the thin and {\it Thick disks} (see Section \ref{sec:4.5.1}). \cite{Comeron2011} concluded these to be comparable. They argued that this
favours the scenario of in situ formation of most of the stars in the {\it Thick disk}. 

The above important study by \cite{Comeron2011} needs to be updated using the thick disk approach (as in Section \ref{sec:4.3}), and the inclusion of the effect of the halo, as discussed next (also, see Section \ref{sec:8}). For the study of vertical structure of a {\it Thick disk}, some of the topics developed in this review would be particularly relevant: such as,  the treatment for a physically thick disk where the $R$ term in the disk Poisson equation needs to be kept in (Section \ref{sec:4.3}); 
 the kinematical effects incorporated through the complete Jeans equations; and the inclusion of effect of non-isothermal velocity dispersion (Section \ref{sec:4.4}).
To model a {\it Thick disk} properly,  a correct, multi-component treatment needs to be done, that includes the thick disk approach (as in Section \ref{sec:4.3}), and the
kinematical effects and non-isothermal velocity dispersion (as in Section \ref{sec:4.4}), as well as the effect of the dark matter halo. 
The vertical structure of {\it Thick disks} needs to be studied systematically. 

The origin of the {\it Thick disk} is not yet well-understood. Various scenarios have been proposed for the formation of the {\it Thick disk}. These include internal mechanisms such as  heating of the thin disk by spiral arms;  radial migration within  the disk; or external mechanisms such as  accretion of stars during a merger with  a satellite galaxy  \cite{vdkF2011,Minchev2015,Vieira2022}.  
However, given the observed structural continuity between the two disks, the actual origin of the {\it Thick disk} may be more complicated; and may involve both internal as well as external mechanisms \cite{Vieira2022}.

The topic of {\it Thick disks} in galaxies remains extremely active in all aspects: namely, its structure, composition, dynamics and evolution. 

\subsection{Dynamical features along the vertical direction\label{sec:7.2}}

\begin{figure}
   \centering
   \includegraphics[height=3.2 in, width= 3.6 in]{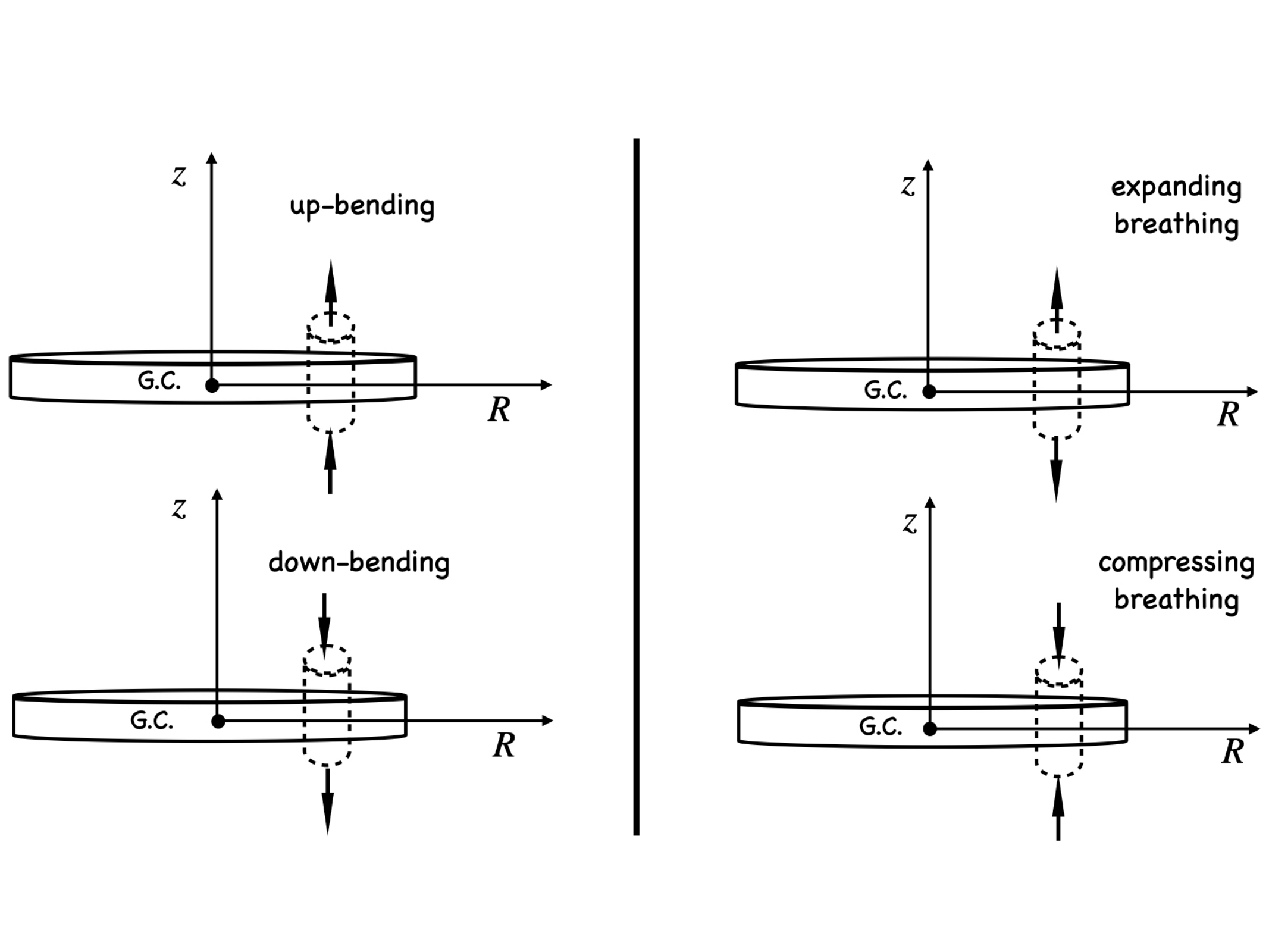}
   \caption{A schematic diagram of bending waves and breathing modes, which illustrates the different bulk motions that characterize these two types of features -- see the text for details.
 $\: $  {\it Source}: S. Ghosh}
       \label{fig.30}
\end{figure}

We have discussed a steady-state distribution in a galactic disk in this review. In addition, the disk also exhibits many complex dynamical features along the vertical direction: such as, warps and 
corrugations (both attributed to bending waves); and  breathing modes -- see Sections \ref{sec:7.2.1} and \ref{sec:7.2.2},  respectively, for details. Both these types of features involve coherent oscillations of stellar disk 
in the direction perpendicular to the mid-plane 
\cite{Widrow2012,Widrow2014,GhoshDeba2022}. These features are collective excitations  
\cite{Weinberg1991,Sellwood1998,Widrow2014}. The associated  vertical motions  
are of two distinct types. The one in which stars on either side of the mid-plane move coherently in the same direction normal to the mid-plane are called bending waves; while the other, in which stars on either side move coherently in opposite directions, namely either compressing towards or expanding away from the mid-plane, are called breathing modes. See Fig. \ref{fig.30} for a schematic diagram which illustrates the different bulk motions shown by these two types of features.

Recent evidence for both these types of features in the Milky Way comes from large-scale, non-zero bulk vertical motions ($\sim$ 10 km s$^{-1}$ magnitude): as seen from \textit{Gaia} (\cite{GaiaColla2018}, see their Fig. 6 c; also, see \cite{Carrillo2018,Lopez2020,Wang2020bulkmotion}; as well as evidence obtained from previous Galactic surveys \cite{Widrow2012,Gomez2012,Carlin2013,Williams2013}. For an axisymmetric disk in a steady-state, the bulk motions along the radial and vertical directions, $v_R$ and $v_z$, respectively, are expected to be zero (e.g., \cite{BT1987,GhoshDeba2022}). Hence, it is important to understand the non-zero bulk motions that are observed and their dynamical implications.
The number density and bulk velocity of stars in the solar neighbourhood show a North-South asymmetry; which indicates a wavelike pattern, probably triggered by a satellite encounter \cite{Widrow2012, Widmark2022}.

These are complex, active  topics and much work has been done on these already.
For example, warps have been long-studied as  features that indicate an m=1 or cos$\phi$ dependence of density along the $z$ direction. 
Early evidence for bending waves such as warps in the Milky Way, and the evidence in other galaxies even now, mainly involves information about the shape  or the spatial distribution of the tracers \cite{Sancisi1976,Bosma1981,ReshetCombes1998}.
Only in recent years, it has become possible to measure the bulk motions associated with warps in the
Galaxy, as discussed above.

In addition to the above two types of features, recent observations from \textit{Gaia} have shown evidence for the existence of a novel, and unexpected, phase-space spiral. This is an extremely active topic of research now (Section \ref{sec:7.2.3}).

\subsubsection{Bending instabilities leading to warps, and corrugations\label{sec:7.2.1}}

  A typical galactic disk exhibits a variety of vertical features w.r.t. the uniform disk distribution, such as warps and corrugations, which can be described in terms of bending instabilities. An axisymmetric disk under an  imposed or external non-axisymmetric potential can produce bending waves w.r.t. the mid-plane (z=0) such that at any location ($\mathrm{R, \phi}$) in the disk, the vertical displacement $\Delta z \propto$ (cos $m \phi$), where $m$ is the azimuthal wavenumber (see \cite{Chaitra2020} for a clear discussion of this topic, and the references therein). If the mode is seen across the entire azimuthal range at a given radius, $R$, then the wavelength of the bending waves satisfies the relation $m \lambda = 2 \pi R$. Here $m=1$
denotes warps; these are commonly observed in galaxies. 
Higher order modes ($m > 1$) have smaller amplitudes and are seen as corrugations of the plane. The study of bending instabilities has a long history, despite which the origin and dynamics of bending waves is not fully understood. Hence the topic is still very active.

A variety of physical mechanisms have been proposed in the literature for the generation of bending waves. The idea that warps are free modes of oscillations was first proposed by \cite{LyndenBell1965}.
The most studied mechanism for the origin of warps is a tidal encounter with another galaxy \cite{HT1969,Weinberg1998,WeinbergBlitz2006,Gomez2013warp}. \begin{figure}
   \centering
   \includegraphics[height=3.2 in, width= 3.5 in]{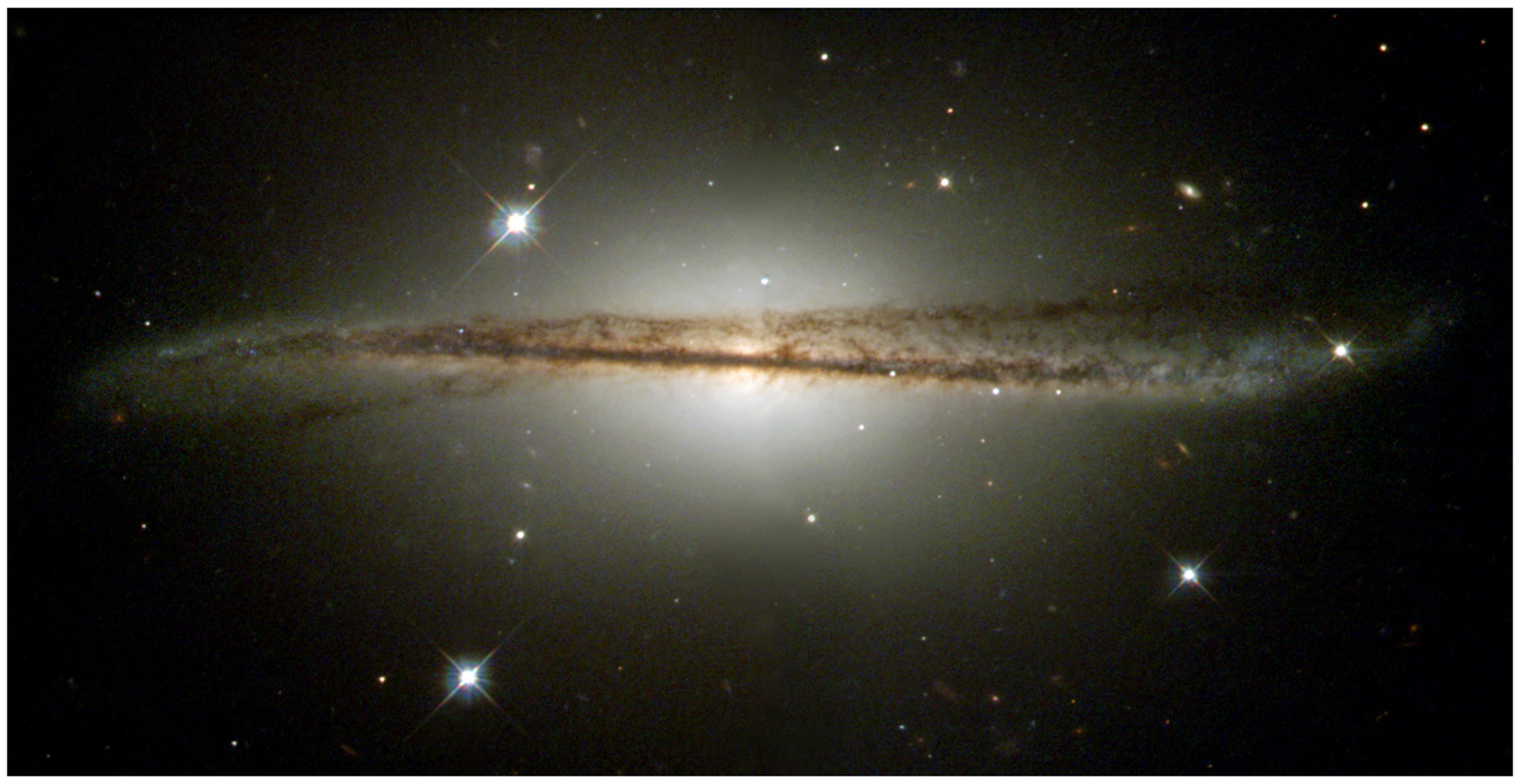}
   \caption{A striking example of a stellar warp in the galaxy 
ESO 510-G13.
The image shows a distribution of stars along $z$ (including the dust lane at the mid-plane) that clearly 
follows an $m=1$ azimuthal variation, indicative of a warp 
in an edge-on galaxy.
 $\: $  {\it Source}: NASA/Hubble Heritage Team}
       \label{fig.31}
\end{figure} 
When triggered by a tidal encounter, if the velocity of the satellite normal to the disk is smaller than that of the stars, the perturbation is taken to be mainly  a bending mode \cite{Widrow2014}. An encounter with a satellite galaxy can cause vertical oscillations, corrugations and flaring of the stellar disk, as shown in the simulations study by \cite{Laporte2019}. The frequency of 
strong warps has been shown to be higher at a high redshift (up to z=2) 
by \cite{Reshetnikov2025} from a study based on the HST and JWST observations. They claim that this is due to a higher frequency of interactions and mergers at high redshift.
The bending instabilities could also be generated by internal processes such as instabilities \cite{Araki1985,Revaz2004warp}; or due to interaction with a bar \cite{Khoperskov2019bar}.
See \cite{Chaitra2020}
for a discussion of other external and internal dynamical processes  for the origin of  bending instabilities. 

\begin{figure}
\centering
 \includegraphics[height= 2.4 in, width= 6.0 in]{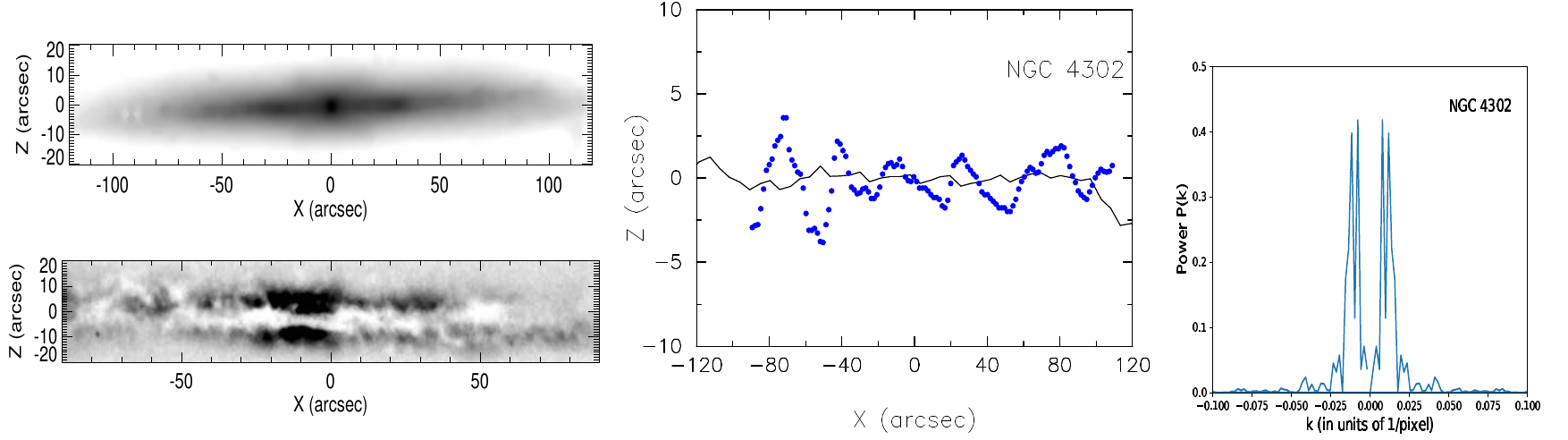}
\caption{Dust corrugations in NGC 4302:  The top left panel shows the Spitzer/IRAC  $\mu$ image; bottom left is the SDSS R-band image given after unsharp masking. The middle panel shows the stellar mid-plane (solid line) and the dust mid-plane (blue filled circles), derived from the left-hand panel images: this clearly shows the dust corrugations. The right-hand panel gives the result of Fourier analysis on the dust mid-plane. The mean wavelength for the two peaks in the power spectrum corresponds to a corrugation of size $\sim 3$ kpc, where 1 arcsec $\sim$ 76 pc. $\: $  {\it Source}: Taken from \cite{Chaitra2020}}
 \label{fig.32}
 \end{figure}

Warps have been the most studied of the bending instabilities: as seen observationally in HI  \cite{Sancisi1976,Bosma1981,Briggs1990,Heald2011}; and in stars \cite{ReshetCombes1998,Sanchez1990warp,Sanchez2003warp,AnnPark2006}, both in external galaxies. See Fig. \ref{fig.31} for a striking example of a stellar warp in the galaxy ESO 510-G13. A Warp have been observed in the Milky Way, using HI as a tracer \cite{Wouter1990,Levine2006,KalberlaD2008}; and using stars as a tracer  (\cite{Han2023warp}, also see references for the stellar warp in the Milky Way in Section \ref{sec:2.1.1}).
 The warps have also been extensively studied theoretically 
\cite{HT1969,NelsonT1995,ChequersW2017}.
 Despite this, however, the origin and dynamics of warps 
 is not yet well-understood.
 
 At the other end of the size spectrum, the higher order (large $m$, or small size and small amplitude) perturbations, known as corrugations, are harder to observe; and hence have not been so well-studied. The higher order bending waves with $m \sim 10$ or larger, known as scalloping, have been observed  in HI at large $R$ in the Milky Way \cite{Kulkarni1982corru,SpickerF1986,Levine2006}.

Corrugations  have been observed in stars in a few external galaxies: in NGC 4244 and NGC 5023 \cite{Florido1991}, and in NGC 5907 \cite{Florido1992warp}. Recently these have been  observed even in dust in a sample of external galaxies \cite{Chaitra2020}, see Fig. \ref{fig.32} for the corrugations in dust seen in NGC 4302. 
Corrugations have been detected in an edge-on superthin LSB galaxy, IC 2233, both, in stars (\cite{MatthewsU2008opt}), and in HI gas (\cite{MatthewsU2008HI}). 

A tidal encounter has been proposed and shown to be the physical mechanism for the origin of corrugations in stars and gas in a galactic disk; see the recent simulations study by \cite{TepperCorru2022}; also, see \cite{Laporte2019}. 

\subsubsection{Breathing motions normal to the galactic disk\label{sec:7.2.2}}

The Galactic stellar disk also shows non-zero vertical bulk motions called breathing modes, with properties as described above; namely, where stars on both sides of the Galactic mid-plane move coherently towards or away from it, see:  (\cite{Widrow2012,Widrow2014}; \cite{GaiaColla2018} - see Fig. 4;  \cite{Carrillo2018,Lopez2020,Wang2020bulkmotion};  and \cite{GhoshDeba2022,Asano2024}, and references therein). Of the above references, \cite{Widrow2012,Widrow2014} are based on the SDSS data; while the rest 
  are mainly based on the detailed kinematic data from the \textit{Gaia} Data Release 2 (\textit{Gaia} DR2) -- except for \cite{Asano2024},  which is based on \textit{Gaia} DR3 data.
  
 The breathing modes could arise directly due to a tidal encounter with a satellite or a galactic sub-halo \cite{Widrow2014,BanikW2022}; or they could arise due to an internal cause such as spiral density waves  \cite{Debattista2014,Faure2014,Monari2016},  or, a bar \cite{Monari2015}. 
 By studying a self-consistent, high-resolution simulation which has prominent spiral arms, \cite{GhoshDeba2022} have shown that  breathing motions are induced by the spiral arms -- see Fig. \ref{fig.33}.
This figure shows the signature of the breathing motions as proposed from their model, namely an increase in the vertical velocity, $v_z$, with $z$ associated with a spiral arm. They show that this result is confirmed by the \textit{Gaia} DR2 data.  
Further, in a follow-up work, \cite{Ankit2022Ghosh} show that the breathing motion is caused by the spiral features generated in the simulation of a fly-by encounter of a satellite. They show that the cause is not the encounter itself as has been argued in the past (e.g., \cite{Widrow2014,BanikW2022}); rather, it is due to the spiral arm triggered by the encounter. 
This point has been confirmed by the recent simulations results by \cite{Asano2025}.

There is observational evidence -- from analysis of \textit{Gaia} DR3 data -- for compressing or expanding breathing motions being aligned w.r.t. various spiral arms in the Galaxy \cite{Widmark2022,Asano2024}. \begin{figure}
 \centering
   \includegraphics[height=3.25 in, width= 3.3in]{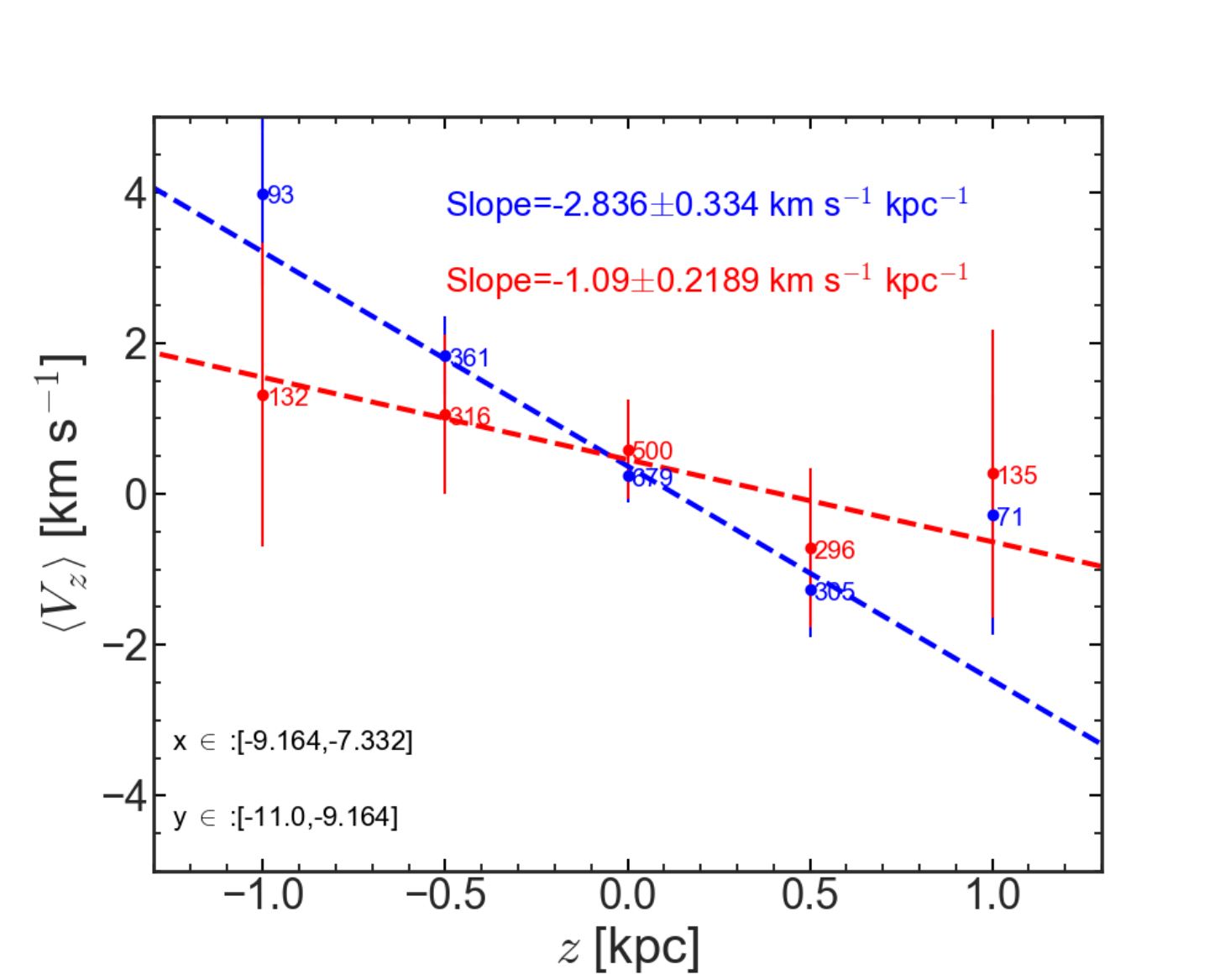}
   \caption{Breathing motions: Plot of mean velocity $<v_z>$ vs. $z$ of stars (between the given x-y range) responding to the spiral pattern in the simulation. The red and blue lines represent the response of the older and younger stars, respectively. This plot clearly shows the breathing modes as predicted by \cite{GhoshDeba2022}: a positive slope would indicate an expanding (positive) breathing motion, while a negative slope (as in this figure) indicates a compressing (negative) breathing motion. The younger stars show a stronger effect. This result is shown to agree with a similar plot obtained using the \textit{Gaia} data for the Milky Way (see Fig. 14 in \cite{GhoshDeba2022}.) $\: $  {\it Source}: Taken from \cite{GhoshDeba2022}}
       \label{fig.33}
\end{figure}
 These studies show that the Local arm and the Outer arm are associated with compressing breathing motions, while the Perseus arm is associated with an expanding breathing motion.
In an interesting study, \cite{Asano2024} show by simulations of an isolated Galaxy, that a transient spiral arm during its growth phase induces compressing breathing motion in the disk stars, while an expanding breathing motion occurs when the arm is getting disrupted. Hence they infer that the Local arm and the Outer arm in the Galaxy are in the growth phase, while the Perseus arm is in the disruption phase. Therefore,  \cite{Asano2024} suggest a general interesting result that the sense of breathing motion (compressing vs. expanding) can be used as a proxy  to decide whether the spiral arm that gave rise to the breathing motion is in a growth phase or a disruption phase, respectively.
The topic of breathing motions is emerging as an active area of research  now.

\subsubsection{Gaia Phase-space spiral\label{sec:7.2.3}}

One of the most enigmatic discoveries from the \textit{Gaia} DR2 data \cite{GaiaColla2018}  is the existence of a prominent spiral feature in the vertical, or $z- v_z$ phase space of the the distribution function of stars, in the solar neighbourhood
\cite{Antoja2018}, see  Fig. \ref{fig.34}. 
Here  $v_z$ is the velocity normal to the disk mid-plane. This has come to be called the \textit{Gaia} phase-space spiral 
or the \textit{Gaia} phase-space snail in the literature \cite{GaiaColla2018,Antoja2018,Widrow2023,
Tremaine2023}. 
 This phase-space feature has a fractional amplitude of a few percent and a complex morphology that depends on the galactic radius, azimuthal location, and angular momentum  \cite{Laporte2019,BH2019phasespace,LiShen2020,
  Antoja2022,Widrow2023}.
 The phase-space feature has a spatial scale of $\sim 200$ pc and a velocity scale of $\sim$ 10 kms$^{-1}$ \cite{Widrow2023}.
A phase-space spiral has also been detected in \textit{Gaia} DR3 data \cite{Antoja2023phase}. For a recent review of this topic, and the Milky Way dynamics in general, in the light of \textit{Gaia} data, see  \cite{Vasiliev2025}.
\begin{figure}
   \centering
   \includegraphics[height=3.1 in, width= 3.8 in]{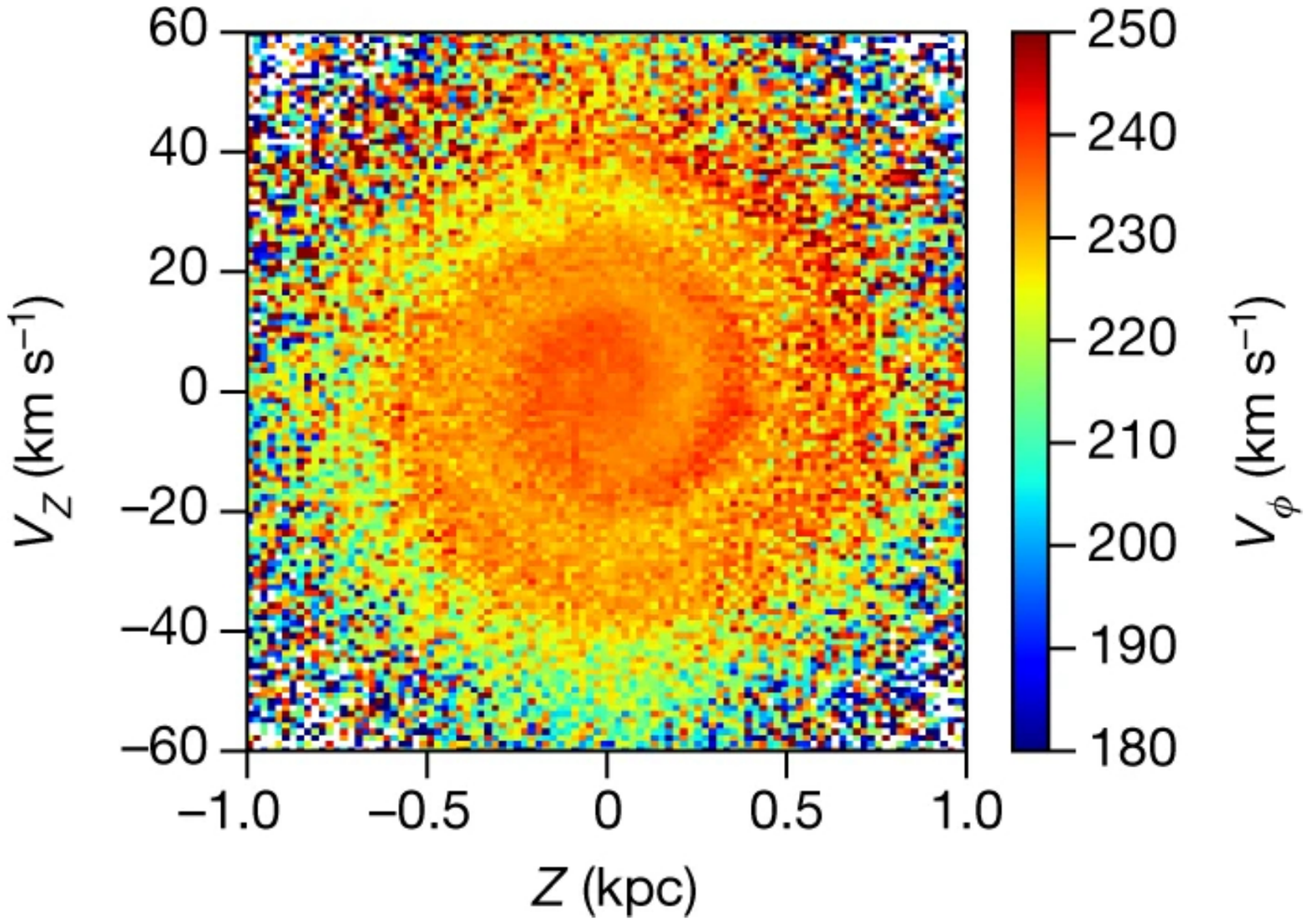}
   \caption{Phase space spiral in the Milky Way  \cite{Antoja2018}:  Plot 
of distribution of stars in the  vertical position-velocity ($\mathrm{z-v_z}$) 
phase space for stars sampled within the galactocentric radial range of 
8.24-8.44 kpc; 
coloured as a function of median azimuthal velocity, $v_{\phi}$, 
in bins of $\Delta z$ = 0.02 kpc and 
$\mathrm{\Delta v_z}$= 1 km s$^{-1}$.   $v_\phi$ is taken to be positive along the direction 
of Galactic rotation. $\: $  {\it Source}: Taken from \cite{Antoja2018}}
       \label{fig.34}
\end{figure}

 The phase-space spiral is believed to be a disturbance in the $\mathrm{z-v_z}$ distribution function from a  past event, such as a tidal  encounter,
that has undergone phase mixing \cite{Tremaine1999,BH2019phasespace,Laporte2019}.
This idea of a kinematical origin leads to an estimated age of $\sim$ 300-900 Myr for the phase-space spiral \cite{Antoja2018}. 
The
Galactic vertical potential near the mid-plane can be represented as an anharmonic oscillator; hence, the  orbits with a smaller $z$ amplitude have a higher vertical oscillation frequency (see Eq. 2 in \cite{Antoja2018}).  Therefore, in this picture,  a perturbation gets sheared into a vertical spiral that gets more tightly wound with time (e.g., \cite{Frankel2023}).
Thus, the phase space spiral as a kinematic feature is shown to get wound up with time.
 \cite{Antoja2018} point out that this estimated age agrees well with the epoch of perturbation due to the last peri-centric passage of the Sagittarius dwarf galaxy, as seen in simulations;  which thus could have triggered  the phase-space spiral.

The origin and dynamics of the phase-space spiral is an  extremely active topic of research now. 
The existence of the phase-space spiral, and the associated observed non-zero bulk motions along radial and vertical directions, $v_R$ and $v_z$, respectively; are taken to be an evidence that the disk is out-of-equilibrium (\cite{Antoja2018,Guo2024}, and references therein). 
This is because a stationary and axisymmetric disk will have zero bulk motions along $R$ and $z$ directions. Alternatively, the existence of phase-space spiral is taken to indicate that the disk is only approximately in a steady-state \cite{Tremaine1999,Widrow2023}; or is under a time-dependent potential \cite{Haines2019} where Jeans equation is no longer valid. 
During simulation studies of a tidal encounter, \cite{Haines2019} find that in low density regions, the estimate of surface density is not given correctly by the usual Jeans equation. Rather, it is an overestimate, which they propose points to a time-dependent potential.
 A tidal encounter with a satellite galaxy has been proposed as one possible physical mechanism for the origin of the phase-space spiral  \cite{Antoja2018,Laporte2019,BH2019phasespace,Haines2019,McMillan2022}. \cite{BanikW2022} propose that a tidal encounter gives rise to both bending and breathing modes in a galactic disk, which then undergo vertical phase mixing which gives rise to local phase-space spirals. 

It has  now been realized that a kinematic phase-space spiral generated by a single event is not adequate to explain all its properties (e.g., \cite{Widrow2023}, and the references therein). For example, N-body simulations have shown \cite{Bennett2022} that an encounter with the Sagittarius dwarf (often applied in this context) cannot match the observed amplitude of the phase-space spiral in the Milky Way.
Another possibility \cite{Tremaine2023} is that the phase-space spirals could repeatedly arise due to a series of perturbations in potential, say, due to galactic sub-halos. These features would then get dissipated due to encounters with the giant molecular clouds in the disk \cite{Tremaine2023}. This would lead to an apparent age or
the net duration of the phase-space spiral to be $\sim$ 0.5 Gyr. Thus, irrespective of the physical mode of origin of the phase-space spiral, it would get erased by scattering off the giant molecular clouds or other small-scale features within $< 1$ Gyr.  
Thus, the existence of a phase-space spiral may not have an origin due to an encounter, and the  kinematic argument for its age would then not be applicable. 

Recently, it has been argued  \cite{Widrow2023} that the phase-space spirals could be swing-amplified perturbations \cite{JT1966swing,GLB1965} in 3-D, where the self-gravity plays a crucial role in the evolution of the features. 
 Further, a co-rotating cloud could excite stationary phase-space spirals. Hence, a simple argument,  based on the epoch of tidal encounter that is 
  used to estimate its age,
 may not be applicable \cite{Widrow2023}. A wake generated in the dark matter halo due to an encounter with a satellite can also give rise to a long-lived phase-space spiral along $z$ \cite{Grand2023wake}.

A phase-space spiral, indicating a deviation from equilibrium, has now also been detected from the \textit{Gaia} DR2 radial velocity catalogue \cite{LiW2023}, by analyzing the residuals of an equilibrium model.
Ripples in the plane have been known for some time \cite{Minchev2009}; also, a phase-space spiral
in the R-$v_R$ phase space has now been discovered \cite{Khanna2019,Hunt2024}. This is a rich and an extremely active subject now. An introduction to the literature and the current ideas on the topic of phase-space spirals is given in this section, citing the relevant papers; so that the reader may get an idea about this fast-developing topic.

\noindent {\bf Phase-space spiral: An out-of-equilibrium feature}

In summary, the  phase-space spiral features are intriguing dynamical entities. They are often taken to be an evidence of a disk that is not in a steady-state or is out-of-equilibrium (e.g. \cite{Antoja2018}; also,
see the discussion above).
However, we emphasise that that the amplitude of the spiral features is only a few \% of the background; 
so the entire disk, or the entire 6-D phase space of the disk, may not be out-of-equilibrium.
 After all, most disks seem to satisfy a density profile 
 as obtained from an equilibrium model (Section \ref{sec:4.2.6}; Appendix A).
 The associated results from such equilibrium models, such as the disk scale heights for stars and gas agree well with observations for the Milky Way \cite{NJ2002}.

Moreover, the origin and life-time of the phase-space spirals is not yet well-understood. These could well be repetitive or generated often; or could be long-lived -as when generated by a co-rotating cloud, as discussed above. Hence, perhaps it is more appropriate to call the phase-space spiral to be an out-of-equilibrium feature, rather than it being an evidence for the entire disk being out-of-equilibrium. This distinction is important for understanding the overall vertical structure and dynamics of the disk. As an interesting example in support of this argument, consider the study by 
 \cite{Widmark2021} who do the analysis of the phase-space spiral, by treating it as a perturbation on the underlying disk potential; and show that this gives information on the galactic potential that is complementary to that given by the bulk of the (undisturbed) matter.
 
Finally, even if the phase-space spiral denotes an out-of-equilibrium feature, so that the disk does not strictly satisfy the vertical Jeans equation; the steady state analysis of a disk in vertical equilibrium as given in this review will provide a useful background to understand the dynamics of the phase-space spiral. It will give the necessary groundwork or the starting point on which to build the dynamical picture of a non-equilibrium feature such as a phase-space spiral. 
We would like to point out that this is analogous to the treatment of a warp or other bending instabilities like corrugations, which are also treated as perturbations on the basic undisturbed, axisymmetric disk; this approach  is routinely taken to study these features (see Section \ref{sec:7.2.1}). . 

Indeed, \cite{LiW2023} have treated the phase-space spiral as a perturbation in an undisturbed disk.
In fact, the dynamics of phase-space spiral has now been used to identify and understand an out-of-equilibrium local region in a disk that is overall in equilibrium \cite{LiW2023,Guo2024}. This is a promising line of research in this topic.

\section{Vertical structure and multi-component disk plus halo model: Future directions\label{sec:8}}

The field of vertical structure of galactic disks promises to remain exciting and active;
driven mainly by new observational data - especially kinematic data, and data for stellar number density; and numerical simulations. A few important  future problems in this field, in particular those  motivated by the multi-component disk plus halo model, are briefly described below.
The model is timely to explain the new data from \textit{Gaia} DR3 and DR4 for the Milky Way, and the data from JWST for high-redshift galaxies. 

\medskip

\noindent {\bf Direct Results and Applications from the Model:}  
  
First, more accurate input parameters for stars including kinematical data are now available for the Milky Way, for example, from \textit{Gaia}. For external galaxies, the various IFU surveys are beginning to provide information on the 2-D kinematics (see Section \ref{sec:2.1.4}), including the stellar velocity dispersion. With better information on the input parameters, more accurate model results for the self-consistent vertical stellar density distribution, and the associated scale heights, can be obtained even for external galaxies. This could, for example, further check and confirm whether a flaring stellar disk is a common feature of galaxies (see Section \ref{sec:5}). For the Milky Way, the theoretical model density profiles can be compared with star counts data for example, from \textit{Gaia}, and LAMOST. 

However, a practical point of concern is as follows. Due to resolution problems, the stellar velocity dispersion measurements from many of the
IFU surveys are still limited to
the inner regions only. For modeling the density distribution, one would like to know the stellar velocity dispersion as a function of radius in the disk. This requires sufficiently broad angular coverage; and also a small sampling distance, or, good resolution. Both these conditions are well-satisfied, for example, in the CALIFA survey, where the measurements cover a distance of $\sim 1-2 R_e$ (see the discussion in \cite{Falcon2017}). The stellar velocity dispersion vs. radius thus measured is given in Fig. 7  in the detailed work by \cite{Falcon2017}; or, see Fig. \ref{fig.35}.   
\begin{figure}
   \centering
   \includegraphics[height= 5.0 in, width= 5.0 in]{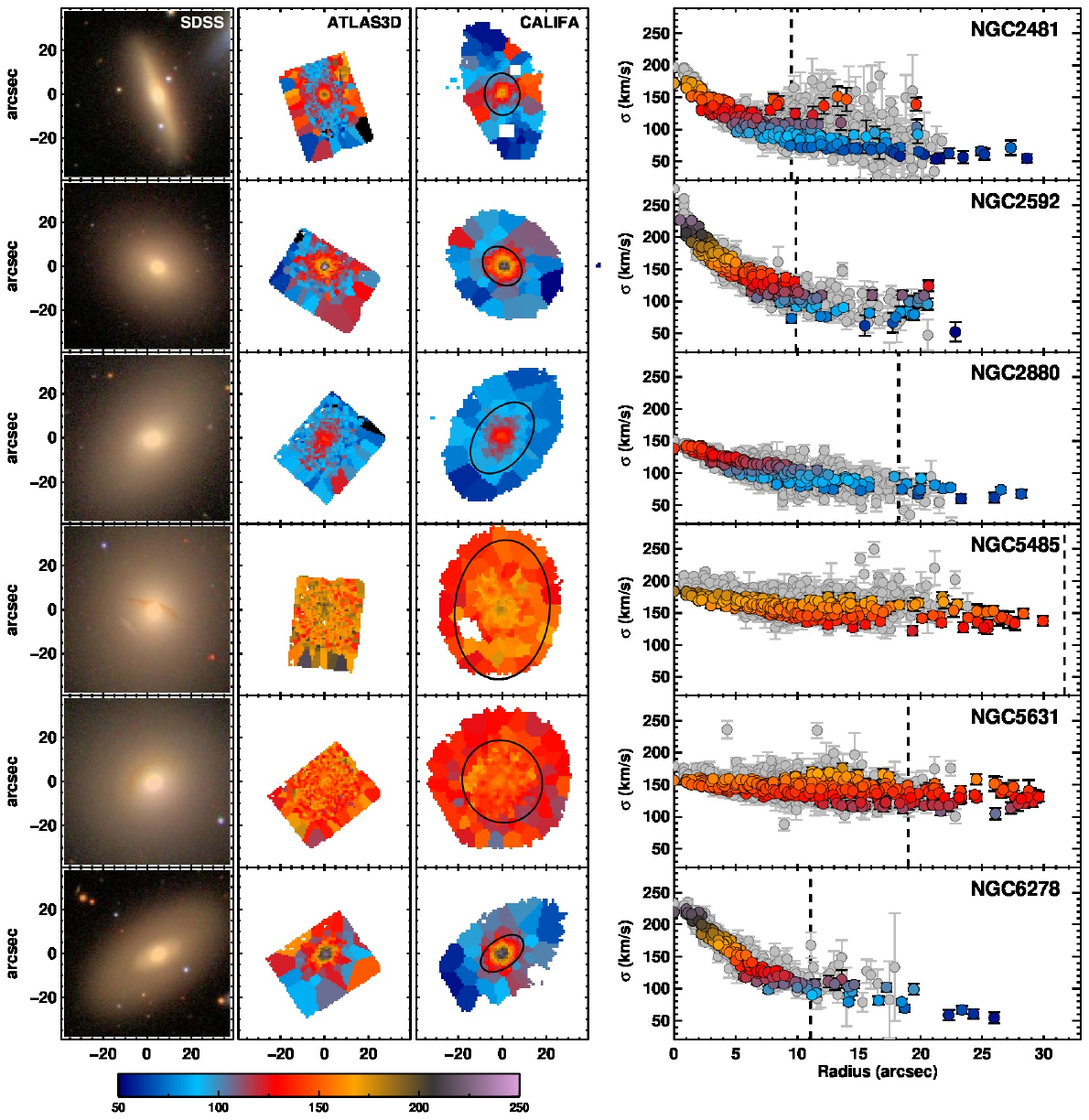}
   \caption{CALIFA Survey results for stellar velocity dispersion in galaxies \cite {Falcon2017}, given in units of km s$^{-1}$. The rows from left to right denote the SDSS images of galaxies, the ATLAS$^{3D}$ velocity dispersion map, the CALIFA velocity dispersion map, and the radial profile for the velocity dispersion (in colour for CALIFA and in grey for ATLAS$^{3D}$. The size of the ellipse shows one effective radius and is indicated by the dashed line in the right-most panel. Thus, the CALIFA survey typically measures values of stellar velocity dispersion beyond one effective radius, $R_e$. $\: $  {\it Source}: Taken from \cite{Falcon2017}}
       \label{fig.35}
\end{figure}
In other surveys, often a single average value for the dispersion, say within the effective radius, is extracted from the data, as done from the SAMI data  by \cite{Oh2022}. A detailed analysis on lines of \cite{Falcon2017}  needs to be done for data from other surveys to extract the values of stellar velocity dispersion as a function of radius.

Another future problem worth doing is as follows. With the detailed information about the stars by the subtype in the Milky Way now available from \textit{Gaia}, it would be possible and worthwhile to formulate a model with a multi-component stellar disk (that is, a disk consisting of various sub-types of stars), and which  also includes the effect of gas and halo (see Section \ref{sec:4.2.1}).

Second, the multi-component disk plus halo model and analysis (Sections \ref{sec:4.2} to \ref{sec:4.4}) can be naturally extended to study the vertical structure of the kinematically and chemically distinct {\it Thick disk} in the Milky Way in a systematic way, as discussed in Section \ref{sec:7.1}.   
To do this, a correct multi-component disk plus halo treatment that includes a thin disk, a {\it Thick disk}, treated in the field of the halo, would need to be formulated in the  (physically) thick disk approach (Section \ref{sec:4.3}). 
If the input parameters for the thin disk and a {\it Thick disk} are known from observations,
as in the Milky Way (see Section \ref{sec:7.1}); then the above procedure can give the density profile, $\rho(z)$, for each disk component. 

On the other hand, when the disk parameters are not known well -- as in external galaxies -- then matching the above resulting model profiles with the observed intensity profiles, the parameters of the thin disk and the 
{\it Thick disk} can be constrained, simultaneously. This would be on lines of what was done by \cite{Comeron2011} except that they had used the simpler thin disk approach; and they did not include the effect of the dark matter halo  (see Section \ref{sec:4.5.1}, and Section \ref{sec:7.1}). Thus, the approach outlined here would be a more realistic treatment of the problem. 
Since {\it Thick disks} are ubiquitous in external galaxies (see e.g. Section \ref{sec:7.1}), this study would provide an insight into the vertical structure of 
{\it Thick disks} in these galaxies.

Third, the multi-component disk plus halo model can now be applied to study the vertical structure of the new regime of high-redshift galaxies.
 This exciting area of study has opened up in the last couple of years with the advent of data from JWST (James Webb Space Telescope), along with ALMA (Atacama Large Millimeter Array). The latter provides information about the gas content of galaxies. The data from JWST, and ALMA,
have revealed surprising new features about the high-redshift galaxies. These galaxies are found to form early on;  are more massive  \cite{Adams2023,Carniani2024,Xiao2024highz};
 and gas-rich \cite{BH2024gasrich} compared to the 
present-day galaxies; and are more evolved than had been expected. These are presumably the precursors of present-day galaxies like the Milky Way. Hence it is important to understand their vertical structure, which could contain evidence about galaxy evolution (see Section \ref{sec:Intro}).

Using the data for a large sample of 181 galaxies between a redshift of z=0 to 5, \cite{LianLuo} have obtained the disk thickness for this sample. This is done by fitting the JWST NIRCAM imaging observational data 
 with a single disk with a sech$^2$ distribution function.
\cite{LianLuo} find that at $z > 1.5$, the disk thickness is comparable to that of the {\it Thick disk} -- and is much larger than the thickness of the thin disk of the Milky Way. They claim that this points to the formation of a 
{\it Thick disk} at early times. A more detailed study involving a 3-D modelling of the  imaging data from CEERS, JADES, and PRIMER surveys using JWST has been done by \cite{Tsukui2024}, where they do a double disk fitting  -- each with a sech$^2$ function -- for a sample of galaxies. They show that a second disk is necessary to explain the observations.

However, both these studies have assumed a sech$^2$ profile -- which is only applicable for a one-component, isothermal disk.  
Recall that the stellar density profile in a real galaxy is typically steeper than sech$^2$, as seen in observations (Section \ref{sec:2}); and also as obtained theoretically for a realistic, multi-component disk plus halo case (Sections \ref{sec:4.2.4} to \ref{sec:4.2.6}, Appendix A). 
Therefore, we suggest the following better and physically more correct approach: namely, obtain physically motivated density profiles by applying
the multi-component disk plus halo model to a coupled thin and a {\it Thick disk} system; and compare the model results with the JWST data. This will be on lines of the application by  \cite{Comeron2011}  to the S$^4$G data;  
except that, here the more general model involving the thick disk approach (Section \ref{sec:4.3}), and other kinematic effects (Section \ref{sec:4.4}), plus the effect of halo, would be included to make the treatment more realistic and complete. This procedure was outlined above in the context of the second future problem given earlier in this Section.
By comparing the model profiles with the JWST data, one can then extract the physical parameters of profiles of the two disks: an important future problem.
This could also help one understand the dynamics and evolution of high redshift galaxies.

\medskip

\noindent{\bf Dynamical implications}

In addition to the above problems involving direct application of the model, 
there are important  implications  of the model results for other dynamical problems. A few of these are outlined below.

First,  the observed HI scale height values and rotation curves  were used as two independent 
constraints on the multi-component disk plus halo model, to get the best-fit values for the shape and density profile of the dark matter halo (Section \ref{sec:6}). 
The 3-D HI data cube for galaxies now available allows an accurate determination of the input parameters: such as the HI
gas scale height values (which were not easy to observe so far), and an accurate rotation velocity measurement as a function of radius (\cite{Teodoro2015}; also, Section \ref{sec:6.5}). 
Hence the above approach can now be applied systematically to a large number of galaxies to obtain their halo properties.			

Second, due to the vertical constraining of a disk component in a coupled system (\cite{NJ2002,BJ2007,SJ2018}; also,  Section \ref{sec:4.2.5}); the disk is more tightly bound closer to the mid-plane. That is,  it takes more energy to move a unit mass by a certain height \cite{SJ2022}. Therefore, a coupled disk is predicted to be more resistant to distortion; say, due to a tidal encounter. 
Detailed numerical simulations are needed to check if a realistic, multi-component disk is indeed more robust to distortion, as predicted by \cite{SJ2022}.
This effect is expected to be higher in high redshift galaxies since these are gas-rich, as seen above.
This could have important implications for the early evolution of galaxies.

Third, the model would have important implications for other dynamical problems in the field, such as warps and other bending instabilities. Typically, in dynamical studies of such features, the disk
vertical profile is assumed to be of sech$^2$ form -- as applicable for a single-component, isothermal disk (e.g., \cite{Widrow2012,Widrow2014,BanikW2022}); or, to have other simple analytical form \cite{Goyari2023}. An early work  that explored the dependence of results on the density profiles \cite{Pranav2010} showed that
realistic vertical stellar density profiles, which are less steep than an exponential, allow warps to set in from a smaller radius, namely,  2-4 R$_D$ -- which lies within the optical disk size (see Appendix A). This explains why optical or stellar warps are ubiquitous (see Section \ref{sec:7.2}).
  The steeper-than-sech$^2$ profile and a vertically constrained distribution obtained in a multi-component disk plus halo model, and as is seen in observed data; and the associated change in the gravitational potential and force; are likely to affect the detailed properties of warps and other bending modes. 
  This general idea needs to be investigated further.

A related point is as follows. In studies of galactic dynamics, typically a simplified analytical form for the potential is used,  for example the popular Miyamoto-Nagai potential -- for its mathematical ease of use \cite{Miyamoto1975,BT1987}. However, while this does give the correct radial profiles of surface density, it does not represent the vertical profiles such as sech$^2$ or an exponential \cite{Dehnen2025}. It has now been realized that the detailed vertical density profile could play an important role in galactic dynamics. Hence a form for potential-density pairs that represent  vertical density profiles such as sech$^2$ or an exponential
has been recently obtained in an important paper by
\cite{Dehnen2025}. This would be useful to study dynamics of disks; for example, for the evolution of a phase-space spiral where the anharmonic term in the potential is important in deciding the winding time \cite{Dehnen2025,Antoja2018}.{\footnote{ 
To put this in  the context of the current review, we emphasize that the self-consistent density profiles resulting from a realistic multi-component disk plus halo model, are more complicated than sech$^2$ or an exponential; and an analytical form for these has not been obtained so far (see Section \ref{sec:4.2.6}). Hence the results for the potential-density pairs from \cite{Dehnen2025} do not cover such a realistic system.}}

\section{Conclusions\label{sec:9}} 

In this review, we have tried to give a cohesive, physical introduction to the vertical structure and related dynamics of a typical, thin galactic disk in hydrostatic equilibrium, while keeping in mind recent observations.
The treatment is comprehensive, starting from the classic, one-component, isothermal disk  \cite{Spitzer1942}; up to the modern picture involving a more realistic, multi-component disk plus halo model
\cite{NJ2002,SJ2018}. This model includes both stars and interstellar gas in a galactic disk on an equal footing. These are taken to be gravitationally coupled, and the disk is in the gravitational field of the dark matter halo. The self-consistent vertical density distribution of each disk component is obtained in terms of the input parameters: namely, the surface density and velocity dispersion of the disk components; and the halo mass distribution.

  The core of the review consists of presenting this model and the results from it. 
The inclusion of gas gravity and dark matter halo is shown to have a crucial constraining effect on the the vertical structure of the stellar disk. 
This model highlights the point that the vertical structure of stellar disk cannot be treated in isolation as has been typically done in the past; rather, the gravitational coupling with gas and the dark matter halo must be taken into account to get the correct physical distribution for each disk component.

The review is written in a pedagogic style, 
stressing the underlying physics and the mathematics in the formulation of the equations, and the treatment to solve these. The development of equations for treating the different physical cases: namely, a thin disk case; the thick/low density disk case (when the $R-z$ coupling is taken into account); and the more general cases including kinematic effects as done by using the complete Jeans equations, and a  non-isothermal velocity dispersion;  has been described clearly, and the difference between these cases is explained. 
The aim is to consider progressively more detailed physics in the problem, and treat realistic cases; starting from the simple, standard isothermal, one-component case that gives a sech$^2$ density profile.  
We hope that in view of the above features, this review will be useful for a reader, to get an easy introduction
to this important and active subject.
 
\subsection{Main results from the multi-component disk plus halo model\label{sec:9.1}}

 1. The most significant result from the model is that 
the self-consistent vertical density distribution of each disk component is constrained to be closer towards the mid-plane: 
such that, the mid-plane density is higher; the disk thickness or the HWHM  of the distribution is smaller; and simultaneously the distribution is steeper  
compared to the one-component, isothermal case. This is due to the inclusion of the additional gravitational force of the other disk components and the dark matter halo in the coupled case (Sections \ref{sec:4.2.5}, \ref{sec:4.2.6}).

2. Due to the lower dispersion of gas, 
it has a smaller scale height than the stars. Hence the gas contributes significantly to the force near the mid-plane. Therefore, gas has a strong constraining effect on the stars despite its small mass fraction in the disk.
 The stellar disk is constrained vertically in the inner galactic disk mainly by the interstellar gas; and  by the dark matter halo in the outer disk.
 Thus, it is necessary to include both gas and the dark matter halo to obtain an accurate vertical distribution of the stellar disk.

We stress that in both these regimes, the stellar distribution is steeper than the sech$^2$ profile obtained for a one-component, isothermal case. If the stellar density profile is represented as sech$^{2/n}$, then $n>1$ in the inner region where the gas has a dominant influence; while $n <1$ in the outer region where the halo is dominant. We point out that the latter case has not been noted earlier in the literature (see Section \ref{sec:4.2.6}). However, the value of $n$ obtained this way is not robust, hence the form sech$^{2/n}$ is not valid for a realistic disk (Section \ref{sec:4.2.6}).

3.  The model simultaneously
gives the self-consistent vertical density distribution in each disk component.
The results from this model explain
the observed scale heights in stars, HI and H$_2$ in the Milky Way very well. We stress that although the focus of this review is to study the stellar vertical distribution; the results for the HI and H$_2$ gas distributions are obtained at the same time, and are an important feature of this model.
The results also naturally explain
 the observed trends in the density distribution of the thin disk such as the steeper-than-sech$^2$ profiles, and broad wings at high $z$ (and high $R$), 
as seen in external galaxies; without  invoking a spurious second disk as is done in the literature to explain the wings.

4. The self-gravitational energy of a multi-component disk is derived using analytical techniques. Surprisingly, the potential energy per unit area for a particular disk component remains the same in the coupled case (Section \ref{sec:4.6}). 
 However, the constrained distribution due to the higher gravitational force  at each $z$ in the coupled case is predicted to make the disk robust against distortion; say, by  a tidal encounter with a satellite galaxy.

5. In the more general versions of the model, additional kinematical features such as  
the tilt of the velocity ellipsoid and the cross terms in the Jeans equations 
are included through the complete Jeans equations (Section \ref{sec:4.3}). Also, the  observed non-isothermal velocity dispersion is taken into account (Section \ref{sec:4.4}). These are shown to  shape the density profile which could be substantially different 
 compared to a sech$^2$ profile that is obtained for a single-component, isothermal case; with the mid-plane density and the disk thickness (HWHM) that differ by $\sim 30-40 \% $. The non-isothermal velocity dispersion causes the density distribution to be more vertically extended and broader at high $z$.  
It may be feasible to compare these predictions with data from future \textit{Gaia} data releases and other data, as well as with high-resolution numerical simulations.

 6. The disk vertical structure is routinely studied assuming an isothermal velocity dispersion, for simplicity. While testing the limits of validity of this assumption, it was shown \cite{SJ2020noniso} that even a small vertical velocity gradient of $ \sim 2- 3 $ km s$^{-1}$ kpc$^{-1}$ results in the mid-plane density being smaller by a sizeable amount $\sim 10-14 \%$
compared to the isothermal, one-component case.  It is possible that such small gradients could arise locally in a galactic disk; indeed, a strictly constant velocity dispersion denotes an idealized case. This result is a warning that the sech$^2$ law that is routinely used for its convenient, analytical form may not be strictly correct even for a single-component disk. 

 7. A generic and robust result is that the typical stellar galactic disk, which is in hydrostatic balance; and when using the observed input parameters, is shown to flare moderately, with the vertical scale height increasing by a factor of 2-3 within the optical disk. This is true even for an isolated galaxy, so an encounter is not necessary for a disk to flare.
We stress that the flaring is not due to the multi-component treatment; rather, it is seen in the stars-alone disk as well (e.g., in the Milky Way - see Section \ref{sec:5.3} for details). In fact, the presence of dark matter halo  reduces the flaring to a moderate value at large radii, as shown in the Milky Way \cite{SJ2018}.
A flaring disk is contrary to the long-held assumption of constant scale height in the literature \cite{vdkS1981a}; however,  it is in agreement with and naturally explains the trend seen in recent observations and simulations (see Section \ref{sec:5}). 
Thus, future studies of galactic structure must take account of the flaring of the stellar disk.

\medskip

\noindent {\bf  Necessity of using the multi-component disk model in future work:}

 In view of the above results, 
 it is clear that the realistic, multi-component disk plus halo model   should be used for an accurate treatment in future studies of disk vertical structure. We stress that such a rigorous, self-consistent calculation is not just of academic interest; rather, it is essential to use this approach to obtain a physically correct vertical density distribution.
 The model results clearly show that  the resulting self-consistent vertical density distribution for stars differs from sech$^2$ and is steeper, 
 in all realistic cases considered;  
 and a flaring stellar disk is shown to be a generic result.
 
In the literature, however, a simple sech$^2$ distribution or its limiting form at high $z$, namely, an exponential function (Section \ref{sec:3.1}); along with a constant scale height for stars, are routinely used for 
convenience (for example, see the references in Section \ref{sec:6.4}).{\footnote{Also, note that the modern text-book titled "Galaxies and Astrophysics of Galaxies", Bovy, J., Princeton University Press (a book under preparation) in fact recommends the usage of an exponential stellar profile and constant scale height.}} The actual model density profile for a realistic disk is more complex than given by either of these profiles (see Section \ref{sec:4.2.6}). 

We caution that the use of a sech$^2$ or an exponential function does not describe the true vertical density distribution for a real disk; and such usage may lead to dynamically incorrect results, 
or may make one miss the underlying physics (e.g., Section \ref{sec:4.5.4}). 

On the other hand, applying the rigorous multi-component disk plus halo model to theoretical problems or to analyze the observed data; though it involves more work, can lead to new physical insights into the disk structure (e.g., \cite{Comeron2011}; also, see Section \ref{sec:4.5.1}). The above model  \cite{NJ2002} used to analyze the S$^4$G data by \cite{Comeron2011} led to the interesting result about the relative masses in the thin and thick disks being comparable, hence the conclusion of an early origin of the thick disk. This new physical result would have been missed if the standard practice of using sech$^2$ for fitting the data had been employed, as noted by the authors. 

Overall, the mass distribution is vertically constrained in a coupled disk -- which will be reflected in the model density profile; and the effect of this constraining on dynamical studies can only be taken into account by using the self-consistent model density profile.

 Note that a simple analytical form to represent the numerical density profile, $\rho(z)$, or the corresponding potential, resulting
from the multi-component disk plus halo model for any of the realistic cases studied,  is not easy to prescribe; and this has not been attempted so far (Sec \ref{sec:4.2.6}, Appendix A),
although it would be convenient to have this.

\subsection{Dynamical implications of the model\label{sec:9.2}} 

As seen above, the multi-component disk plus halo model significantly changes the self-consistent vertical density distribution, and the thickness of each disk component.
The associated gravitational potential  
and the net vertical force are also different compared to the one-component case. 
Thus, apart from being interesting in themselves, the results for the disk vertical structure have a direct 
bearing on the various dynamical processes in the disk. 

Various examples of the dynamical applications have already been discussed in the previous sections (in particular, see Sections \ref{sec:4.5}, \ref{sec:5} and \ref{sec:6} for details): namely, star formation; disk stability; the feasibility and evolution of spiral arms; and, the constraining of the density profile and shape of the dark matter halo. Thus the results from this model have broader implications for various fundamental properties of galaxies.

\subsection{Unique features of study of disk vertical structure\label{sec:9.3}}

The study of vertical structure of the disk has some unique features.
In keeping with the pedagogic nature of this review, it is worth summarising these features, which were discussed in this review.

First, the self-consistent vertical density distribution in a disk depends only on the local conditions at a given $R$. Thus the study of disk vertical structure reduces to a local, one-dimensional problem along z. 
This is true for a thin disk: for a one-component case (Section \ref{sec:3.1}), and for a multi-component disk (Section \ref{sec:4.2.1}); as well as for a physically thick disk if it is in a rotational equilibrium (Section \ref{sec:4.3.1}).
Thus, interestingly, the vertical structure allows one to trace the local gravitational potential, unlike in the planar case (see Section \ref{sec:3}).

On the other hand, the entire vertical density distribution along the z direction at a given radius has to be determined self-consistently by jointly solving the Poisson equation and the equation of hydrostatic equilibrium. Thus the determination of the vertical structure is a harder problem than that of the galactic mass model  where a test particle approach can be used to determine the planar mass distribution from the observed rotation curve 
(see Section \ref{sec:3} for details).

Second, due to the open nature of disk geometry, each disk component responds to the potential of the other disk components, even those lying at larger $z$ values (see Appendix A in \cite{JogSolomon1984}). For example, the gas -- despite having a smaller scale height than stars --  can respond to the stellar potential (Section \ref{sec:4.2.1}). This makes it possible to formulate a multi-component disk plus halo model. 

Third, since the disk vertical structure  can be phrased as a local, one-dimensional problem along $z$; the self-consistent density distribution in a disk component in the coupled case is only affected by the local properties of the other disk components and the halo, at a given $R$. Thus, the stellar density distribution is affected by H$_2$ gas in the inner disk where most of the molecular gas is observed, and by the HI gas in the middle and outer Galaxy. This trend will be true in all galaxies. 

Fourth, the potential energy per unit area for a disk component is unchanged even in a coupled case. The interaction acts as an internal force, merely redistributing the energy along $z$. However, it takes more energy to raise a unit mass by a certain distance along $z$ in the coupled case, or the gravitational potential is higher. Hence, the disk component is more tightly bound and harder to distort than the one-component case (see Section \ref{sec:4.6}).

Finally, the disk vertical structure contains clues regarding
the formation and evolution of the disk (Section \ref{sec:Intro}).

\subsection{Future problems and trends\label{sec:9.4}} 

There are a number of important, open problems in this field, covering  the range from direct model applications to dynamical implications of the model.
This topic was discussed in some detail, giving the motivation and the detailed outline of each problem, in Section \ref{sec:8}. 
A few of these are briefly summarized here, starting with the direct model applications:

 1. First, crucial input parameters such as stellar velocity dispersion values in external galaxies can now be extracted from the IFU data, available from various surveys. This can yield more accurate model results for the vertical density profiles. This will also allow one to check and confirm whether the stellar disk flares in general, see Section \ref{sec:5}.
Alternatively, the model results could be taken as a template to be compared with new data such as \textit{Gaia} DR3, or other data that become available from future observations. 

 2. Second, the model could be applied to understand the vertical structure of the kinematically and chemically distinct {\it Thick disk} component. To do this, the multi-component disk plus halo model has to applied in the physically thick disk approach (see Sections \ref{sec:7.1}, and \ref{sec:8} for details).

 3. The data from JWST has revealed the surprising result that disk galaxies were common, and even dominant, at a redshift as high as 4-6 \cite{Ferreira2023,Kartal2023,Robertson2023}. These even showed spiral features, which indicates an underlying rotating disk at a redshift of 4 (\cite{Kuhn2024}; also, see \cite{Yan2024}). This is an emergent area of research now.

Thus, the time is now ripe for a systematic study of the vertical structure of disks in  high-redshift galaxies. Physically motivated density profiles for a thin and a {\it Thick disk} can be obtained by applying the multi-component disk plus halo model to such a system. Then, by comparing these density profiles with the data from JWST, the best-fit parameters for the thin and {\it Thick disk} can be extracted (see Section \ref{sec:8} for details).
Additionally, a study of disk vertical structure could shed light on the dynamics and evolution of  high redshift galaxies, which is not well-understood. 

\medskip

In addition, the overall constrained vertical distribution in a coupled case can have various dynamical implications, which are worth pursuing: 

1. First, it would be interesting 
to check by simulations, if the vertical constraining that is obtained in a realistic, multi-component disk makes the disk more robust against tidal distortion  - as was predicted by \cite{SJ2022}. This would be particularly important for the early evolution of high-redshift, gas-rich galaxies. 

 2. Second, the steeper-than-sech$^2$ profile and a vertically constrained distribution, as resulting from a multi-component disk plus halo model, is likely to affect the properties of warps (see e.g., \cite{Pranav2010}) and other bending modes. This topic should be investigated in detail. 

 3. Third, the 3-D HI data cube now available for galaxies  can yield more accurate observed scale height values and rotation curves. Hence, the new approach developed (Section \ref{sec:6}), where the above two are used as independent constraints on the model, can now be applied systematically to a large number of galaxies; so as to better constrain the shape and density profile of their dark matter halos. 

\subsection {Summary and outlook\label{sec:9.5}}

In the end, we would like to stress that the study of disk vertical structure is  an ongoing progress: with many outstanding unsolved problems, mainly triggered by new observations; and a growing realization that a multi-component disk plus halo model is necessary 
to treat a realistic galactic disk.
This model, with the rigorous and self-consistent  approach embodied in it,
would provide the necessary theoretical framework to physically understand and interpret the  new observations from \textit{Gaia}, JWST, and other upcoming future observatories. 
These include data on high redshift galaxies, which have surprising and unexpected properties, as discussed above.
This is a new, relatively unexplored field, and the multi-component disk plus halo model discussed in this review is a timely tool to analyze this data.
This  will help us better understand the vertical structure of galactic disks, and thus also help shed light on the dynamics and early  evolution of galaxies.

The
past experience from   \textit{Gaia} DR2 and RAVE shows that these  led to the discovery of unexpected features such as non-isothermal velocity dispersion (Section \ref{sec:4.4.2}), and the enigmatic phase-space spirals (Section \ref{sec:7.2.3}). 
The latter indicate that a fraction of the disk could be out-of-equilibrium, which is an active topic of research now.
  We can similarly expect that data from JWST and other future observations are likely to reveal more surprises, 
 that will open up new theoretical topics in the field.  The study of disk vertical structure is thus an exciting topic of study now, and promises to remain so in the coming years. 

We hope that this review will provide an in-depth introduction to the important topic of the vertical structure and dynamics of a galactic disk; along with an outline of some of the open, challenging problems in it. This would, hopefully, enthuse more people to work in this field and to explore its rich physics further.

\bigskip

\noindent {\bf Acknowledgments:}

During the past two decades, I have had the great pleasure of working with many students, and other collaborators, on the topic of the vertical structure of a galactic disk and related dynamics, and in particular, on developing the multi-component disk plus halo model and working out its various applications.
 It is a pleasure to thank my students; in chronological order of our work together, they are: Chaitra Narayan, Kanak Saha, Pratyush Pranav, Arunima Banerjee, Rathulnath R.,  Soumavo Ghosh, and Suchira Sarkar. 
 Special thanks are due to Chaitra and Suchira  for extensive collaboration; also to Chaitra for being there at the beginning of this study, and to Suchira  for several useful discussions during the writing of this review.  I also thank the various collaborators; in chronological order, they are:  Leo Blitz, Evan Levine, Lynn Matthews, Elias Brinks, Ioannis Bagetakos, Nirupam Roy, and Meera Nandakumar.  Finally, I would  like to thank friends who are physicists but not astronomers: namely, (the late) Rohini Godbole for appreciating the results from this model over the years; and Suman Iyer for giving a patient hearing during the writing of this review.
 
I am happy to acknowledge the support from INSA, New Delhi  as an INSA Senior Scientist for the past two years (2023-2025), and also from DST, New Delhi for a J.C. Bose National Fellowship for an extended period in the past during a large part of the work reported here.

\newpage

\begin{table}
\caption{Glossary of important terms used:}
\label{table:1}
\medskip
\centering
  \begin{tabular}{l l}
\hline 

Symbol used $\: \:$  & Description of the physical parameter \\
 &   \\
\hline
    &  \\

$\mathrm{R, \phi, z}$ & Galactic cylindrical coordinates, where $R$ is the\\
    & Galactocentric radius and $z$ is normal to the Galactic plane\\

$\mathrm{{\rho(z)}_i}$ & Mass density as a function of $z$, with the subscript\\
              &  $i$=s,HI,H$_2$,h corresponding to stars, interstellar HI, and H$_2$ \\
              & gas, and dark matter halo, respectively\\

$\mathrm{\Sigma_i (R)}$ & Surface density at a given $R$ for a disk component, with\\ 
               & the subscript $i$=s,HI,H$_2$, as defined above.\\

$\mathrm{{\Phi}_i}$ & Gravitational potential, with the subscript $i$=s,HI,H$_2$,h as\\
           & defined above.\\

$ \mathrm{\langle{(v_z^2)}_i\rangle^{1/2} =(\sigma_z)_i}$ & R.M.S. velocity dispersion along $z$, with the\\
                                                    & subscript $i$=s,HI,H$_2$ as defined above. \\
          
$\mathrm{(K_z)_i}$ & Force per unit mass along the $z$ direction, with the subscript\\
         & $i$=s,HI,H$_2$,h as defined above. \\

$\mathrm{R_D,R_v}$ & The scale length for  exponential fall-off with radius  of  stellar\\ 
     & disk surface density and vertical velocity dispersion, respectively \\

$\mathrm{\rho_0, z_0}$ & The mid-plane disk density and the parameter indicating  \\
                 & the scale height in an isothermal, one-component, thin disk\\
                 & distribution given by sech$^2$(z) \\

$\rho_{0d}$ & The mid-plane disk density in a general, coupled system \\

HWHM  & The half-width at half maximum (HWHM) of the vertical \\
       & density distribution, also referred to as the scale height\\
       & of the vertical density distribution\\

$\mathrm{{\rho}_{0h}}$, R$_c$ & The mass density of the dark matter halo at the centre of a\\
                    & galaxy, and the core radius of the halo, respectively for a\\
                    & psuedo-isothermal density distribution\\  

$ \mathrm{q = c/a}$          & Shape of the spheroid, where $c$ and $a$ correspond to the \\
                    & the vertical and the radial major axis of the spheroid, \\
                    & respectively. $q >1, <1$ corresponds to a prolate and an \\
                    & oblate spheroid respectively. $q=1$ represents a sphere with\\ 
                    & $a=b=c$ where $b$ is the semi-minor axis in the disk plane.\\
$\mathrm{W}$       & Gravitational potential energy per unit area for a disk\\
  &  \\
\hline
\end{tabular}
   \end{table}

\newpage

\noindent {\bf Acronyms}

\noindent CALIFA Survey: Calar Alto Legacy Integral Field Area Survey

\noindent ISM: Interstellar Medium

\noindent IFU: Integral Field Unit

\noindent JWST: James Webb Space Telescope

\noindent LAMOST: The Large Sky Area Multi-Object Fiber Spectroscopic Telescope

\noindent SDSS: Sloane Digital Sky Survey

\noindent S$^4$G:  {\it Spitzer} Survey of Spiral Structure in Galaxies

\newpage

\section {Appendix A. Thickness of a self-gravitating disk and disk vertical density profile \label{App.A} }

\noindent {\bf A.1 Thickness of a self-gravitating disk}

For a disk mass distribution, its vertical density profile, $\rho(z)$, and thickness are important physical properties that characterize the disk and also provide a handy way of comparing various disks. However, for a gravitating system, neither its radial extent nor its thickness is sharp or well-defined. 
Though this is an important point, this is not often mentioned clearly  
in the literature. Further, a different definition of thickness 
is employed for different cases. The aim of this Appendix is to point this out and bring some clarity and uniformity to the definition of disk thickness that has been employed in the literature for 
different density profiles.

The radial extent of disk of an external galaxy, 
is taken to be given by the  radius of the optical disk or the visible extent of the stellar disk. The radius of the optical disk is typically defined to be the radius of an isophote of a given limiting magnitude, such as 26.5 mag arc sec$^{-2}$ (known as the Holmberg radius, R$_H$); or 25 mag arc sec$^{-2}$, corrected for inclination (known as the de Vaucouleurs radius), see \cite{BT1987}. For the typical values of central intensity of a spiral galaxy, and the sky brightness,  $\mathrm{R_H \sim 4-5 R_D}$ (e.g., \cite{NJ2003}).{\footnote{The above definition for the radius of the optical disk depends on the sky brightness as well as the technological capabilities of detecting faint regions \cite{Trujillo2020}. 
The actual radial extent of the mass distribution in a galactic disk could be much larger. This is confirmed by the detection of faint, highly extended outer regions of some galaxies using  modern sensitive telescopes:
for example, the detection by the Dragonfly Telescope Array of outer, faint disks up to $\sim 18 R_D$ in M101 \cite{DragonflyM101} and up to $\sim 23 R_D$ in NGC 2841 \cite{Dragonfly2841}.}}

It is clear that the functional form of the vertical density profile, or specifically the rate of fall-off of the density distribution, would be important in defining its vertical extent as well as what can be taken to be its thickness. So the density profile and the thickness are directly coupled.

For example, the observed stellar distribution (based on the star counts) close to the Galactic mid-plane is typically fitted by an exponential distribution and hence the rate of exponential fall-off of density,$z_{exp}$, is taken to denote the disk thickness or scale height; and its value in $\sim 300$ pc in the solar neighbourhood (e.g., \cite{MihalasRoutly}). 
Another way to estimate the characteristic thickness is to divide the total surface density by the mid-plane mass density of stars; and this gives a value of 350 pc in the solar neighbourhood \cite{BT1987}.
The interstellar HI gas typically is taken to obey a Gaussian distribution, and the half-width-at-half-maximum (HWHM) of this distribution is taken to define the gas thickness  \cite{Lockman1984,DL1990}. For HI gas, this is measured to be $\sim 150 pc$ in the solar neighbourhood  \cite{Wouter1990,DL1990}.

Theoretically, a gravitating, one-component, isothermal, thin disk distribution has a form sech$^2 (z/z_0)$ along $z$ (\cite{Spitzer1942}; also, see Section \ref{sec:3.1}). Here $z_0$ is defined to be the scale parameter, as per the notation used by \cite{Spitzer1942,Camm1950}. Later, in the literature, the quantity $z_0$ has often been refereed to as the "scale height" \cite{vdkF2011}, and is taken to be an indicator of disk thickness.

\medskip

\noindent {\bf A.2 HWHM of vertical density distribution as the thickness}

In general, the density distribution of stars in a galactic disk deviates from the sech$^2$ function predicted for an isothermal, one-component disk. 
This is seen observationally for external galaxies (e.g., \cite{vdk1988}); and, also  found theoretically for the self-consistent density distribution resulting from a multi-component disk plus halo model (see Sections \ref{sec:4.1} to \ref{sec:4.4}, in particular, see \cite{SJ2018}). For each disk component in the coupled case, the vertical density distribution is more concentrated towards the mid-plane, and is steeper, than the sech$^2$ distribution obtained for the corresponding one-component case.  Thus, for a constant surface density (i.e., at a given $R$), the additional gravitational force due to gas and the halo results in a steeper profile, and simultaneously a smaller disk thickness for stars (see Section \ref{sec:4.2.5}). The non-isothermal velocity dispersions tends to have the opposite effect of making the density profile broader than sech$^2$,
especially at high $z$- the latter appears as a wing
(Section \ref{sec:4.4.2}).

Thus, even though each isothermal disk component by itself would satisfy a sech$^2$ distribution; the resulting stellar vertical density distribution in the coupled system is completely different from sech$^2$, see Section \ref{sec:4.2.6} for details. In these realistic cases, it is not meaningful to use $z_0$ as  the indicator of disk thickness. Hence, instead, the HWHM (Half Width at Half Maximum) of the vertical density distribution of a disk component was introduced
as an indicator of the scale height disk thickness: by \cite{complex}  while treating the effect of a cloud complex on the disk; and by \cite{NJ2002} while treating a multi-component disk plus halo model. The use of HWHM was in analogy to what is routinely done in the study of ISM  \cite{DL1990}.
The HWHM was also refereed to as the scale height (or disk thickness)   
in the above two papers and in other subsequent papers on this topic. It should be pointed out that
the definition of the HWHM takes account of the density variation with $z$; but it does not explicitly 
depend on  the actual functional form of the density distribution. Hence HWHM is a robust indicator of disk thickness. In addition, it is easy to determine its value  from the model results as well as from the observed data.

The subsequent detailed work on the multi-component disk plus halo model showed that in all realistic cases the resulting density profile, $\rho(z)$,  differs from sech$^2$; and could instead be attempted to be specified by a more general functional form, sech$^{2/n}$ (\cite{vdk1988}; or, see Eq. \ref{eq4.7} ). However, it is found that here the exponent $n$ is not robust; rather, it varies with $\Delta z$, the range of $z$ over which the density distribution is fitted by the above functional form (see Section \ref{sec:4.2.6}). Thus $n$ as defined by \cite{vdk1988} is not a useful indicator of the steepness of the vertical density profile; and 
Eq. (\ref{eq4.7})
 as proposed by \cite{vdk1988} is not valid for a realistic galactic disk.

This behaviour was shown for the various physical cases considered: namely, the multi-component disk plus halo model, for a thin disk 
as applied to the Milky way \cite{SJ2018}; and for a thick disk as applied to UGC 7321 \cite{SJ2019};
and also for one-component thick disk, with kinematic effects included (Model B respectively from \cite{SJ2020Jeans}); and the non-isothermal case \cite{SJ2020noniso}. Therefore, defining a scale height based on the functional form of the density distribution (analogous to $z_0$ for the sech$^2$ distribution) would not be feasible. Hence, the use of HWHM introduced and used by \cite{complex,NJ2002} to denote the scale height or disk thickness makes good practical sense.  This usage was then also adopted in the above-mentioned subsequent papers.
The HWHM from the above physically obtained density distributions is a  robust indicator of disk thickness or scale height. 

Therefore, we warn that in most realistic cases, simply assuming a sech$^2$ vertical density distribution, or  taking the high $z$ limit of this -- namely, an exponential distribution -- and taking the associated $z_0$ parameter (see Eq.(\ref{eq3.5})) as the scale height, and further assuming it to be constant, as is routinely done in the literature,
does not represent the true, detailed physics in the problem. Such usage would  lead to  dynamically incorrect results.

Finally, we comment on a point made by
 (\cite{Tsukui2024}); namely, that the parameter $n \neq 1$ in \cite{vdk1988} ($n$ as defined by Eq. (\ref{eq4.7})
  corresponds to a non-isothermal case. We point out that this is not strictly true.  As shown above, any deviation from a one-component, isothermal case (including the isothermal, multi-component case as well as
a non-isothermal, one-component case) gives rise to a density profile that has $n \neq 1$. 
In any case, as described above, the parameter $n$ obtained for a general case is not robust. This is because 
Eq. (\ref{eq4.7}) proposed by \cite{vdk1988}
is not adequate to explain the results obtained  for a realistic disk (see the discussion in Section \ref{sec:4.2.6}). 

As discussed above, most realistic cases give a density profile different than sech$^2$. We point out that a simple analytical form for the actual density profile
resulting from the multi-component disk plus halo model for any of the realistic cases studied  is not easy to prescribe, and has not been attempted so far (Sec \ref{sec:4.2.6}); although such an expression would be useful.  

\noindent {\bf Vertical profile of gas density distribution}

A distribution of any self-gravitating disk component will be more constrained closer to the mid-plane in the coupled case, and would show a distribution steeper than sech$^2$ (Section \ref{sec:4.2.5}). The HWHM of the density distribution  is used to denote the thickness for each of the disk components, namely, stars, HI and H$_2$ \cite{NJ2002}, also see the above discussion.

\cite{Patra2019} states that the gas distribution in the coupled case is a Gaussian (see Fig. 7 in \cite{Patra2019}). However, that paper has not done a mathematical fit to check this or to see whether the fit to $n$ (in sech$^{2/n}$) is a function of $z$. In any case, a fit close to a Gaussian for gas is not surprising. 
It is well-known (e.g., \cite{Rohlfsdensity}) that in the potential of the main mass component of the disk, namely the stars, the gas (when treated as massless particles) has a Gaussian distribution (as given by Eq.(\ref{eq3.6})). 
In the study of ISM, the HI distribution is routinely fitted by a Gaussian (\cite{DL1990};also, see Section \ref{sec:3.2}).
However, in the more realistic, coupled case, since it includes gas gravity, the gas distribution would be expected to be steeper than this. 
Certainly the HI gas distribution in the field of dark matter halo and stars 
is much steeper than the gas-alone case - the latter under its own self-gravity (as seen in Section \ref{sec:4.2.5}, Fig. \ref{fig.15}). A quantitative measurement of the steepness of gas vertical density distribution in the coupled case needs to be done, to ensure a full understanding.

\medskip

\noindent {\bf A.3 Half-mass scale height, $z_{1/2}$, as an indicator of disk thickness:}

Another indicator of disk thickness was introduced by \cite{SJ2018}, namely, $z_{1/2}$, which they called the half-mass scale height,
measured from the mid-plane. This is defined to contain half the mass in a column of unit area at a given radius, or half the total stellar surface density at a given radius. This will also implicitly depend on
the shape of the density profile at larger z values, unlike the value
of HWHM. Hence a smaller $z_{1/2}$ indicates a distribution that is
concentrated closer to the mid-plane. This parameter has not been noted or used much in the literature; however, see \cite{Sotillo2023} who independently defined a similar quantity to define thickness during the analysis of TNG50 simulations data. From an observational viewpoint, $z_{1/2}$  can be thought of as an indicator of the half-light
scale-height of an edge-on galaxy with M/L considered to be
constant. That is, $z_{1/2}$  is the height that contains half the light from a disk in a column density at a given radius.
The two definitions (HWHM and $z_{1/2}$) give values of heights that are comparable, with the HWHM being somewhat larger than the $z_{1/2}$ values (see Tables 2 and 3 from \cite{SJ2018}).

\medskip

\noindent {\bf A.4 Some special cases of density profiles:}

In some special cases, a simple analytical function is a reasonably good fit to the vertical density distribution, and is used for simplicity. The conversion between the HWHM and the scale height corresponding to a particular functional form, for some typical cases, is given next.
For example, consider  a sech$^2$ distribution (proportional to sech$^2 (z/z_0))$); a sech distribution (proportional to sech$(z/z_{sech})$), and an exponential distribution (proportional to exp$(-z/z_{exp})$).
In these cases, the relation between the corresponding scale-factors and the HWHM is given respectively as:  z$_0$ = HMHM/0.88; z$_{sech}$ = HWHM/0.759 ; and z$_{exp}$ = HWHM/0.693.

\newpage

\bibliographystyle{elsarticle-num-names}
\bibliography{main}

\end{document}